\documentclass[twocolumn]{aastex701}

\usepackage{graphicx}
\usepackage{amsmath}
\usepackage{amssymb}
\usepackage{float}
\usepackage{subcaption}
\begin{document}

\title{Revisiting the Supernova Engines in the 3C~397 and W49B Supernova Remnants}

\author[orcid=0009-0005-5637-6538,gname='Cole',sname='Treyturik']{Cole Treyturik}
\affiliation{Department of Physics and Astronomy, University of Manitoba, Winnipeg, MB R3T 2N2, Canada}
\email[show]{treyturc@myumanitoba.ca}

\author[orcid=0000-0001-5581-486X]{Chelsea Braun} 
\affiliation{Department of Physics and Astronomy, University of Manitoba, Winnipeg, MB R3T 2N2, Canada}
\affiliation{Fredonia State University of New York, 280 Central Avenue, Fredonia, NY, 14063, USA}
\email{braunc@fredonia.edu}

\author[orcid=0000-0001-6189-7665]{Samar Safi-Harb}
\affiliation{Department of Physics and Astronomy, University of Manitoba, Winnipeg, MB R3T 2N2, Canada}
\email{samar.safi-harb@umanitoba.ca}

\author[orcid=0000-0003-2624-0056]{Christopher L. Fryer}
\affiliation{Center for Nonlinear Theory, Los Alamos National Laboratory, Los Alamos, NM 87545, USA}
\affiliation{Department of Physics, The George Washington University, Washington, DC 20052, USA}
\email{fryer@lanl.gov}

\author[0000-0002-4231-8717]{Gilles Ferrand}
\affiliation{Department of Physics and Astronomy, University of Manitoba, Winnipeg, MB R3T 2N2, Canada}
\affiliation{RIKEN Center for Interdisciplinary Theoretical and Mathematical Sciences (iTHEMS), Wak\={o}, Saitama 351-0198 Japan}
\email{gilles.ferrand@umanitoba.ca}

\begin{abstract}

The nature of the supernova remnants (SNRs) 3C~397 and W49B has long been a subject of debate, with prior studies offering conflicting interpretations between thermonuclear and core-collapse scenarios. To help settle this debate, we present a systematic, spatially resolved, spectroscopic analysis of both remnants using \textit{XMM-Newton}. By applying multi-component thermal models, we derive key physical properties including elemental abundances, ejecta temperatures, ambient densities, and explosion energetics. We compare the inferred metal abundance ratios to a wide range of core-collapse and thermonuclear nucleosynthesis models, including new models whose explosion energies differ from the canonical value of 10$^{51}$~ergs. We find that the observed Fe/Si and Ca/Si ratios in both SNRs are best matched by certain thermonuclear models. However, no model fully reproduces the complete set of observed abundance patterns. In 3C~397, high Fe enrichment and spatial abundance variations suggest interaction with a dense progenitor environment, and W49B’s composition is overall consistent with a thermonuclear origin; however both require a low energy ($\sim$10$^{50}$~erg) supernova explosion. We additionally map the Fe K$\alpha$ line centroid energies and find a spread, with W49B falling within the core-collapse region -- highlighting both environmental complexity and the limitations of this diagnostic for supernova classification. Our results highlight the need for caution in relying on any single diagnostic or nucleosynthesis model for supernova typing, underscore the need for improved nucleosynthesis models, and motivate future  high-resolution, high-throughput observations.

\end{abstract}


\section{Introduction}
\label{sec:intro}
Supernova (SN) explosions come in many different forms, but two main pathways exist on which these forms are expected to fall: the collapse and subsequent explosion of a massive ($\gtrsim$8M$_{\odot}$) star, or the thermonuclear deflagration and/or subsequent detonation of a white dwarf. These pathways, known respectively as core-collapse (CC) and Type Ia (Ia), are the main cases used to classify SN explosions, as they vary fundamentally in their progenitors. In both scenarios, the explosion results in a shock wave which expands outwards, interacting with, shocking, and heating the surrounding interstellar medium (ISM). A reverse shock runs backward into the SN ejecta heating the shocked ejecta as well to X-ray emitting temperatures.

The swept-up material forms a nebula referred to as a 
supernova remnants (SNR). SNRs, as with their SN progenitors, can be classified as either Ia or CC. While the mechanisms behind the creation of these different types vary greatly, it can sometimes be not so straightforward to assign one of these types to a given SNR. X-ray emission holds clues to inferring the SN explosion properties through the SNR properties (e.g., \cite{2023MNRAS.525.6257B}).

3C~397 is a bright and middle-aged Galactic SNR ($\sim$1.3--5.3~kyr at an estimated distance of 8--10~kpc; \citealt{2000ApJ...545..922S, leahy_2016})\footnote{\url{http://snrcat.physics.umanitoba.ca/SNRrecord.php?id=G041.1m00.3}} whose explosion mechanism remains uncertain. Originally, 3C~397 was considered a CC remnant, owing to its unusual morphology,  proximity to the Galactic plane, association with a molecular cloud, and enhanced intermediate-mass element (IME)  abundances consistent with core-collapse nucleosynthesis \citep{2000ApJ...545..922S, safi-harb_2005}. However, subsequent analyses have suggested a Type Ia origin based on the iron-group elements and centroid energy of the Fe-K line \citep{yamaguchi_2015, yamaguchi_2014}. In particular, the discovery of ejecta clumps highly enriched in neutron-rich elements, including  Cr, Mn, Fe, and Ni, has been interpreted as favoring the scenario of a near-Chandrasekhar-mass white dwarf progenitor \citep{yamaguchi_2015, 2021ApJ...913L..34O}. Recent re-analysis of the Suzaku data also favors a Type Ia origin \citep{martinez-rodriguez_2020}, though some of these arguments remain contested (e.g., \citealt{siegel_2020}).

Another well known SNR with a highly debated origin is W49B. Like 3C~397, W49B is a bright, middle-aged Galactic remnant ($\sim$2.9--6~kyr at a distance of $\sim$7.5--11.3~kpc; \citealt{hwang_2000, zhou_2018, sun+chen_2020}).\footnote{\url{http://snrcat.physics.umanitoba.ca/SNRrecord.php?id=G043.3m00.2}} The X-ray emission from W49B is dominated by the emission from its center, for which it was classified as a mixed-morphology SNR \citep{rho_petre_1998}, and initially presumed to host a pulsar wind nebula \citep{pye_1984}. This scenario was later ruled out due to prominent Fe-K line emission \citep{smith_1985}. 

The jet-like and apparent asymmetry have led to the speculation of a hypernova origin \citep{keohane_2007}, but the observed abundances better match yields from more traditional supernova models \citep{miceli_2006}. Due to this, as well as its mixed-morphology classification, W49B was assumed to be a CC SNR \citep{hwang_2000}. This conclusion was further supported by possible evidence of a black hole \citep{lopez_2013b}. Yet more recent work has proposed a type Ia origin based on abundance ratios and metal distribution within the SNR \citep{zhou_2018}.

In this paper, we revisit the explosion origins of both 3C~397 and W49B. We perform a systematic, spatially-resolved spectroscopic study of both objects using X-ray observations from \textit{XMM-Newton}, and for the first time compare the fitted abundance patterns to a broad suite of core-collapse and Type Ia nucleosynthesis models. Our goal is to provide updated constraints on the explosion mechanisms and motivate improvements to theoretical models of stellar explosions.

This paper is organized as follows: in Section \ref{sec:obs}, we summarize the observation data, as well as the steps taken for data preparation and the method of region selection. In Section \ref{sec:spectroscopy}, we describe our \textit{XMM-Newton} spectroscopic study of the objects, and compare our findings to those of previous studies. In Section \ref{sec:results}, we discuss our derived analysis, including the X-ray and supernova explosion properties, an analysis of the Fe~K$\alpha$ line centroids, and compare our spectroscopic results to a wide range of models of supernova nucleosynthesis yields. In Section \ref{sec:discussion}, we discuss our findings. Finally, in Section \ref{sec:con}, we summarize our results and their implications.

\section{Observations and Data Preparation}

\begin{table}
	\centering
	\caption{Observation data used in the study. All SNRs use \textit{XMM-Newton} data. Dates for the observations are detailed in Section \ref{sec:obs_3C397} and Section \ref{sec:obs_W49B} for 3C~397 and W49B, respectively.}
	\label{tbl:ExpTime}
    \begin{tabular}{cccc} 
		Source      &
		OBS ID      & 
		Detector    & 
		\begin{tabular}[c]{@{}c@{}} 
			Exp. Time \\ (ks) 
		\end{tabular} \\
		\hline
        3C 397  & 0085200301 & MOS, pn & 17.06 \\
        & 0085200401 & MOS, pn & 20.12 \\
        & 0085200501 & MOS, pn & 23.37 \\
        & 0830450101 & MOS, pn & 140.0 \\
        W49B    & 0084100401 & MOS, pn & 18.95 \\
        & 0084100501 & MOS, pn & 18.96 \\
        & 0724270101 & MOS, pn & 118.5 \\
        & 0724270201 & MOS, pn & 71.20 \\ \hline
	\end{tabular}
\end{table}

\label{sec:obs}
All observations used in this analysis made use of archival data obtained by the \textit{XMM-Newton} X-ray telescope (see Table \ref{tbl:ExpTime}), with observation dates for 3C~397 and W49B detailed in Section \ref{sec:obs_3C397} and Section \ref{sec:obs_W49B}, respectively. Observations were performed using the European Photon Imaging Camera (EPIC), using both Metal Oxide Semi-conductor (MOS) CCD cameras, as well as the pn camera, which together afford an energy range of 0.15 to 15 keV, with a spectral resolution of roughly 0.1 keV \citep{turner_2001, struder_2001}. 

\begin{figure*}
    \begin{center}
        \subfloat[(a) 3C 397 (17 Regions)]{\includegraphics[angle=0,width=0.4\textwidth]{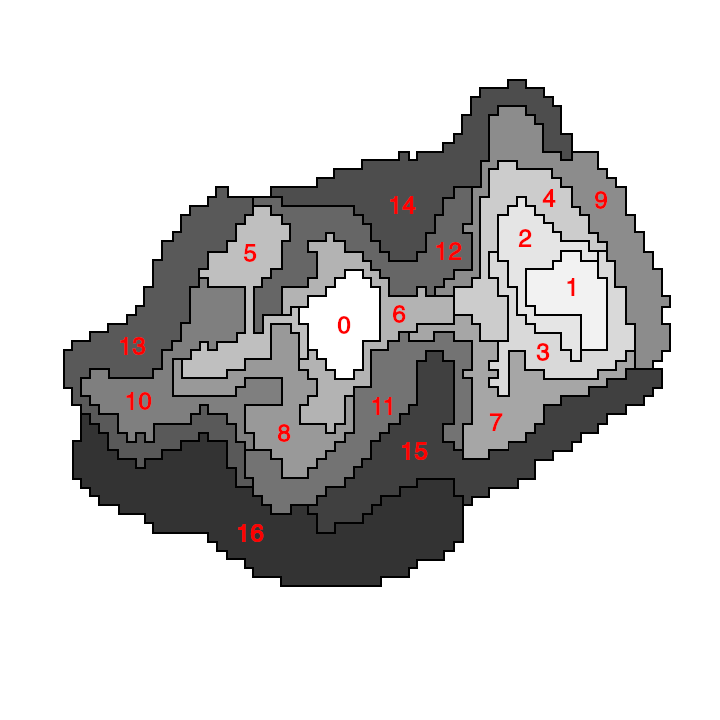}}
        \hspace{1cm}
        \subfloat[(b) W49B (19 Regions)]{\includegraphics[angle=0,width=0.4\textwidth]{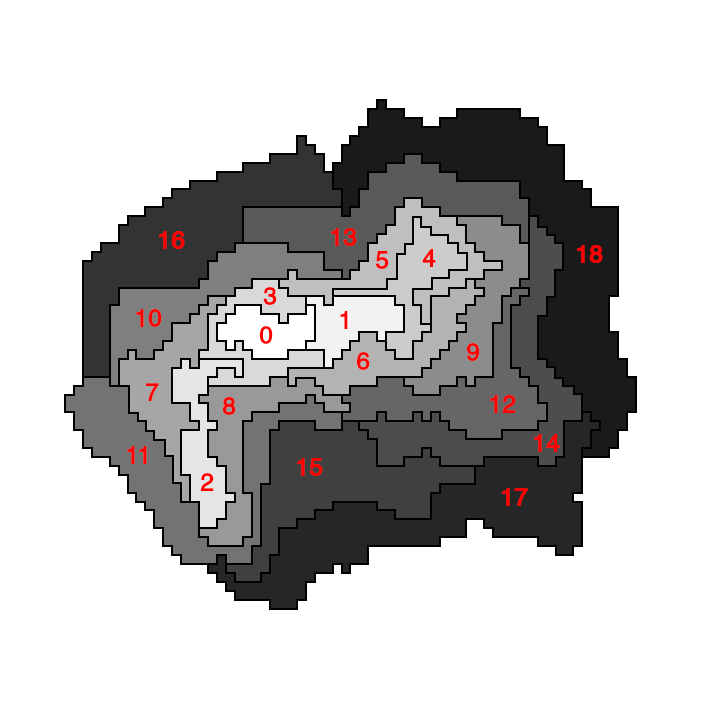}}
        \caption{Selected regions for 3C 397 (left) and W49B (right), as generated by the \textit{contbin} algorithm. Every color represents a different region. Regions are labeled in order of decreasing flux of their highest-flux pixel.}
        \label{fig:regions}
    \end{center}
\end{figure*}

For all observations, calibration and filtering were performed using the \textit{XMM-Newton} Science Analysis System\footnote{\url{https://www.cosmos.esa.int/web/xmm-newton/sas}} (SAS) v19.1.0, with the latest calibration files being used. All observation files were reprocessed using the SAS tasks \textit{emproc} and \textit{epproc} and filtered for good time intervals (GTI), bad pixels, and out-of-time events, and all data was checked for any potential photon pile-up using the command \textit{epatplot}. The MOS data was filtered to retain patterns 0 -- 12 in the 200 -- 12000 eV range and using the \textit{\#XMMEA\_EM} flag, while the pn data was filtered to retain patterns 0 -- 4 in the 200 -- 15000 eV range and using the \textit{\#XMMEA\_EP} flag. 

For our spectroscopic study, we generated region maps for both objects using the \textit{contbin}\footnote{\url{https://www-xray.ast.cam.ac.uk/papers/contbin/}} algorithm \citep{sanders_2006}, an adaptively smoothed binning program which generates regions based on a set of input parameters. This allowed for regions to be generated which seamlessly encompassed the entirety of the SNR whilst maintaining the morphology of the SNR. Regions generated via this method are intended to have the same (or, at least, quite similar) amounts of counts per region, and are generated based on surface brightness, due to the general correlation found between spectral properties and changes in surface brightness. Due to the use of multiple sets of observations for each object, we opted not to run the algorithm on each observation. Instead, broadband (0.1 -- 10 keV) images from all observations were first merged to create a single image, from which a mask was created to better constrain the algorithm to the SNR. The results were then binned to an appropriate bin size before running the algorithm. This was done to avoid the regions being biased towards the surface brightness of any particular observation. The final number of regions (17 and 19, for 3C 397 and W49B, respectively) was chosen to allow for sufficient statistics for each region during the spectral fitting process, whilst still allowing for the examination of any arcsecond-scale structure within the object. These regions are shown in Figure~\ref{fig:regions}. Therein, each color represents a different region, and each region is labeled according to the output of the \textit{contbin} algorithm -- that is to say, in order of descending flux of their highest-flux pixel.

\subsection{Observations of 3C~397}
\label{sec:obs_3C397}
The observations of the SNR 3C~397 by \textit{XMM-Newton} EPIC were taken during two epochs, and all pointings were done using the thin filter and full frame mode. For the first epoch, three observations were taken -- on 2003 October 15, 2004 April 23, and 2004 April 25 (OBS ID: 0085200301, 0085200401, and 0085200501, respectively). The second epoch used a single pointing, taken on 2018 October 22 (OBS ID: 0830450101).

\begin{figure*}
    \begin{center}
        \subfloat[(a) 3C 397 Region 00 Spectrum]{\includegraphics[angle=0,width=0.45\textwidth]{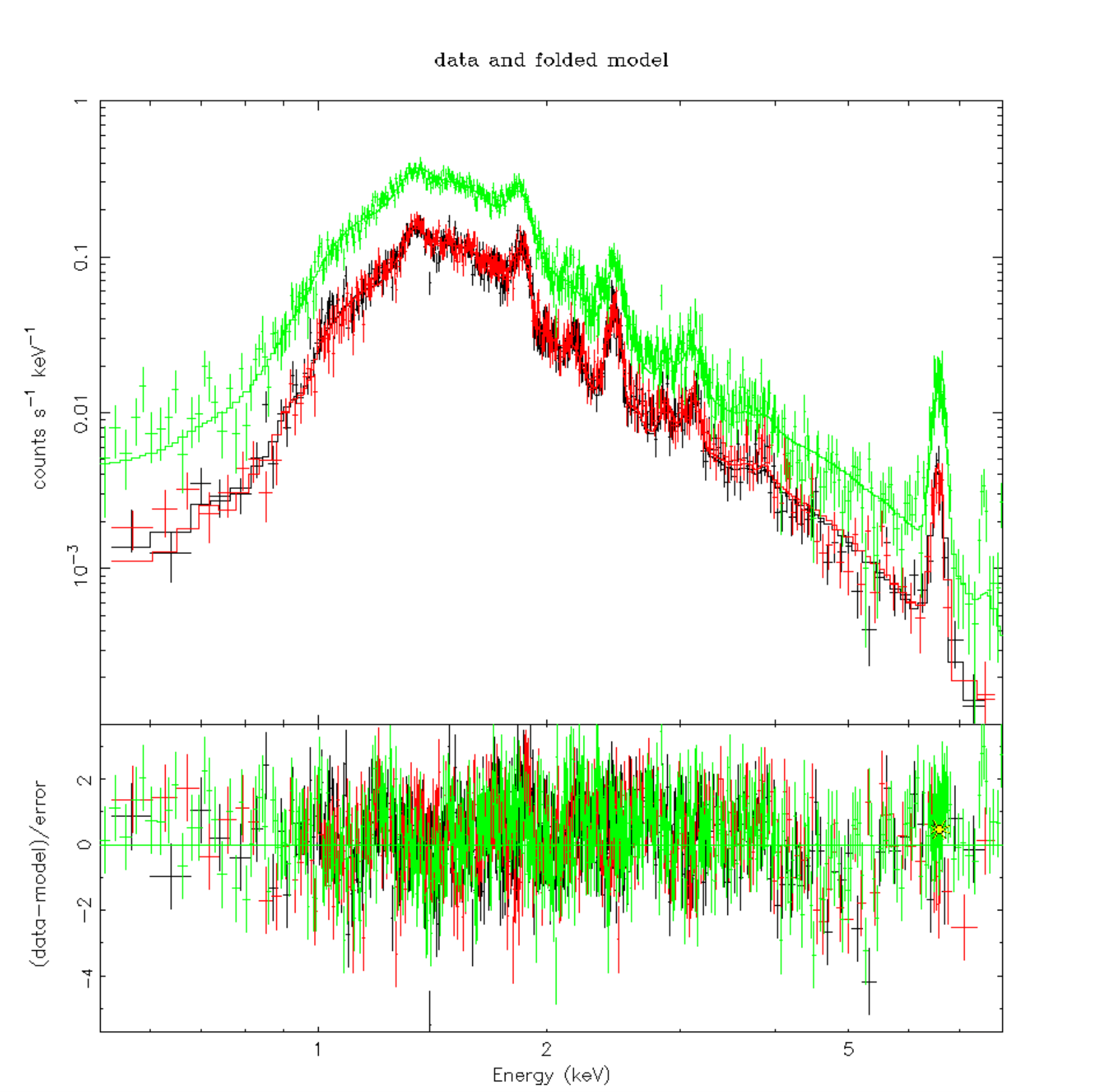}}
        \hspace{1cm}
        \subfloat[(b) W49B Region 00 Spectrum]{\includegraphics[angle=0,width=0.45\textwidth]{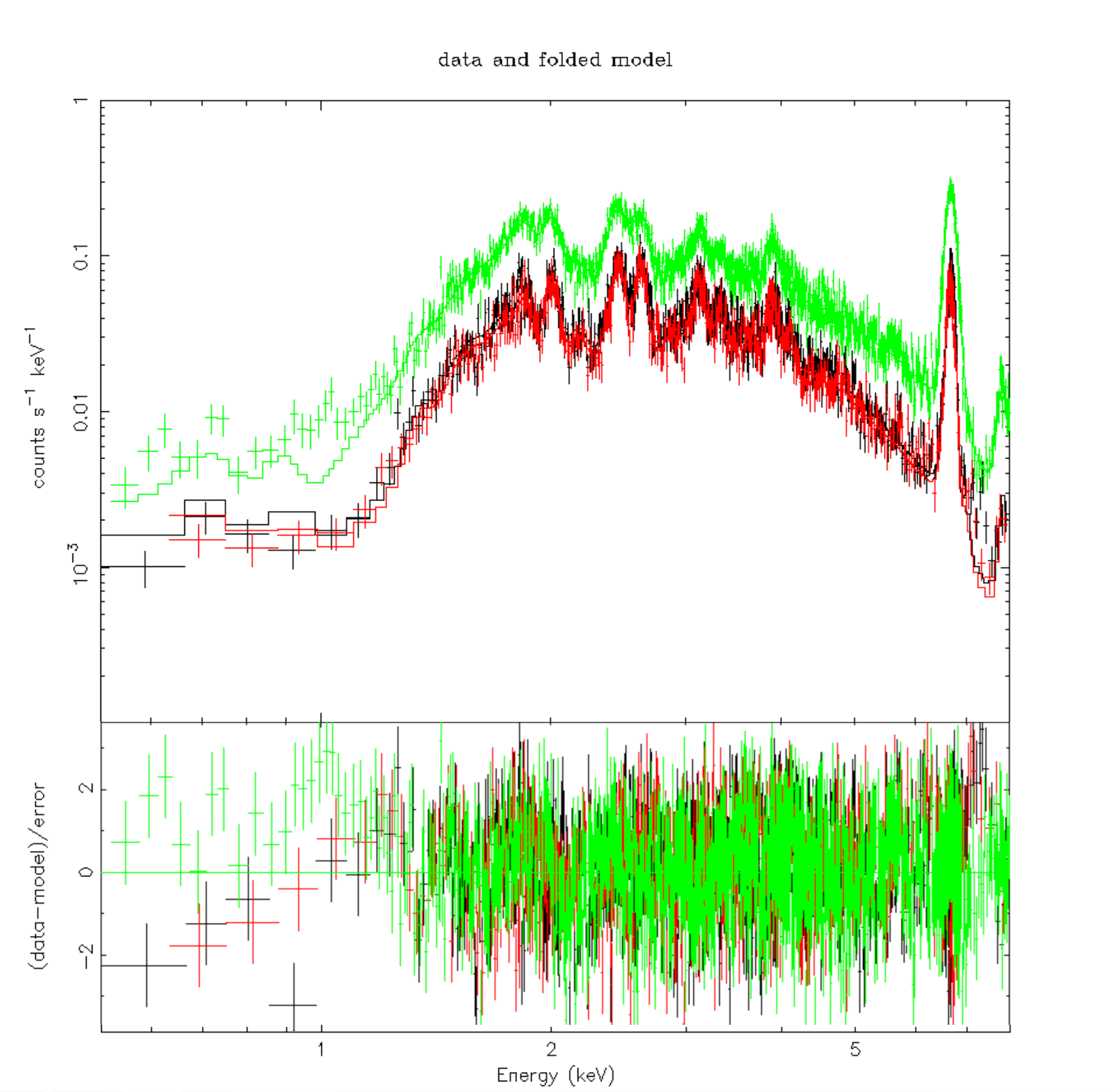}}
        \caption{Sample fit spectra for both 3C 397 (left) and W49B (right). Both spectra were taken from the Region 00 of the respective SNR, and both show only the data for the longest observation used in each object (OBS ID 0830450101 and OBS ID 0724270101 for 3C 397 and W49B, respectively). In both figures, black represents MOS1 data, red represents MOS2 data, and green represents pn data.}
        \label{fig:sample_spectra}
    \end{center}
\end{figure*}

The spectral analysis was performed using XSPEC version 12.11.1, and the spectra were binned using a minimum of 20 counts bin$^{-1}$. For our spectroscopic study, our \textit{contbin} regions were generated using a minimum signal-to-noise ratio of 125 and a signal-to-noise smoothing value of 30. For our spectral analysis of the entire SNR, we subtracted a background region from a circular ring in the source-free region surrounding the SNR. For our spatially resolved spectroscopic study, we divided the remnant up into three sections of roughly equal size. Each spectroscopic region was assigned to a section, and each section was assigned a circular background region from the source-free zone surrounding the SNR, the location of which corresponded to the section's location in the remnant.

\subsection{Observations of W49B}
\label{sec:obs_W49B}
The \textit{XMM-Newton} EPIC observations of the SNR W49B used were taken during two epochs. The first epoch was divided into three pointings, taken on 2004 April 3, 2004 April 5, and 2004 April 13 (OBS ID: 0084100401, 0084100501, and 0084100601, respectively). The second epoch was divided into two pointings, taken on 2014 April 18 and 2014 April 19 (OBS ID: 0724270101 and 0724270201, respectively). All observations for both epochs were made using the medium filter and full frame modes. Due to significant contamination from proton flaring events, observation 0084100601 was excluded from our analysis. 

The spectral analysis was again performed using XSPEC version 12.11.1, and the spectra were binned using a minimum of 20 counts bin$^{-1}$. For our spectroscopic study, our \textit{contbin} regions were generated using using a minimum signal-to-noise ratio of 275 and a signal-to-noise smoothing value of 30. For our spectral analysis of the entire SNR, we subtracted a background region from a circular ring in the source-free region surrounding the SNR. For our spatially resolved spectroscopic study, we used the same method as in the case of SNR 3C~397, save that the remnant was divided into four sections, rather than three.

\section{Spectroscopy}
\label{sec:spectroscopy}
We performed a spatially-resolved spectroscopic study on both 3C~397 and W49B, using the regions generated by the \textit{contbin} algorithm, as defined in Section \ref{sec:obs}. Data were modelled using the software XSPEC version 12.11.1 \citep{arnaud_1996}. Abundance tables were set to use the values as defined by Wilms \citep{wilms_2000}, using the XSPEC command \textit{abund wilm}, while soft X-ray absorption was accounted for by use of the Tuebingen-Bolder ISM absorption model, \texttt{TBABS}, which is characterized by the molecular hydrogen column density, defined therein by the parameter $N_H$.

Previous studies have shown that single-component models are too simplistic to produce adequate fits to the spectra of either 3C~397 \citep{2000ApJ...545..922S, safi-harb_2005} or W49B \citep{miceli_2008, zhou_2018}, even when looking at arcsecond-scale regions, and that at least two components are needed. As a result, we opted to fit all of our regions using a variety of two-component models, using both collisional ionization equilibrium (CIE) models as well as non-equilibrium ionization (NEI) models. The models tested included:

\begin{enumerate}
    \item \texttt{APEC/VAPEC/VVAPEC}, a CIE model for diffuse gas, characterized by a constant electron temperature.
    \item \texttt{PSHOCK/VPSHOCK/VVPSHOCK}, an NEI model for a plane-parallel shocked plasma with a constant temperature.
    \item \texttt{NEI/VNEI/VVNEI}, an NEI collisional plasma model, which assumes a constant temperature and a single ionization parameter.
    \item \texttt{RNEI/VRNEI/VVRNEI}, an NEI collisional plasma model for recombining plasmas. This model assumes that the plasma started in a CIE state, and includes an initial temperature parameter.
\end{enumerate}

All of these models are characterized by a temperature, kT; an ionization timescale, n$_\text{e}$t, where n$_\text{e}$ is the post-shock electron density and t is the time since the passage of the shock; a normalization constant $K=\frac{10^{-14}}{4\pi D^2}\int n_en_HdV$; and variable abundances for several elements, including Mg, Si, S, Ar, Ca, Fe, and Ni. The statistical quality of the results was based on reduced chi-squared statistics, $\chi^2_\nu$, where a $\chi^2_\nu<2$ is considered a good fit, with values closer to 1 being preferable. The exception to this was the global fit to the SNR W49B, for which a $\chi^2_\nu<2$ was not possible with our two-component models. The errors for individual parameters were obtained using the \textit{Markov Chain Monte Carlo} method, as described by the XSPEC manual. For this, we made use of the Goodman-Weare algorithm with 8 walkers, a total chain length of 200,000 steps, and a burn-in phase length of 100,000 steps.

\subsection{Spectroscopy of 3C 397}

\begin{figure*}
	\begin{center}
		\subfloat[(a) N$_{H}$]{\includegraphics[angle=0,width=0.3\textwidth,scale=0.5]{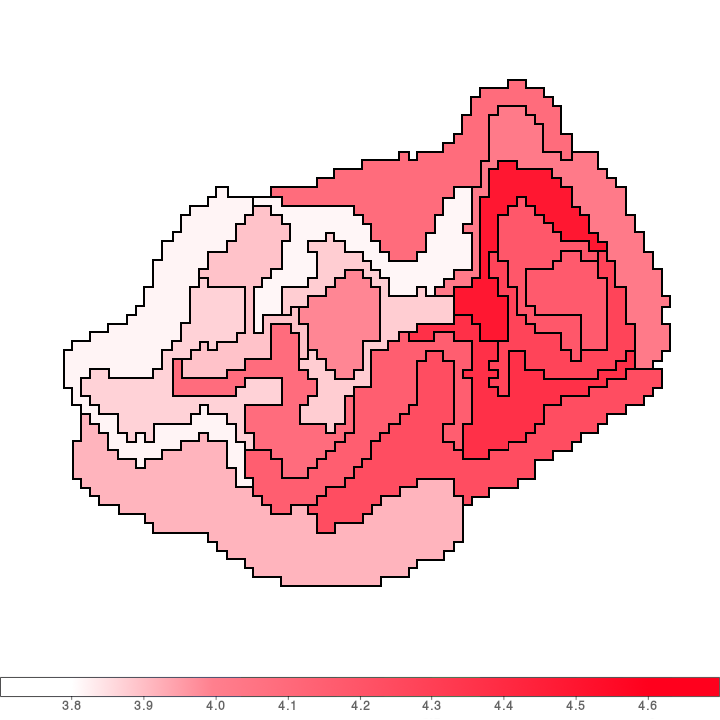}}
		\subfloat[(a) kT$_{h}$]{\includegraphics[angle=0,width=0.3\textwidth,scale=0.5]{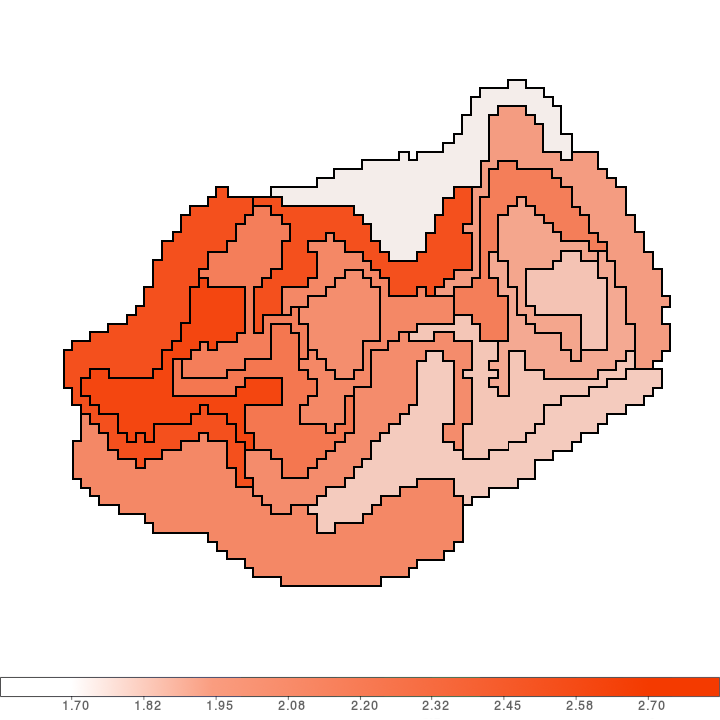}}
		\subfloat[(c) kT$_{c}$]{\includegraphics[angle=0,width=0.3\textwidth]{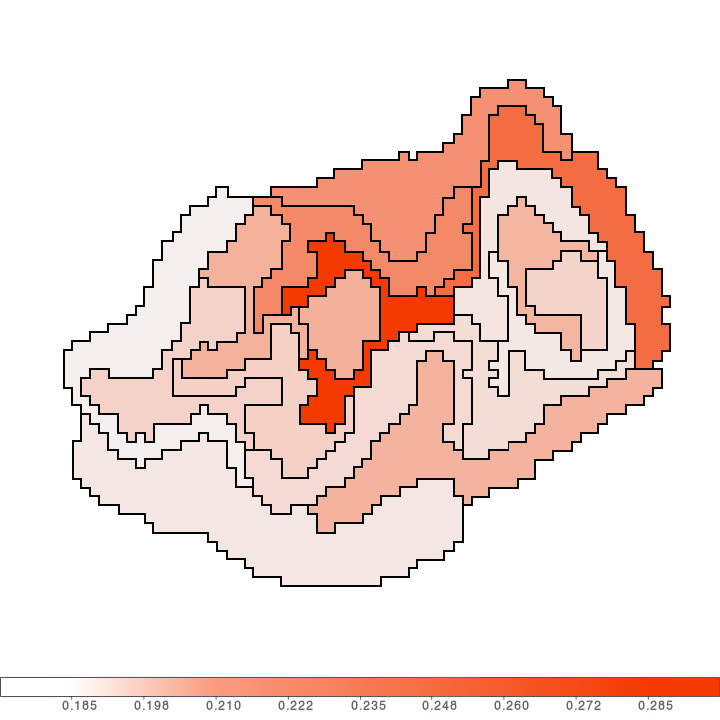}} \\
		\subfloat[(d) Mg]{\includegraphics[angle=0,width=0.3\textwidth]{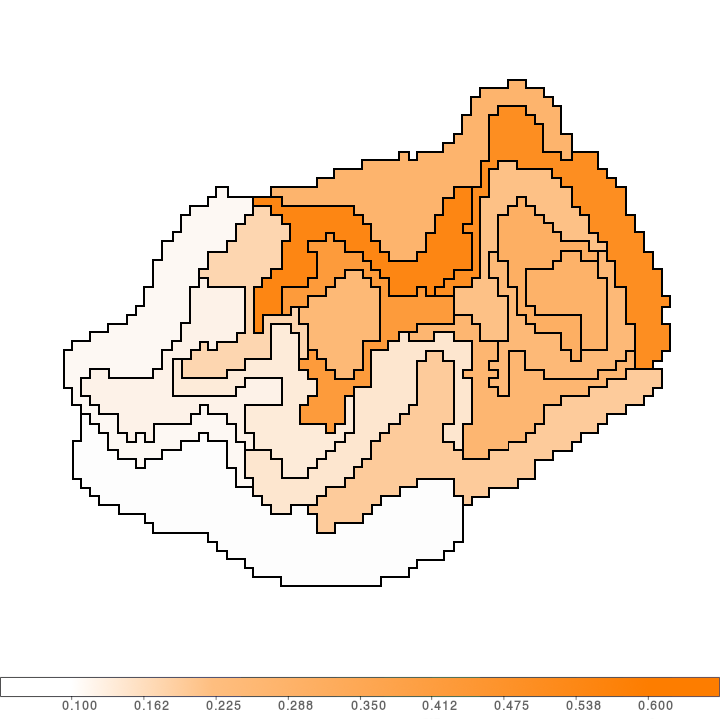}}
		\subfloat[(e) Si]{\includegraphics[angle=0,width=0.3\textwidth,scale=0.5]{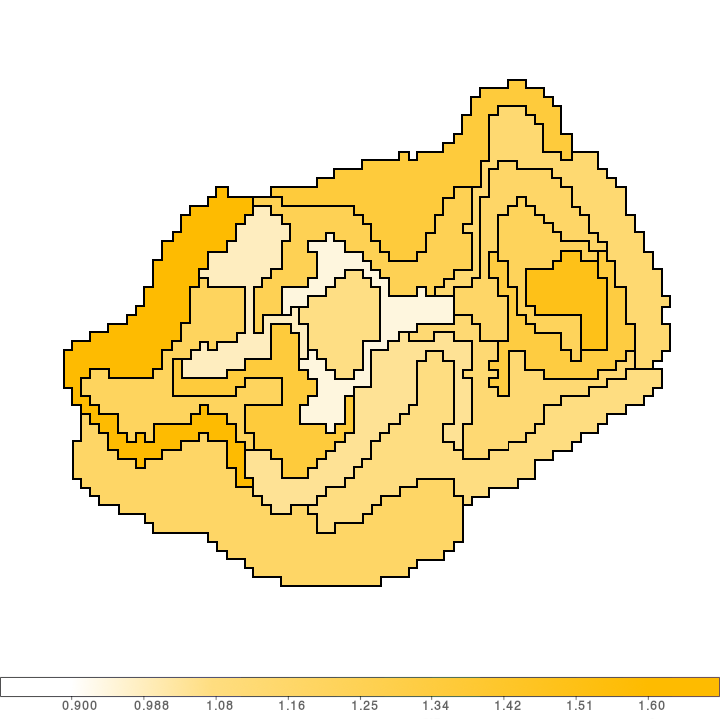}}
		\subfloat[(f) S]{\includegraphics[angle=0,width=0.3\textwidth]{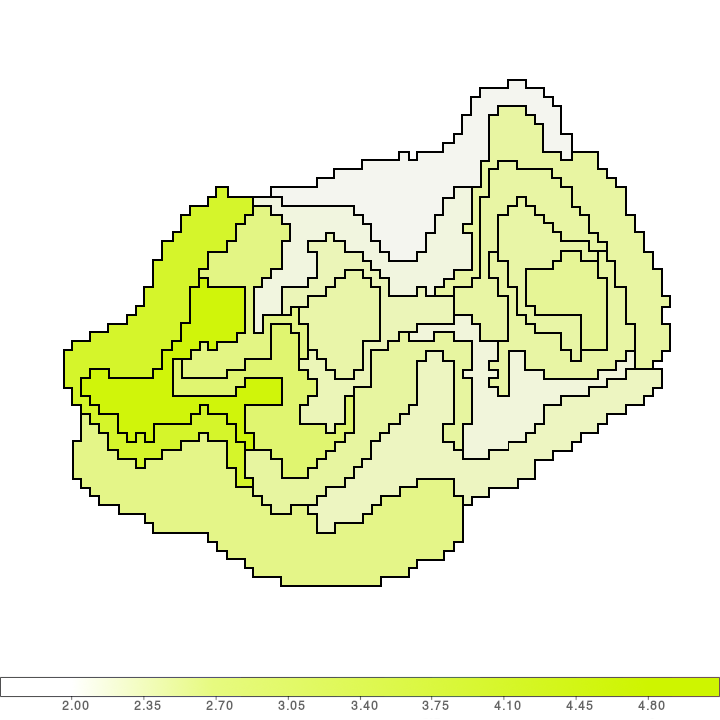}} \\
		\subfloat[(g) Ar]{\includegraphics[angle=0,width=0.3\textwidth]{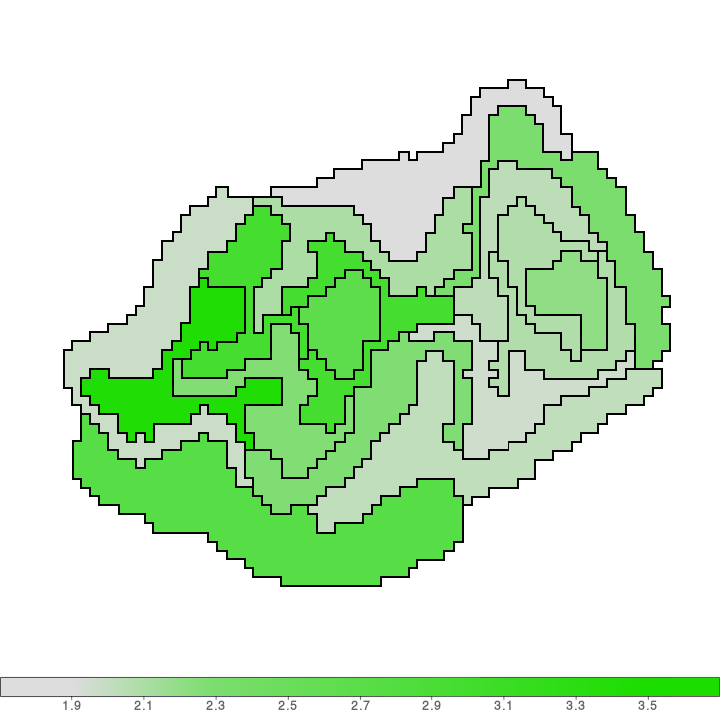}}
		\subfloat[(h) Ca]{\includegraphics[angle=0,width=0.3\textwidth,scale=0.5]{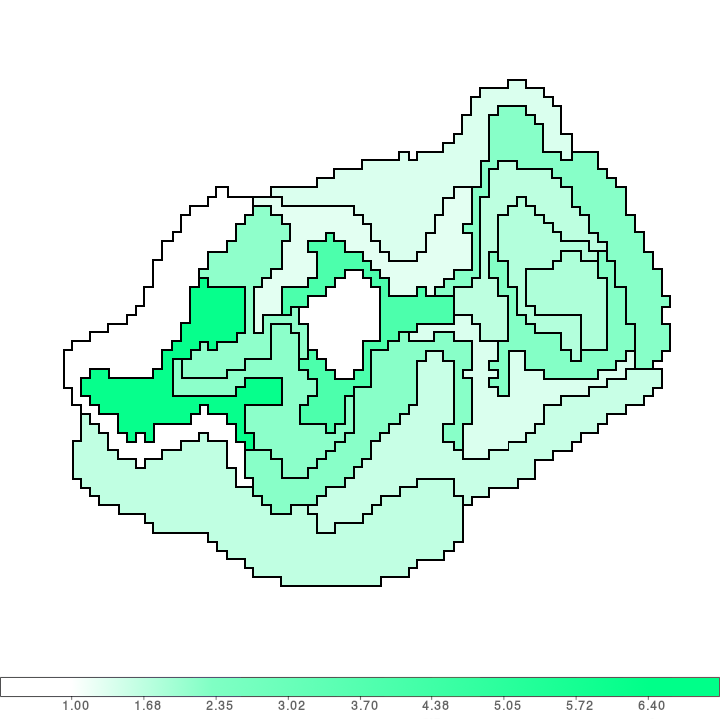}}
		\subfloat[(i) Fe]{\includegraphics[angle=0,width=0.3\textwidth]{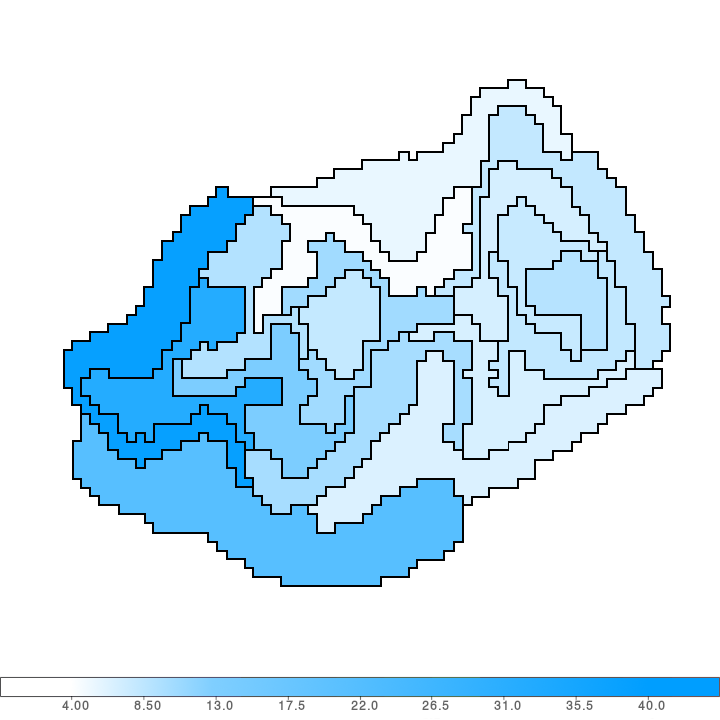}}
		
		\caption{3C 397: Distribution maps for the given parameters. All maps are in a linear scale. N$_{H}$ is in units of $10^{22}$ cm$^{-2}$, kT$_{h}$ and kT$_{c}$ are in units of keV, and all other elements are in terms of solar abundances. The hue of each map is arbitrary.}
		\label{fig:3C397_distribution_maps}	
	\end{center}
\end{figure*}

The analysis of 3C 397 used four data sets, taken across two epochs. All data sets were extracted and then fit simultaneously within the energy range of 0.5 -- 8~keV. Initial testing revealed that our arcsecond-scale regions were best fit with a two-component \texttt{VNEI}+\texttt{VAPEC} model, where the \texttt{VNEI} and \texttt{VAPEC} models represented the hot- and cold-components, respectively. For the initial step of the fitting process, the hydrogen column density, hot-component temperature and normalization, and ionization timescale were freed and fit. The abundances in the hot-component were then freed and fit one at a time in the order of Mg, Si, S, Ar, Ca, then Fe (with Ni tied to Fe). The cold-component temperature and normalization were then freed and allowed to fit. \added{While previous studies -- such as \cite{yamaguchi_2015} or \cite{oshiro_2021} -- were able to examine the emission from elements such as Cr and Mn, they made use of data that was possessed of a higher sensitivity than that used in this study. As a result, while we did attempt to isolate the emission from these elements using the models with extra abundance parameters (\texttt{VVAPEC/VVPSHOCK/VVNEI/VVRNEI}), the lack of sensitivity resulted in these elements ultimately not being included in our investigation.} After testing, it was found that \added{the abundances of Mg and Si in the hot component with consistent with solar values, and were unable to reproduce the observed emission lines, yielding little improvement to the fit. As a result, we opted to freeze these parameters to solar, and allow the abundances of Mg and Si in the cold component to vary instead. This improved the fit, and allowed us to reproduce the observed emission lines}. In region 13, allowing the Ca abundance to vary did not improve the fit, and it was instead frozen to solar.

The spectral fits for the individual regions can be found in Table~\ref{tbl:3C397_data}, a sample spectrum can be found in Figure \ref{fig:sample_spectra}, and distribution maps for each parameter can be seen in Figure \ref{fig:3C397_distribution_maps}. Each region was well fit ($\chi^{2}_{\nu} < 1.4$) with the \texttt{VNEI}+\texttt{VAPEC} model described above. The column density ranged from 3.83--4.53$\times 10^{22}$~cm$^{-2}$, with the lowest column densities being associated with the north-eastern portions of the SNR and the highest column densities being associated with the interior regions in the SNR's western half. Temperatures for the hot component, associated with the \texttt{VNEI} model, range from 1.74--2.63 keV, with a distribution somewhat opposite that of the column density, with the highest temperatures being in the north-eastern portions of the SNR. The soft component, associated with the \texttt{VAPEC} model, displayed temperatures that range from 0.19--0.29 keV, with the higher temperatures being associated with the northern portions of the SNR. The abundances of Mg, Si, S, Ar, Ca, and Fe within the SNR were enhanced above solar values except for Mg, which was subsolar in all cases. Most of these abundances appear to be highest in the eastern half of the SNR; Mg is an exception, in that it appears more concentrated in the western half of the SNR.

There are two previous studies to which we can compare our results. The first is a \textit{Chandra} study by \cite{safi-harb_2005}. This was a spatially-resolved study in which an absorbed two-component \texttt{VPSHOCK}+\texttt{VPSHOCK} model was used. They found column densities between $2.27-4.10\times10^{22}$ cm$^{-2}$, which are slightly lower than those determined by in our study, although this can be explained by the use of different absorption models (e.g. their study made use of the \texttt{WABS} model, while our work uses the \texttt{TBABS} model). They found hot component temperatures between 1.4--3.4 keV, and cold component temperatures between 0.15--0.26 keV, depending on which region, which are comparable to our results. They also found subsolar abundances for Mg, near-solar abundances for Si, enhanced abundances of S and Ca, and significantly enhanced abundances of Fe, all of which agree with our findings. The second study is from \cite{martinez-rodriguez_2020}, which looked at \textit{Suzaku} data. They performed a global fit of the SNR using an absorbed three-component \texttt{VVNEI}+\texttt{VNEI}+\texttt{NEI} model, using the solar abundances from \cite{wilms_2000}. They determined a column density of $3.49\times10^{22}$ cm$^{-2}$, and a hottest component temperature of 1.89 keV, both slightly lower than our own. They did, however, find a cold component temperature of 0.22 keV, which matches our findings. Additionally, they found generally enhanced abundances for Si, S, Ar, and Ca, and significantly enhanced iron-group abundances, with the enhanced abundances being present in their higher-temperature components, all of which matches our results quite well.

\subsection{Spectroscopy of W49B}

\begin{figure*}
	\begin{center}
		\subfloat[(a) N$_{H}$]{\includegraphics[angle=0,width=0.3\textwidth,scale=0.5]{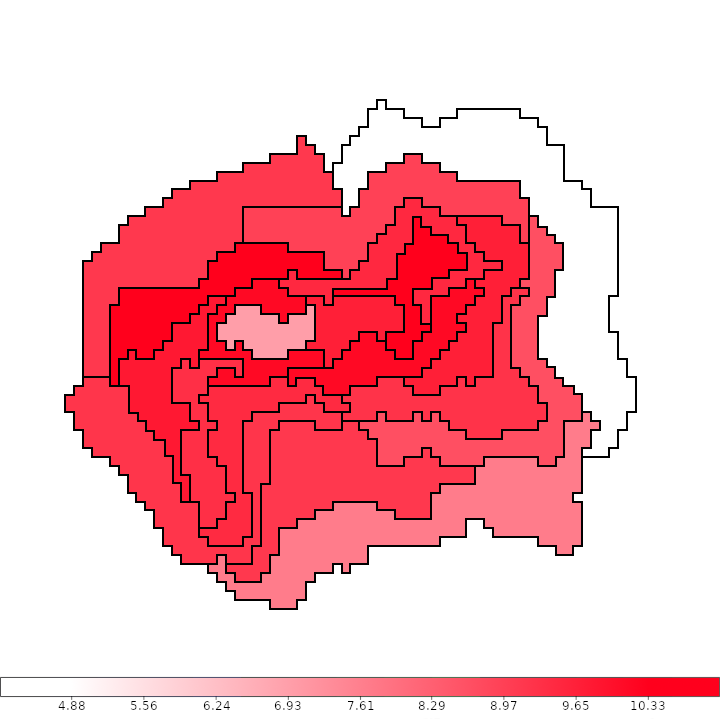}}
		\subfloat[(a) kT$_{h}$]{\includegraphics[angle=0,width=0.3\textwidth,scale=0.5]{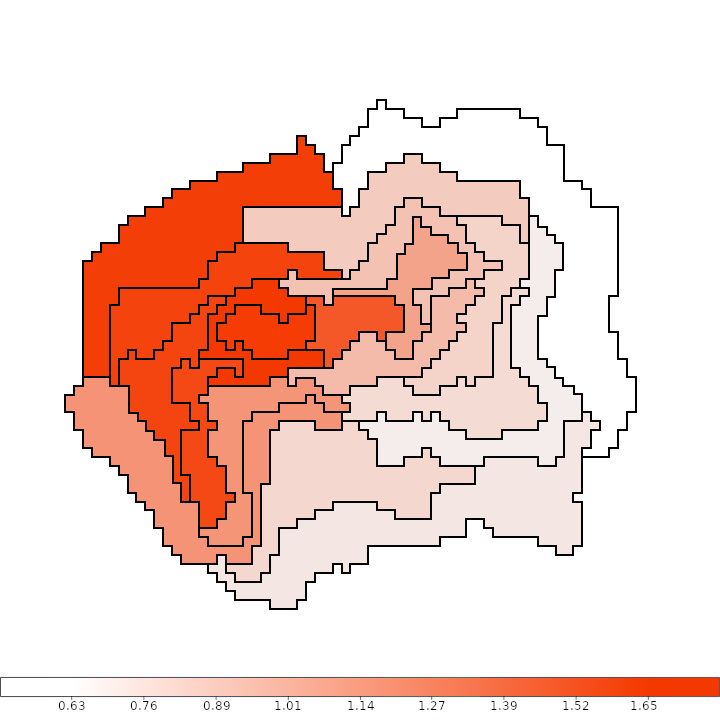}}
		\subfloat[(c) kT$_{c}$]{\includegraphics[angle=0,width=0.3\textwidth]{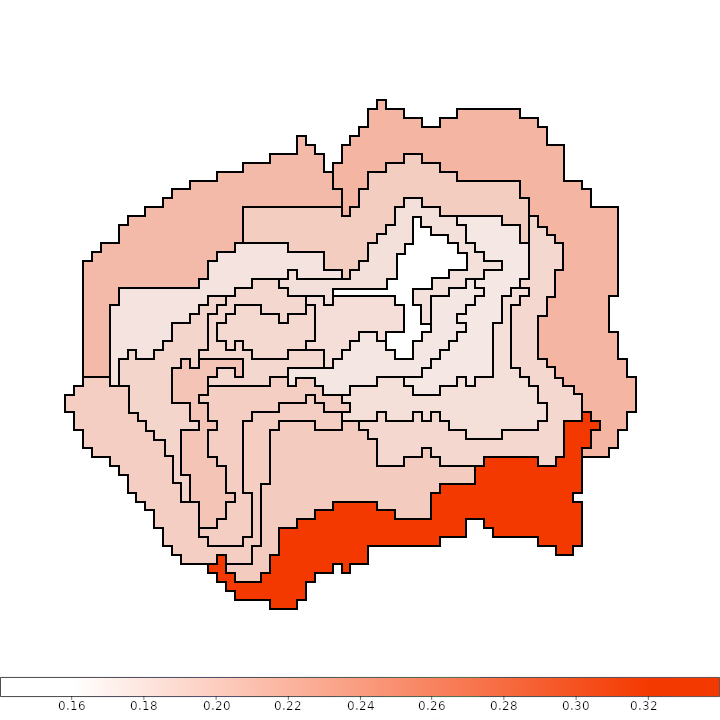}} \\
		\subfloat[(d) Mg]{\includegraphics[angle=0,width=0.3\textwidth]{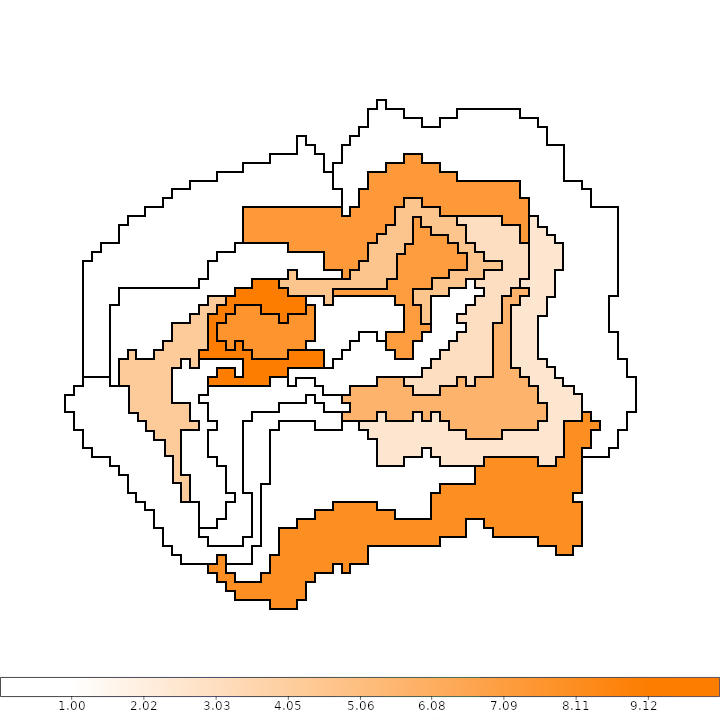}}
		\subfloat[(e) Si]{\includegraphics[angle=0,width=0.3\textwidth,scale=0.5]{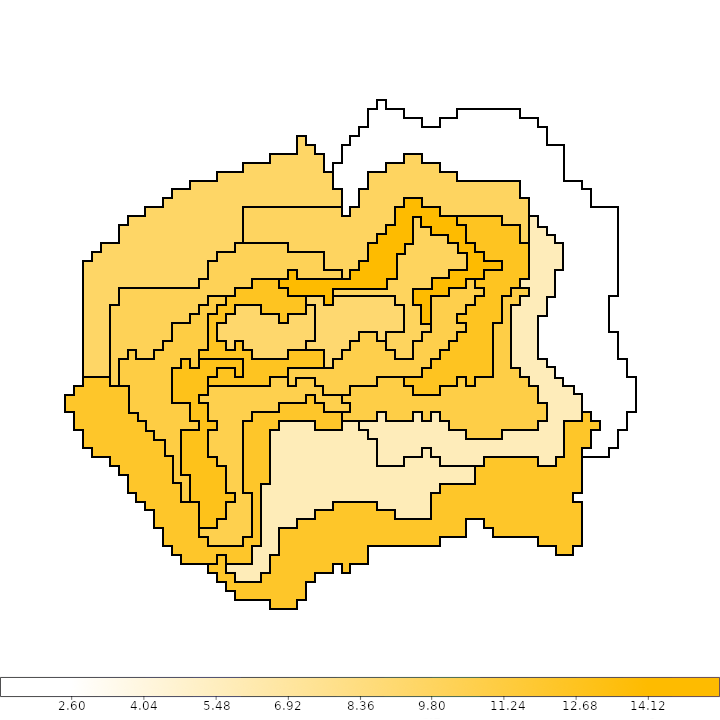}}
		\subfloat[(f) S]{\includegraphics[angle=0,width=0.3\textwidth]{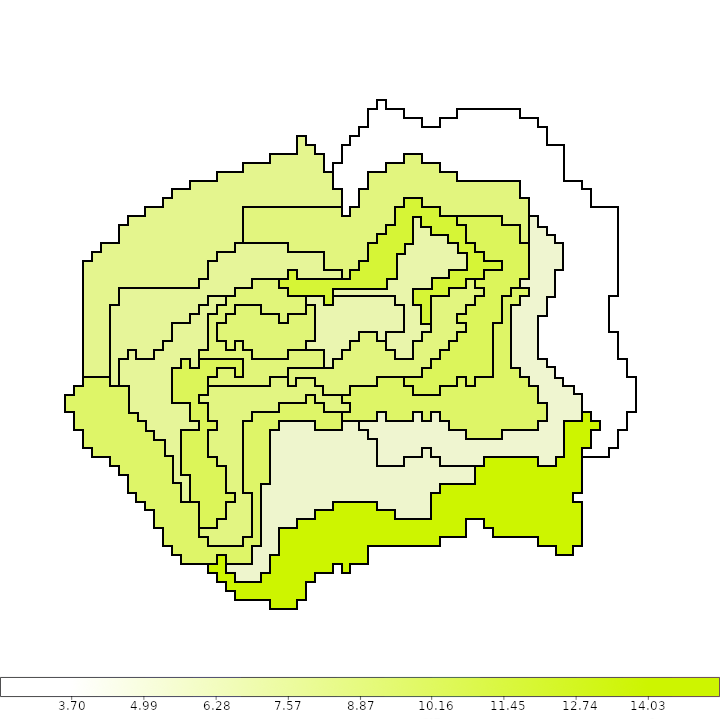}} \\
		\subfloat[(g) Ar]{\includegraphics[angle=0,width=0.3\textwidth]{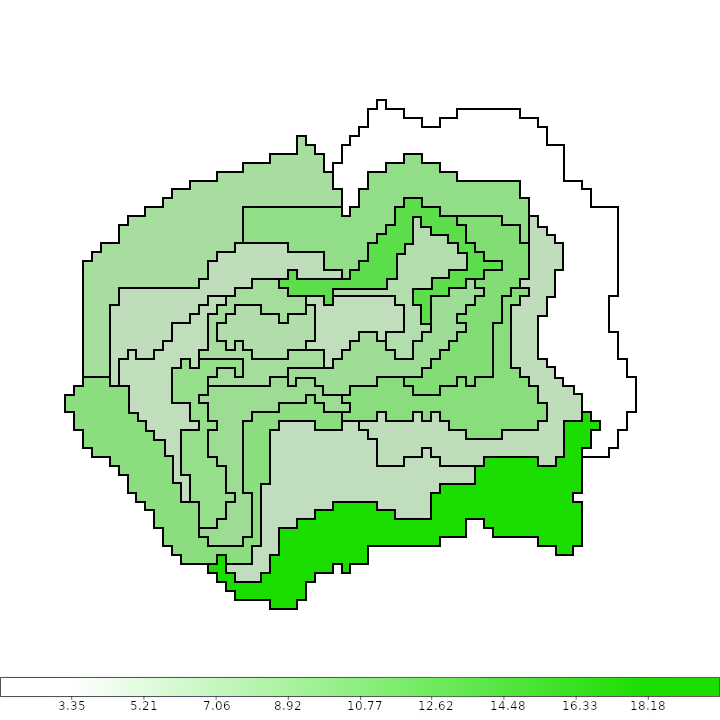}}
		\subfloat[(h) Ca]{\includegraphics[angle=0,width=0.3\textwidth,scale=0.5]{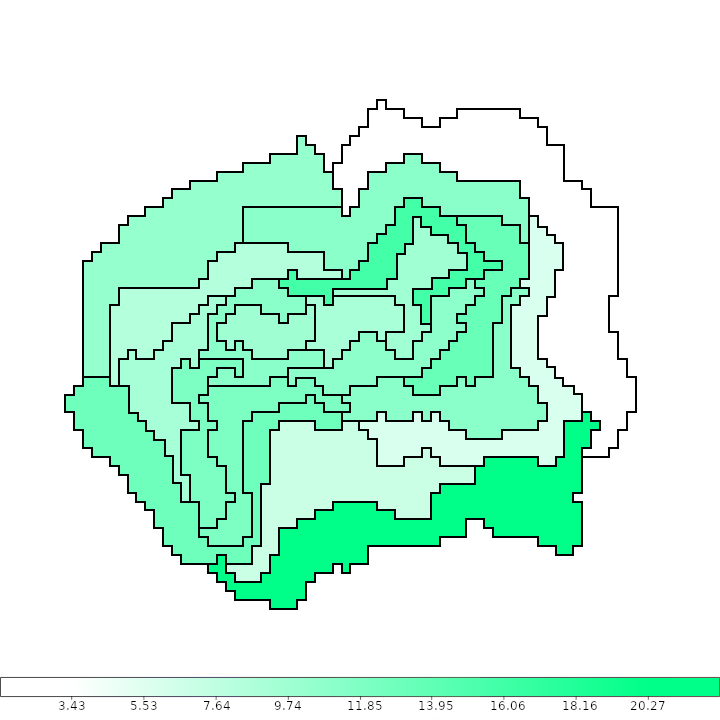}}
		\subfloat[(i) Fe]{\includegraphics[angle=0,width=0.3\textwidth]{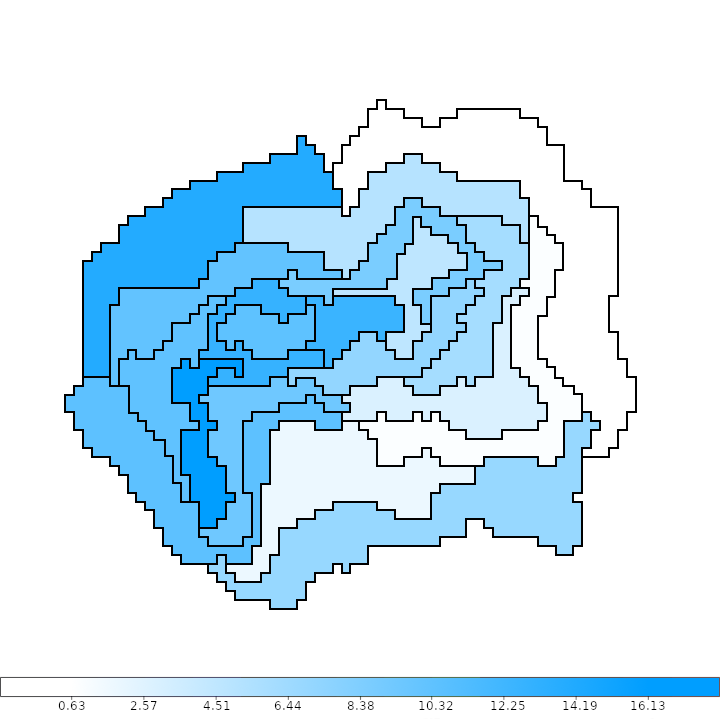}}
		\caption{W49B: Distribution maps for the given parameters. All maps are in a linear scale. N$_{H}$ is in units of $10^{22}$ cm$^{-2}$, kT$_{h}$ and kT$_{c}$ are in units of keV, and all other elements are in terms of solar abundances. The hue of each map is arbitrary.}
		\label{fig:W49B_distribution_maps}
	\end{center}
\end{figure*}

Our analysis of W49B made use of four data sets, which were extracted and then fit simultaneously over the energy range of 0.5 -- 8~keV. After initial testing, it was found that each region was fit best by using a two-component \texttt{VRNEI}+\texttt{VAPEC} model, with the \texttt{VRNEI} component acting as the hot-component and the \texttt{VAPEC} component acting as the cold-component. During the fitting process, the hydrogen column density, both plasma temperatures and normalizations, and the ionization timescale were initially freed and fit. The abundances in the \texttt{VRNEI} component were then freed one at a time in the order of Si, S, Fe, Ni, Ar, Ca, Mg, and allowed to fit. \added{Similarly to 3C 397, previous studies -- such as \cite{sawada_2025} -- have detected emission from Cr and Mn in W49B. These studies made use of data from the \textit{Suzaku} telescope, which possesses higher sensitivity than the data used in this study. As a result, we did not investigate the presence of these elements in W49B.} W49B is known to suffer from heavy interstellar absorption, which can negatively affect the accuracy with which one can detect and identify emission lines below $\sim 2.0$ keV, such as that of Mg, which is typically prominent between $1.3 - 1.5$ keV. This difficulty is reflected in our findings, in which the Mg abundances tend towards having larger uncertainties than other abundances, and in that for some regions, we were unable to isolate the Mg line and thus froze its abundance at solar. \added{As a significant velocity was recently measured in the ejecta \citep{w49b_xrism}, we allowed the redshift to vary freely to account for the bulk velocity of the ejecta.} The fit in certain regions was not improved by freeing one or more of the abundances; in these regions, these abundances were frozen at solar values.

The spectral fits for the individual regions can be found in Table~\ref{tbl:w49b_data}, a sample spectrum can be found in Figure \ref{fig:sample_spectra}, and distribution maps for each parameter can be seen in Figure \ref{fig:W49B_distribution_maps}. Each region is well fit ($\chi^{2}_{\nu} < 1.25$) with the \texttt{VRNEI}+\texttt{VAPEC} model described above. We find that the column density is quite high across the SNR, with values ranging from 4.88$\times 10^{22}$~cm$^{-2}$ to 10.33$\times 10^{22}$~cm$^{-2}$, with the outer portions of the SNR displaying lower column densities on average. Temperatures for the hot component, associated with the \texttt{VRNEI} model, range from 0.64--1.65 keV, with the north-eastern portions of the SNR displaying the highest values thereof. Temperatures for the cold component, associated with the \texttt{VAPEC} model, showed a smaller range of variation, from 0.18--0.32 keV. The lower temperatures emerge in the SNR's interior -- particularly, in the western half of the SNR -- while the higher temperatures appear in the outer portions of the SNR. The abundances within the SNR -- those of Mg, Si, S, Ar, Ca, and Fe -- were all consistently enhanced above solar values, with all of these abundances appearing highest in the north-eastern portion of the SNR.

We can compare our results to several previous studies. One such study is that of \cite{keohane_2007}, in which \textit{Chandra} data was used to perform a spatially resolved study of the SNR. An absorbed, single temperature \texttt{VMEKAL} model was used, yielding a global column density of 5.18$\times 10^{22}$~cm$^{-2}$ -- slightly lower than our average value, but comparable to that found in some of our spectroscopic regions -- and a plasma temperature of 1.58 keV. Enhanced abundances for Si, S, Ar, Ca, Fe, and Ni were found, in agreement with our findings. \cite{lopez_2013}, using \textit{Chandra} data, fit 136 small-scale regions using an absorbed single-component CIE model with the solar abundances of \cite{asplund_2009}. They found values for N$_{H}$ in the range of $4-12 \times 10^{22}$~cm$^{-2}$ and kT from $0.7 - 2.5$ keV, both of which are wider ranges than our findings, but are not dissimilar. Though they did not report abundance values for each region of their analysis, they did report the average values, which we find to be notably lower than our findings, albeit with similar ratios between elements. \cite{zhou_2018} similarly used \textit{Chandra} data and the solar abundances of \cite{asplund_2009} to fit an absorbed \texttt{VRNEI}+\texttt{VAPEC} model to a selection of 177 arcsecond-scale regions. For this, they froze the value of N$_{H}$ to $8 \times 10^{22}$~cm$^{-2}$, which yielded values for kT$_{h}$ between 0.7 and 2.2 keV, a mean value for kT$_{c}$ of 0.27 keV, and ionization timescales ranging between $1 - 10 \times 10^{11}$~s cm$^{-3}$. They too found lower average abundances values for Si, S, Ar, Ca, and Fe than our findings; the ratios between most of these elements were similar than ours, with the notable exception of Si, which appears lower relative to the other elements in their findings than in ours.

We can also compare our results to those of \cite{holland-ashford_2020}, who used \textit{XMM-Newton} data, the solar abundances of \cite{asplund_2009}, and a spatially-resolved binning method in which the SNR was divided into 46 regions of equal size. They fit these regions using an absorbed three-component model, resulting in either a \texttt{VAPEC}+\texttt{VVRNEI}+\texttt{VVRNEI} or \texttt{VAPEC}+\texttt{VVAPEC}+\texttt{VVRNEI} model, dependent on the region in question. They found a range of column densities between $7.2 - 8.5 \times 10^{22}$~cm$^{-2}$, which is close to our findings. The plasma temperature for the \texttt{APEC} component averaged $0.18$ keV, with little variation across the remnant, while the plasma temperatures for their hotter components ranged between $0.31 - 0.76$ keV and $0.87 - 1.62$ keV for the \texttt{VVAPEC}/\texttt{VVRNEI} and \texttt{VVRNEI} component, respectively. The abundances reported are quite similar to our own findings, with theirs being slightly higher on average. Finally, we can compare to \cite{siegel_2020}, an \textit{XMM-Newton} study using observation ID 0724270101. The authors used a smoothed particle inference (SPI) technique to study the entire SNR, using two models: an absorbed \texttt{VMEKAL} model with two temperatures components, as well as an absorbed \texttt{VRNEI}+\texttt{APEC} model. Both were reported in terms of the solar abundances reported in \cite{anders_1989}. The authors reported a column density and temperature of 7.0$\times 10^{22}$~cm$^{-2}$ and 0.98 keV, respectively, both of which are notably lower than our findings. However, they also reported enhanced abundances of Si, S, Ar, Ca, Fe, and Ni, with Ni being poorly constrained, all of which were comparable to our results.

\section{Results}
\label{sec:results}
We have performed spatially resolved spectroscopy on both of the SNRs involved in our study. In the following, we estimate the physical properties and supernova explosion parameters of both 3C~397 and W49B using the spectral fit parameters from the \texttt{VNEI}+\texttt{VAPEC} and \texttt{VRNEI}+\texttt{VAPEC} models summarized in Tables~\ref{tbl:3C397_data} and~\ref{tbl:w49b_data}, respectively.

\subsection{X-ray Properties}
\label{sec:dis_xray}
It is possible to derive useful information, such as several explosion properties as well as the progenitor mass, from the X-ray properties of the remnants. To do so, we first need the distance to the SNR in question, as well as its physical size. We define the volume, V, of the X-ray emitting regions as the area of the region as generated by \textit{contbin}, multiplied by the radius of the SNR, $R_{s}$, on account of the line-of-sight depth of the objects. Because the physical radius of the SNR is highly dependent upon the measured distance, we introduce a scaling factor, defined as $D_{3.1} = D / 3.1$ kpc, such that the calculations in which the distance value is involved can be easily rescaled should an updated value for the distance become available.

For a given volume V, the amount of plasma available to produce the observed flux is known as the emission measure (EM). This is given as $\text{EM} = \int n_en_HdV \sim fn_en_HV$ where $n_e$ is the post-shock electron density, $n_H$ is the proton density, and the parameter $f$ is a filling factor between 0 and 1, where 0 corresponds to the given volume not contributing to the observed flux, and 1 corresponds to the whole volume having a contribution. This factor differs between the hard and soft components of our SNR models, and we distinguish these as $f_h$ and $f_s$, respectively. Filling factors may differ based on the element being examined; however, we opt to assume that they are all the same, regardless of the chosen element. The Rankine-Hugoniot jump conditions relate parameters between pre- and post-shock conditions. For plasmas with solar abundances, the electron density $n_e$ is approximately related to the proton density $n_H$ through the relation $n_e=1.2n_H$. Additionally, the ambient density, $n_0$, can be estimated from the electron density, through $n_e=4.8n_0$. Here, $n_0$ includes only hydrogen \citep{borkowski_2001, 2000ApJ...545..922S}. Lastly, the normalization factor, $K=\frac{10^{-14}}{4\pi D^2}\int n_en_HdV$, can be obtained directly from the spectral fits to the plasma models used, and can then be used to derive the EM.

While XSPEC \added{users may tend to} assume that the ratio between electron density and hydrogen density is a constant, for the ejecta-dominated plasmas found in an SNR, this is not the case, and the actual value can be significantly smaller. \cite{leahy_2023} defines the electron density ratio $r_{e} = n_{e}/(1.2 \, n_{H,0})$, where $n_{H,0}$ is a fiducial hydrogen density. The exact value of this ratio varies with energy and with the composition of the plasma in question, but can be smaller than the value assumed by XSPEC by a factor of 0.1 to 0.001. This ratio represents the ratio of the electron density to the XSPEC value, and serves as a corrective factor for the electron density and all parameters derived therefrom.

As there are several evolutionary phases in which an SNR can fall, which phase the SNR is in plays an important factor in estimating the age of the remnant. By assuming that the SNR is in the early free expansion phase of its evolution, we are able to estimate a lower limit on its age. By assuming an initial expansion velocity of $V_0\sim5000$~km~s$^{-1}$ -- a value which is consistent for young SNRs \citep{reynolds_2008} -- and given the radius of the SNR, we are able to estimate the lower age limit as $t=R_s/V_0$. The Sedov-Taylor phase, during which the supernova shockwave can no longer propagate freely into the surrounding medium, is entered into when the swept-up mass (M$_{\text{sw}}$) well exceeds the mass of the ejecta (M$_{\text{ej}}$). By assuming that the SNR is expanding into a uniform ambient medium, we can estimate the swept-up mass as $\text{M}_{\text{sw}}=1.4m_pn_0\times(4/3\pi $R$^3_sf)$, where $m_p$ is the mass of a proton. We are able to derive an upper limit on the age of the SNR, as $t_{\text{SNR}}=\eta R_s/V_s$, where $\eta=0.4$ for a shock in the Sedov phase \citep{sedov}. When the SNR transitions to the Sedov-Taylor phase, the shock will have been slowed down through interactions with the surrounding medium, and the previously assumed shock velocity of $V_0\sim5000$~km~s$^{-1}$ will no longer be applicable. At this point, we can derive the shock speed V$_S$ from the shock temperature T$_S$ obtained from the Rankine-Hugoniot jump conditions for a strong shock as $\text{V}_s=(16k_B\text{T}_s/3\mu m_h)^{1/2}$, where $\mu=0.604$ is the mean mass per free particle of a fully ionized plasma, $k_B=1.38\times 10^{-16}$~erg~K$^{-1}$ is the Boltzmann's constant, and $m_h$ is the mass of a hydrogen atom. Assuming a uniform ambient medium with density $n_0$, in the Sedov blast wave model the explosion energy E$_*$ is given by E$_*=(1/2.02)R^5_Sm_nn_0t^{-2}_{\text{SNR}}$. Here, $m_n=1.4m_p$ is the mean mass of the nuclei, $m_p$ is the mass of a proton, and t$_{\text{SNR}}$ is the age of the SNR. Typically, explosion energy for a supernova is expected to be of the order of $10^{51}$~ergs. In cases where the shock failed to achieve full electron-ion equilibriation during the Sedov phase, the electron temperature (and hence the shock velocity, $V_s$) will act as a lower limit on the shock temperature $T_s$, which will result in an underestimation of the SNR's explosion energy.

Additionally, we estimated the ejecta mass for each element following the process detailed in \cite{zhou_2016}. The mass for a given element $Z$ is estimated via $M_{Z} = A_{Z} M_{ej} (M_{Z_{\odot}}/M_{\odot})$, where $A_{Z}$ is the abundance value, $M_{ej} = 1.4 n_{H} m_{H} V$ is the mass of the model's shock-heated ejecta component, and $(M_{Z_{\odot}}/M_{\odot})$ is the solar mass fraction of the element $Z$ as determined from the used solar abundances (here, those of \citealt{wilms_2000}). While the filling factor suggests that these ejecta masses act as upper limits on ejecta yields, the inclusion of the corrective factor $r_{e}^{-1/2}$ suggests instead that these ejecta masses act as lower limits. As such, we stress that these ejecta masses are highly uncertain, and are not meant as strong determiners of progenitor type.

\subsection{Nucleosynthesis Models and Progenitor Mass}
\label{sec:dis_nucsyn}
One additional tool which we can make use of is the analysis of nucleosynthesis yields, which can be used to narrow down the range of possible progenitor class for a given supernova remnant. To do so, we compare the nucleosynthesis yields from multiple models -- each of which features a different description of the supernova progenitor -- to the abundance values of the ejecta component obtained from our fitted X-ray data. Here, the abundance ratios used are with respect to Si, and are given by ($X$/Si)/($X$/Si)$_{\odot}$, with $X$ being the measured ejecta mass of either Mg, S, Ar, Ca, or Fe with respect to the solar values of \cite{wilms_2000}.

Because the origins of both of the SNRs involved in this study have historically been debated and remain unclear, in this analysis, we made use of 12 sets of models for comparison. Five of these models are representative of a CC origin: one of the first developed nucleosynthesis models of spherical explosions \citep{woosley_weaver_1995}, hereafter labeled WW95; a set of bipolar explosion models \citep{Maeda_2003}, which we now refer to as M03; a set of hypernova models \citep{Nomoto_2006}, now referred to as N06; a set of spherical explosions for a variety of progenitor masses \citep{sukhbold_2016}, now referred to as S16; and a set of explosion models which makes use of 3 progenitors, but which features a broad range of explosion energies and includes a mix of explosion energies per mass, so as to better simulate asymmetrical explosions \citep{2018ApJ...856...63F}, now referred to as F18.

The other seven models are representative of a Ia scenario: a trio of two-dimensional, asymmetric, models: two delayed-detonation, and one pure deflagration \citep{maeda_2010}, referred hereafter as M10; a set of three-dimensional delayed-detonation models \citep{seitenzahl_2013}, now referred to as S13; a set of three-dimensional, asymmetric, pure deflagration models \citep{fink_2014}, hereafter labeled F14; a collection of two-dimensional DDT models \citep{leung_nomoto_2018}, now labeled LN18; a unique three-dimensional smoothed particle hydrodynamics simulation that looks at the dynamically-driven double-degenerate double-detonation (D$^{6}$) model \citep{tanikawa_2018}, now referred to as T18; a wide set of one-dimensional models exploring both sub-Chandrasekhar-mass scenarios as well as Chandraskehar-mass scenarios, and which explores the reaction rate of C+O \citep{bravo_2019} using the standard C+O reaction rate, as well as the same rate scaled down by a factor ($1-\xi_{CO}$) with $\xi_{CO}=0.9$, now referred to as B19; and a set of two-dimensional double-detonation sub-Chandrasekhar-mass and pure deflagration Chandrasekhar-mass models \citep{leung_nomoto_2020a, leung_nomoto_2020b}, hereafter referred to as LN20. 

\begin{figure*}
    \begin{center}
        \subfloat[(a) 3C 397 Best Fit Ia Models]{\includegraphics[angle=0,width=0.45\textwidth]{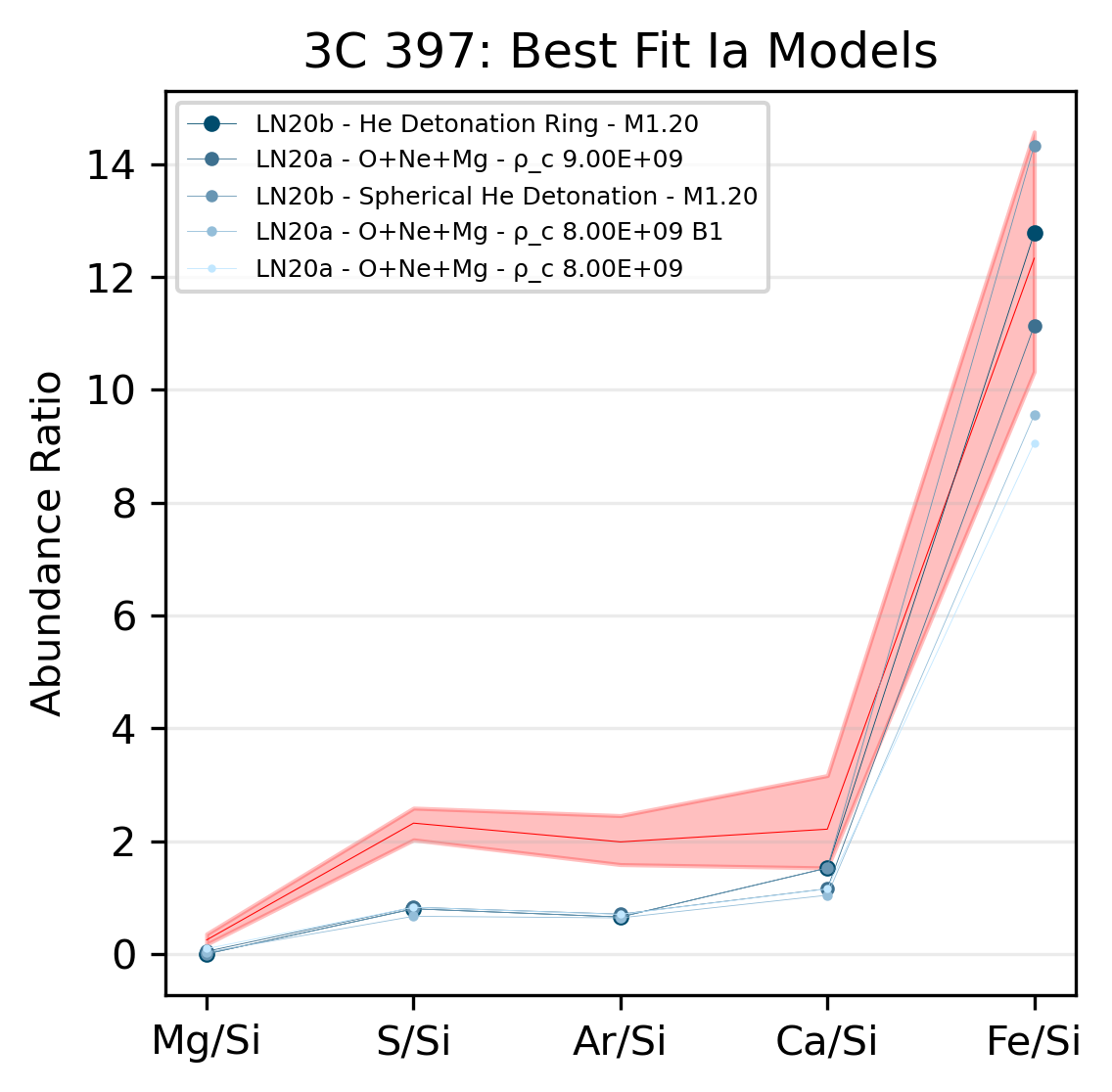}}
        \hspace{1cm}
        \subfloat[(b) 3C 397 Best Fit CC Models]{\includegraphics[angle=0,width=0.45\textwidth]{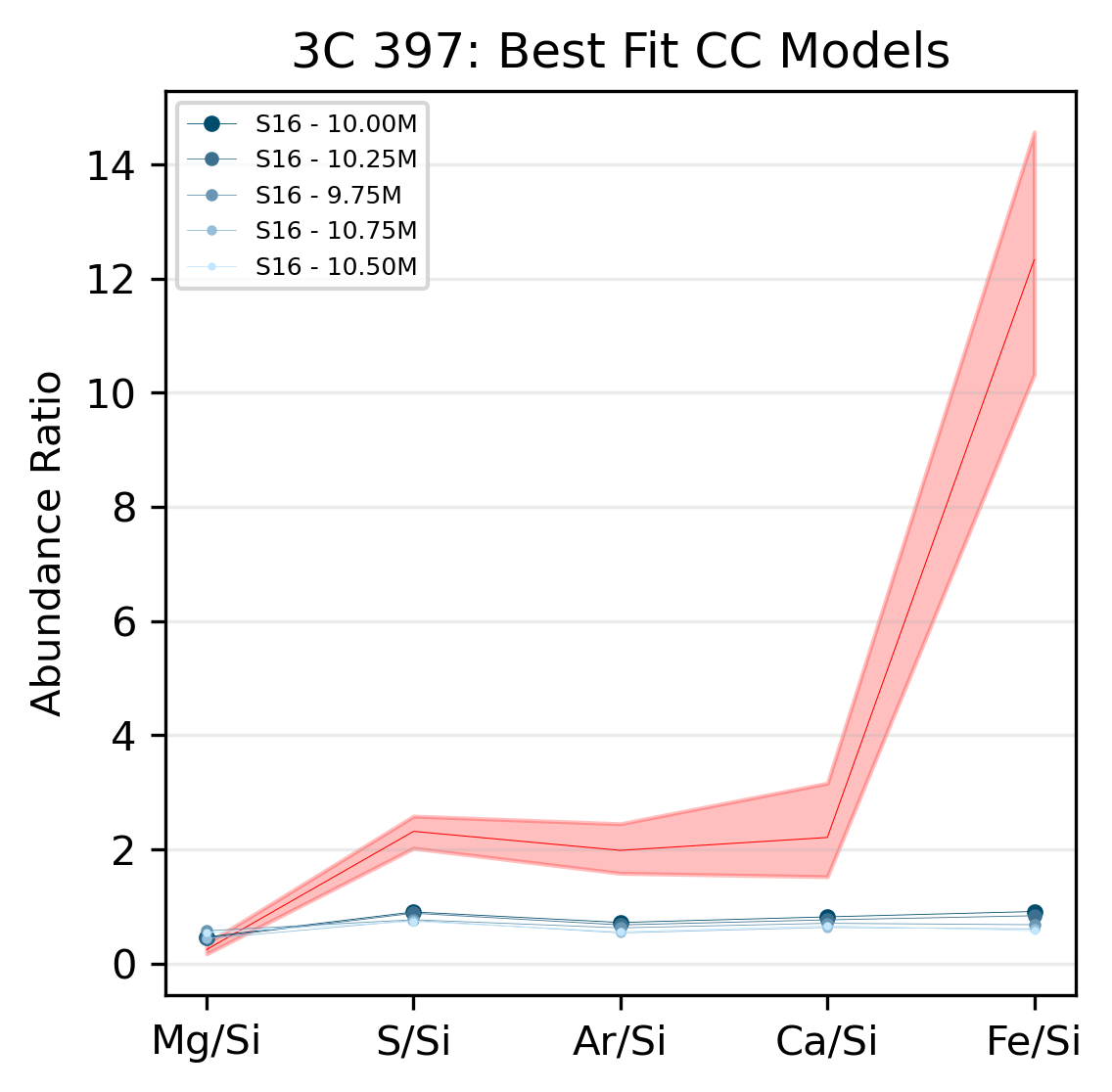}}
        \caption{Best-fit nucleosynthesis models for 3C 397, comparing to both Ia and CC supernova nucleosynthesis yields. See text for details of the models.}
        \label{fig:3C397_best_fits}
    \end{center}
\end{figure*}

\subsection{3C 397}
\label{sec:dis_3C397}
For 3C 397, we assumed a distance of 8.5 kpc \citep{martinez-rodriguez_2020}, which yields a scaling factor of $D_{8.5}~=~D / 8.5$ kpc. We measured a radius of 150$''$, from which we obtained a physical size of $R_s~=~1.91\times10^{19}$~D$_{8.5}$~cm. We used the average values from our regional fits of the SNR to acquire the derived explosion properties. In doing so, we acquired the swept-up mass, M$_\text{SW}~=~3.05\pm{0.03}~f_s^{1/2}$D$_{8.5}^{5/2}$~M$_\odot$; the shock velocity V$_S$~=~$1.21\pm{0.005}\times 10^{3}~$km~s$^{-1}$; the free expansion age estimate t$_\text{SNR}~=~1160$~D$_{8.5}$~yr; the Sedov age estimate t$_\text{SNR}~=~2010\pm{8}~$D$_{8.5}$~yr; and the explosion energy  E$_*~=~7.31^{+0.08}_{-0.10}\times 10^{49}~f_s^{-1/2}$~D$_{8.5}^{5/2}$~erg. The derived X-ray properties, as described in Section \ref{sec:dis_xray}, can be found in Table~\ref{tbl:3c397_xray_properties}, and we estimate the ejecta mass for each element as: 
\begin{itemize}
    \item M$_\text{Mg}=0.062_{-0.008}^{+0.010}~f_h^{1/2} \, r_{e}^{-1/2}$ D$_{8.5}^{5/2}$~M$_\odot$,
    \item M$_\text{Si}=0.26\pm{0.01}~f_h^{1/2} \, r_{e}^{-1/2}$ D$_{8.5}^{5/2}$~M$_\odot$,
    \item M$_\text{S}=0.016\pm{0.001}~f_h^{1/2} \, r_{e}^{-1/2}$ D$_{8.5}^{5/2}$~M$_\odot$,
    \item M$_\text{Ar}=0.0035\pm{0.0003}~f_h^{1/2} \, r_{e}^{-1/2}$ D$_{8.5}^{5/2}$~M$_\odot$,
    \item M$_\text{Ca}=0.0023_{-0.0003}^{+0.0004}~f_h^{1/2} \, r_{e}^{-1/2}$ D$_{8.5}^{5/2}$~M$_\odot$,
    \item M$_\text{Fe}=0.31\pm{0.02}~f_h^{1/2} \, r_{e}^{-1/2}$ D$_{8.5}^{5/2}$~M$_\odot$,
    \item M$_\text{Ni}=0.014\pm{0.001}~f_h^{1/2} \, r_{e}^{-1/2}$ D$_{8.5}^{5/2}$~M$_\odot$.
\end{itemize}

\begin{table*}
\movetableright=-15mm
	\caption{Derived X-ray properties for 3C 397. The subscripts 'h' and 's' refer to the hard and soft components, respectively.}
	\label{tbl:3c397_xray_properties} 
	\begin{tabular}{cccccc}
	Region & Emission Measure, EM$_h$ & Electron Density, n$_e{_h}$ & Emission Measure, EM$_s$ & Electron Density, n${_e}_s$ & Ambient Density, n${_0}_s$ \\  		  
	& ($\times 10^{57}$~$f_h \, r_{e}^{1/2}$~D$_{8.5}^{-2}$~cm$^{-3}$) &
    ($f_h^{-1/2} \, r_{e}$~D$_{8.5}^{-1/2}$~cm$^{-3}$) &
    ($\times 10^{60}$~$f_s \, r_{e}^{1/2}$~D$_{8.5}^{-2}$~cm$^{-3}$) &
    ($f_s^{-1/2} \, r_{e}$~D$_{8.5}^{-1/2}$~cm$^{-3}$) &
    ($f_s^{-1/2}$~D$_{8.5}^{-1/2}$~cm$^{-3}$) \\ \hline
	\textbf{0} & $3.15_{-0.13}^{+0.29}$ & $3.07_{-0.06}^{+0.14}$ & $1.80_{-0.14}^{+0.25}$ & $73.4_{-2.9}^{+5.0}$ & $15.3_{-0.6}^{+1.1}$ \\
\textbf{1} & $3.05_{-0.26}^{+0.18}$ & $3.52_{-0.15}^{+0.11}$ & $2.31_{-0.26}^{+0.27}$ & $97.1_{-5.4}^{+5.7}$ & $20.2_{-1.1}^{+1.2}$ \\
\textbf{2} & $4.42_{-0.38}^{+0.22}$ & $3.77_{-0.16}^{+0.09}$ & $2.31_{-0.23}^{+0.30}$ & $86.2_{-4.3}^{+5.6}$ & $18.0_{-0.9}^{+1.2}$ \\
\textbf{3} & $4.96_{-0.28}^{+0.40}$ & $3.62_{-0.10}^{+0.15}$ & $6.46_{-0.91}^{+0.68}$ & $130_{-9}^{+7}$ & $27.2_{-1.9}^{+1.4}$ \\
\textbf{4} & $3.12_{-0.20}^{+0.14}$ & $2.76_{-0.09}^{+0.06}$ & $3.42_{-0.51}^{+0.37}$ & $91.5_{-6.9}^{+5.0}$ & $19.1_{-1.4}^{+1.0}$ \\
\textbf{5} & $2.44_{-0.22}^{+0.25}$ & $2.27_{-0.10}^{+0.11}$ & $1.78_{-0.17}^{+0.35}$ & $61.2_{-3.0}^{+6.0}$ & $12.8_{-0.6}^{+1.3}$ \\
\textbf{6} & $1.87_{-0.16}^{+0.03}$ & $1.95_{-0.08}^{+0.02}$ & $0.35_{-0.12}^{+0.09}$ & $26.7_{-4.6}^{+3.6}$ & $5.6_{-1.0}^{+0.7}$ \\
\textbf{7} & $4.70_{-0.25}^{+0.08}$ & $2.98_{-0.08}^{+0.02}$ & $3.87_{-0.47}^{+0.39}$ & $85.6_{-5.2}^{+4.3}$ & $17.8_{-1.1}^{+0.9}$ \\
\textbf{8} & $2.48_{-0.21}^{+0.02}$ & $2.16_{-0.09}^{+0.01}$ & $2.85_{-0.35}^{+0.38}$ & $73.5_{-4.6}^{+4.9}$ & $15.3_{-0.9}^{+1.0}$ \\
\textbf{9} & $2.05_{-0.77}^{+0.54}$ & $1.62_{-0.30}^{+0.21}$ & $0.57_{-0.47}^{+0.45}$ & $27.0_{-11}^{+11}$ & $5.6_{-2.3}^{+2.2}$ \\
\textbf{10} & $1.37_{-0.11}^{+0.12}$ & $1.52_{-0.06}^{+0.07}$ & $2.41_{-0.28}^{+0.33}$ & $63.9_{-3.7}^{+4.4}$ & $13.3_{-0.8}^{+0.9}$ \\
\textbf{11} & $2.73_{-0.13}^{+0.08}$ & $1.98_{-0.05}^{+0.03}$ & $3.16_{-0.34}^{+0.34}$ & $67.3_{-3.6}^{+3.7}$ & $14.0_{-0.8}^{+0.8}$ \\
\textbf{12} & $4.49_{-0.54}^{+0.34}$ & $2.59_{-0.16}^{+0.10}$ & $0.89_{-0.09}^{+0.10}$ & $36.6_{-1.9}^{+2.1}$ & $7.6_{-0.4}^{+0.4}$ \\
\textbf{13} & $1.13_{-0.03}^{+0.06}$ & $1.12_{-0.01}^{+0.03}$ & $2.80_{-0.40}^{+0.38}$ & $55.8_{-4.0}^{+3.8}$ & $11.6_{-0.8}^{+0.8}$ \\
\textbf{14} & $4.06_{-0.37}^{+0.28}$ & $2.15_{-0.10}^{+0.07}$ & $0.80_{-0.10}^{+0.08}$ & $30.1_{-1.9}^{+1.5}$ & $6.3_{-0.4}^{+0.3}$ \\
\textbf{15} & $4.65_{-0.15}^{+0.23}$ & $2.08_{-0.03}^{+0.06}$ & $2.32_{-0.16}^{+0.14}$ & $46.4_{-1.6}^{+1.4}$ & $9.7_{-0.3}^{+0.3}$ \\
\textbf{16} & $2.25_{-0.24}^{+0.08}$ & $1.15_{-0.06}^{+0.02}$ & $3.98_{-0.45}^{+0.28}$ & $48.4_{-2.8}^{+1.7}$ & $10.1_{-0.6}^{+0.3}$ \\ \hline 	
	\end{tabular}
\end{table*}

The nucleosynthesis models that most closely compare to our results can be seen in Figure \ref{fig:3C397_best_fits}; comparisons to all tested models can be found in Appendix \ref{apx:plots}, in Figures \ref{fig:3c397_ia} and \ref{fig:3c397_cc}. Comparing to the nucleosynthesis models mentioned above, we find that while no one model is capable of reproducing our results, the Ia models fare significantly better at doing so than the CC models. This is largely due to the particularly high Fe/Si ratio found in the SNR: only a few Ia models are capable of reproducing this, with the best results coming from the He detonation and O+Ne+Mg WD models of LN20. The Mg/Si ratio is reasonably well fit for most of the models tested, with the CC models tending to produce slightly higher abundances than observed. The S/Si ratio is not well reproduced by any model, while the Ar/Si ratio only manages to be reproduced by the D$^{6}$ model of T19, although the 11~M$_\odot$ model of WW95 comes quite close. The Ca/Si ratio is well reproduced by a number of Ia models: the models which can reproduce the Fe/Si ratio can also reproduce the Ca/Si ratio, and several of the models found in B19 can do so as well.

\subsection{W49B}
\label{sec:dis_W49B}
For W49B, a distance of 9.3 kpc was used \citep{sun+chen_2020}, yielding a scaling factor of $D_{9.3} = D / 9.3$ kpc. We measured a radius of 132$''$, from which we obtained a physical size of $R_s = 1.84\times10^{19}$~D$_{9.3}$~cm. The explosion properties were derived from the averages of the regional fits. Values obtained from this include the swept-up mass, M$_\text{SW} = 10.8_{-0.3}^{+0.1}~f_s^{1/2}$D$_{9.3}^{5/2}$~M$_\odot$; the shock velocity V$_S$~=~$8.68^{+0.13}_{-0.02}\times 10^{2}~$km~s$^{-1}$; the free expansion age estimate t$_\text{SNR}=1160$~D$_{9.3}$~yr; the Sedov age estimate t$_\text{SNR}=2680^{+41}_{-9}~$D$_{9.3}$~yr; and the explosion energy  E$_*=1.20\pm{0.04}\times 10^{50}~f_s^{-1/2}$~D$_{9.3}^{5/2}$~erg. The derived X-ray properties, as described in Section \ref{sec:dis_xray}, can be found in Table~\ref{tbl:w49b_xray_properties}, and we estimate the mass of the ejecta for each element to be:
\begin{itemize}
    \item M$_\text{Mg} = 0.131^{+0.052}_{-0.019}~f_h^{1/2} \, r_{e}^{-1/2}$ D$_{9.3}^{5/2}$~M$_\odot$,
    \item M$_\text{Si} = 0.270^{+0.020}_{-0.025}~f_h^{1/2} \, r_{e}^{-1/2}$ D$_{9.3}^{5/2}$~M$_\odot$,
    \item M$_\text{S} = 0.179^{+0.015}_{-0.014}~f_h^{1/2} \, r_{e}^{-1/2}$ D$_{9.3}^{5/2}$~M$_\odot$,
    \item M$_\text{Ar} = 0.047^{+0.003}_{-0.005}~f_h^{1/2} \, r_{e}^{-1/2}$ D$_{9.3}^{5/2}$~M$_\odot$,
    \item M$_\text{Ca} = 0.036^{+0.005}_{-0.003}~f_h^{1/2} \, r_{e}^{-1/2}$ D$_{9.3}^{5/2}$~M$_\odot$,
    \item M$_\text{Fe} = 0.543^{+0.057}_{-0.058}~f_h^{1/2} \, r_{e}^{-1/2}$ D$_{9.3}^{5/2}$~M$_\odot$,
    \item M$_\text{Ni} = 0.075^{+0.021}_{-0.015}~f_h^{1/2} \, r_{e}^{-1/2}$ D$_{9.3}^{5/2}$~M$_\odot$.
\end{itemize}

\begin{figure*}
    \begin{center}
        \subfloat[(a) W49B Best Fit Ia Models]{\includegraphics[angle=0,width=0.45\textwidth]{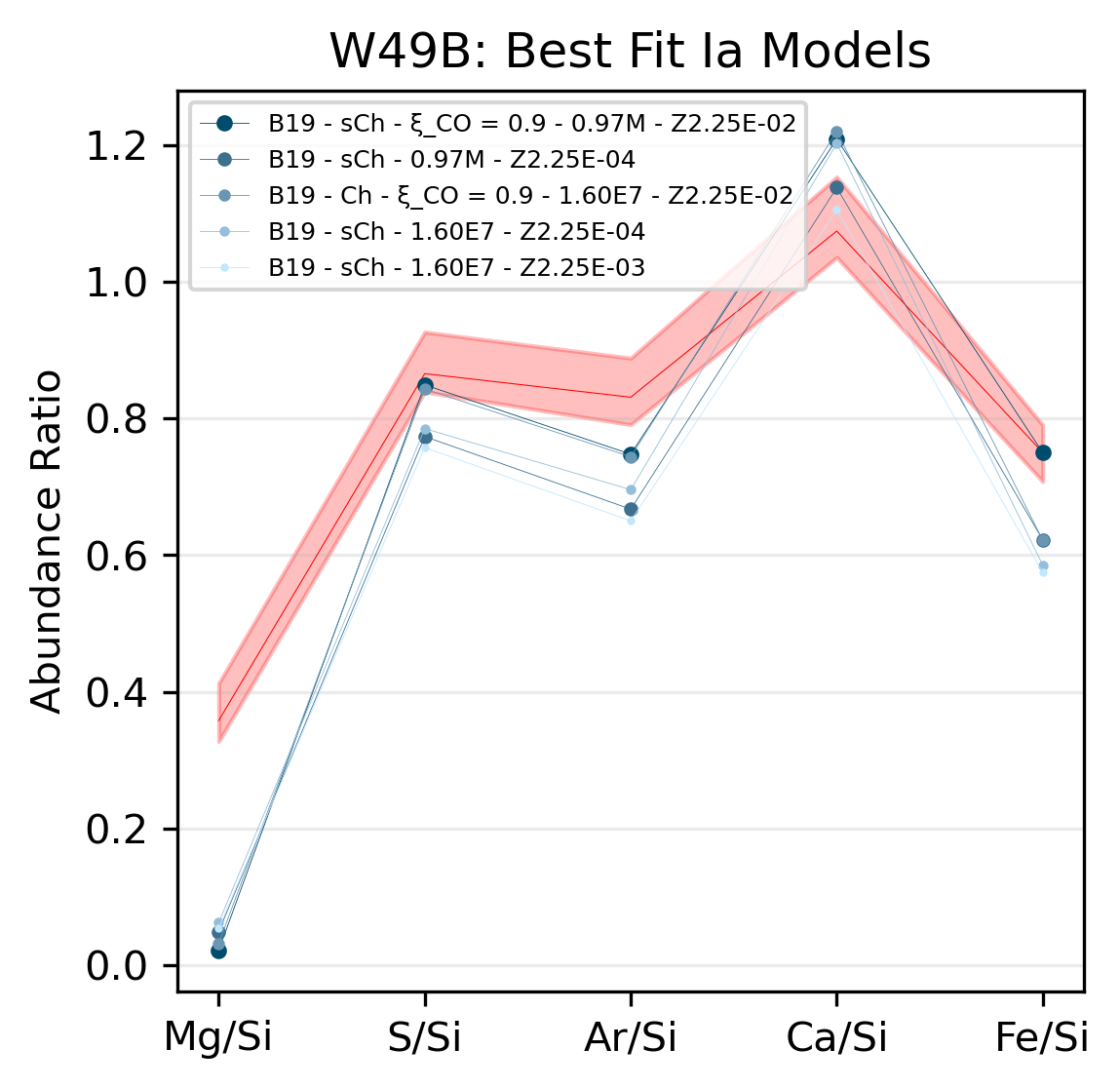}}
        \hspace{1cm}
        \subfloat[(b) W49B Best Fit CC Models]{\includegraphics[angle=0,width=0.45\textwidth]{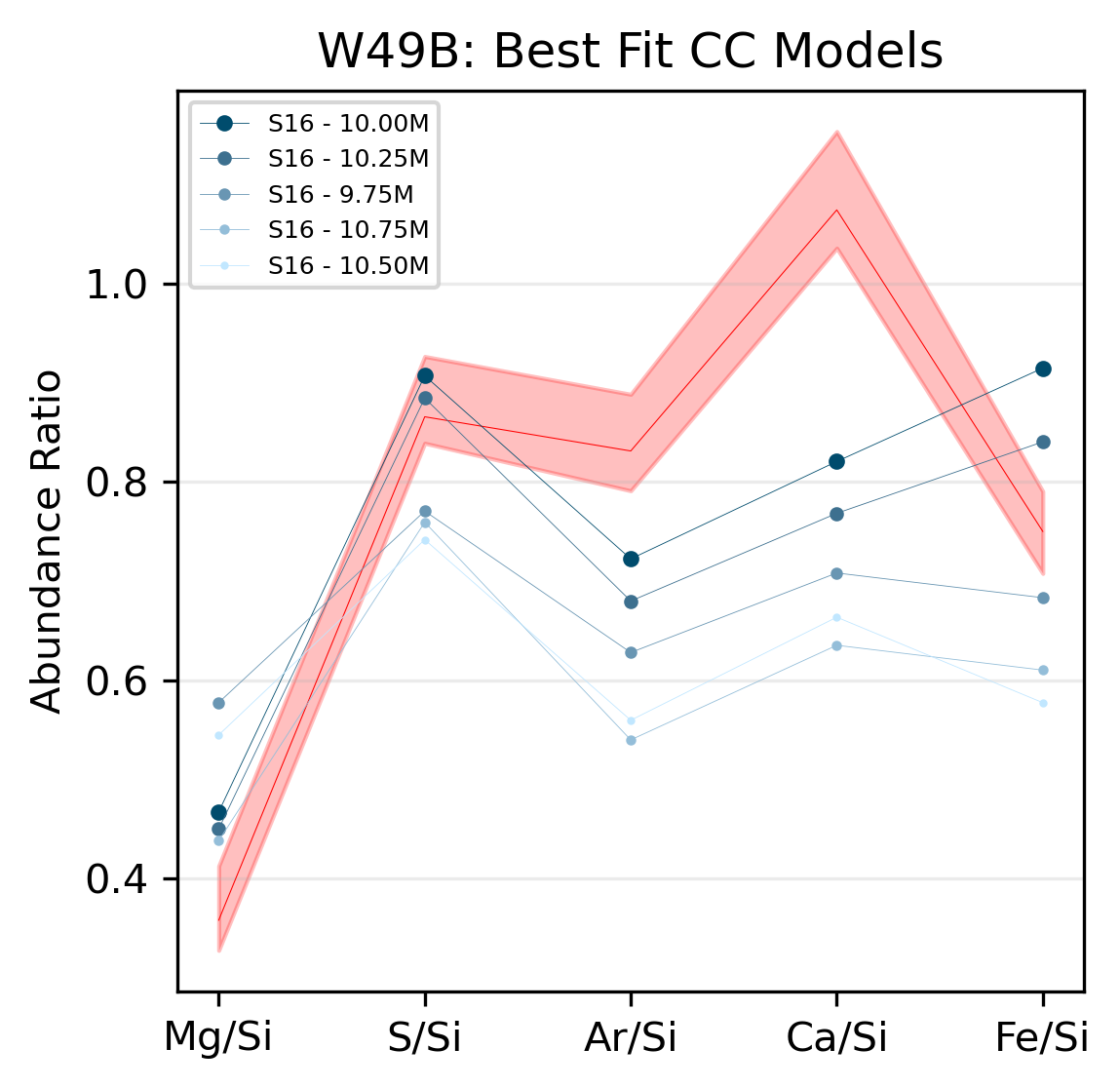}}
        \caption{Best-fit nucleosynthesis models for W49B, comparing to both Ia and CC supernova nucleosynthesis yields. See text for details of the models.}
        \label{fig:W49B_best_fits}
    \end{center}
\end{figure*}

\begin{table*}
\movetableright=-15mm
    \caption{Derived X-ray properties for W49B. The subscripts ’h’ and ’s’ refer to the hard and soft components, respectively.}
    \label{tbl:w49b_xray_properties}
    \begin{tabular}{cccccc}
    Region & Emission Measure, EM$_h$ & Electron Density, n$_e{_h}$ & Emission Measure, EM$_s$ & Electron Density, n${_e}_s$ & Ambient Density, n${_0}_s$ \\  		  
	& ($\times 10^{58} f_{h} r_{e}^{1/2} D_{9.3}^{-2}$~cm$^{-3}$) &
    ($f_{h}^{-1/2} r_{e} D_{9.3}^{-1/2}$~cm$^{-3}$) &
    ($\times 10^{61} f_{s} r_{e}^{1/2} D_{9.3}^{-2}$~cm$^{-3}$) &
    ($f_{s}^{-1/2} r_{e} D_{9.3}^{-1/2}$~cm$^{-3}$) &
    ($f_{s}^{-1/2} D_{9.3}^{-1/2}$~cm$^{-3}$) \\
    \hline
    \textbf{0}  & $2.33_{-0.26}^{+0.16}$ & $9.50_{-0.53}^{+0.32}$ & $3.26_{-0.81}^{+0.65}$ & $356_{-44}^{+35}$     & $74.1_{-9.2}^{+7.3}$   \\
\textbf{1}  & $2.70_{-0.22}^{+0.07}$ & $9.61_{-0.38}^{+0.13}$ & $3.79_{-0.61}^{+2.21}$ & $360_{-29}^{+105}$    & $75.0_{-6.4}^{+21.9}$  \\
\textbf{2}  & $2.18_{-0.37}^{+0.01}$ & $8.14_{-0.69}^{+0.02}$ & $2.16_{-2.15}^{+0.01}$ & $256_{-127}^{+1}$     & $53.3_{-26.6}^{+0.1}$  \\
\textbf{3}  & $1.88_{-0.69}^{+0.01}$ & $7.10_{-1.31}^{+0.01}$ & $3.43_{-3.43}^{+0.01}$ & $303_{-152}^{+1}$     & $63.2_{-31.6}^{+0.1}$  \\
\textbf{4}  & $1.31_{-0.01}^{+0.01}$ & $5.97_{-0.01}^{+0.01}$ & $4.36_{-0.03}^{+0.03}$ & $344_{-1}^{+1}$       & $71.6_{-0.2}^{+0.3}$   \\
\textbf{5}  & $3.29_{-0.35}^{+0.12}$ & $8.48_{-0.45}^{+0.15}$ & $4.47_{-0.66}^{+1.25}$ & $313_{-23}^{+44}$     & $65.1_{-4.8}^{+9.1}$   \\
\textbf{6}  & $4.61_{-0.44}^{+0.23}$ & $10.3_{-0.5}^{+0.3}$   & $8.60_{-1.58}^{+0.83}$ & $444_{-41}^{+21}$     & $92.5_{-8.5}^{+4.5}$   \\
\textbf{7}  & $4.11_{-0.28}^{+0.08}$ & $9.63_{-0.33}^{+0.09}$ & $6.82_{-0.11}^{+0.97}$ & $392_{-3}^{+28}$      & $81.8_{-0.7}^{+5.8}$   \\
\textbf{8}  & $3.26_{-0.33}^{+0.24}$ & $8.78_{-0.45}^{+0.32}$ & $2.78_{-0.62}^{+0.49}$ & $256_{-29}^{+23}$     & $53.4_{-6.0}^{+4.7}$   \\
\textbf{9}  & $3.65_{-2.54}^{+0.05}$ & $8.56_{-2.97}^{+0.06}$ & $5.00_{-5.00}^{+0.05}$ & $316_{-158}^{+2}$     & $65.9_{-33.0}^{+0.3}$  \\
\textbf{10} & $2.30_{-0.17}^{+0.01}$ & $5.82_{-0.22}^{+0.01}$ & $7.06_{-2.14}^{+1.10}$ & $323_{-49}^{+25}$     & $67.2_{-10.2}^{+5.2}$  \\
\textbf{11} & $2.12_{-0.17}^{+0.17}$ & $4.61_{-0.19}^{+0.18}$ & $1.93_{-0.33}^{+0.27}$ & $139_{-12}^{+10}$     & $29.0_{-2.5}^{+2.1}$   \\
\textbf{12} & $7.25_{-0.53}^{+0.50}$ & $10.3_{-0.4}^{+0.4}$   & $4.97_{-0.73}^{+1.21}$ & $269_{-20}^{+33}$     & $56.0_{-4.1}^{+6.8}$   \\
\textbf{13} & $3.47_{-0.39}^{+0.15}$ & $6.09_{-0.34}^{+0.13}$ & $1.64_{-0.28}^{+0.19}$ & $132_{-11}^{+7}$      & $27.6_{-2.3}^{+1.6}$   \\
\textbf{14} & $11.2_{-3.5}^{+0.1}$   & $12.2_{-1.9}^{+0.1}$   & $2.26_{-0.35}^{+0.03}$ & $172_{-13}^{+1}$      & $36.0_{-2.8}^{+0.2}$   \\
\textbf{15} & $6.20_{-0.39}^{+0.66}$ & $7.82_{-0.25}^{+0.41}$ & $1.56_{-0.33}^{+0.28}$ & $124_{-13}^{+11}$     & $25.9_{-2.7}^{+2.3}$   \\
\textbf{16} & $2.06_{-0.13}^{+0.10}$ & $4.16_{-0.13}^{+0.10}$ & $1.09_{-0.07}^{+0.56}$ & $95.8_{-3.0}^{+24.7}$ & $20.0_{-0.6}^{+5.1}$   \\
\textbf{17} & $3.05_{-0.43}^{+0.30}$ & $4.58_{-0.32}^{+0.22}$ & $0.07_{-0.02}^{+0.02}$ & $22.6_{-2.8}^{+3.0}$  & $4.71_{-0.59}^{+0.63}$ \\
\textbf{18} & $7.20_{-0.45}^{+0.58}$ & $6.11_{-0.19}^{+0.25}$ & $0.21_{-0.04}^{+0.04}$ & $32.9_{-3.3}^{+3.3}$  & $6.86_{-0.69}^{+0.69}$ \\ \hline
    \end{tabular}
\end{table*}

Figure \ref{fig:W49B_best_fits} shows a comparison between our results and the nucleosynthesis models that most closely reproduce them; comparisons to all tested models can be found in Appendix \ref{apx:plots}, in Figures \ref{fig:w49b_ia} and \ref{fig:w49b_cc}. Comparing to the nucleosynthesis models, while we again find that no tested models are able to fully recreate the data, we do note much better agreement than in the case of 3C~397. We find that the Mg/Si ratio cannot be reproduced by any of the tested Ia models, all of which produce less Mg than found. CC models, comparatively, tend to produce larger amounts of Mg, resulting in a greater agreement with our results, with models such as the WW95 13M or 25M, or the S16 20.5M or 26.0M being able to reproduce our findings. The S/Si and Ar/Si ratios can be accurately reproduced through a variety of both Ia and CC models, while the Ca/Si and Fe/Si ratios are generally only able to be reproduced by Ia models, as the CC models tend to produce too little of these elements. Overall, the best fit is achieved by a 0.97 M$_{\odot}$ progenitor of B19, with $\xi_{CO}=0.9$. However, from the results of our mass estimation, the total ejecta mass within the SNR should be at least $\sim$ 1.28 M$_{\odot}$, which is of course incompatible with a 0.97 M$_{\odot}$ progenitor. We thus suggest that the most likely progenitor for this SNR is a Ch-mass progenitor with a reduced C+O reaction, also from B19.

\subsection{Fe K\texorpdfstring{$\alpha$}{-Alpha} Line Centroids}
\label{sec:dis_fe-k}

\begin{figure*}
    \begin{center}
        \subfloat[(a) 3C 397]{\includegraphics[angle=0,width=0.4\textwidth]{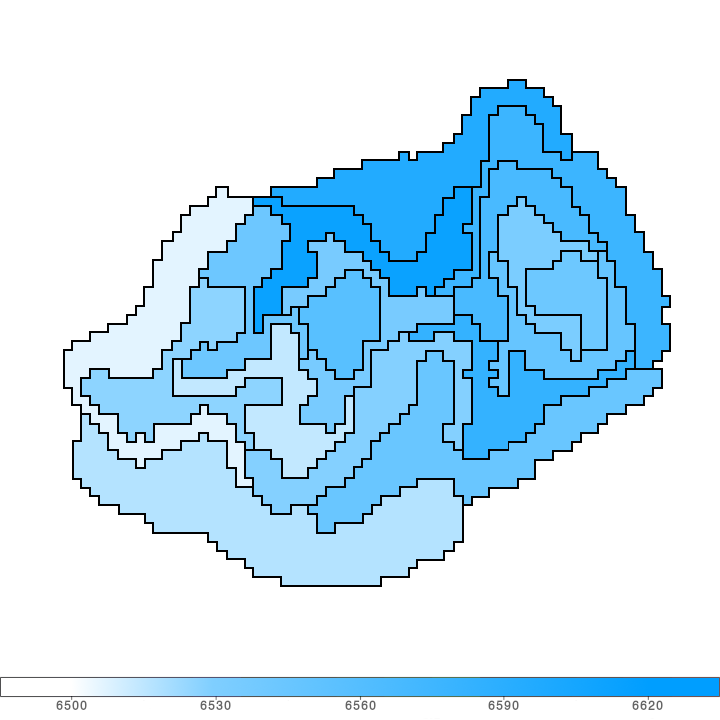}}
        \hspace{1cm}
        \subfloat[(b) W49B]{\includegraphics[angle=0,width=0.4\textwidth]{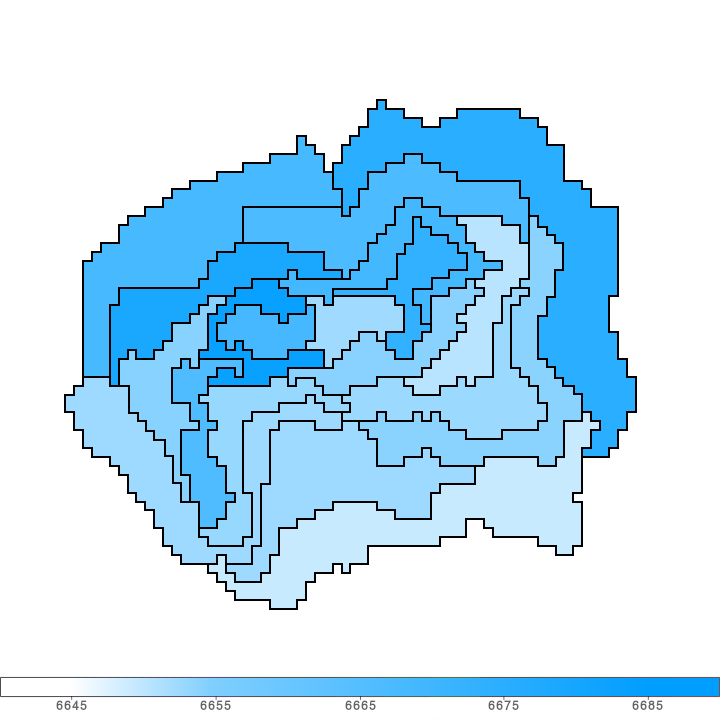}}
        \caption{Fe K$\alpha$ centroid distribution maps for 3C 397 (left) and W49B (right). Line energies are given in eV.}
        \label{fig:fe-k_maps}
    \end{center}
\end{figure*}

\begin{table}
    \caption{Energies of the Fe K$\alpha$ centroid for each of our arcsecond-scale regions within 3C 397 (left) and W49B (right)}
    \label{tbl:fe_energy}
    \centering
    \begin{tabular}{cc|cc}
    \multicolumn{2}{c}{\textbf{3C 397}} & \multicolumn{2}{c}{\textbf{W49B}} \\
\multicolumn{1}{c}{Region} & \multicolumn{1}{c}{Centroid Energy} & \multicolumn{1}{c}{Region} & \multicolumn{1}{c}{Centroid Energy} \\
 & \multicolumn{1}{c}{(eV)} &  & \multicolumn{1}{c}{(eV)} \\
    \hline
    \textbf{0} & $6578_{-10}^{+12}$ & \textbf{0} & $6674_{-2}^{+2}$ \\
\textbf{1} & $6569_{-11}^{+12}$ & \textbf{1} & $6660_{-7}^{+7}$ \\
\textbf{2} & $6562_{-11}^{+12}$ & \textbf{2} & $6673_{-2}^{+2}$ \\
\textbf{3} & $6572_{-9}^{+9}$ & \textbf{3} & $6684_{-2}^{+2}$ \\
\textbf{4} & $6586_{-11}^{+11}$ & \textbf{4} & $6677_{-3}^{+3}$ \\
\textbf{5} & $6569_{-13}^{+15}$ & \textbf{5} & $6674_{-4}^{+1}$ \\
\textbf{6} & $6564_{-10}^{+11}$ & \textbf{6} & $6664_{-2}^{+2}$ \\
\textbf{7} & $6596_{-9}^{+9}$ & \textbf{7} & $6664_{-3}^{+2}$ \\
\textbf{8} & $6529_{-9}^{+9}$ & \textbf{8} & $6661_{-3}^{+2}$ \\
\textbf{9} & $6592_{-10}^{+10}$ & \textbf{9} & $6656_{-3}^{+4}$ \\
\textbf{10} & $6554_{-10}^{+10}$ & \textbf{10} & $6681_{-2}^{+2}$ \\
\textbf{11} & $6558_{-7}^{+7}$ & \textbf{11} & $6660_{-1}^{+7}$ \\
\textbf{12} & $6615_{-15}^{+16}$ & \textbf{12} & $6661_{-3}^{+2}$ \\
\textbf{13} & $6513_{-13}^{+12}$ & \textbf{13} & $6673_{-3}^{+3}$ \\
\textbf{14} & $6604_{-20}^{+20}$ & \textbf{14} & $6663_{-3}^{+4}$ \\
\textbf{15} & $6571_{-8}^{+9}$ & \textbf{15} & $6660_{-6}^{+6}$ \\
\textbf{16} & $6535_{-8}^{+8}$ & \textbf{16} & $6674_{-2}^{+2}$ \\
 &  & \textbf{17} & $6654_{-1}^{+3}$ \\
 &  & \textbf{18} & $6678_{-4}^{+16}$ \\ \hline
    \end{tabular}
\end{table}

One method by which we can determine the explosion type of these SNRs is through observing the location of the centroid of their Fe~K$\alpha$ emission lines. This method, which was proposed in \cite{yamaguchi_2014}, suggests that there is a clear divide in the Fe~K$\alpha$ centroid energy at $\sim$6550 eV, with the centroids of Ia SNRs appearing below this value, and the centroids of CC SNRs appearing above it. This distinction is attributed to the ambient medium into which the SNRs were expanding, with core-collapse SNRs typically expanding into higher density media, owing to mass loss from their progenitor star, while Type Ia SNRs were typically expanding into lower density media. 

\begin{figure*}
    \begin{center}
        \subfloat[(a) 3C 397 Fe K$\alpha$ Line Centroid Energies By Region]{\includegraphics[angle=0,width=0.45\textwidth]{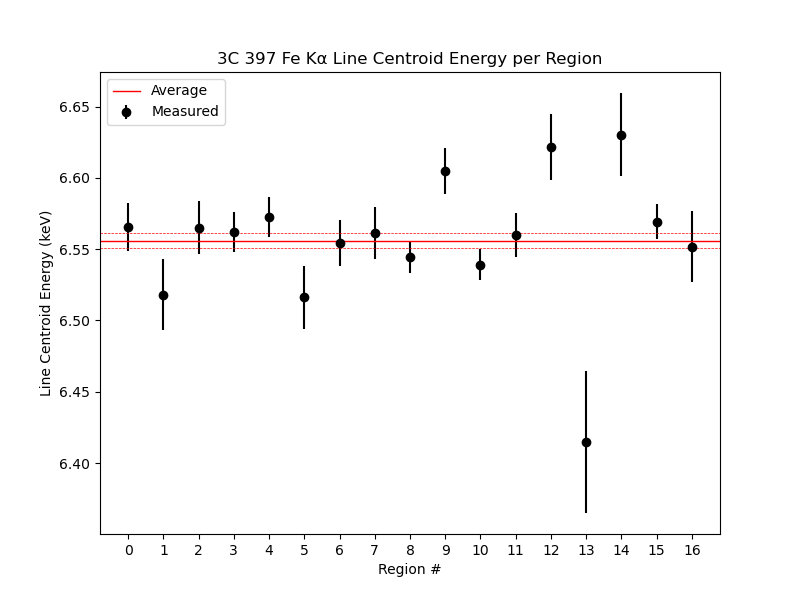}}
        \hspace{1cm}
        \subfloat[(b) W49B Fe K$\alpha$ Line Centroid Energies By Region]{\includegraphics[angle=0,width=0.45\textwidth]{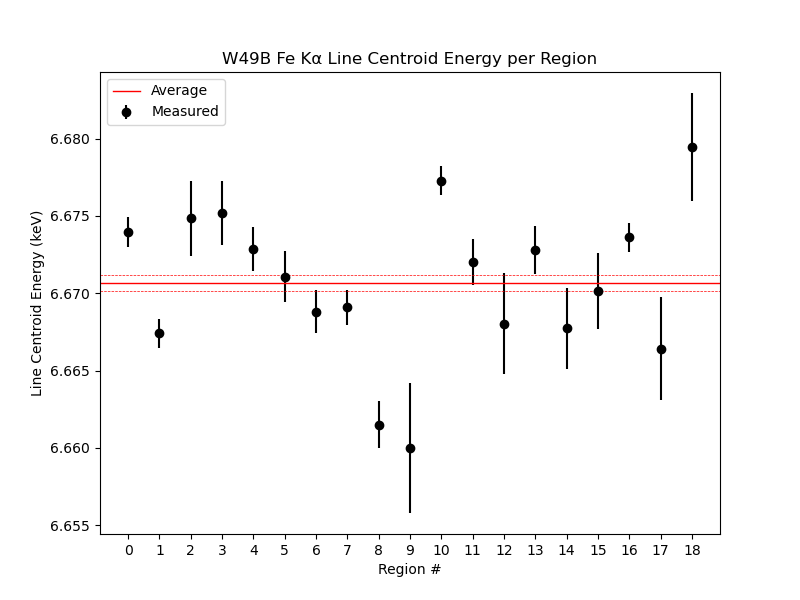}}
        \caption{Energies of the Fe K$\alpha$ line centroid for each region used in our spatially resolved study for 3C 397 (left) and W49B (right).}
        \label{fig:fe-k_plot}
    \end{center}
\end{figure*}

To determine the location of these centroids, we extracted the 5-10 keV regime from the pn observations of both SNRs, and fit them each with an absorbed power-law model to account for the continuum, plus a Gaussian to account for the Fe~K$\alpha$ emission. Additionally, we performed the same analysis for the arcsecond-scale regions used in our spatially-resolved spectroscopic analysis, using the same methods as noted above. The results are summarized in Table~\ref{tbl:fe_energy}, and are represented visually in Figures~\ref{fig:fe-k_maps} (to highlight the spatial distribution of values) and \ref{fig:fe-k_plot} (to show the range of energies). From the figures, it is clear that there is a variation in the Fe~K$\alpha$ centroid energy across each remnant.

For W49B, our global fit yielded a centroid energy of $6669_{-1}^{+1}$ eV, compared to $6663_{-1}^{+1}$~eV from \cite{yamaguchi_2014} and $6665_{-1}^{+1}$~eV from \cite{siegel_2021}. From our spatially-resolved analysis, we found a variation of roughly 30 eV across the SNR, with values ranging from $6654_{-1}^{+3}$ eV in region 17 up to $6684_{-2}^{+2}$ eV in region~3. Correspondingly, we noticed the highest centroid energies in the northern and eastern parts of the SNR. All centroid energies -- both from the global spectrum and from the spatially-resolved analysis -- lie within the regime associated with a CC designation, and are consistent with the findings of both \cite{yamaguchi_2014} and \cite{siegel_2021}.

For 3C~397, our global fit determined a centroid energy of $6567_{-4}^{+2}$~eV, compared to the values of $6556_{-3}^{+4}$~eV from \cite{yamaguchi_2014}, $6552_{-2}^{+3}$~eV from \cite{martinez-rodriguez_2020}, and $6578_{-6}^{+7}$~eV from \cite{siegel_2021}. From our spatially-resolved analysis, we note a rather large variation in the centroid energies of roughly 100 eV across the SNR, with the highest value of $6615_{-15}^{+16}$~eV found in region 12, and the lowest value of $6513_{-13}^{+12}$~eV found in region 13. In general, the highest centroid energies were found in the northern and western portions of the remnant. However, 3C~397 presents an interesting situation: not only is the variation of the centroid energies large, but so too do they exist on both sides of the dividing line proposed by \cite{yamaguchi_2014}. This is not entirely unexpected: as mentioned above, the environment into which as SNR expands is expected to play an important role in determining the energy of its Fe K$\alpha$ line centroid, and 3C~397 is believed to be expanding into and interacting with a molecular cloud along its northern and western border \citep{safi-harb_2005, kilpatrick_2016}. Both of these peculiarities mirror the findings of \cite{siegel_2021}.

Based on these results, we must echo the findings of \cite{siegel_2021} in suggesting that caution should be taken when attempting to use the Fe K$\alpha$ centroid energy as a determining property of supernova progenitor classification. While -- based on previous studies -- there certainly appears to be a broad correlation between Fe K$\alpha$ centroid energy and progenitor class, the dividing line may be more nebulous than previously expected, with W49B being a good example as to how the results of such a study seems to stand at odds with the results of a more conventional spectroscopic approach. Further, environmental conditions may play a greater role than previously thought, with spatial variation in centroid energies across an SNR being potentially significant, as evidenced by 3C~397. It would thus be prudent to use the results of such an analysis as a supportive piece of evidence, rather than as the sole indicator, and to ensure that one makes use of spatially-resolved analysis whenever possible.

\section{Discussion}
\label{sec:discussion}
As mentioned above, we find that for both objects, nucleosynthesis models of Type Ia supernovae produce better fits to the data than do models of CC supernovae. Despite this, we find that our results cannot be fully reproduced by any one model of a given class. Here, we discuss both what worked and what did not, some of the possible reasoning behind this discrepancy, as well as some possible avenues for further investigation based on our results.

\subsection{Thermonuclear Supernovae}
The greatest discrepancy between our results and the theoretical models is largely due to the Fe abundances found in the remnants, as CC models tend to predict lower abundances of Fe relative to that of Si. For W49B, several CC models -- notably, those models in the $9-11$ M$_{\odot}$ range -- are able to come close to reproducing the observed Fe/Si ratio, but for 3C~397, no such model exists. The CC model that comes closest is the $10.0$ M$_{\odot}$ model from S16, with an Fe/Si ratio of $0.91$, which falls notably short of 3C~397's observed Fe/Si ratio of $12.3$.

The IMEs examined are less consistent than Fe, but still tend towards favoring a Ia origin. Ca behaves similar to Fe, in that the Ca/Si ratio produced by Ia models is in general higher than that produced by CC models, and this is echoed in our results for both objects, with only Ia models being able to reproduce the observed Ca/Si ratios. Ar and S are less decisive, with Ia and CC models faring equally well when compared to W49B, and equally poorly when compared to 3C~397. The main exception to this trend is Mg, which is consistently underproduced by all Ia models for both 3C~397 and W49B. CC models fare somewhat better in this regard, with several models being able to closely match the abundance ratios found for both objects. The general trend, however, is that Ia models produce less Mg than observed, while CC models produce more.

Additionally, there is some degree of inconsistency between what is observed in the SNRs, and what is implied by the models. As an example, W49B is generally considered to be an asymmetric SNR (see e.g., \cite{lopez_2013}). This, one might assume, would suggest an explosion with a similar property. However, when comparing to the models of S13 -- a series of models primarily parametrized by the number of ignition kernels -- we find much the opposite: the best fit arises from models with higher numbers of ignition kernels (and thus, lower degrees of asymmetry), rather than those with a lower number of kernels.

\subsection{Core-Collapse Supernovae}
The Fe/Si ratio produced in core-collapse supernovae is set by the structure of the star and the explosion energy. The Fe production increases with explosion energy both because stronger explosion energies limit the amount of supernova fallback and the more powerful shocks by these explosions produce higher temperatures that synthesize more iron-peak elements. However, there are limitations to this increase in production. Most of the shock-heated Fe production occurs in the relatively dense Si layer. As the structure of the star becomes more diffuse, even strong supernova shocks are unable to produce conditions to burn C/O material into iron-peak elements as the blast wave propagates through the star. For most supernova progenitors ($<20-25M_\odot$), it is difficult to produce more than 0.1-0.2M$_\odot$ of Fe~\citep{2018ApJ...856...63F,2020ApJ...890...35A}.

\begin{figure*}
    \begin{center}
    \subfloat[]{\includegraphics[angle=0,width=0.40\textwidth,scale=0.5]{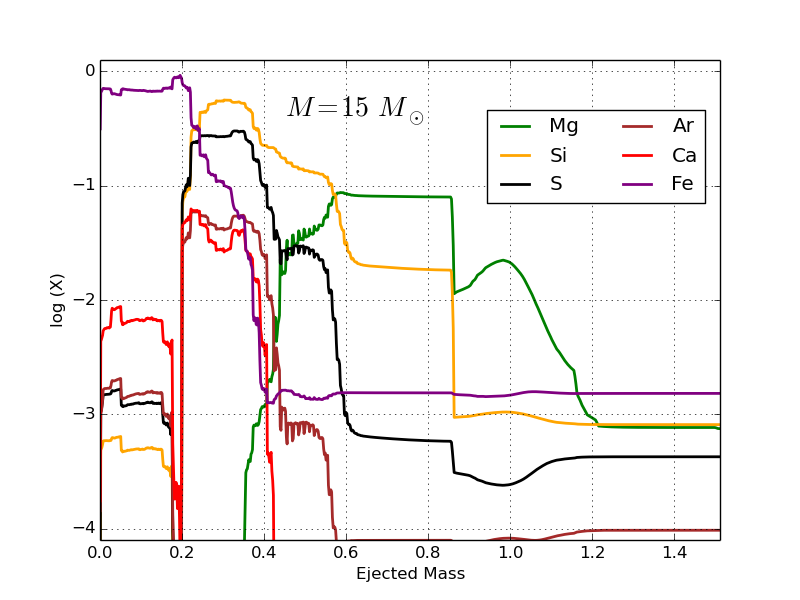}}
    \hspace{1cm}
    \subfloat[]{\includegraphics[angle=0,width=0.40\textwidth,scale=0.5]{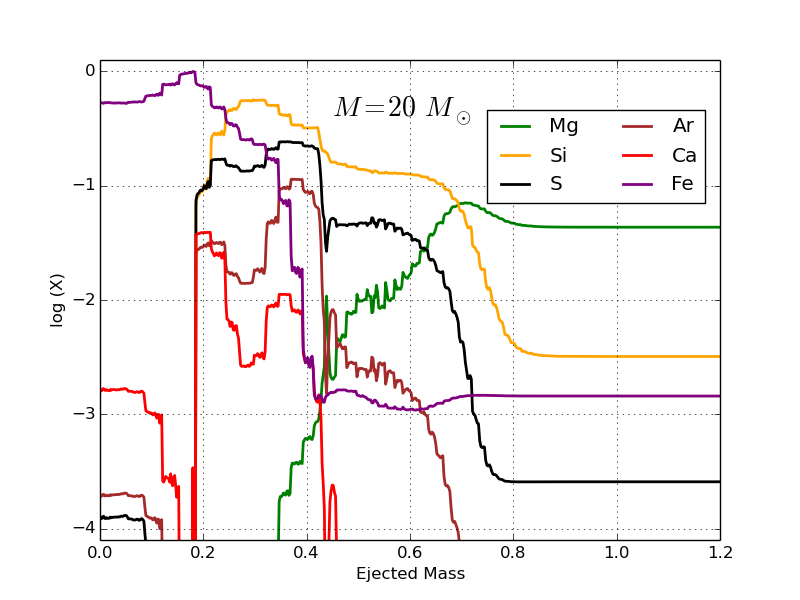}}
    \hspace{1cm}
    \subfloat[]{\includegraphics[angle=0,width=0.40\textwidth,scale=0.5]{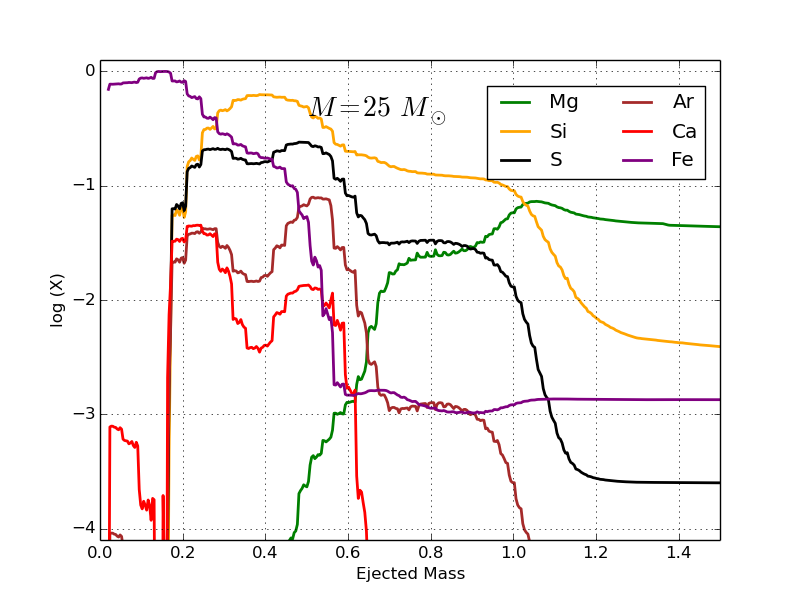}}
    \caption{Mass fraction of the labeled elements versus enclosed mass, for strong explosions of a 15, 20, and 25\,M$_\odot$ star.}
    \label{fig:simmf}
    \end{center}
\end{figure*}

To study the abundances in detail, we use the yields from ~\cite{2020ApJ...890...35A}. This paper studied a 15, 20, and 25\,$M_\odot$ with a broad range of explosion properties. Figure~\ref{fig:simmf} shows the mass fractions of the innermost ejecta from strong explosions (few times $10^{51}\,{\rm erg}$) of a 15, 20, and 25\,$M_\odot$ progenitor stars. Fe is concentrated in the innermost ejecta. In this region, there is very little Si, Ar, Ca, and S. In these strong explosions, we have nearly maximized the Fe production with the total Fe production in this inner solar mass of ejecta equal to 0.18, 0.19, 0.24\,$M_\odot$. In these models the total Fe/Si ratio can exceed 1, explaining the total Fe yields for W49B. However, increasing the explosion energy ultimately also increases the region where Si is produced (fusion in the C/O layer). 

\begin{figure*}
    \begin{center}
        \subfloat[(a) 3C 397 vs F18]{\includegraphics[angle=0,width=0.45\textwidth]{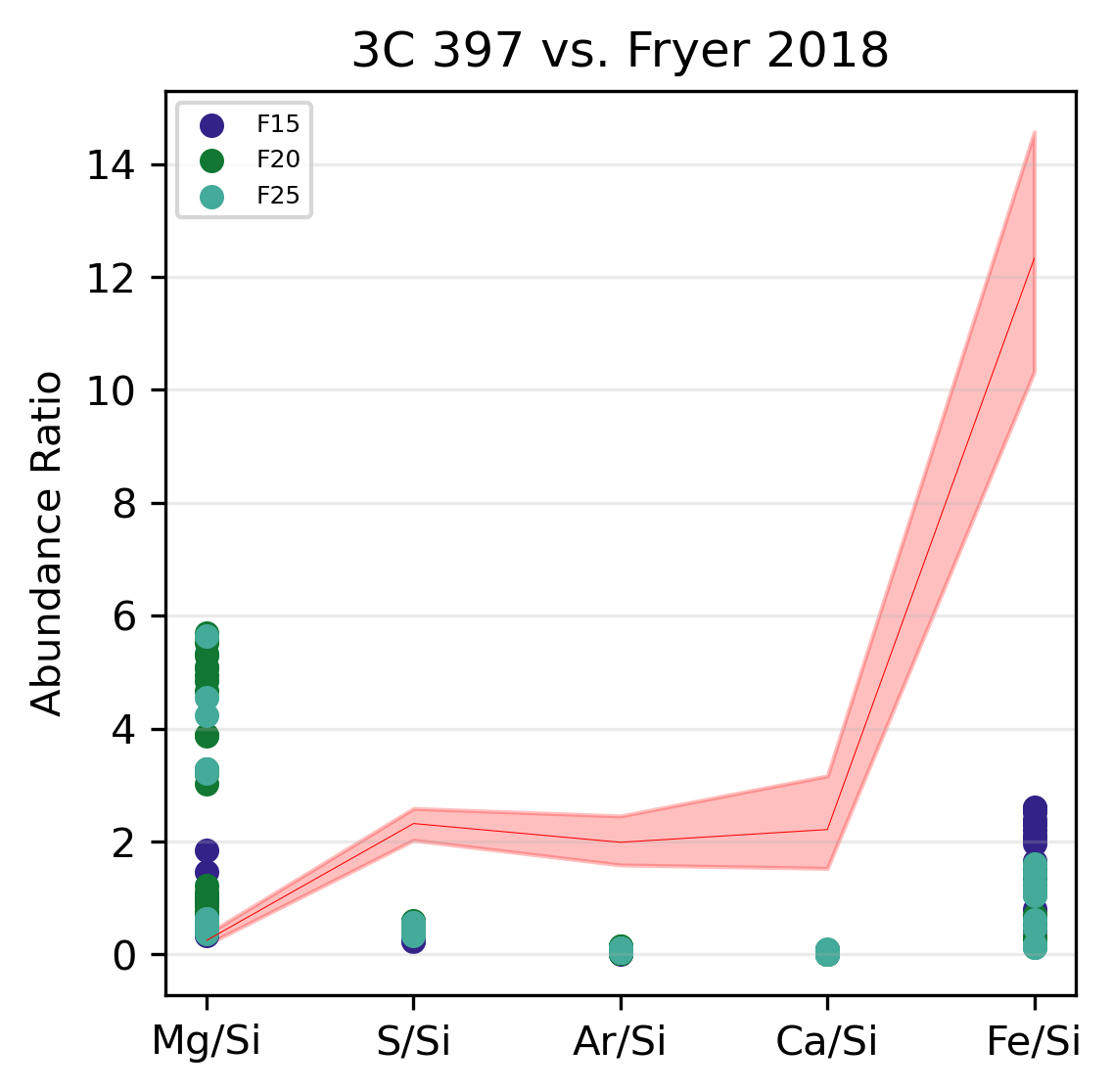}}
        \hspace{1cm}
        \subfloat[(b) W49B vs F18]{\includegraphics[angle=0,width=0.45\textwidth]{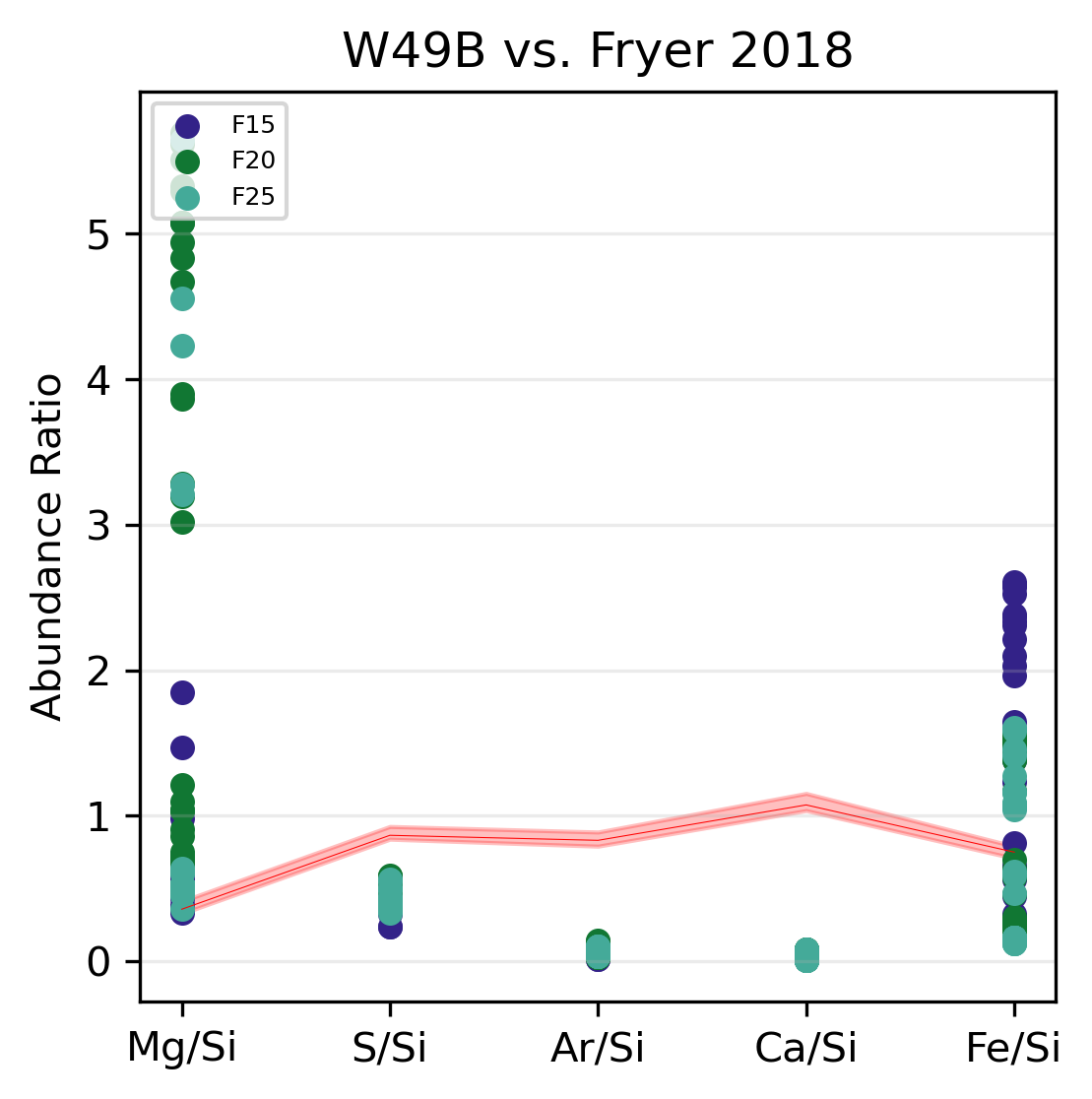}}
        \caption{Best-fit abundances for Mg, S, Ar, Ca, and Fe relative to Si relative to solar values from the solar abundances from \protect\cite{wilms_2000}, plotted over the predicted relative abundances [X/Si]/[X/Si] for the model of F18. Labeled masses are in units of M$_\odot$.}
        \label{fig:fryer_models}
    \end{center}
\end{figure*}

Increasing the explosion energy further does not significantly increase the iron production but does increase the Si production~\citep{2023MNRAS.525.6257B}. With this set of progenitors, explaining the high Fe/Si ratio of 3C~397 is not possible with any explosion energy (see Figure~\ref{fig:fryer_models}). However, it may be that current observations are only focused on the innermost iron-rich ejecta. The innermost $\sim 0.2\,M_\odot$ of the ejecta in these explosions is predominantly iron. If we are only observing the innermost $0.2-0.3\,M_\odot$ of the ejecta, the Fe/Si ratio can be above 10 for our models.

Even if we explain the Fe/Si ratio in 3C~397 by arguing that the observations are biased toward the innermost the ejecta, we can not explain the relative abundances of the intermediate elements (S, Ar, Ca) produced in the Si layer. In our models, we see from Figure~\ref{fig:simmf} that throughout the Si layer, Si production exceeds the S production and S production exceeds both Ar and Ca. Typically S production is 2--3 times lower than Si and the Ar and Ca production is nearly 10 times lower than that of Si. Ar and Ca are close, but typically Ar production is higher than Ca. These mass ratios are consistent across all explosion energies for the 15, 20, and 25M$_\odot$ models from \cite{2020ApJ...890...35A}. And because these four elements are all produced in the same region, arguing that the observations are only measuring selection portions of the ejecta can not explain the observed Ca/Si and Ar/Si ratios.

The low inferred explosion energies for these remnants may suggest an alternate core-collapse scenario for these remnants.  Many studies of core-collapse explosions focus on stars above 12$_\odot$ with energies above a few times $10^{50} {\rm \, erg}$.  But the inferred explosion energies for our two remnants, $\sim 7.3\times10^{49}, 1.2\times10^{50} {\rm erg}$ for 3C~397, W49B respectively, suggest that we are not studying the right progenitors and explosions for these remnants. Such low explosion energies are expected in low mass stars with zero-age main sequence star masses below 12\,M$_\odot$ or from very massive stars that barely explode~\citep{2001ApJ...554..548F,2018ApJ...852...40W,2019ApJ...882..170J,2022ApJ...926..147B}. We will discuss both scenarios in this section.

Based on standard Initial Mass Functions, we expect nearly half of all stellar collapses to arise from stars below 12\,M$_\odot$ and these progenitors are the most likely candidates for low energy explosions.  For these low-mass core-collapse progenitors, the binding energy is well below $10^{50}~{\rm erg}$ and the inferred supernova energy from the supernova is nearly identical to the explosion energy when the shock is launched.  When the shock is launched, it consists of the material that is processed in the explosive region itself (hereafter, we refer to this ejecta as the ``innermost'' ejecta).  In this study, we will assume this innermost ejecta ($M_{\rm ejecta}$) can range from $<0.01-0.3\,M_\odot$:  compare the results of \cite{2018ApJ...852...40W} to those of \cite{1999ApJ...516..892F} and \cite{2019ApJ...882..170J}. The work of \cite{2019ApJ...882..170J} argues that strong explosion burning during the collapse can lead to a large amount of engine-driven ejecta.  This ejecta is initially moving outward at a shock velocity ($v_{\rm shock}$):
\begin{equation}
    v_{\rm shock} = (2 E_{\rm Exp}/M_{\rm ejecta})^{1/2}.
\end{equation}
Using an ejecta mass of  $0.1\,M_\odot$, the corresponding initial shock velocities for our explosions are $8.5-10\times10^8 {\rm \, cm \, s^{-1}}$ for 3C 397 and W49B respectively.

Supernovae produce new elements either in this innermost ($<0.01-0.3\,M_\odot$) ejecta that is processed in the explosive engine or when a strong shock hits the stellar material (shock-heated nucleosynthesis).  The shock heated nucleosynthesis for these low mass stars will be minimal.  The amount of shock-heated nucleosynthesis  depends on the shock temperatures.  For the low shock velocities of these weak explosions, we expect the mass of shock-heated nucleosynthetic material to be low.  To estimate this shock-heated nucleosynthesis we must a) determine how much the stellar ejecta compresses before it is hit by the supernova shock and b) determine the temperatures achieved when the shock hits the stellar material. Since low-mass star explosions are rapid~\citep{2001ApJ...554..548F,2022ApJ...926..147B}, there is very little time for the bulk of the stellar material above the explosive region to compress significantly before it is hit by the shock from the innermost ejecta.  If we assume no compression, we can then estimate the temperature of the shock-heated material utilizing analysis of the strong shock conditions for a radiation dominated gas~\citep{2020ApJ...898..123F},(Fryer et al. in prep):
\begin{equation}
    T_{\rm Shock} = \left(\frac{f_{\rm eff} \rho v_{\rm shock}^2}{2 a}\right)^{1/4}
\end{equation}
where $f_{\rm eff}$ is an efficiency factor, $\sim 0.2$~(Fryer et al., in prep) and $a=7.5675\times10^{-15} {\rm \, erg \, cm^{-3} \, K^{-4}}$.  With this formula and an understanding of the structure of the star from \cite{2016ApJ...821...38S}, we can infer the shock heated temperature of the shock-heated ejecta.  To calculate the peak temperature of the shock heated ejecta, we study the pre-shock conditions of the material just above the "innermost" ejecta that is processed through the explosive engine (see previous paragraph).  This mass coordinate depends on the mass of the neutron star (we vary this mass from 1.2-1.4\,M$_\odot$ corresponding to gravitational masses of $\sim 1.1 - 1.3\,M_\odot$) and the mass of the innermost ejecta (we assume $0.1\,M_\odot$).  Shock-heated ejecta lies at the mass coordinate just above the neutron star mass plus this innermost ejecta mass and the material lying just above this mass coordinate corresponds to highest-temperature shock-heated material.  

Table~\ref{tbl:shockheating} shows the expected shock-heated temperatures compared to the temperature in the star prior to explosion.  If this temperature is not significantly higher than the stellar temperature, the shock will not drive nuclear burning and the yields will be set by the progenitor (including the nature of its initial metallicity).  For the $9\,M_\odot$ progenitor, these weak shocks do increase the temperature of the stellar material, but the peak temperatures are low (below $2-3\times10^8 K$).  The weak shocks in the $12\,M_\odot$ progenitor do not increase the temperature demonstrably and will not drive any additional nuclear burning.

\begin{figure*}
	\begin{center}
		\includegraphics[angle=0,width=0.45\textwidth,scale=0.5]{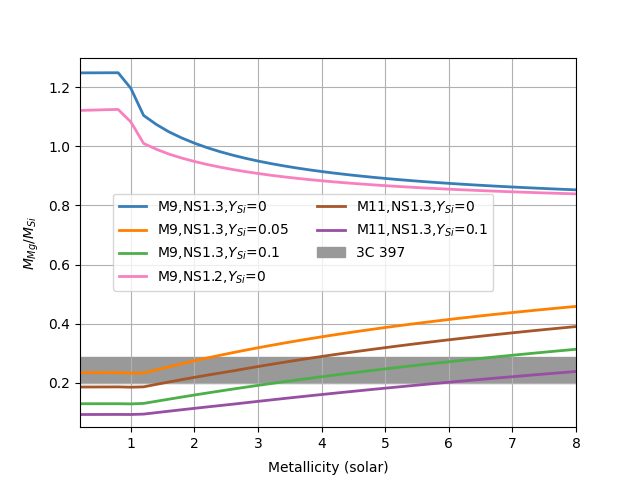}
        \includegraphics[angle=0,width=0.45\textwidth,scale=0.5]{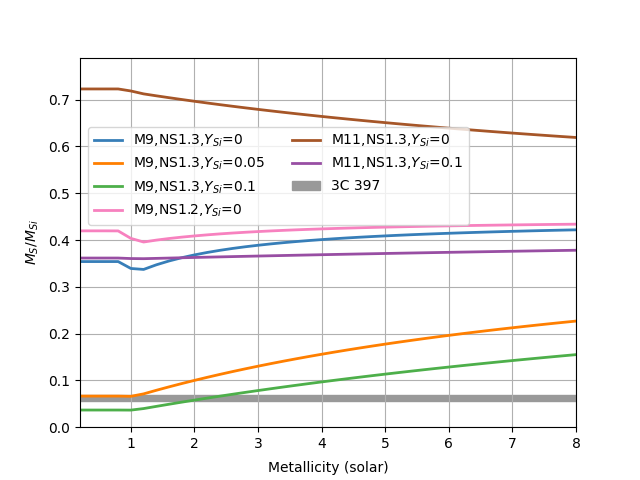}
        \includegraphics[angle=0,width=0.45\textwidth,scale=0.5]{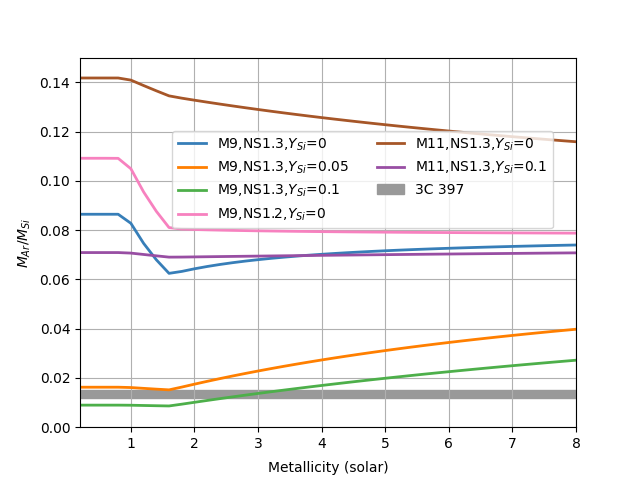}
        \includegraphics[angle=0,width=0.45\textwidth,scale=0.5]{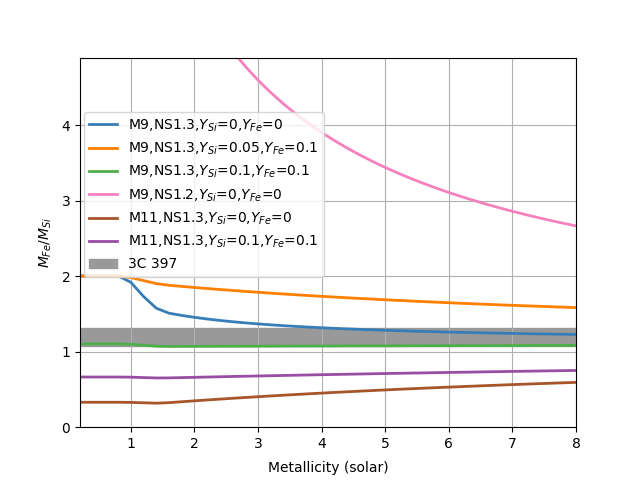}
		\caption{Yield ratios of different elements to Silicon as a function of stellar metallicity for a variety of progenitors, compact remnant masses and composition of the innermost ejecta.  These yields from low-energy explosions of low-mass progenitors are compared to 3C 397 and a rough fit is achieved at metallicities between 1-3 times solar.}
		\label{fig:lowEyield3C397}	
	\end{center}
\end{figure*}

\begin{figure*}
	\begin{center}
		\includegraphics[angle=0,width=0.45\textwidth,scale=0.5]{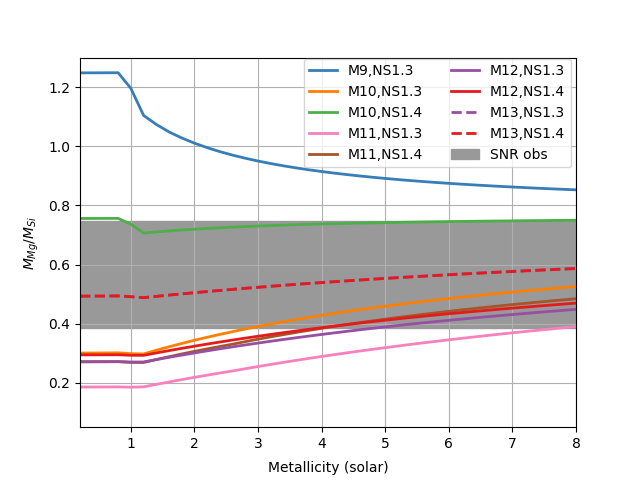}
        \includegraphics[angle=0,width=0.45\textwidth,scale=0.5]{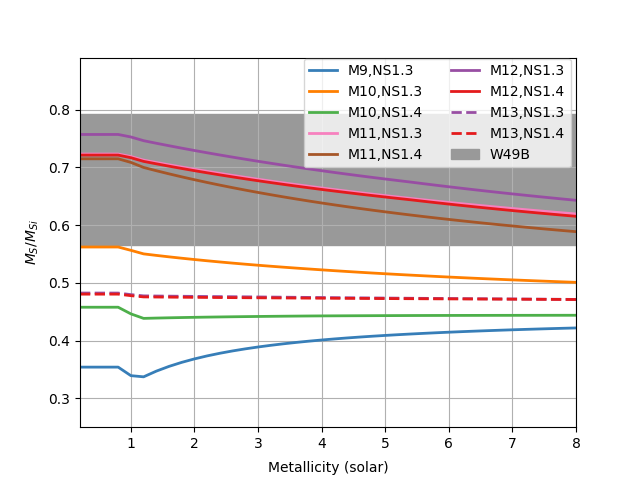}
        \includegraphics[angle=0,width=0.45\textwidth,scale=0.5]{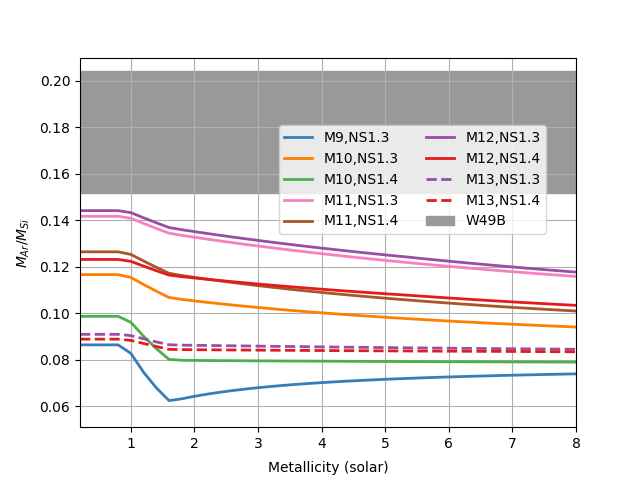}
        \includegraphics[angle=0,width=0.45\textwidth,scale=0.5]{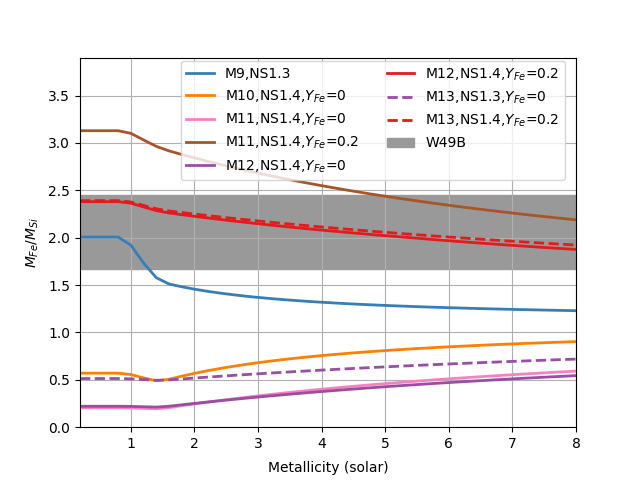}
		\caption{Yield ratios of different elements to Silicon as a function of stellar metallicity for a variety of progenitors, compact remnant masses and composition of the innermost ejecta.  These yields from low-energy explosions of low-mass progenitors are compared to W49B and, except for Ar, a rough fit is achieved at high metallicities.}
		\label{fig:lowEyieldW49B}	
	\end{center}
\end{figure*}

All of this analysis indicates that shock heating doesn't affect the yields for the weak SN explosion energies observed in the 3C 397 and W49B remnants.  This means that the yield of these stars will be limited to the yields from the innermost ejecta processed in the engine and the composition of the star at formation (which only depends on its metallicity).  To calculate the yields from these supernovae we assume the yields are a combination of the metallicity of the nascent star and engine-synthesized innermost ejecta.  Figures~\ref{fig:lowEyield3C397},~\ref{fig:lowEyieldW49B} show the expected yields varying the progenitors, the compact remnant mass, and metallicity of the stellar progenitor.  Our default model assumes the mass of the engine-synthesized innermost ejecta is small, but we have models that both increase the Si and Fe yields.  With these parameters, we can produce marginal fits to the abundance ratio data for stars that have metallicities slightly higher than solar.

However, it is important to note that the total masses of the observed elements in these explosions are much lower than those inferred from the data.  We are under-producing the observed total ejecta masses by a factor of 3--10 in our models.  These masses will be extremely difficult to produce with low-energy, low-mass progenitors.  It may be that explosive nuclear burning can alter this ejecta, as is seen in models of electron capture supernovae~\citep{2019A&A...622A..74J,2019ApJ...882..170J}.  Although the electron capture supernova models do not match the observed abundances, these models are exceptionally stochastic and it may be possible to tune them to match the observations.

\begin{table*}
\movetableright=-10mm
\caption{Shock-heating conditions from Weak Explosions in Low-Mass Stars.  The temperature and density is determined at the base of the star just above the innermost ejecta processed in the engine.  These temperatures reflect the hottest temperatures achieved through shock heating.}
\label{tbl:shockheating}
\begin{tabular}{cccccc}
Mass$_{\rm Progenitor}$ & Mass$_{\rm compact remnant}$  & Explosion Energy ($10^{50} {\rm \, erg}$) & T$_{\rm star}$ (K) & $\rho_{\rm star} ({\rm g \, cm^{-3}})$ & T$_{\rm star, post-shock}$ (K)\\ 
\hline
   9.0 & 1.2 & 1.2 & 210 & $2.0\times10^8$ & $2.4\times10^8$ \\
   9.0 & 1.3 & 1.2 & 0.27 & $1.5\times10^7$ & $4.5\times10^7$ \\
   9.0 & 1.4 & 1.2 & $10^{-5}$ & $5.7\times10^6$ & $3.5\times10^7$ \\
   9.0 & 1.2 & 0.75 & 210 & $2.0\times10^8$ & $2.1\times10^8$ \\
   9.0 & 1.3 & 0.75 & 0.27 & $1.5\times10^7$ & $4.0\times10^7$ \\
   9.0 & 1.4 & 0.75 & $10^{-5}$ & $5.7\times10^6$ &  $3.0\times10^7$ \\
   12.0 & 1.2 & 1.2 & $1.1\times10^7$ & $3.6\times10^9$ & $3.6\times10^9$ \\
   12.0 & 1.3 & 1.2 & $1.1\times10^6$ & $2.4\times10^9$ & $2.0\times10^9$ \\
   12.0 & 1.4 & 1.2 & $3.2\times10^5$ & $1.6\times10^9$ & $1.5\times10^9$ \\
   12.0 & 1.2 & 0.75 & $1.1\times10^7$ & $3.6\times10^9$ & $3.2\times10^9$ \\
   12.0 & 1.3 & 0.75 & $1.1\times10^6$ & $2.4\times10^9$ & $1.8\times10^9$ \\
   12.0 & 1.4 & 0.75 & $3.2\times10^5$ & $1.6\times10^9$ & $3.6\times10^9$ \\
   \hline
\end{tabular}
\end{table*}

For weak explosions in high mass stars, most of the innermost ejecta falls back onto the compact remnant. None of the shock-heated material that could have undergone further nuclear burning is ejected from the star. It is possible to develop an extremely asymmetric model that will eject some innermost engine-synthesized material.  But it is very unlikely that these explosions can match the high iron yields seen in the data.  Given these difficulties and the relative rarity of these events, we believe the low-mass progenitor the more likely core-collapse candidate, requiring that the inferred total-mass yields from these remnants is 3-10 times too high.

\subsection{Caveats}
\label{sec:caveats}
Above, we have shown that while there is agreement in some aspects between our fitted models and the suite of nucleosynthesis models used in this study, no available model from the literature was able to fully reproduce our findings. Here, we elaborate on some caveats in our analysis that may account for some of the discrepancies observed.

First, the ejecta components from which we have inferred our metal abundances are quite complex, and the degree of this complexity likely varies from region to region. While we have attempted to account for this complexity through the use of two-component models for our spectral fits, it is very possible that these models are still too simplistic in their assumptions, and that more components could be required to accurately represent the data. Second, while our analysis has been performed on the shock-heated ejecta present within these SNRs, it is possible that not all of the ejecta has been shock-heated, or is otherwise not visible in the X-ray band. Third, while this study was spatially-resolved in nature, our comparisons were produced by the averaging out of ejecta yields across the SNRs. It is possible that some regions within these objects would be a better representation of the expected nucleosynthetic yields, given the likelihood of asymmetry in the explosions. Fourth, the spectral resolutions of the \textit{XMM-Newton} CCDs are not so fine as to be able to resolve every possible emission line in the observed energy range, which may introduce some degree of degeneracy in the parameters, leading to a less accurate measurement of the plasma properties. And fifth, there exist systematic errors in the telescope calibration and plasma models (in this case, ATOMDB) used in this study. 

\section{Conclusions}
\label{sec:con}
We have presented a systematic X-ray study of the two unusual SNRs 3C~397 and W49B using the widest selection of nucleosynthesis models available to date, with the goal to address the ongoing question of their SN progenitors. This was done through use of spatially resolved spectroscopy, by first fitting the ejecta component within the various regions of these remnants to determine their metal abundances, then comparing these abundances to models of nucleosynthesis yields from 12 models of supernova explosions, which are labeled and referenced as WW95, M03, N06, M10, S13, F14, S16, F18, LN18, T18, B19, and LN20. We find that the observed abundances do not fully match the yields for any of the tested explosion models.

While we found that our observations were unable to directly reproduce the yields of any of the tested explosion models, in both cases, a greater degree of agreement was found when comparing to models of Type Ia supernovae rather than to those of CC supernovae. For 3C~397, this was largely driven by the particularly high Fe/Si ratio, which only a select few Ia models were able to account for, suggesting a high-density, near Ch-mass WD progenitor, with the possibility of an ONeMg WD. For W49B, this agreement was more general, owing to the overall pattern of abundances observed in the SNR, and suggesting a low-metallicity, subCh-mass WD progenitor.

We also considered the energy of the centroids of the Fe~K$\alpha$ lines of both SNRs. For W49B, the results of this would place it firmly within the realm of CC SNRs, standing at odds with the results of our spectroscopic study. For 3C 397, the results are less certain, lying almost entirely on the dividing line between the two classifications. However, for both SNRs, we note a significant degree of spatial variation in the energies of this line; this is of particular note for 3C 397, in which this variation results in centroid energies on both sides of the dividing line. We thus urge that caution be used in the employ of this method to ensure that such spatial variations are accounted for.

The results of this study highlight the need for improvements in both the observational data and in the models of supernova nucleosynthesis. Observationally, given the caveats listed above, a future study will benefit from the high-resolution X-ray spectroscopy available to the recently-launched \textit{XRISM} mission and the future \textit{NewAthena} \citep{2025hsa..conf..148C, 2025NatAs...9...36C}, as this would would result in a reduction of degeneracies in the parameter space, allowing for a more accurate measurement of the chemical properties of these objects. Sensitive, high-resolution and sensitive imaging instruments -- such as those aboard the proposed \textit{AXIS} Probe mission \citep{2023SPIE12678E..1ER, 2023arXiv231107673S} -- would also strongly benefit future studies, allowing for the better isolation and localization of ejecta.

From a theoretical perspective, these results underscore the need for both the continued refinement of existing models and the development of new ones that probe underexplored regions of parameter space. While properties such as progenitor mass, ambient density, and metallicity have been extensively modelled, other key parameters—such as explosion energy (particularly values below the canonical 10$^{51}$~ergs value) and explosion asymmetry -- have received comparatively less attention, yet may be equally critical in determining the final outcome of a supernova explosion.

\section*{Acknowledgments}
This research is primarily supported by the Natural Sciences and Engineering Research Council of Canada (NSERC) through the Discovery Grants and Canada Research Chairs program.  The work of CLF  was supported by the US Department of Energy through the Los Alamos National Laboratory. Los Alamos National Laboratory is operated by Triad National Security, LLC, for the National Nuclear Security Administration of U.S.\ Department of Energy (Contract No.\ 89233218CNA000001).  We made use of NASA's Astrophysics Data System (ADS) and of data and software provided by the High Energy Astrophysics Science Archive Research Center (HEASARC), which is a service of the Astrophysics Science Division at NASA/GSFC.  This work also made use of data obtained with XMM--Newton, a European Space Agency science mission with instruments and contributions directly funded by ESA Member States and NASA. Literature studies of both remnants made use of SNRcat, the online catalogue of supernova remnants\footnote{\url{http://snrcat.physics.umanitoba.ca}} \citep{2012AdSpR..49.1313F} hosted at the University of Manitoba. We thank Vikram Dwarkadas for discussions relevant to this work. \added{We thank the referee for their careful reading of the manuscript, and for their insightful comments and suggestions.}

\facilities{XMM-Newton}

\software{HEAsoft \citep{heasoft}, XMM-SAS \citep{xmm-sas}, CIAO \citep{2006SPIE.6270E..1VF}, SAOImage~DS9 \citep{2003ASPC..295..489J}, XSPEC v12.11.1 \citep{1996ASPC..101...17A}}

\bibliographystyle{aasjournalv7}
\bibliography{bibliography}

\begin{table*}
    \movetabledown=30mm
    \begin{rotatetable*}
        \caption{\textbf{3C397}: Spectral properties for regions fit with a \texttt{VNEI}+\texttt{VAPEC} model with  S, Ar, Ca, and Fe (Ni tethered to Fe) allowed to vary for the hard (H) \texttt{VNEI} component, and where the Mg and Si were allowed to vary for the soft (S) \texttt{VAPEC} component. Missing abundances, as well as all other abundances not mentioned, were frozen at solar values. Abundances are in solar units from the abundances table of \protect\cite{wilms_2000}. Data was grouped with a minimum of 20 counts per bins.}
        \label{tbl:3C397_data}
        \begin{center}
            \renewcommand{\arraystretch}{1.5}
            \begin{tabular}{cccccccccccc}
                  Region &
  N$_{H}$ &
  kT$_{c}$ &
  Mg &
  Si &
  kT$_{h}$ &
  S &
  Ar &
  Ca &
  Fe &
  $\tau_{h}$ &
  $\chi_{\nu}^{2}$ (DoF) \\
  &
  ($\times 10^{22}$ cm$^{-2}$)&
  (keV)&
  &
  &
  (keV)&
  &
  &
  &
  &
  ($\times 10^{10}$ s cm$^{-3}$)&
  \\ \hline
0 &
  $4.18_{-0.02}^{+0.06}$ &
  $0.22_{-0.01}^{+0.01}$ &
  $0.37_{-0.06}^{+0.02}$ &
  $1.24_{-0.09}^{+0.07}$ &
  $2.25_{-0.14}^{+0.07}$ &
  $2.87_{-0.22}^{+0.26}$ &
  $2.95_{-0.70}^{+0.30}$ &
  - &
  $12.3_{-1.0}^{+1.8}$ &
  $5.80_{-0.01}^{+0.01}$ &
  $1.17$ ($2199$) \\
1 &
  $4.32_{-0.04}^{+0.03}$ &
  $0.20_{-0.01}^{+0.01}$ &
  $0.39_{-0.08}^{+0.04}$ &
  $1.53_{-0.09}^{+0.08}$ &
  $1.96_{-0.06}^{+0.07}$ &
  $3.09_{-0.21}^{+0.16}$ &
  $2.53_{-0.34}^{+0.27}$ &
  $2.72_{-1.32}^{+0.53}$ &
  $14.2_{-1.7}^{+1.9}$ &
  $6.18_{-0.01}^{+0.01}$ &
  $1.24$ ($1976$) \\
2 &
  $4.33_{-0.06}^{+0.04}$ &
  $0.22_{-0.01}^{+0.01}$ &
  $0.40_{-0.03}^{+0.07}$ &
  $1.35_{-0.10}^{+0.06}$ &
  $2.12_{-0.04}^{+0.08}$ &
  $2.90_{-0.14}^{+0.25}$ &
  $2.27_{-0.54}^{+0.51}$ &
  $2.61_{-0.57}^{+1.02}$ &
  $11.8_{-1.6}^{+1.3}$ &
  $6.70_{-0.01}^{+0.01}$ &
  $1.15$ ($1974$) \\
3 &
  $4.39_{-0.02}^{+0.02}$ &
  $0.19_{-0.01}^{+0.01}$ &
  $0.36_{-0.05}^{+0.02}$ &
  $1.42_{-0.05}^{+0.07}$ &
  $2.10_{-0.06}^{+0.04}$ &
  $2.93_{-0.19}^{+0.25}$ &
  $2.27_{-0.33}^{+0.42}$ &
  $3.60_{-0.82}^{+0.88}$ &
  $12.6_{-0.5}^{+1.0}$ &
  $7.54_{-0.01}^{+0.01}$ &
  $1.25$ ($2132$) \\
4 &
  $4.53_{-0.07}^{+0.04}$ &
  $0.19_{-0.01}^{+0.01}$ &
  $0.34_{-0.06}^{+0.03}$ &
  $1.32_{-0.14}^{+0.10}$ &
  $2.33_{-0.12}^{+0.06}$ &
  $2.93_{-0.08}^{+0.14}$ &
  $2.17_{-0.30}^{+0.31}$ &
  $2.40_{-0.67}^{+0.97}$ &
  $9.95_{-0.85}^{+1.94}$ &
  $7.53_{-0.01}^{+0.01}$ &
  $1.18$ ($1822$) \\
5 &
  $3.99_{-0.02}^{+0.03}$ &
  $0.22_{-0.01}^{+0.01}$ &
  $0.25_{-0.06}^{+0.04}$ &
  $1.07_{-0.05}^{+0.14}$ &
  $2.33_{-0.11}^{+0.03}$ &
  $3.29_{-0.29}^{+0.25}$ &
  $3.15_{-0.60}^{+0.72}$ &
  $3.38_{-1.98}^{+1.83}$ &
  $14.7_{-1.7}^{+3.1}$ &
  $6.87_{-0.01}^{+0.01}$ &
  $1.18$ ($1735$) \\
6 &
  $3.96_{-0.09}^{+0.07}$ &
  $0.29_{-0.01}^{+0.02}$ &
  $0.48_{-0.08}^{+0.18}$ &
  $0.99_{-0.02}^{+0.10}$ &
  $2.28_{-0.08}^{+0.01}$ &
  $2.69_{-0.48}^{+0.10}$ &
  $3.16_{-0.50}^{+0.43}$ &
  $4.79_{-0.91}^{+1.47}$ &
  $17.3_{-1.7}^{+1.3}$ &
  $6.23_{-0.60}^{+0.60}$ &
  $1.28$ ($2087$) \\
7 &
  $4.45_{-0.03}^{+0.04}$ &
  $0.20_{-0.01}^{+0.01}$ &
  $0.38_{-0.03}^{+0.04}$ &
  $1.28_{-0.11}^{+0.08}$ &
  $1.95_{-0.05}^{+0.03}$ &
  $2.30_{-0.17}^{+0.11}$ &
  $2.02_{-0.36}^{+0.49}$ &
  $1.79_{-0.31}^{+0.41}$ &
  $9.63_{-0.55}^{+0.79}$ &
  $11.2_{-0.01}^{+0.01}$ &
  $1.16$ ($2237$) \\
8 &
  $4.26_{-0.03}^{+0.04}$ &
  $0.20_{-0.01}^{+0.01}$ &
  $0.17_{-0.04}^{+0.07}$ &
  $1.44_{-0.12}^{+0.10}$ &
  $2.37_{-0.03}^{+0.11}$ &
  $3.52_{-0.23}^{+0.21}$ &
  $2.66_{-0.51}^{+0.64}$ &
  $3.44_{-0.54}^{+1.08}$ &
  $22.5_{-3.7}^{+2.2}$ &
  $4.48_{-0.01}^{+0.01}$ &
  $1.14$ ($2156$) \\
9 &
  $4.22_{-0.32}^{+0.27}$ &
  $0.26_{-0.09}^{+0.10}$ &
  $0.53_{-0.32}^{+0.46}$ &
  $1.29_{-0.16}^{+0.02}$ &
  $2.17_{-0.13}^{+0.01}$ &
  $2.94_{-0.67}^{+0.32}$ &
  $2.71_{-0.09}^{+1.00}$ &
  $3.53_{-0.48}^{+2.53}$ &
  $12.4_{-2.3}^{+4.5}$ &
  $6.83_{-2.18}^{+2.18}$ &
  $1.24$ ($2044$) \\
10 &
  $3.94_{-0.03}^{+0.02}$ &
  $0.20_{-0.01}^{+0.01}$ &
  $0.14_{-0.05}^{+0.05}$ &
  $1.34_{-0.08}^{+0.13}$ &
  $2.63_{-0.12}^{+0.16}$ &
  $4.68_{-0.33}^{+0.46}$ &
  $3.46_{-0.81}^{+0.87}$ &
  $6.26_{-0.88}^{+0.98}$ &
  $35.0_{-3.5}^{+3.9}$ &
  $4.21_{-0.01}^{+0.01}$ &
  $1.14$ ($1943$) \\
11 &
  $4.31_{-0.04}^{+0.03}$ &
  $0.20_{-0.01}^{+0.01}$ &
  $0.19_{-0.02}^{+0.05}$ &
  $1.19_{-0.04}^{+0.08}$ &
  $2.26_{-0.04}^{+0.09}$ &
  $2.97_{-0.21}^{+0.17}$ &
  $2.66_{-0.46}^{+0.21}$ &
  $3.53_{-0.76}^{+0.78}$ &
  $16.5_{-1.6}^{+1.1}$ &
  $5.34_{-0.01}^{+0.01}$ &
  $1.19$ ($2383$) \\
12 &
  $3.83_{-0.04}^{+0.03}$ &
  $0.24_{-0.01}^{+0.01}$ &
  $0.56_{-0.01}^{+0.06}$ &
  $1.37_{-0.06}^{+0.16}$ &
  $1.85_{-0.06}^{+0.08}$ &
  $2.25_{-0.14}^{+0.23}$ &
  $2.31_{-0.27}^{+0.43}$ &
  $1.58_{-0.58}^{+0.37}$ &
  $4.71_{-0.69}^{+0.63}$ &
  $15.5_{-0.01}^{+0.01}$ &
  $1.31$ ($1938$) \\
13 &
  $3.84_{-0.04}^{+0.05}$ &
  $0.20_{-0.01}^{+0.01}$ &
  $0.12_{-0.06}^{+0.06}$ &
  $1.60_{-0.13}^{+0.14}$ &
  $2.58_{-0.11}^{+0.10}$ &
  $4.31_{-0.65}^{+0.28}$ &
  $2.06_{-0.85}^{+0.94}$ &
  - &
  $39.2_{-7.9}^{+5.0}$ &
  $3.81_{-0.01}^{+0.01}$ &
  $1.17$ ($1561$) \\
14 &
  $4.26_{-0.08}^{+0.04}$ &
  $0.24_{-0.01}^{+0.01}$ &
  $0.38_{-0.08}^{+0.09}$ &
  $1.43_{-0.13}^{+0.07}$ &
  $1.74_{-0.02}^{+0.08}$ &
  $2.07_{-0.24}^{+0.25}$ &
  $1.91_{-0.37}^{+0.28}$ &
  $1.78_{-0.84}^{+1.36}$ &
  $7.11_{-0.94}^{+0.70}$ &
  $8.03_{-0.01}^{+0.01}$ &
  $1.21$ ($1697$) \\
15 &
  $4.36_{-0.03}^{+0.03}$ &
  $0.22_{-0.01}^{+0.01}$ &
  $0.29_{-0.04}^{+0.06}$ &
  $1.24_{-0.08}^{+0.07}$ &
  $1.92_{-0.06}^{+0.07}$ &
  $2.59_{-0.09}^{+0.07}$ &
  $2.14_{-0.30}^{+0.08}$ &
  $2.17_{-0.40}^{+0.33}$ &
  $9.10_{-0.80}^{+0.59}$ &
  $8.82_{-0.01}^{+0.01}$ &
  $1.28$ ($2735$) \\
16 &
  $4.03_{-0.02}^{+0.02}$ &
  $0.19_{-0.01}^{+0.01}$ &
  $0.06_{-0.05}^{+0.05}$ &
  $1.32_{-0.08}^{+0.11}$ &
  $2.28_{-0.08}^{+0.02}$ &
  $3.24_{-0.18}^{+0.26}$ &
  $3.00_{-0.28}^{+0.70}$ &
  $2.33_{-0.29}^{+1.50}$ &
  $27.7_{-1.0}^{+3.9}$ &
  $4.49_{-0.01}^{+0.01}$ &
  $1.19$ ($2340$) \\ \hline
            \end{tabular}
        \end{center}
    \end{rotatetable*}
\end{table*}

\begin{table*}
    \movetabledown=60mm
    \begin{rotatetable*}
        \caption{\textbf{W49B} Spectral properties for regions fit with a \texttt{VRNEI}+ \texttt{VAPEC} model with Si, S, Ar, Ca, Fe, and Ni allowed to vary for the hard (H) \texttt{VRNEI} component, and where the Mg was allowed to vary for the soft (S) \texttt{VAPEC} component. Missing abundances, as well as all other abundances not mentioned, were frozen at solar values. Abundances are in solar units from the abundances table of \protect\cite{wilms_2000}. Data was grouped with a minimum of 20 counts per bins.}
        \label{tbl:w49b_data}
        \begin{center}
            \renewcommand{\arraystretch}{1.5}
            \begin{tabular}{cccccccccccccc}
                  Region &
  N$_{H}$ &
  kT$_{c}$ &
  kT$_{h}$ &
  Mg &
  Si &
  S &
  Ar &
  Ca &
  Fe &
  Ni &
  $\tau_{h}$ &
  Redshift &
  $\chi_{\nu}^{2}$ (DoF) \\
&
  ($\times 10^{22}$ cm$^{-2}$) &
  (keV) &
  (keV) &
  &
  &
  &
  &
  &
  &
  &
  ($\times 10^{11}$ s cm$^{-3}$) &
  ($\times10^{-3}$) &
  \\ \hline
0 &
  $6.95_{-0.16}^{+0.05}$ &
  $0.18_{-0.01}^{+0.01}$ &
  $1.63_{-0.01}^{+0.06}$ &
  $7.66_{-2.00}^{+4.22}$ &
  $9.23_{-0.60}^{+1.27}$ &
  $9.03_{-0.58}^{+1.08}$ &
  $6.74_{-0.57}^{+0.58}$ &
  $9.74_{-0.66}^{+1.09}$ &
  $10.4_{-0.8}^{+1.1}$ &
  $32.5_{-4.8}^{+9.32}$ &
  $9.67_{-0.60}^{+1.89}$ &
  $-2.72_{-0.04}^{+0.29}$ &
  $1.17$ ($4277$) \\
1 &
  $9.68_{-0.11}^{+0.30}$ &
  $0.18_{-0.01}^{+0.01}$ &
  $1.48_{-0.08}^{+0.01}$ &
  - &
  $9.06_{-0.26}^{+1.74}$ &
  $6.61_{-0.10}^{+1.18}$ &
  $5.58_{-0.21}^{+0.95}$ &
  $9.11_{-0.32}^{+1.79}$ &
  $12.6_{-0.4}^{+1.8}$ &
  $40.9_{-5.2}^{+17.0}$ &
  $7.46_{-1.19}^{+0.25}$ &
  $-2.81_{-0.01}^{+0.03}$ &
  $1.25$ ($4001$) \\
2 &
  $9.33_{-0.04}^{+0.01}$ &
  $0.20_{-0.01}^{+0.01}$ &
  $1.56_{-0.01}^{+0.40}$ &
  - &
  $12.9_{-0.1}^{+3.0}$ &
  $10.3_{-0.1}^{+0.6}$ &
  $8.78_{-0.36}^{+0.01}$ &
  $12.0_{-0.3}^{+0.1}$ &
  $16.1_{-4.3}^{+0.1}$ &
  $66.5_{-64.4}^{+0.4}$ &
  $9.75_{-0.01}^{+9.33}$ &
  $-2.78_{-0.01}^{+0.20}$ &
  $1.24$ ($3856$) \\
3 &
  $10.15_{-2.19}^{+0.06}$ &
  $0.18_{-0.01}^{+0.01}$ &
  $1.65_{-0.01}^{+0.51}$ &
  $9.12_{-0.03}^{+5.93}$ &
  $12.7_{-5.4}^{+0.1}$ &
  $8.90_{-0.01}^{+3.84}$ &
  $7.62_{-6.30}^{+0.00}$ &
  $11.1_{-0.1}^{+9.1}$ &
  $12.9_{-6.9}^{+0.1}$ &
  $31.4_{-0.2}^{+10.3}$ &
  $9.85_{-1.42}^{+0.18}$ &
  $-2.92_{-0.05}^{+0.08}$ &
  $1.16$ ($4243$) \\
4 &
  $10.33_{0.00}^{+0.01}$ &
  $0.16_{-0.01}^{+0.00}$ &
  $1.07_{-0.01}^{+0.01}$ &
  $7.10_{-0.01}^{+0.01}$ &
  $9.87_{-0.01}^{+0.01}$ &
  $6.83_{-0.02}^{+0.01}$ &
  $6.63_{-0.02}^{+0.02}$ &
  $9.71_{0.00}^{+0.00}$ &
  $4.55_{0.00}^{+0.00}$ &
  $11.4_{-0.1}^{+0.1}$ &
  $5.79_{-0.01}^{+0.01}$ &
  $-2.47_{-0.01}^{+0.02}$ &
  $1.19$ ($4260$) \\
5 &
  $9.48_{-0.13}^{+0.17}$ &
  $0.17_{-0.01}^{+0.01}$ &
  $0.90_{-0.03}^{+0.04}$ &
  $4.56_{-1.39}^{+3.05}$ &
  $14.1_{-0.3}^{+2.0}$ &
  $11.8_{-0.5}^{+1.3}$ &
  $13.2_{-1.0}^{+1.4}$ &
  $15.9_{-0.6}^{+2.0}$ &
  $8.69_{-0.76}^{+1.38}$ &
  $25.8_{-5.1}^{+9.1}$ &
  $4.37_{-0.11}^{+0.21}$ &
  $-2.13_{-0.15}^{+0.08}$ &
  $1.11$ ($4500$) \\
6 &
  $10.14_{-0.22}^{+0.07}$ &
  $0.17_{-0.01}^{+0.01}$ &
  $0.94_{-0.04}^{+0.02}$ &
  - &
  $10.7_{-0.5}^{+1.1}$ &
  $8.69_{-0.30}^{+0.82}$ &
  $8.18_{-0.07}^{+1.47}$ &
  $11.0_{-0.1}^{+1.9}$ &
  $7.22_{-0.61}^{+1.13}$ &
  $23.1_{-6.9}^{+6.8}$ &
  $4.75_{-0.25}^{+0.14}$ &
  $-1.25_{-0.19}^{+0.32}$ &
  $1.15$ ($4034$) \\
7 &
  $9.81_{-0.07}^{+0.15}$ &
  $0.18_{-0.01}^{+0.01}$ &
  $1.58_{-0.04}^{+0.01}$ &
  $4.22_{-1.50}^{+2.74}$ &
  $11.0_{-0.2}^{+1.4}$ &
  $7.59_{-0.11}^{+0.81}$ &
  $5.59_{-0.16}^{+1.09}$ &
  $9.30_{-0.20}^{+1.00}$ &
  $10.3_{-0.2}^{+0.9}$ &
  $26.4_{-2.1}^{+6.9}$ &
  $11.2_{-0.5}^{+0.1}$ &
  $-1.93_{-0.25}^{+0.24}$ &
  $1.21$ ($4303$) \\
8 &
  $9.44_{-0.17}^{+0.13}$ &
  $0.19_{-0.01}^{+0.01}$ &
  $1.15_{-0.04}^{+0.04}$ &
  - &
  $10.6_{-0.9}^{+1.3}$ &
  $8.54_{-0.54}^{+0.91}$ &
  $8.31_{-0.84}^{+0.96}$ &
  $12.0_{-1.1}^{+1.3}$ &
  $9.40_{-0.91}^{+1.81}$ &
  $39.4_{-11.1}^{+10.7}$ &
  $5.99_{-0.31}^{+0.35}$ &
  $-1.10_{-0.82}^{+0.07}$ &
  $1.25$ ($3629$) \\
9 &
  $9.68_{-0.33}^{+0.01}$ &
  $0.17_{-0.01}^{+0.01}$ &
  $0.81_{-0.04}^{+0.06}$ &
  $2.96_{-2.90}^{+7.19}$ &
  $12.6_{-2.9}^{+7.1}$ &
  $10.2_{-0.4}^{+7.4}$ &
  $10.4_{-3.4}^{+6.5}$ &
  $13.4_{-3.9}^{+8.3}$ &
  $6.09_{-2.93}^{+3.82}$ &
  $17.7_{-0.2}^{+19.6}$ &
  $4.56_{-0.58}^{+0.01}$ &
  $-1.06_{-0.09}^{+0.07}$ &
  $1.14$ ($4018$) \\
10 &
  $10.33_{-0.26}^{+0.08}$ &
  $0.17_{-0.01}^{+0.01}$ &
  $1.59_{-0.02}^{+0.06}$ &
  - &
  $10.2_{-0.6}^{+1.3}$ &
  $7.59_{-0.39}^{+0.73}$ &
  $5.60_{-0.39}^{+0.86}$ &
  $8.38_{-0.67}^{+1.09}$ &
  $10.2_{-0.6}^{+0.9}$ &
  $17.2_{-3.3}^{+5.3}$ &
  $10.4_{-0.1}^{+1.0}$ &
  $-2.78_{-0.03}^{+0.03}$ &
  $1.16$ ($4542$) \\
11 &
  $9.21_{-0.14}^{+0.14}$ &
  $0.19_{-0.01}^{+0.01}$ &
  $1.16_{-0.04}^{+0.03}$ &
  - &
  $12.2_{-1.1}^{+1.3}$ &
  $9.66_{-0.92}^{+1.30}$ &
  $9.58_{-1.01}^{+1.38}$ &
  $12.9_{-1.3}^{+1.5}$ &
  $10.5_{-0.8}^{+1.5}$ &
  $37.3_{-8.1}^{+12.7}$ &
  $6.34_{-0.48}^{+0.30}$ &
  $-2.72_{-0.08}^{+0.32}$ &
  $1.20$ ($4105$) \\
12 &
  $9.24_{-0.11}^{+0.27}$ &
  $0.17_{-0.01}^{+0.01}$ &
  $0.76_{-0.02}^{+0.02}$ &
  $5.70_{-1.34}^{+1.84}$ &
  $10.9_{-0.9}^{+1.3}$ &
  $9.70_{-0.61}^{+1.07}$ &
  $9.79_{-0.81}^{+0.81}$ &
  $11.2_{-0.8}^{+1.4}$ &
  $2.91_{-0.33}^{+0.41}$ &
  $8.43_{-1.94}^{+1.28}$ &
  $4.60_{-0.10}^{+0.18}$ &
  $-1.00_{-0.08}^{+0.11}$ &
  $1.08$ ($3851$) \\
13 &
  $8.94_{-0.10}^{+0.10}$ &
  $0.19_{-0.01}^{+0.01}$ &
  $0.85_{-0.02}^{+0.04}$ &
  $7.34_{-1.56}^{+1.87}$ &
  $9.76_{-0.77}^{+1.17}$ &
  $8.72_{-0.73}^{+1.08}$ &
  $9.11_{-0.76}^{+1.53}$ &
  $11.2_{-1.2}^{+1.1}$ &
  $5.05_{-0.45}^{+0.95}$ &
  $11.0_{-2.9}^{+5.5}$ &
  $4.88_{-0.18}^{+0.13}$ &
  $-1.95_{-0.02}^{+0.02}$ &
  $1.15$ ($4351$) \\
14 &
  $8.64_{-0.13}^{+0.48}$ &
  $0.18_{-0.01}^{+0.01}$ &
  $0.67_{-0.01}^{+0.01}$ &
  $2.47_{-0.01}^{+0.58}$ &
  $5.68_{-0.05}^{+2.64}$ &
  $5.24_{-0.01}^{+2.34}$ &
  $5.61_{-0.03}^{+1.71}$ &
  $5.92_{-0.01}^{+1.51}$ &
  $0.87_{-0.01}^{+0.35}$ &
  $2.08_{-0.02}^{+4.90}$ &
  $4.88_{-0.03}^{+0.08}$ &
  $-1.93_{-0.01}^{+0.04}$ &
  $1.11$ ($3599$) \\
15 &
  $9.13_{-0.17}^{+0.18}$ &
  $0.19_{-0.01}^{+0.01}$ &
  $0.78_{-0.03}^{+0.02}$ &
  - &
  $5.77_{-0.63}^{+0.45}$ &
  $5.39_{-0.49}^{+0.30}$ &
  $5.53_{-0.58}^{+0.34}$ &
  $6.95_{-0.72}^{+0.49}$ &
  $1.81_{-0.29}^{+0.26}$ &
  $9.79_{-2.89}^{+2.32}$ &
  $5.14_{-0.12}^{+0.14}$ &
  $-2.13_{-0.44}^{+0.21}$ &
  $1.12$ ($3437$) \\
16 &
  $9.14_{-0.05}^{+0.24}$ &
  $0.21_{-0.01}^{+0.01}$ &
  $1.62_{-0.05}^{+0.05}$ &
  - &
  $9.48_{-0.49}^{+1.36}$ &
  $8.06_{-0.59}^{+0.94}$ &
  $7.43_{-0.89}^{+1.14}$ &
  $10.4_{-1.45}^{+1.55}$ &
  $14.0_{-1.2}^{+1.6}$ &
  $36.0_{-7.5}^{+10.2}$ &
  $9.02_{-0.82}^{+1.49}$ &
  $-2.79_{-0.03}^{+0.05}$ &
  $1.19$ ($4619$) \\
17 &
  $7.68_{-0.14}^{+0.10}$ &
  $0.32_{-0.02}^{+0.03}$ &
  $0.64_{-0.03}^{+0.01}$ &
  $8.06_{-2.47}^{+2.26}$ &
  $12.3_{-1.1}^{+2.3}$ &
  $14.0_{-1.1}^{+2.3}$ &
  $18.2_{-1.2}^{+3.2}$ &
  $20.6_{-1.4}^{+3.5}$ &
  $6.92_{-0.99}^{+1.38}$ &
  $34.5_{-9.9}^{+14.5}$ &
  $4.39_{-0.14}^{+0.27}$ &
  $-0.58_{-0.11}^{+0.06}$ &
  $1.17$ ($3358$) \\
18 &
  $4.88_{-0.08}^{+0.06}$ &
  $0.21_{-0.01}^{+0.01}$ &
  $0.63_{-0.02}^{+0.02}$ &
  - &
  $2.60_{-0.30}^{+0.17}$ &
  $3.70_{-0.26}^{+0.19}$ &
  $3.35_{-0.33}^{+0.26}$ &
  $3.43_{-0.41}^{+0.33}$ &
  $0.63_{-0.06}^{+0.09}$ &
  $0.46_{-0.29}^{+0.74}$ &
  $4.91_{-0.16}^{+0.12}$ &
  $-1.67_{-0.73}^{+0.31}$ &
  $1.11$ ($3781$) \\ \hline
            \end{tabular}
        \end{center}
    \end{rotatetable*}
\end{table*}

\appendix
\section{Nucleosynthesis Plots}
\label{apx:plots}
Herein, we present plots comparing the results of our spatially-resolved spectroscopic study of both 3C 397 and W49B to 11 of the 12 models of supernova nucleosynthesis used in this study. Four of these are of CC supernovae: WW95, M03, N06, and S16, while the remaining seven are of Ia supernovae: M10, S13, F14, LN18, T18, B19, and LN20. For 3C 397, the Ia results can be seen in Figure \ref{fig:3c397_ia}, while the CC results can be seen in Figure \ref{fig:3c397_cc}. For W49B, the Ia results can be seen in Figure \ref{fig:w49b_ia}, while the CC results can be seen in Figure \ref{fig:w49b_cc}.

\begin{figure*}
	\begin{center}
		\subfloat{\includegraphics[angle=0,width=0.40\textwidth,scale=0.5]{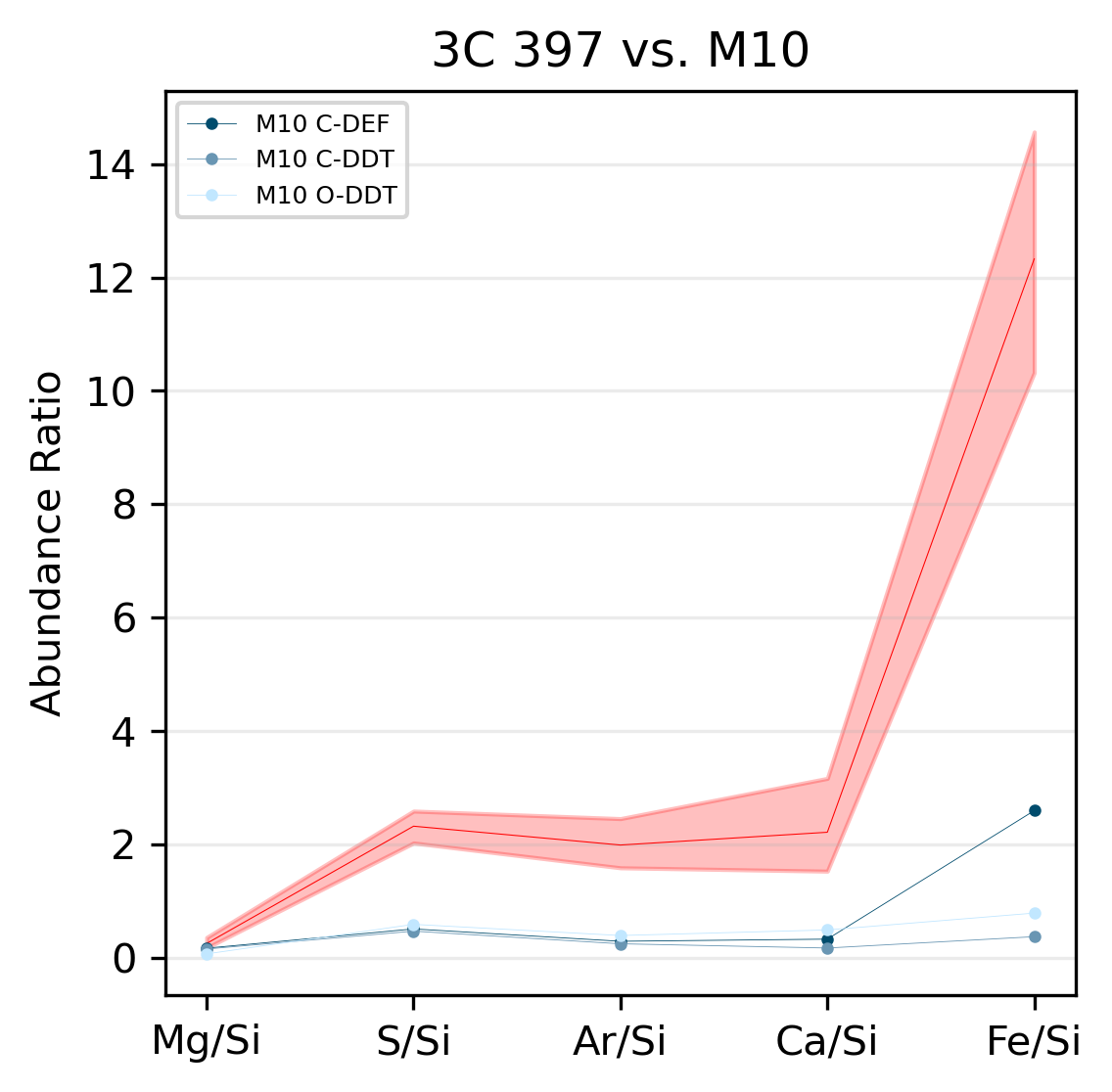}}
		\subfloat{\includegraphics[angle=0,width=0.40\textwidth,scale=0.5]{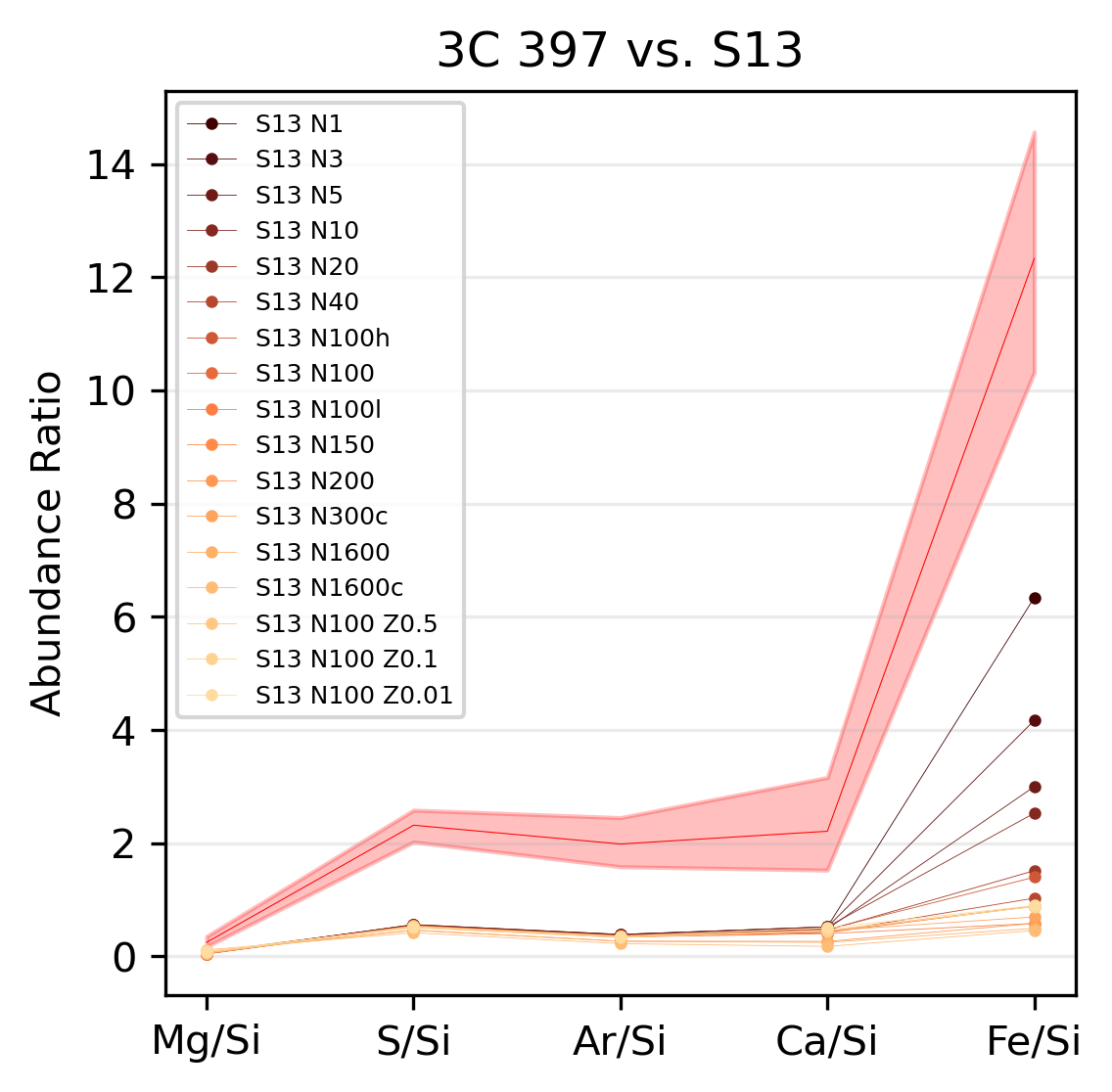}} \\
		\subfloat{\includegraphics[angle=0,width=0.40\textwidth]{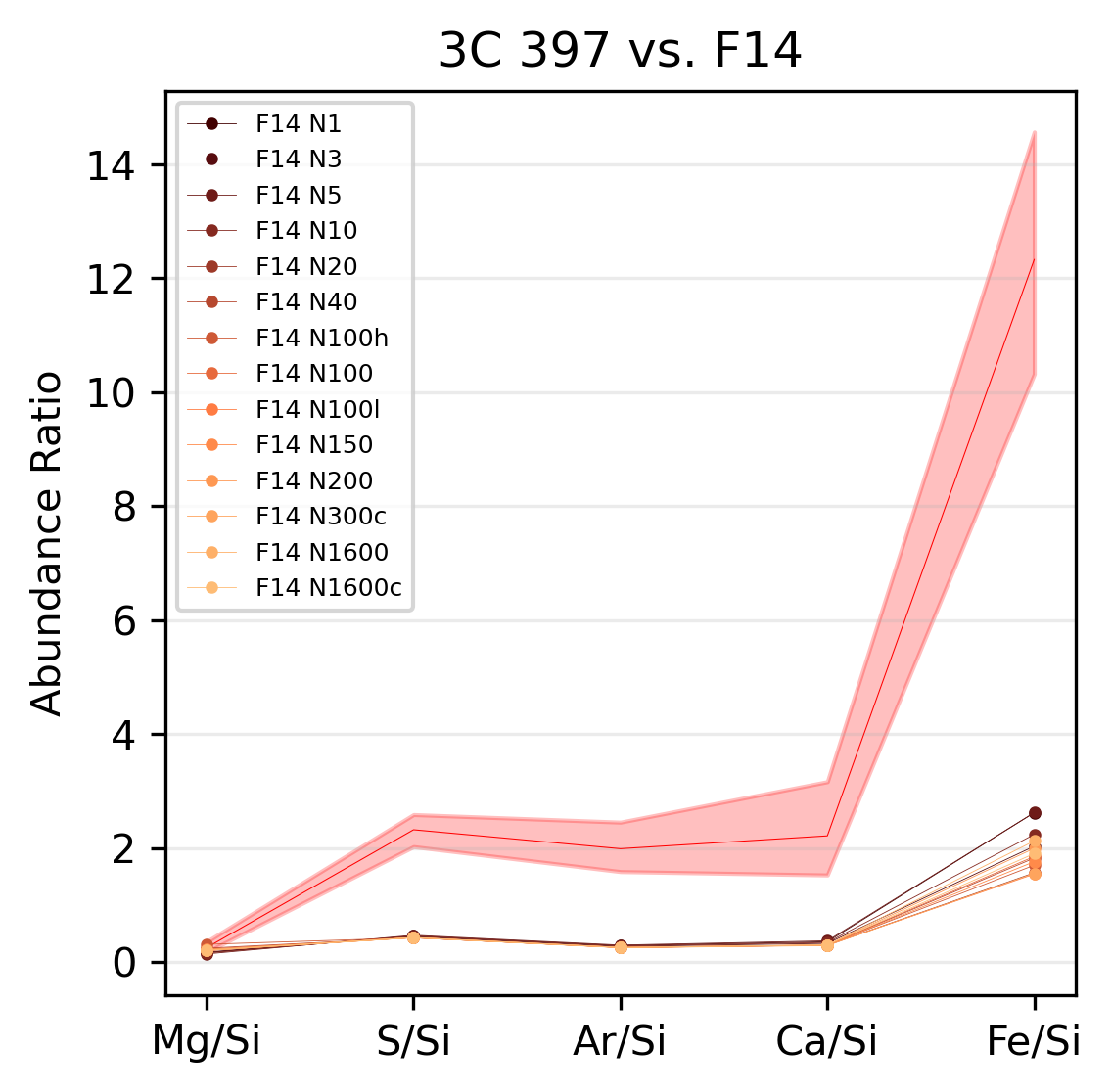}}
		\subfloat{\includegraphics[angle=0,width=0.40\textwidth]{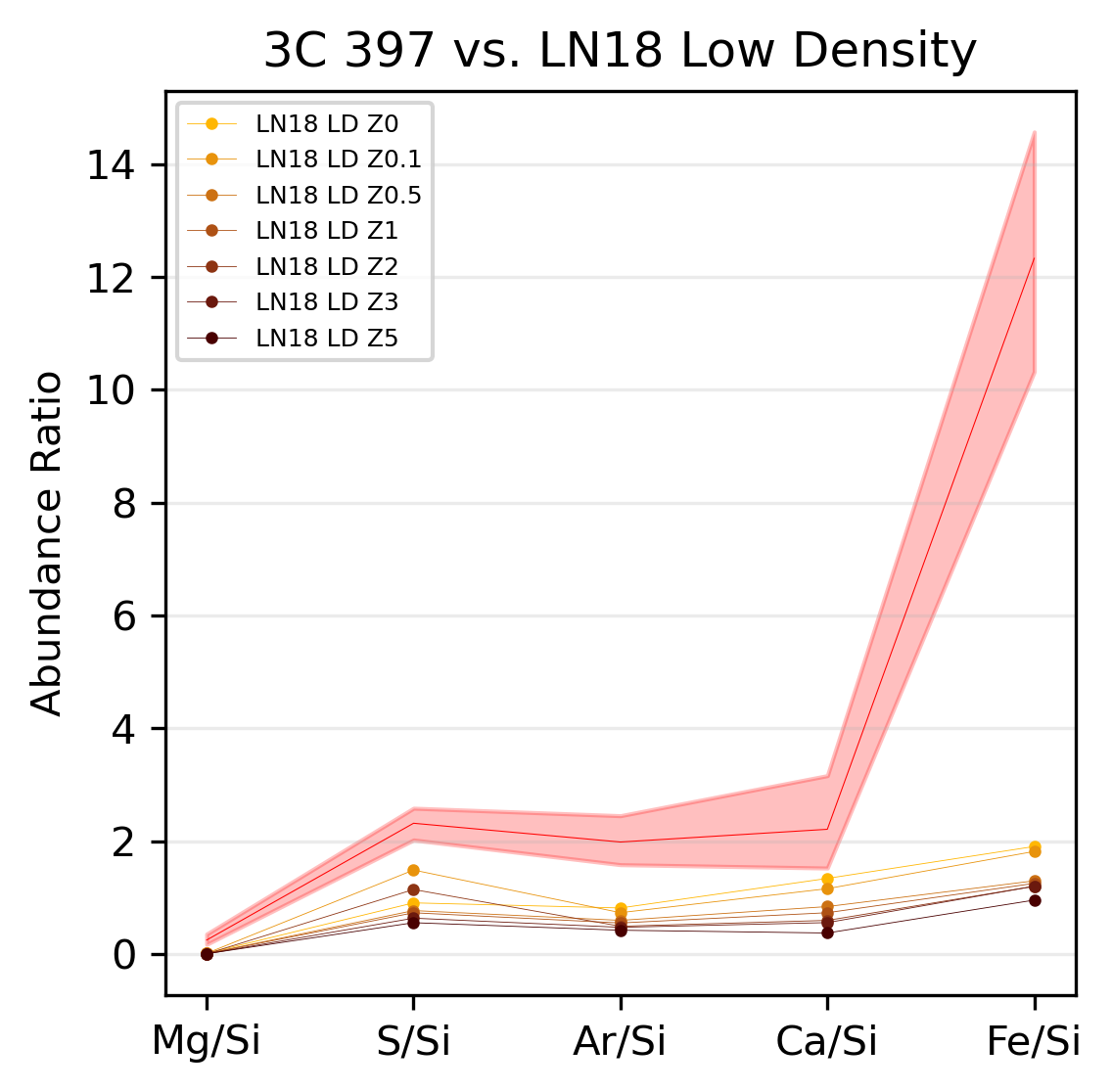}} \\
		\subfloat{\includegraphics[angle=0,width=0.40\textwidth,scale=0.5]{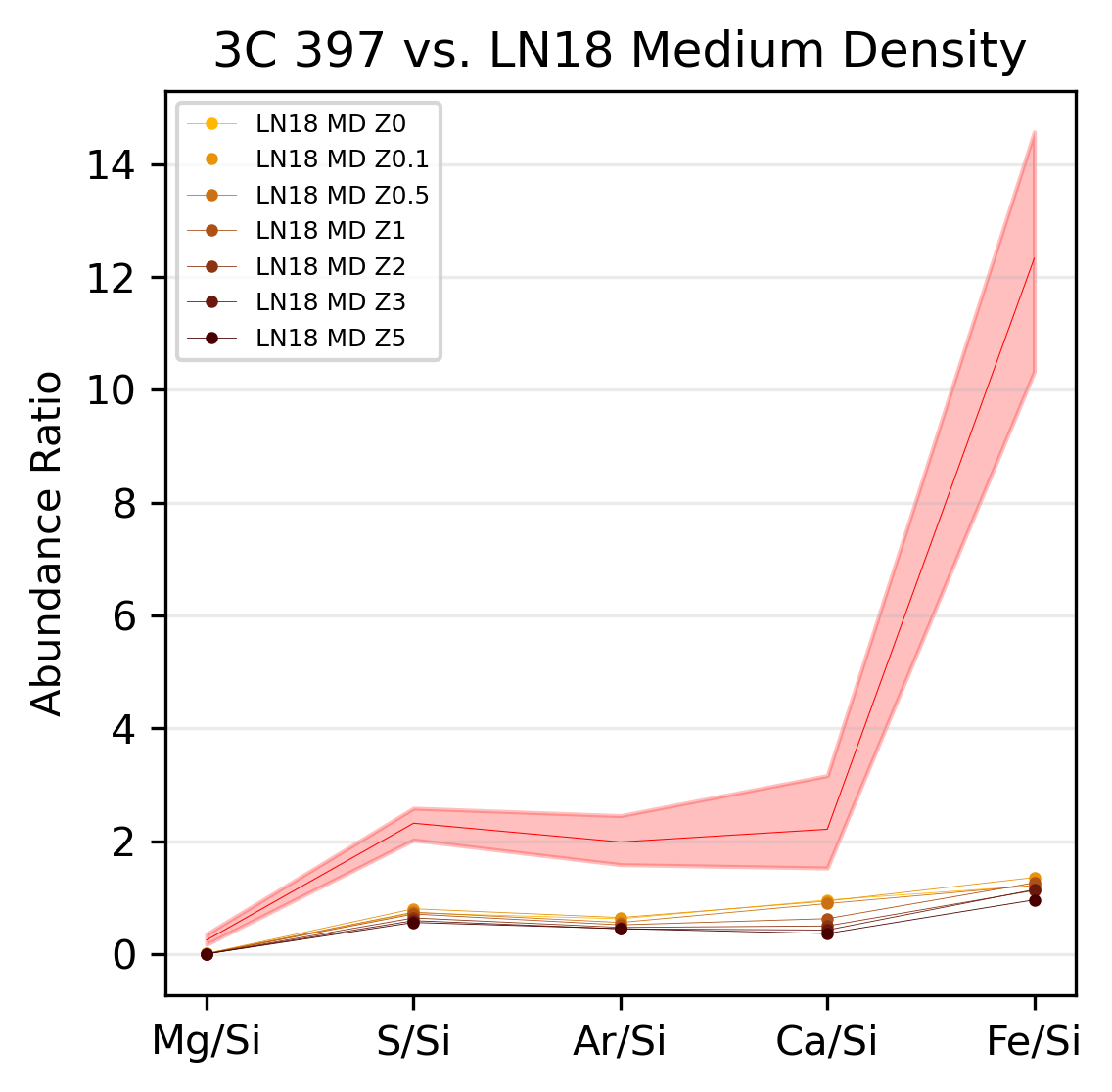}}
		\subfloat{\includegraphics[angle=0,width=0.40\textwidth]{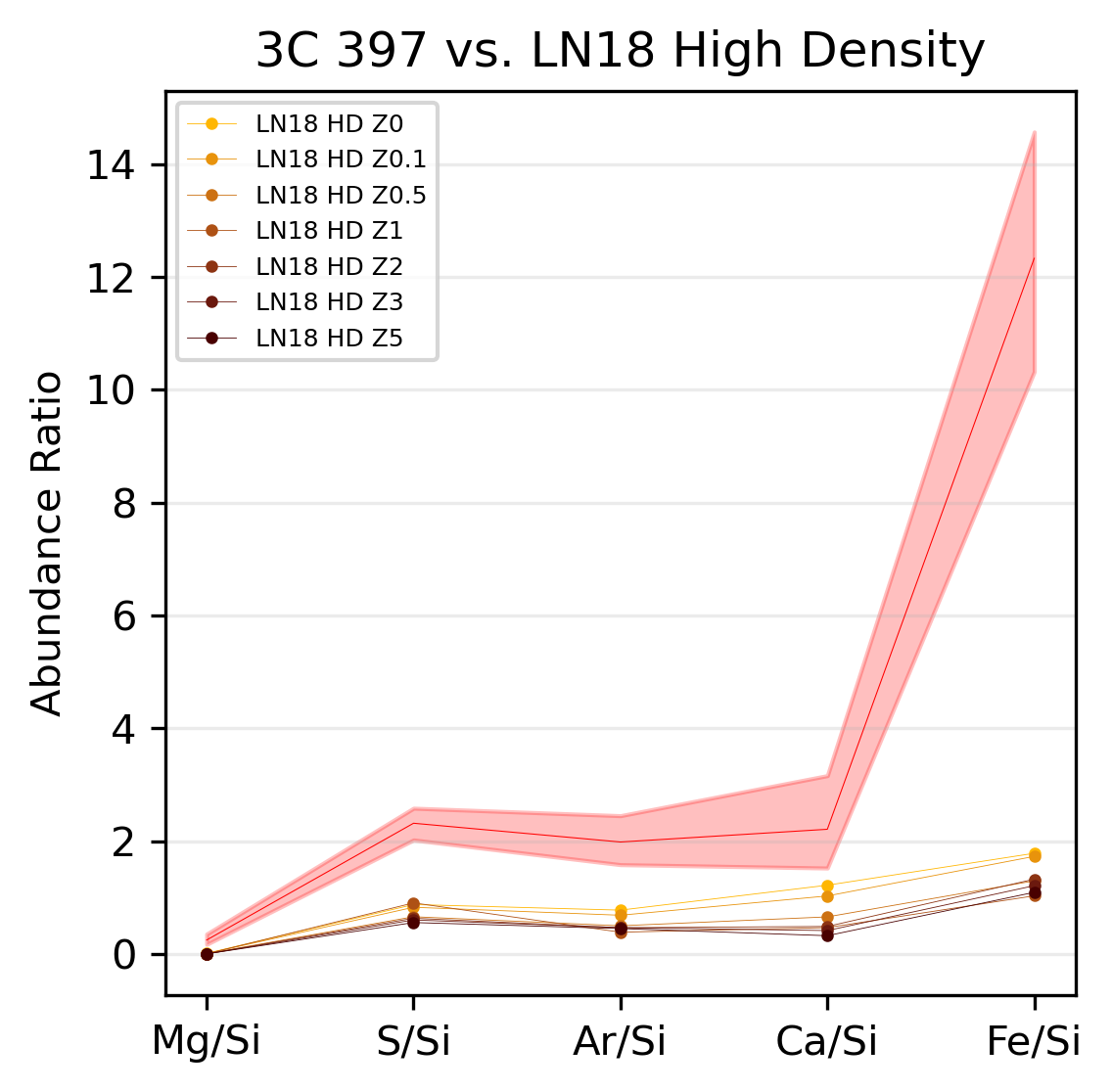}}
	\end{center}
    \caption{Nucleosynthesis comparisons between 3C 397 and the tested Ia models.}
    \label{fig:3c397_ia}
\end{figure*}

\begin{figure*}\ContinuedFloat
	\begin{center}
		\subfloat{\includegraphics[angle=0,width=0.40\textwidth,scale=0.5]{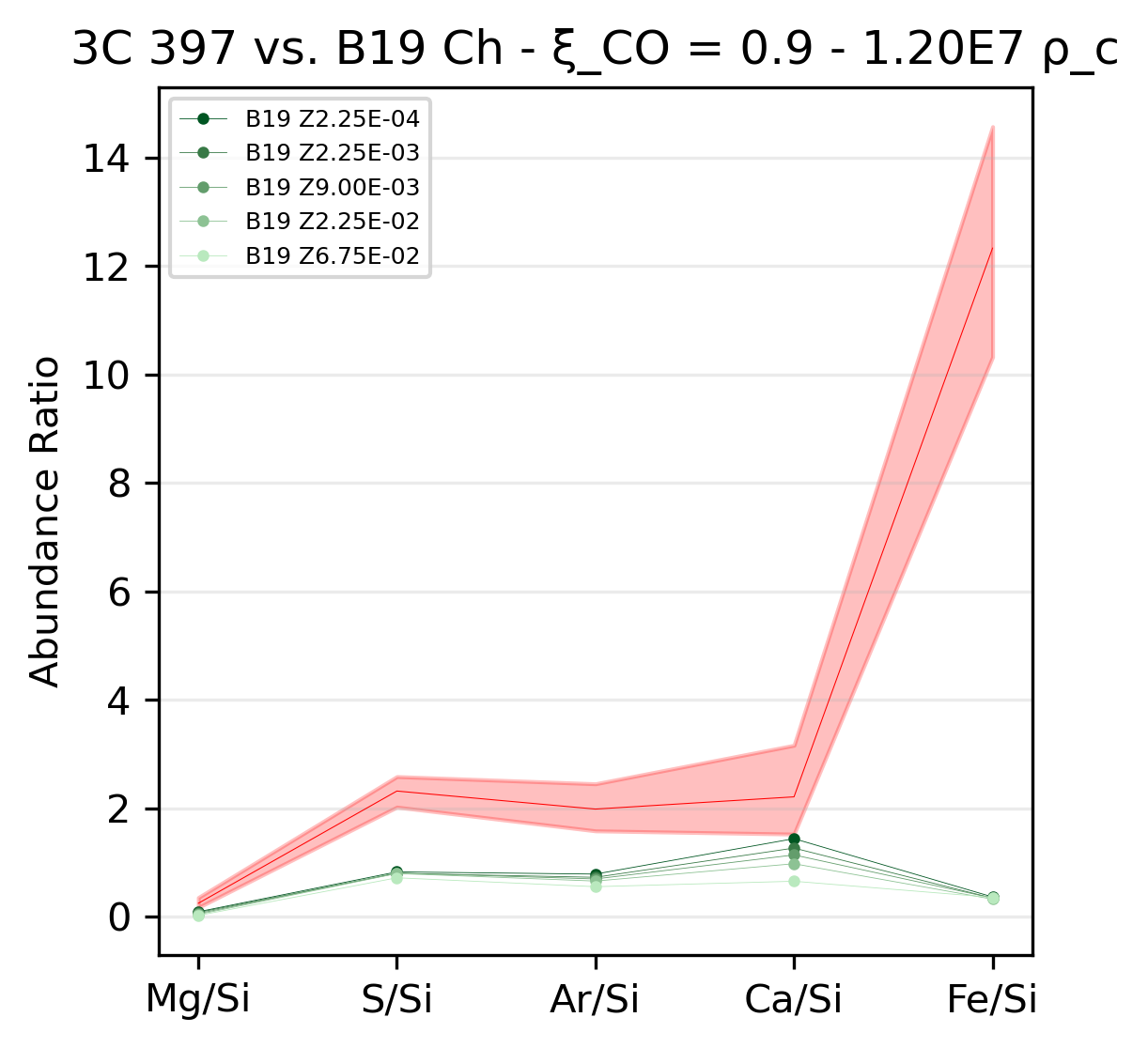}}
		\subfloat{\includegraphics[angle=0,width=0.40\textwidth,scale=0.5]{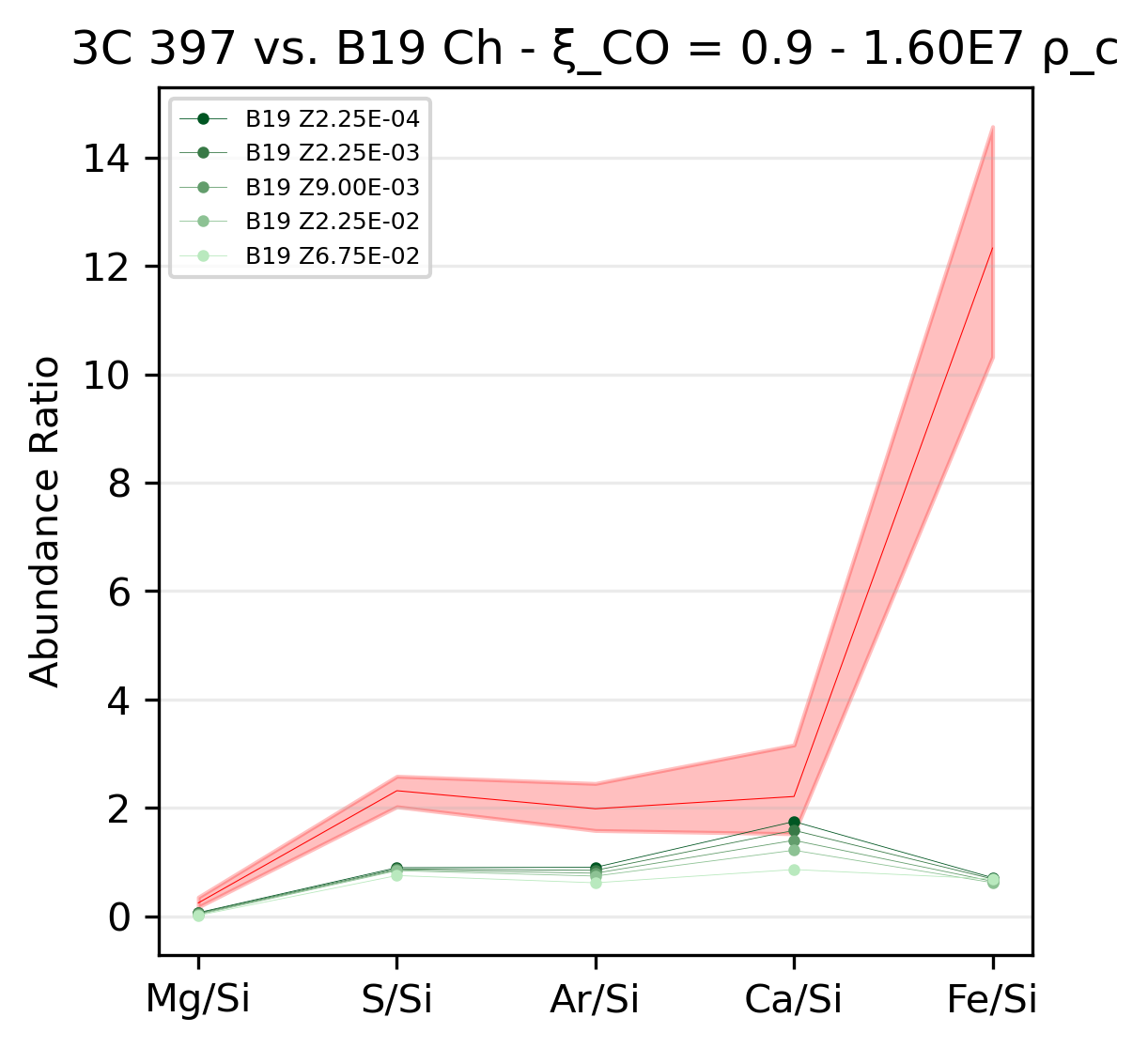}} \\
		\subfloat{\includegraphics[angle=0,width=0.40\textwidth]{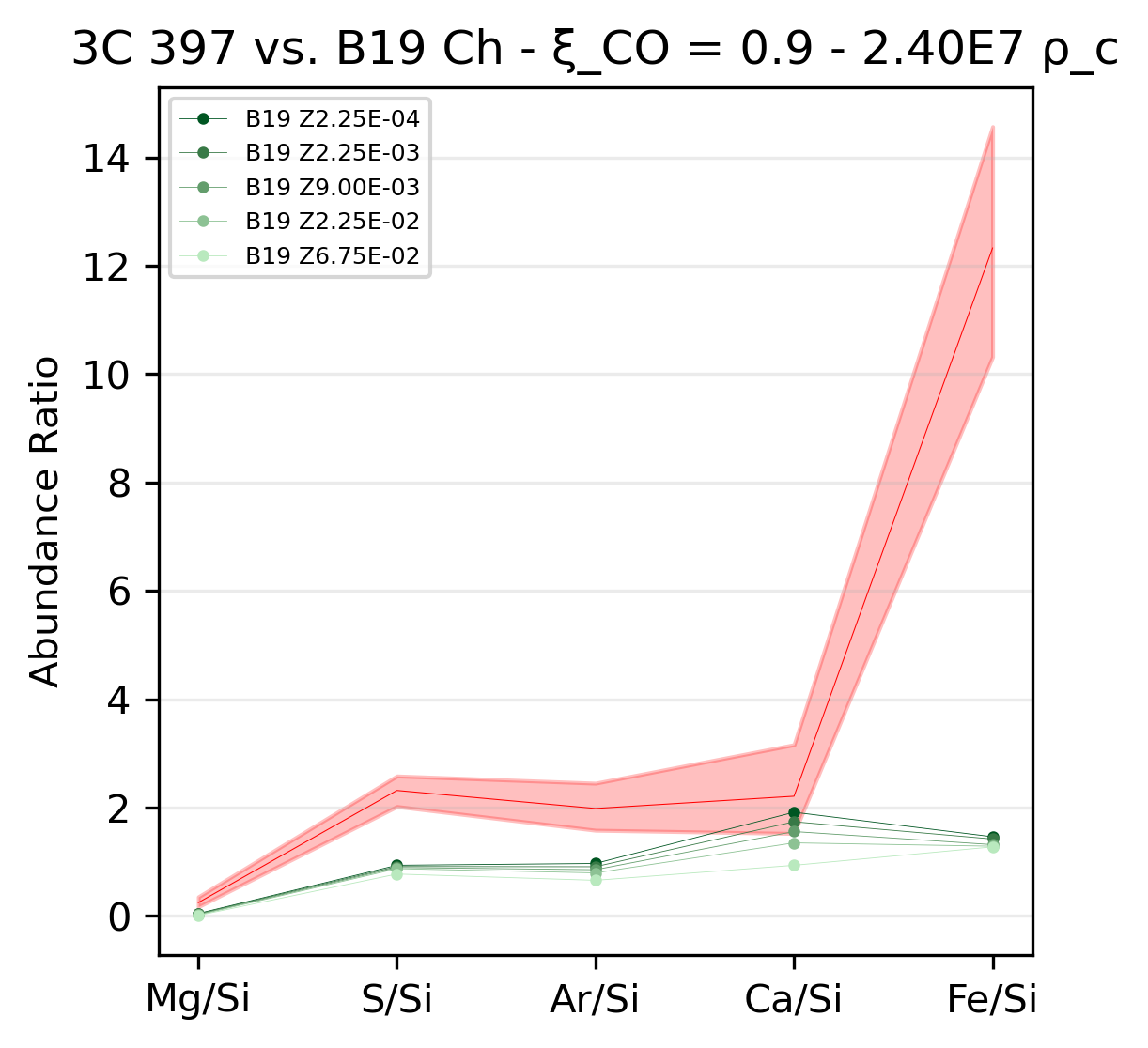}}
		\subfloat{\includegraphics[angle=0,width=0.40\textwidth]{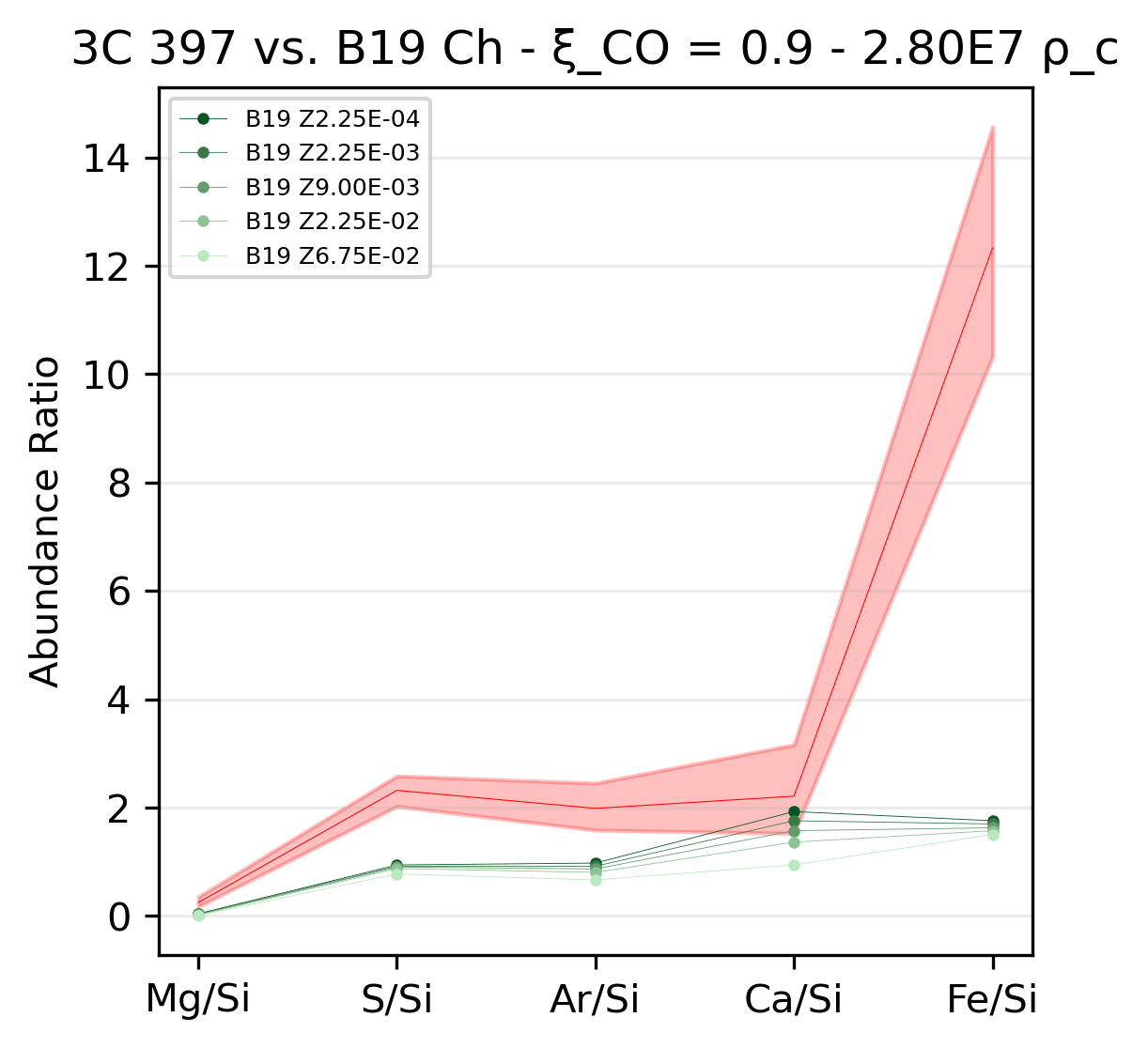}} \\
		\subfloat{\includegraphics[angle=0,width=0.40\textwidth,scale=0.5]{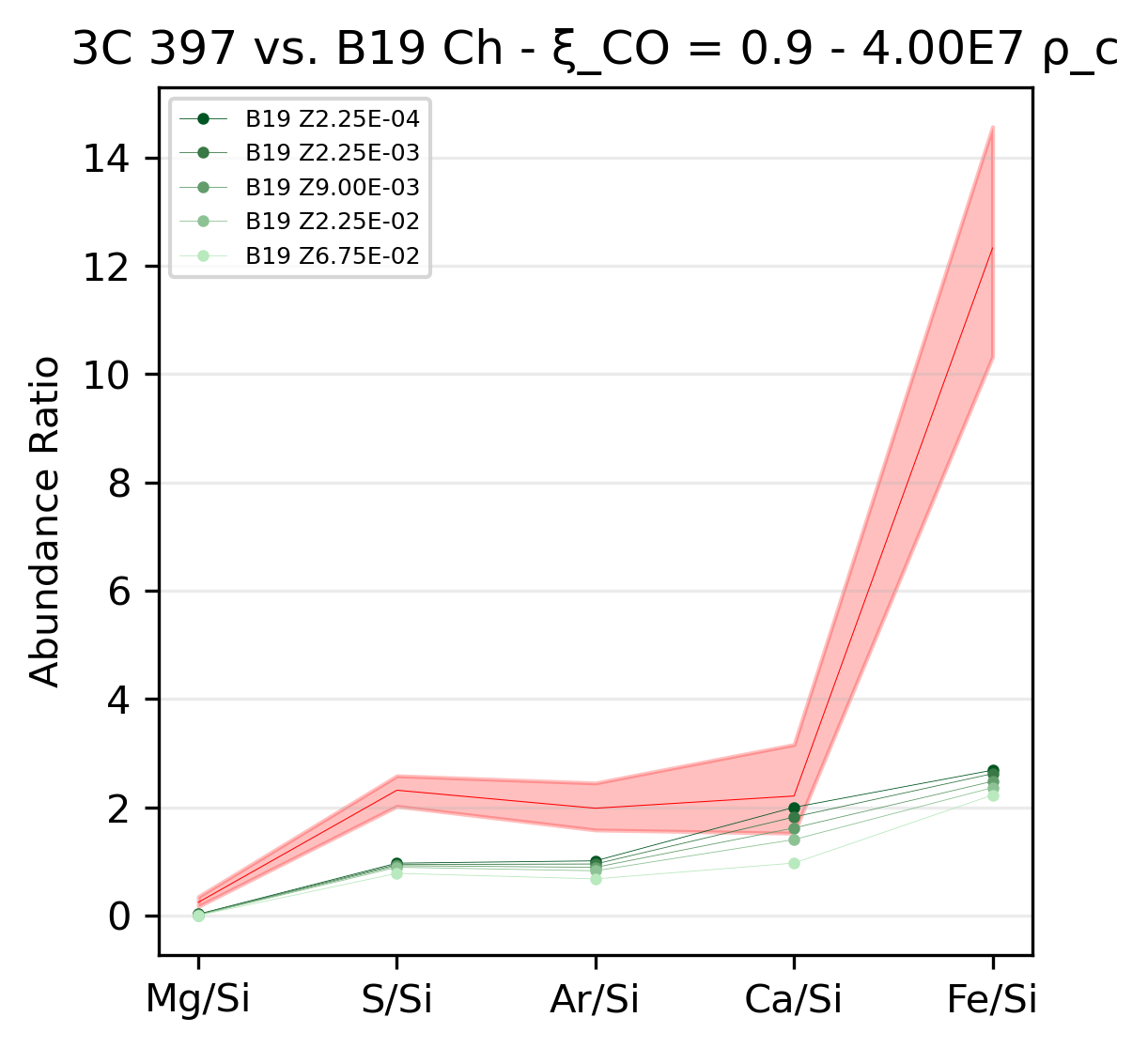}}
		\subfloat{\includegraphics[angle=0,width=0.40\textwidth]{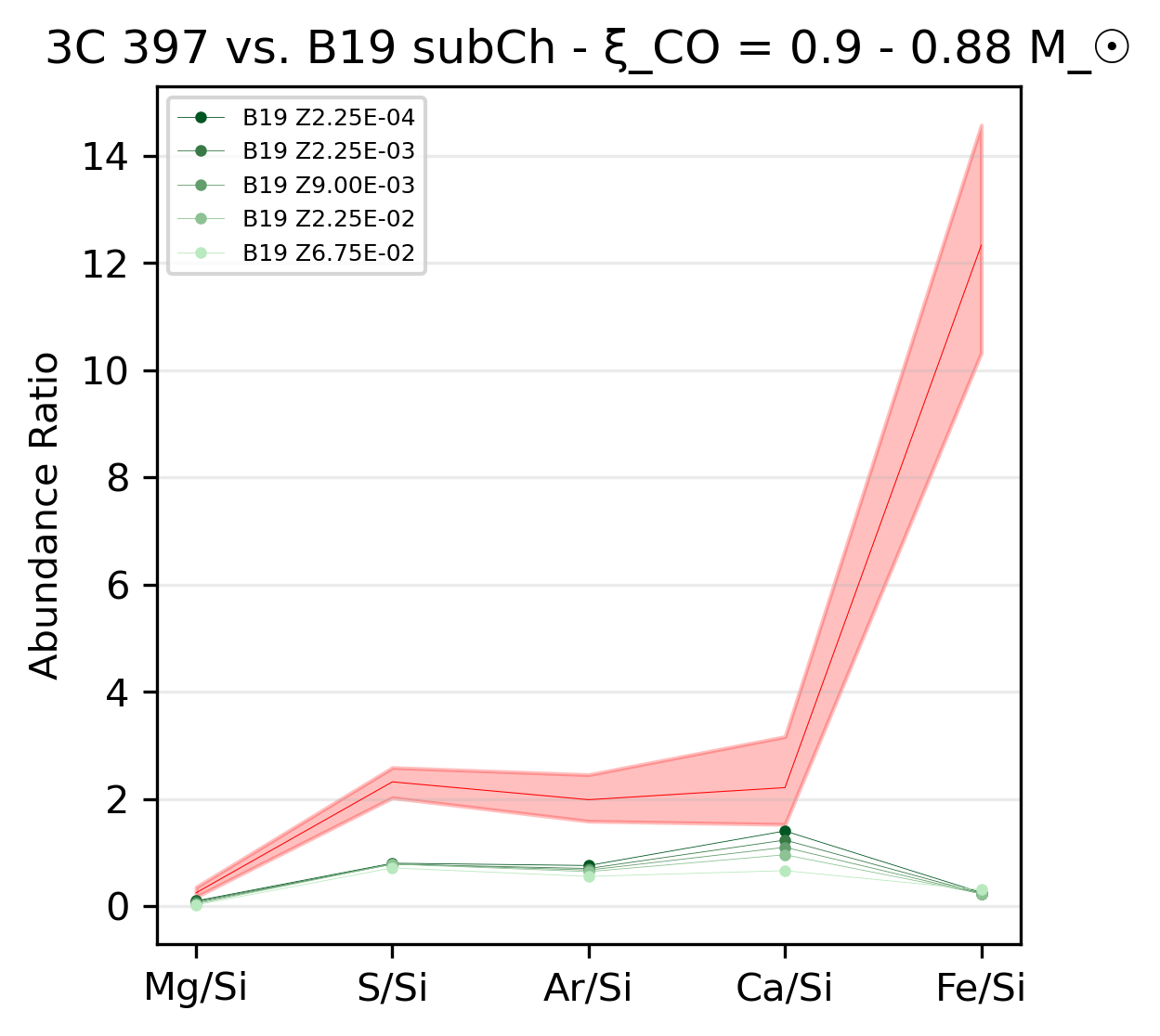}}
	\end{center}
    {Continued from above.}
\end{figure*}

\begin{figure*}\ContinuedFloat
	\begin{center}
		\subfloat{\includegraphics[angle=0,width=0.40\textwidth,scale=0.5]{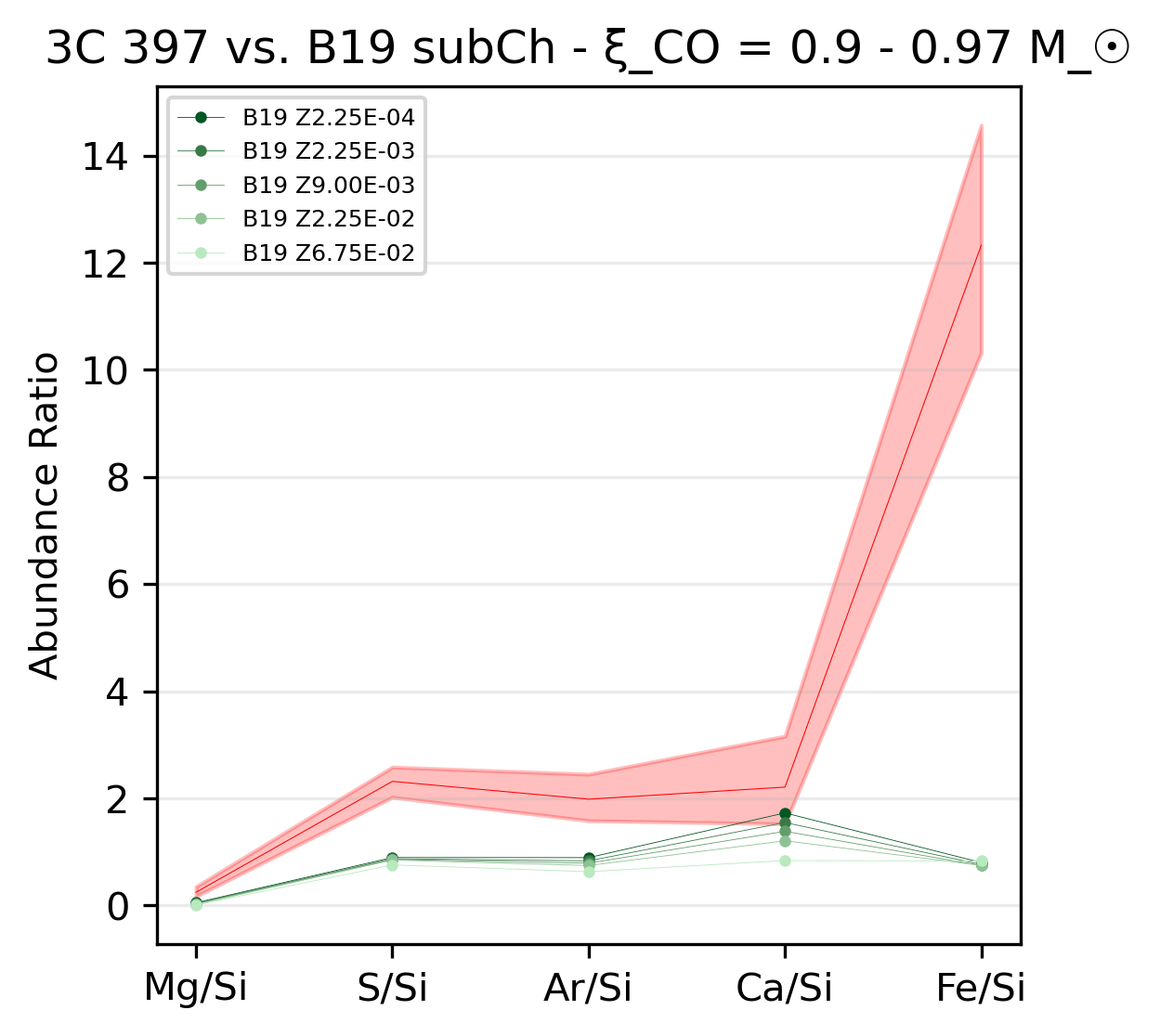}}
		\subfloat{\includegraphics[angle=0,width=0.40\textwidth,scale=0.5]{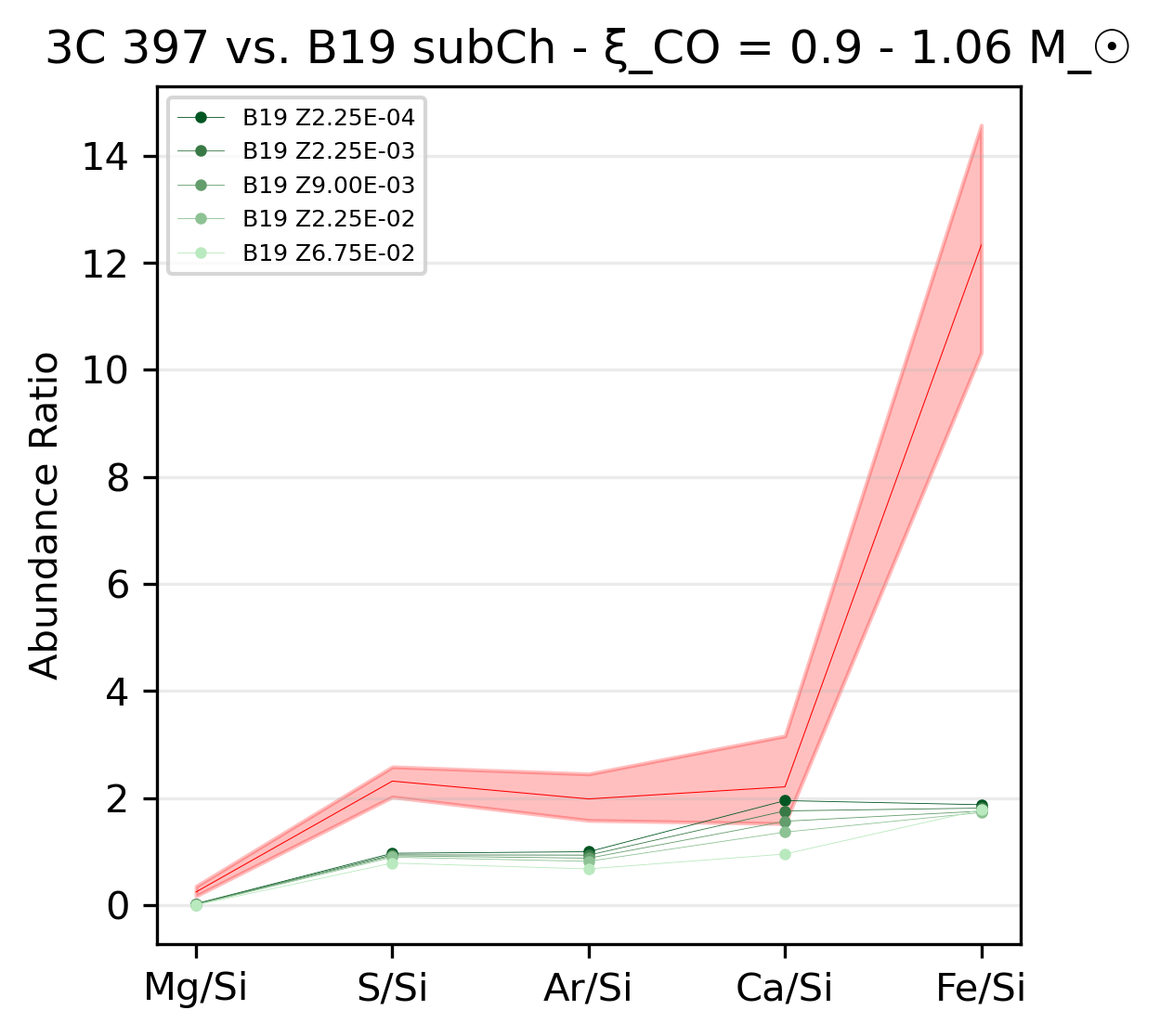}} \\
		\subfloat{\includegraphics[angle=0,width=0.40\textwidth]{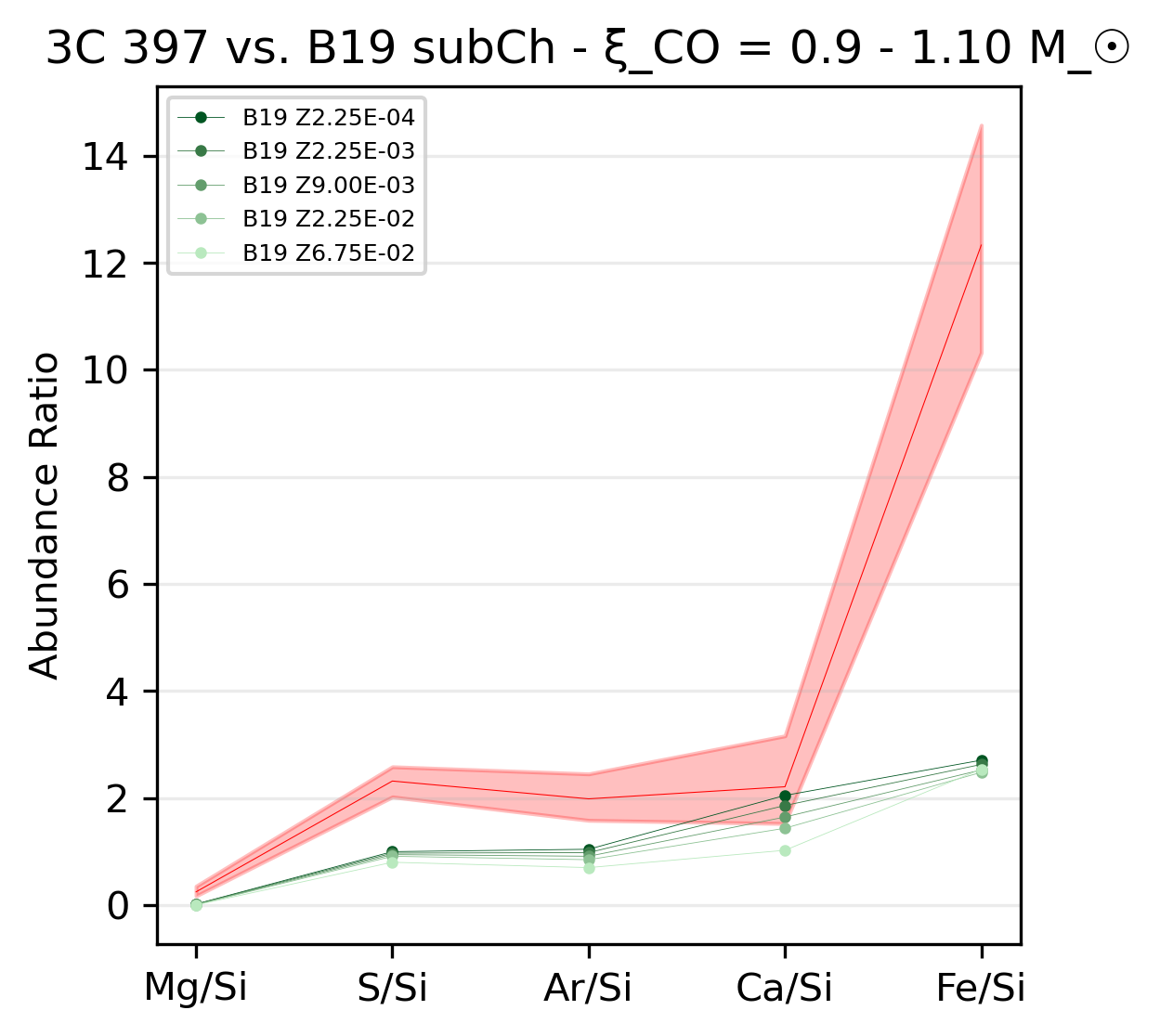}}
		\subfloat{\includegraphics[angle=0,width=0.40\textwidth]{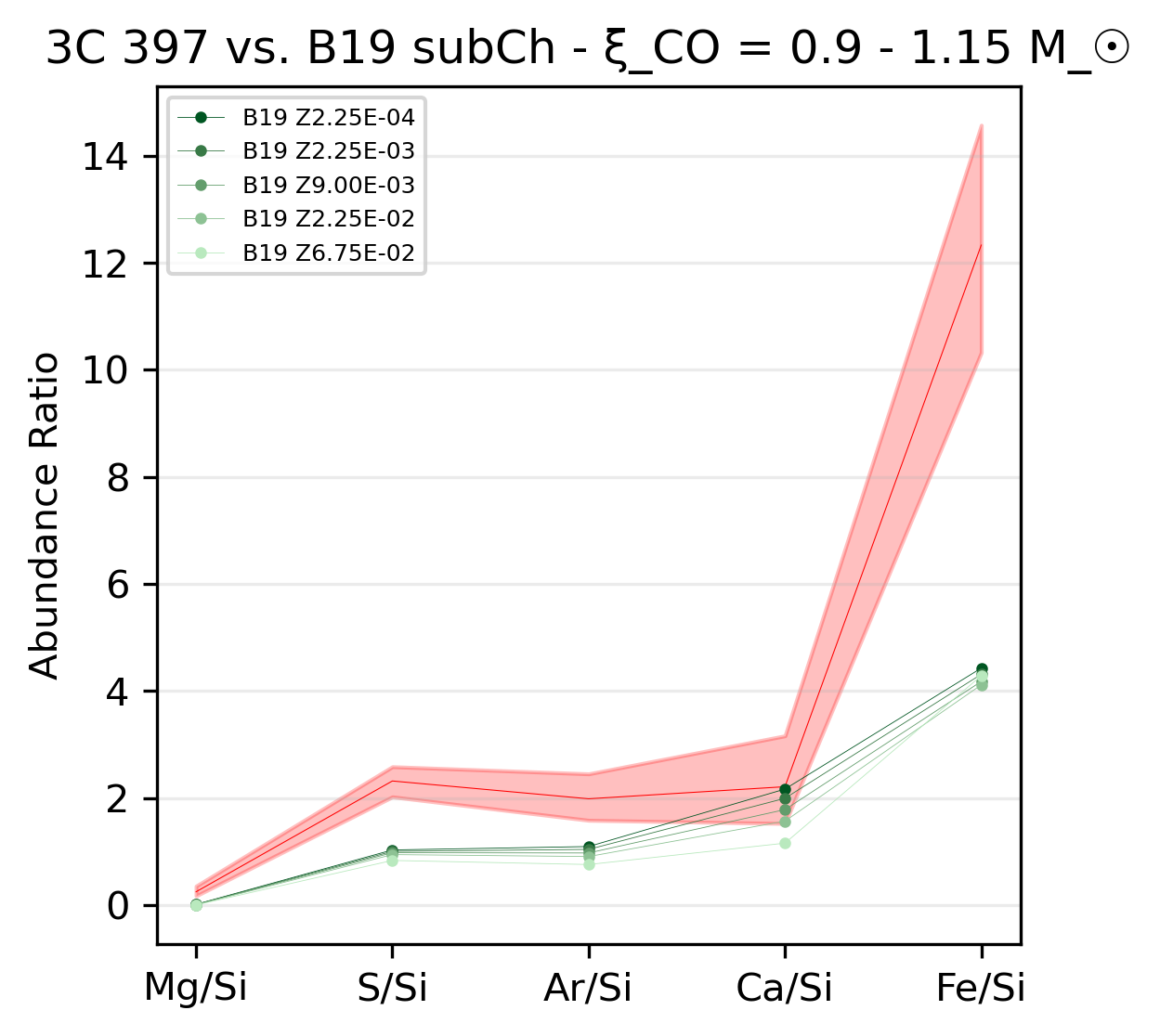}} \\
		\subfloat{\includegraphics[angle=0,width=0.40\textwidth,scale=0.5]{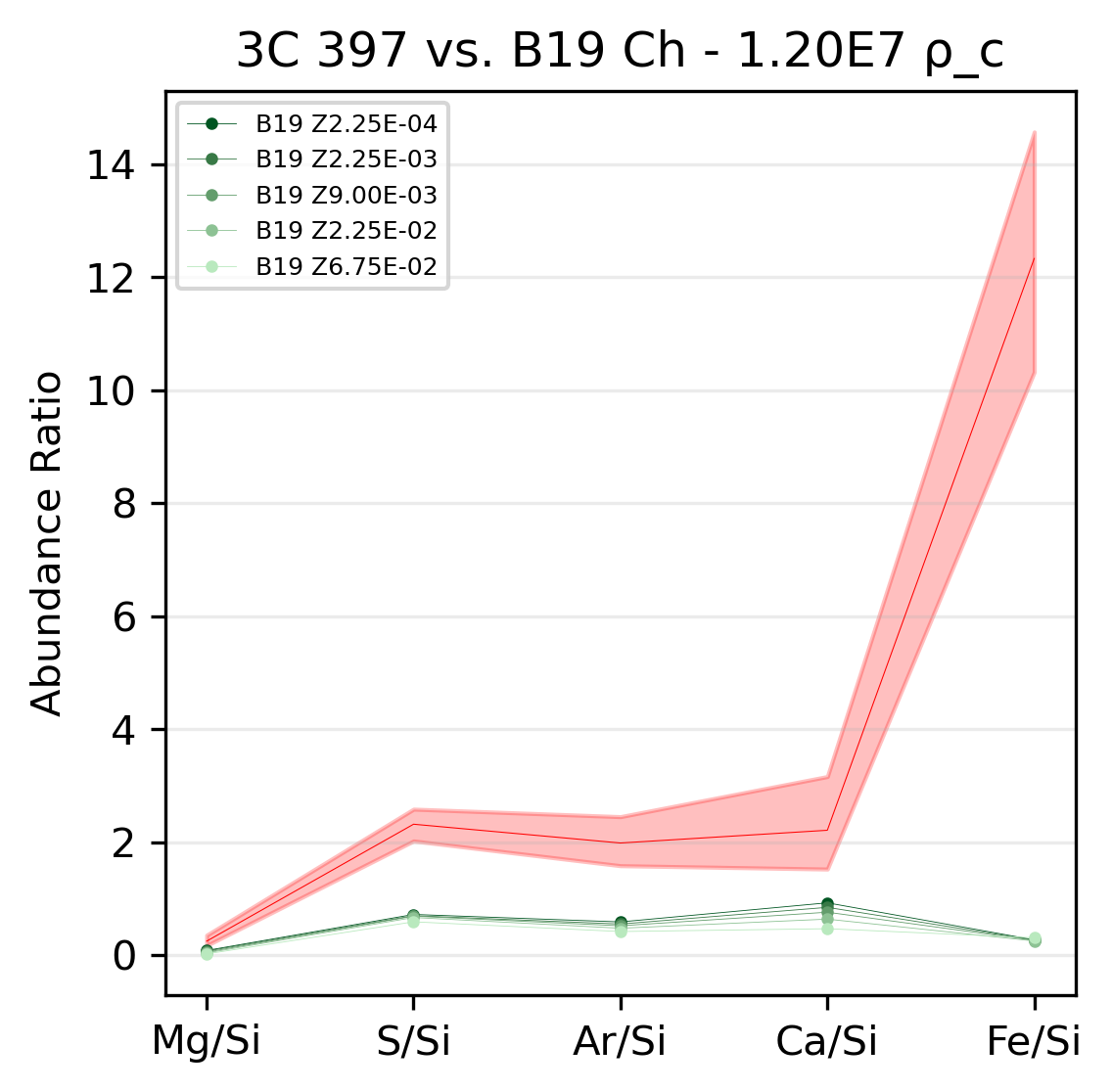}}
		\subfloat{\includegraphics[angle=0,width=0.40\textwidth]{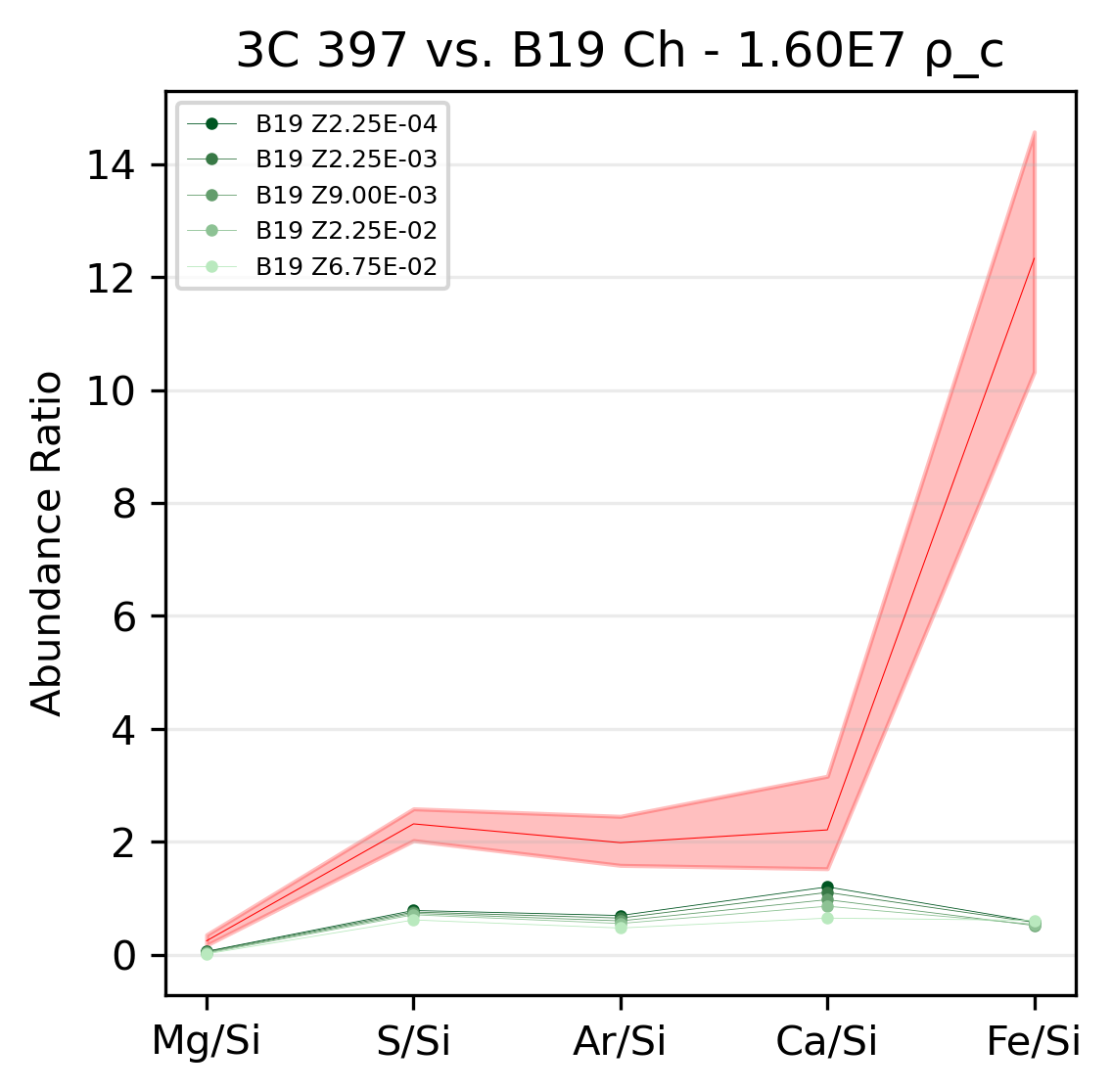}}
	\end{center}
    {Continued from above.}
\end{figure*}

\begin{figure*}\ContinuedFloat
	\begin{center}
		\subfloat{\includegraphics[angle=0,width=0.40\textwidth,scale=0.5]{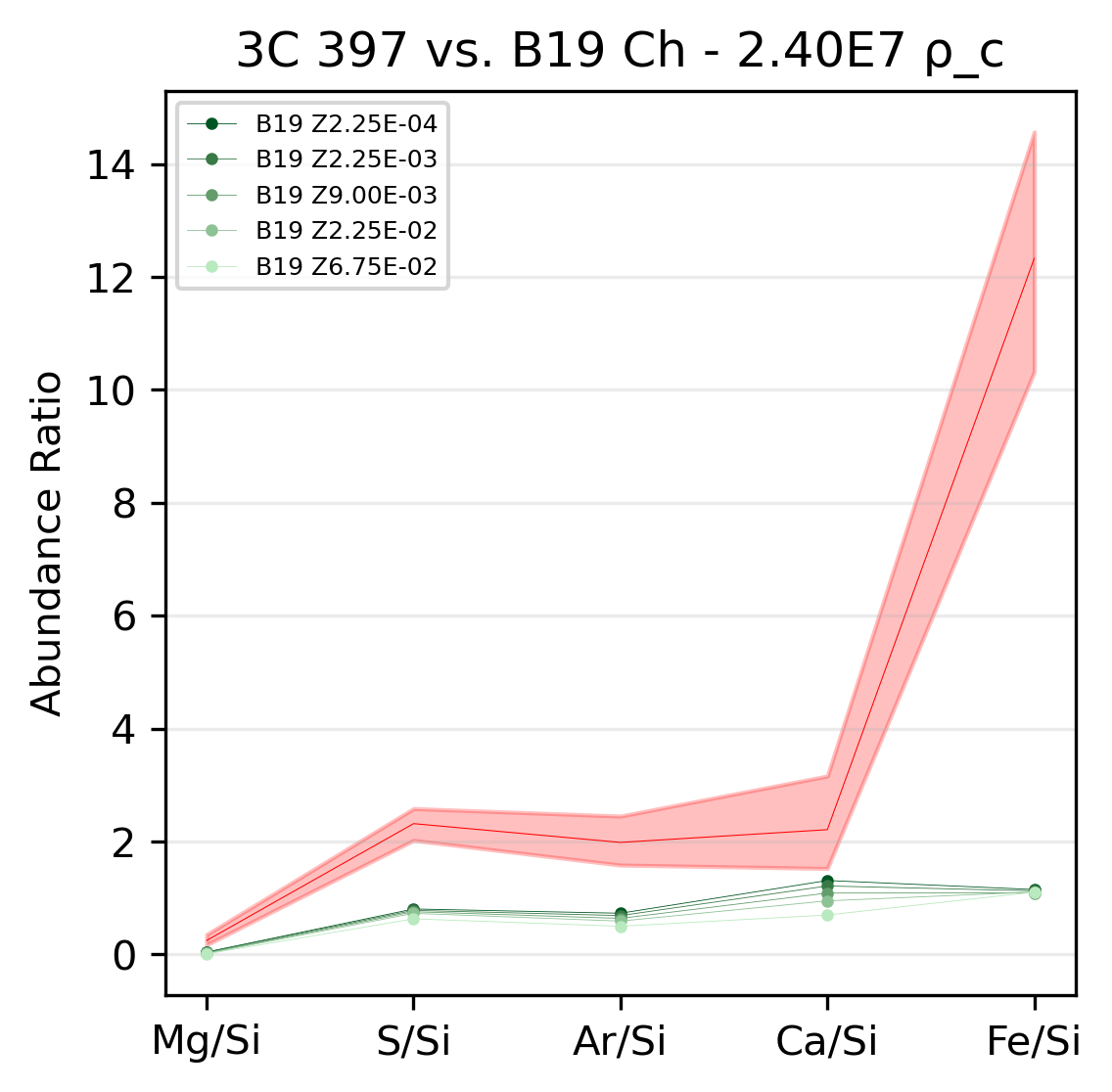}}
		\subfloat{\includegraphics[angle=0,width=0.40\textwidth,scale=0.5]{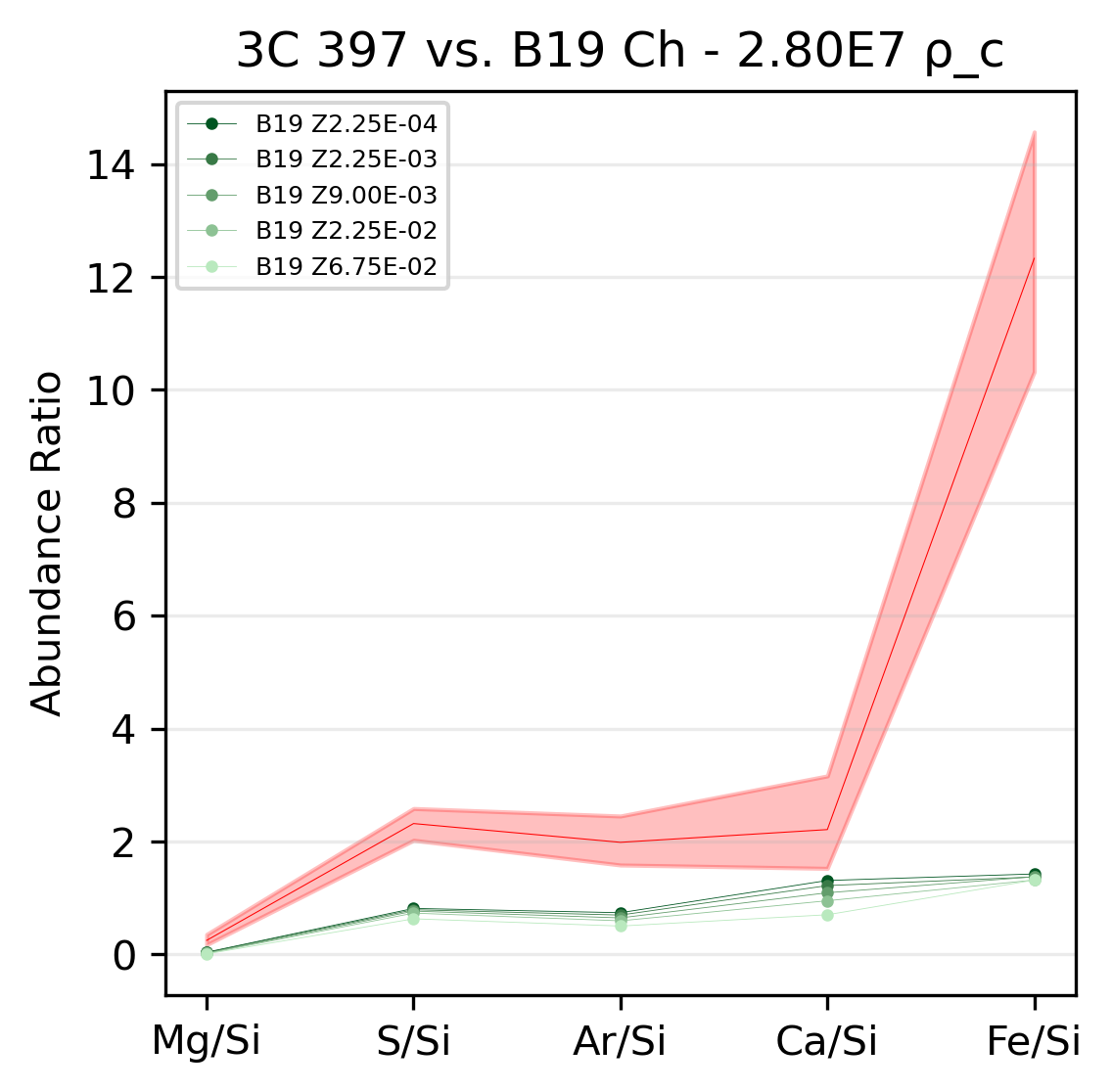}} \\
		\subfloat{\includegraphics[angle=0,width=0.40\textwidth]{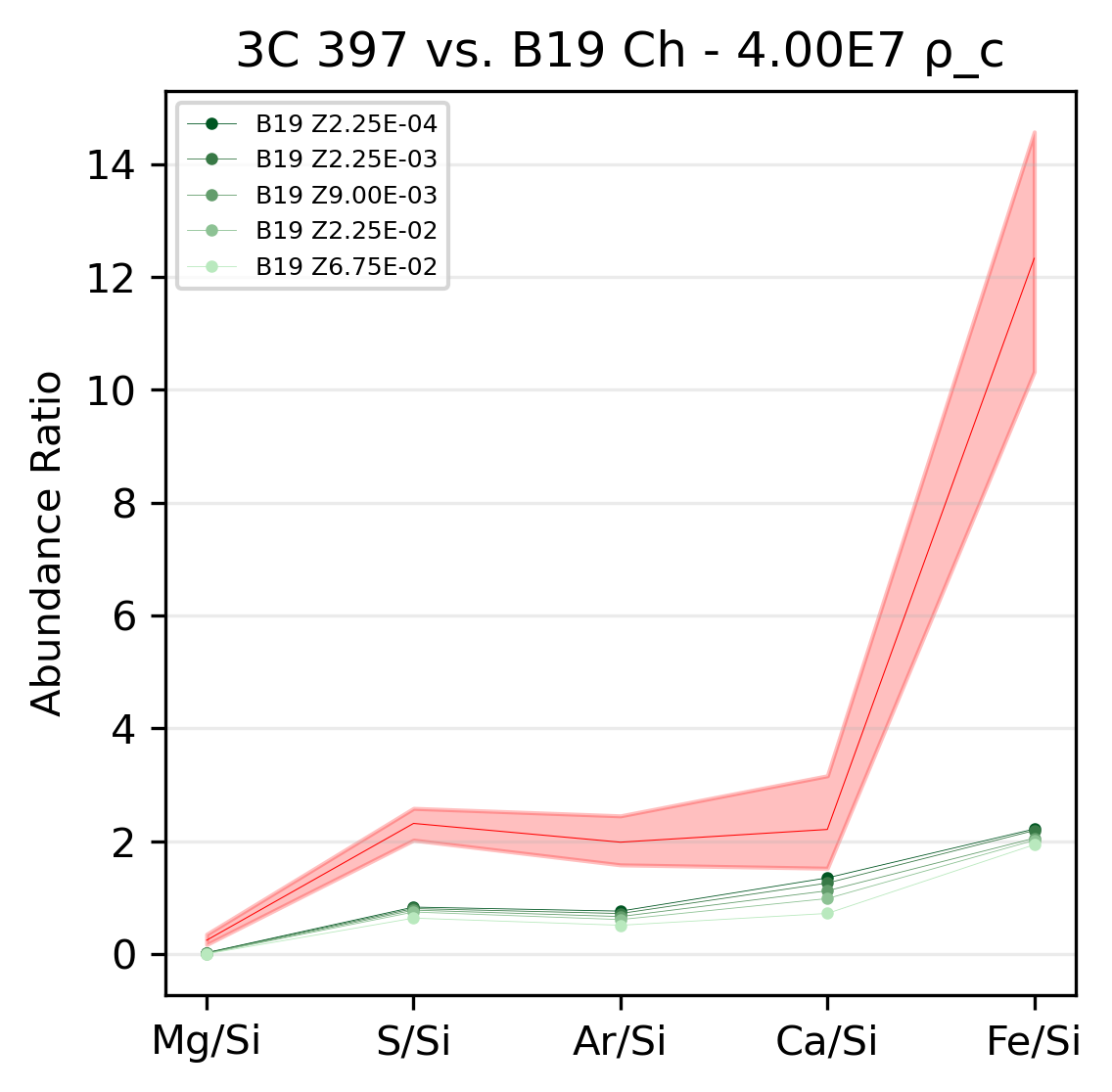}}
		\subfloat{\includegraphics[angle=0,width=0.40\textwidth]{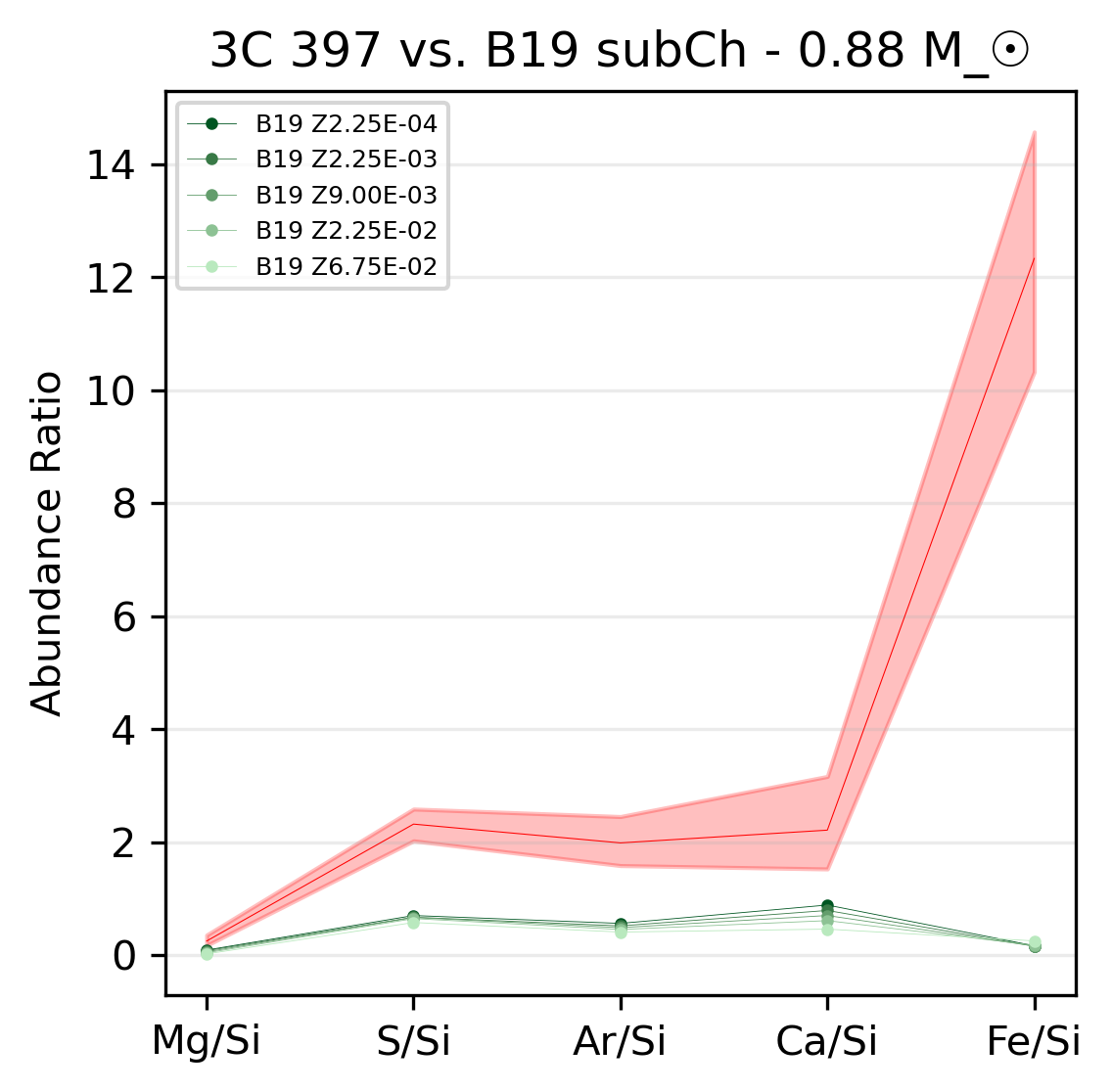}} \\
		\subfloat{\includegraphics[angle=0,width=0.40\textwidth,scale=0.5]{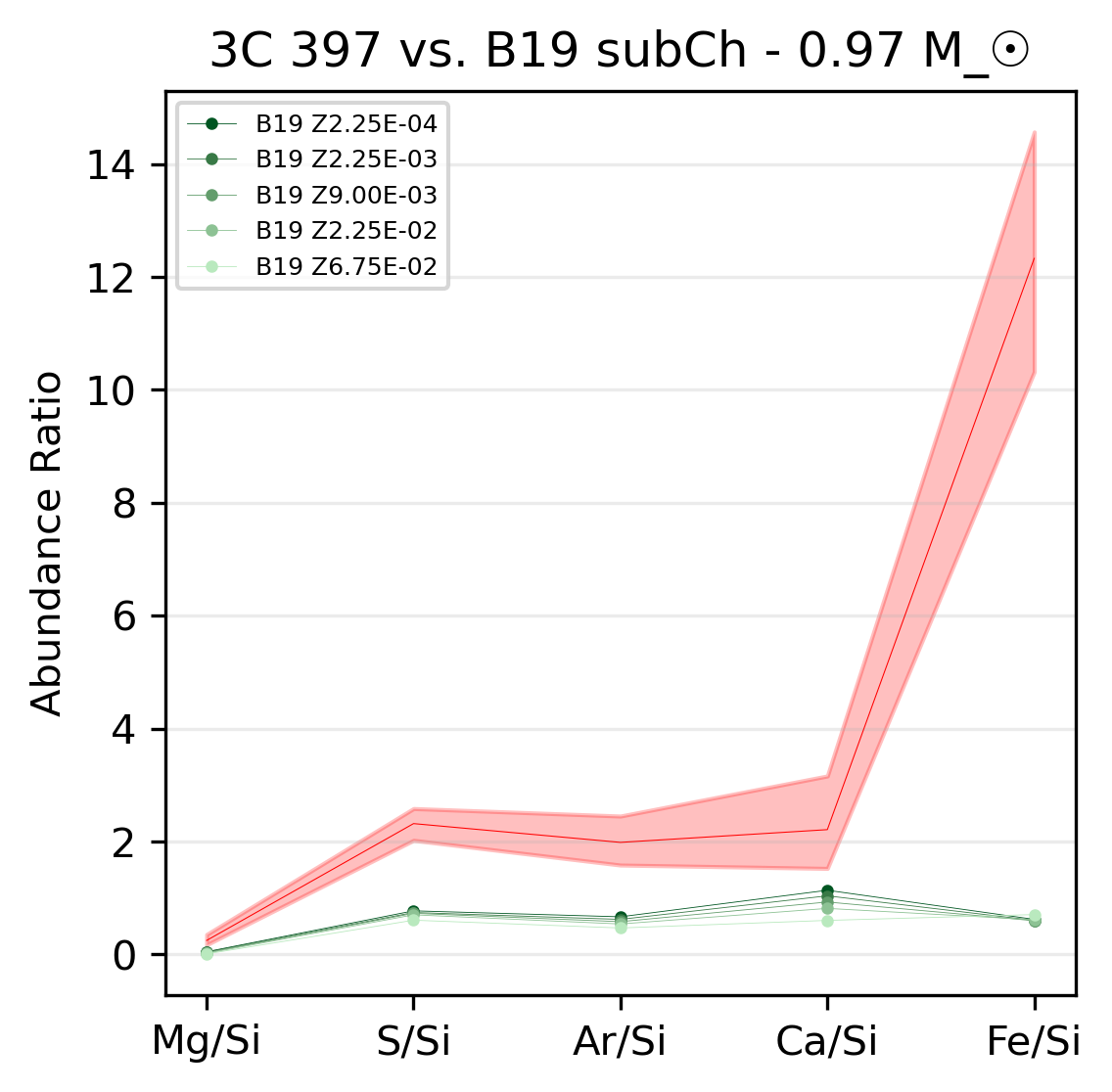}}
		\subfloat{\includegraphics[angle=0,width=0.40\textwidth]{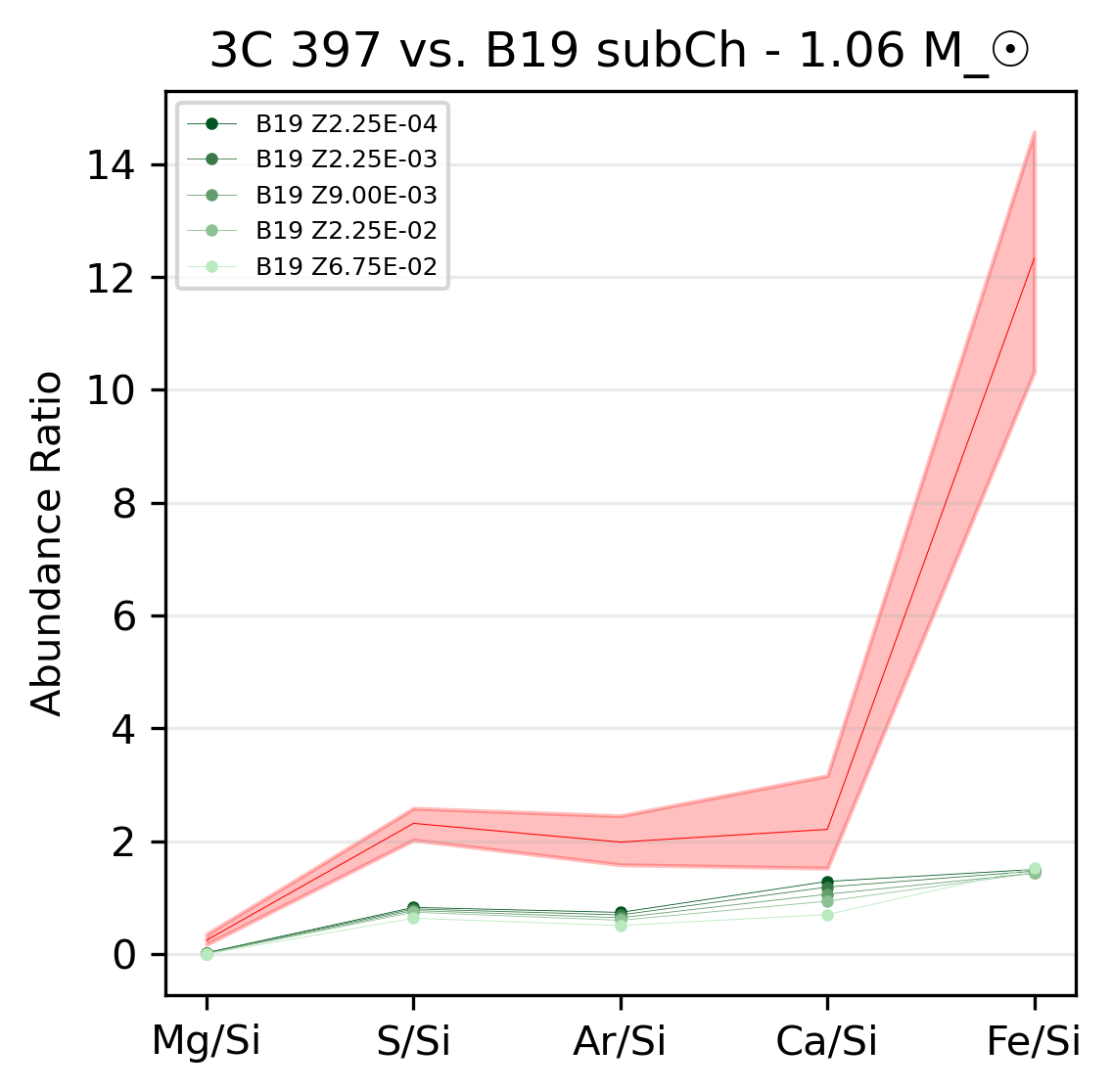}}
	\end{center}
    {Continued from above.}
\end{figure*}

\begin{figure*}\ContinuedFloat
	\begin{center}
		\subfloat{\includegraphics[angle=0,width=0.40\textwidth,scale=0.5]{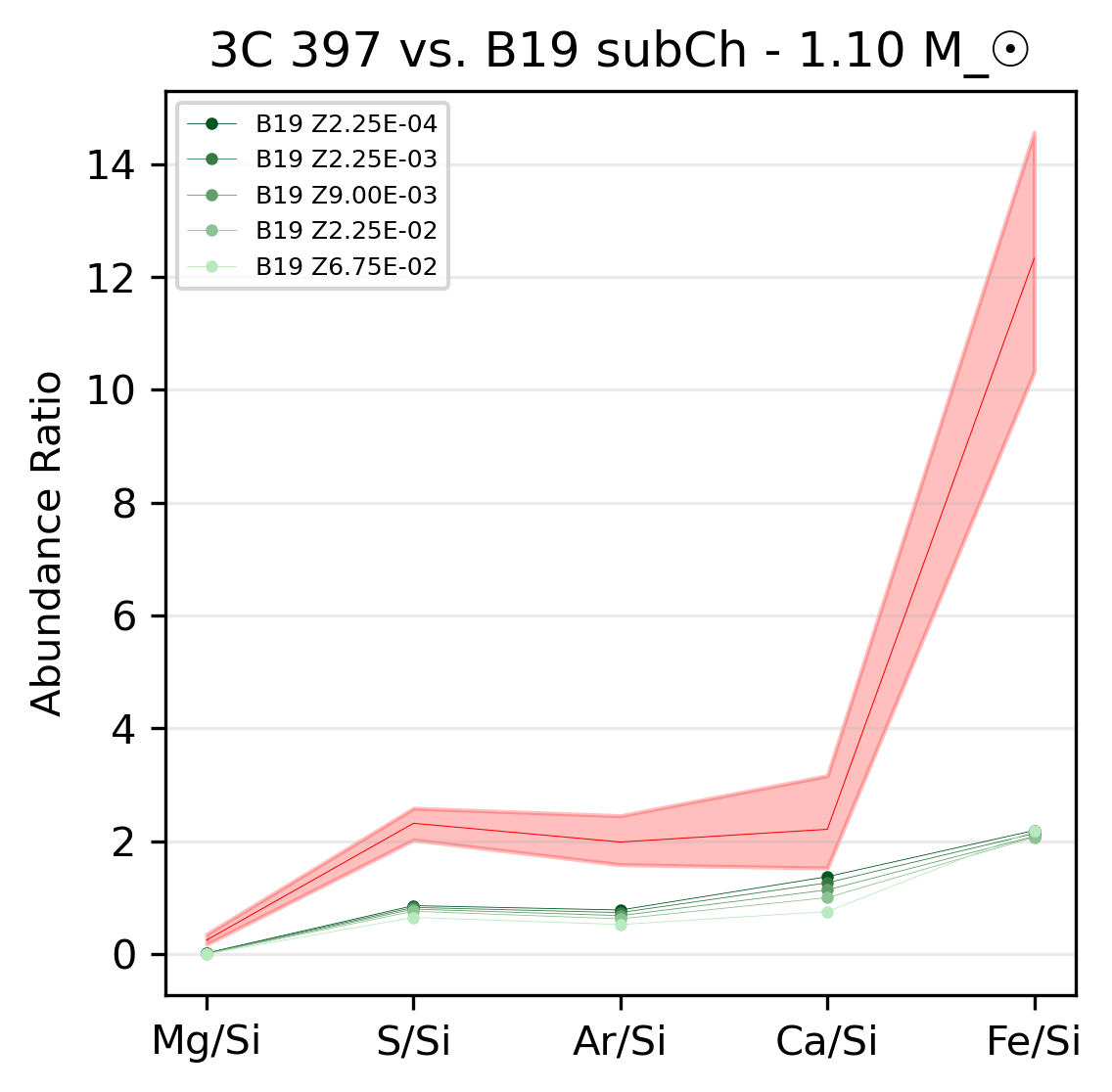}}
		\subfloat{\includegraphics[angle=0,width=0.40\textwidth,scale=0.5]{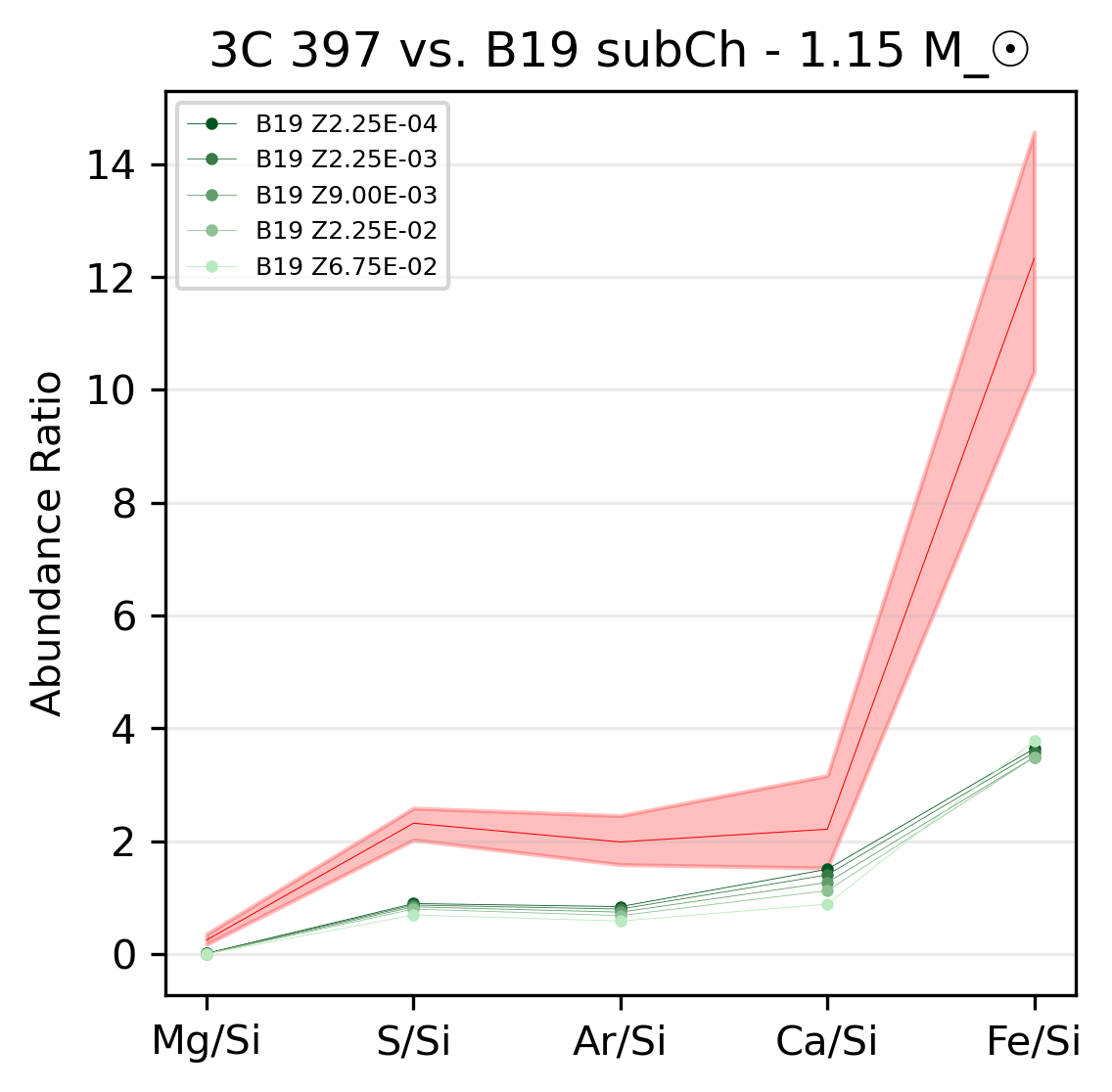}} \\
		\subfloat{\includegraphics[angle=0,width=0.40\textwidth]{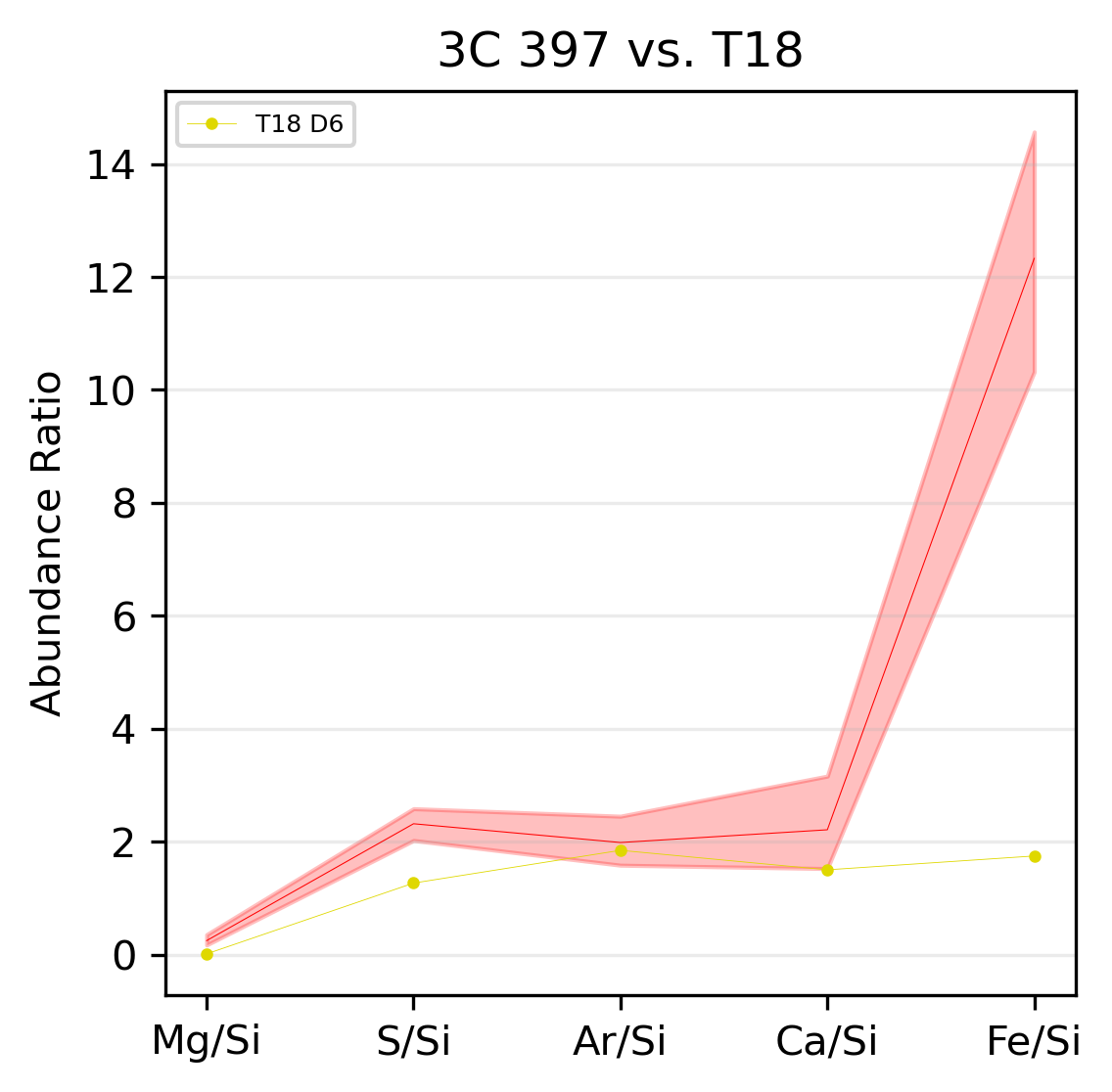}}
		\subfloat{\includegraphics[angle=0,width=0.40\textwidth]{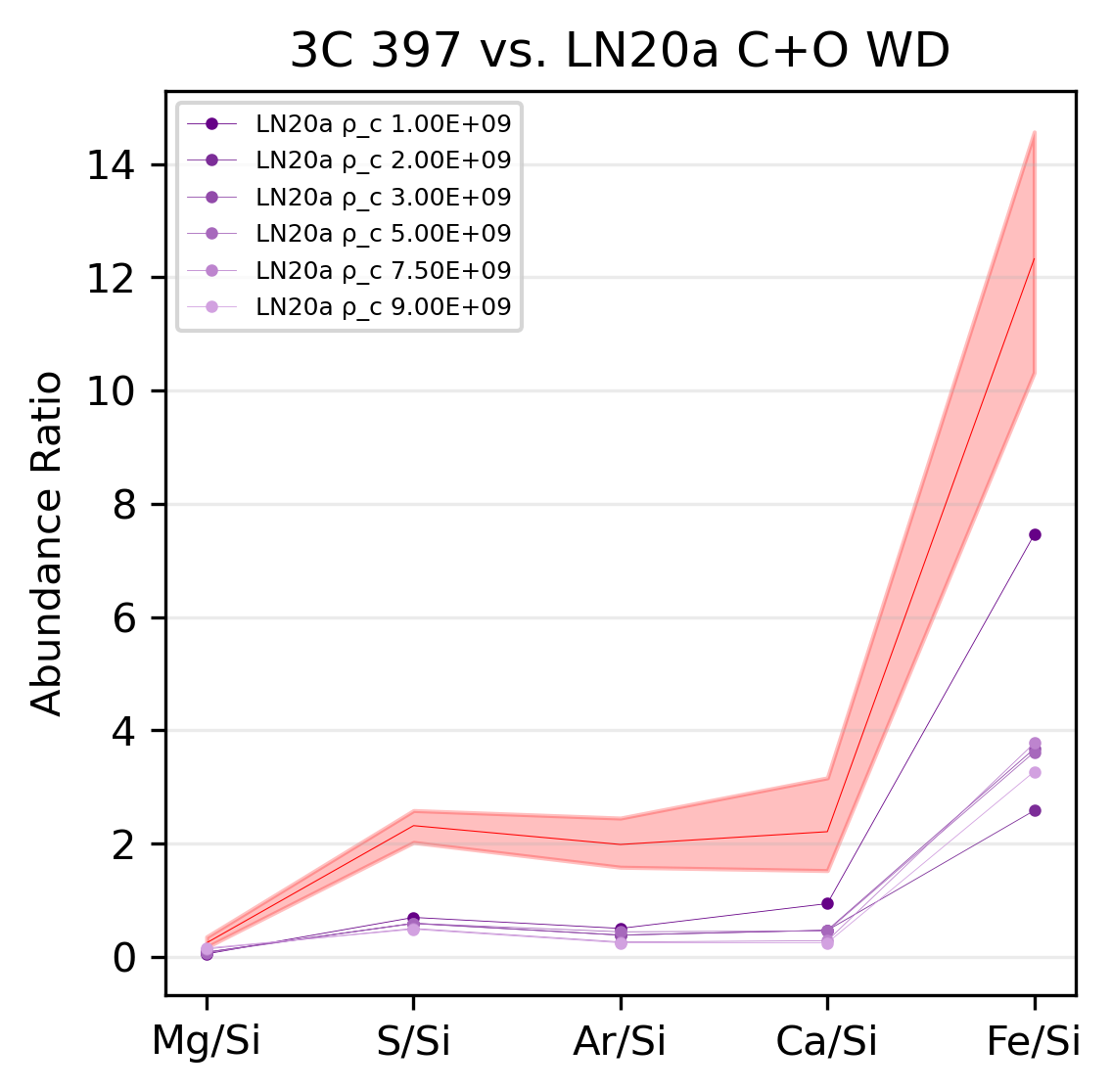}} \\
		\subfloat{\includegraphics[angle=0,width=0.40\textwidth,scale=0.5]{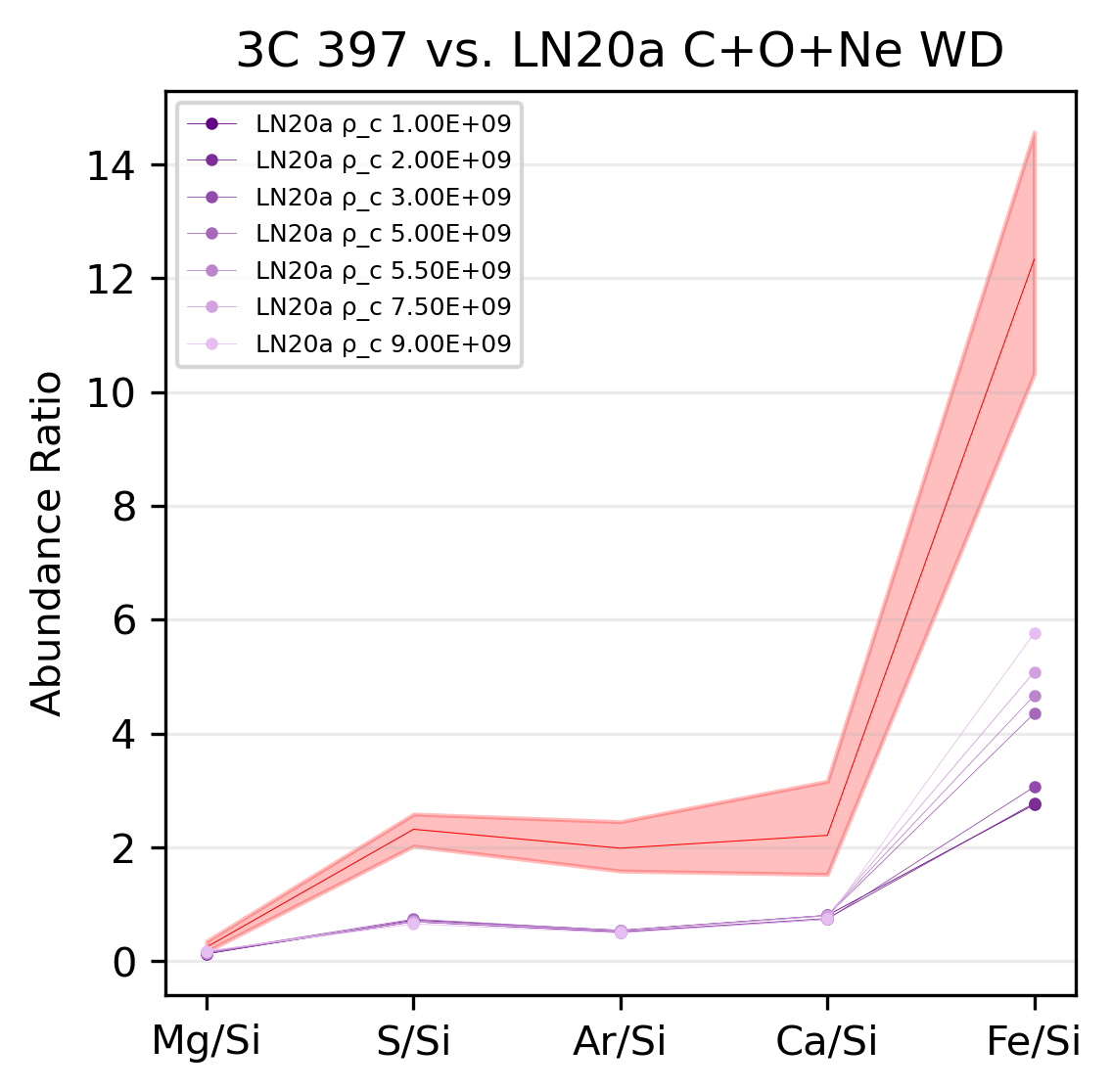}}
		\subfloat{\includegraphics[angle=0,width=0.40\textwidth]{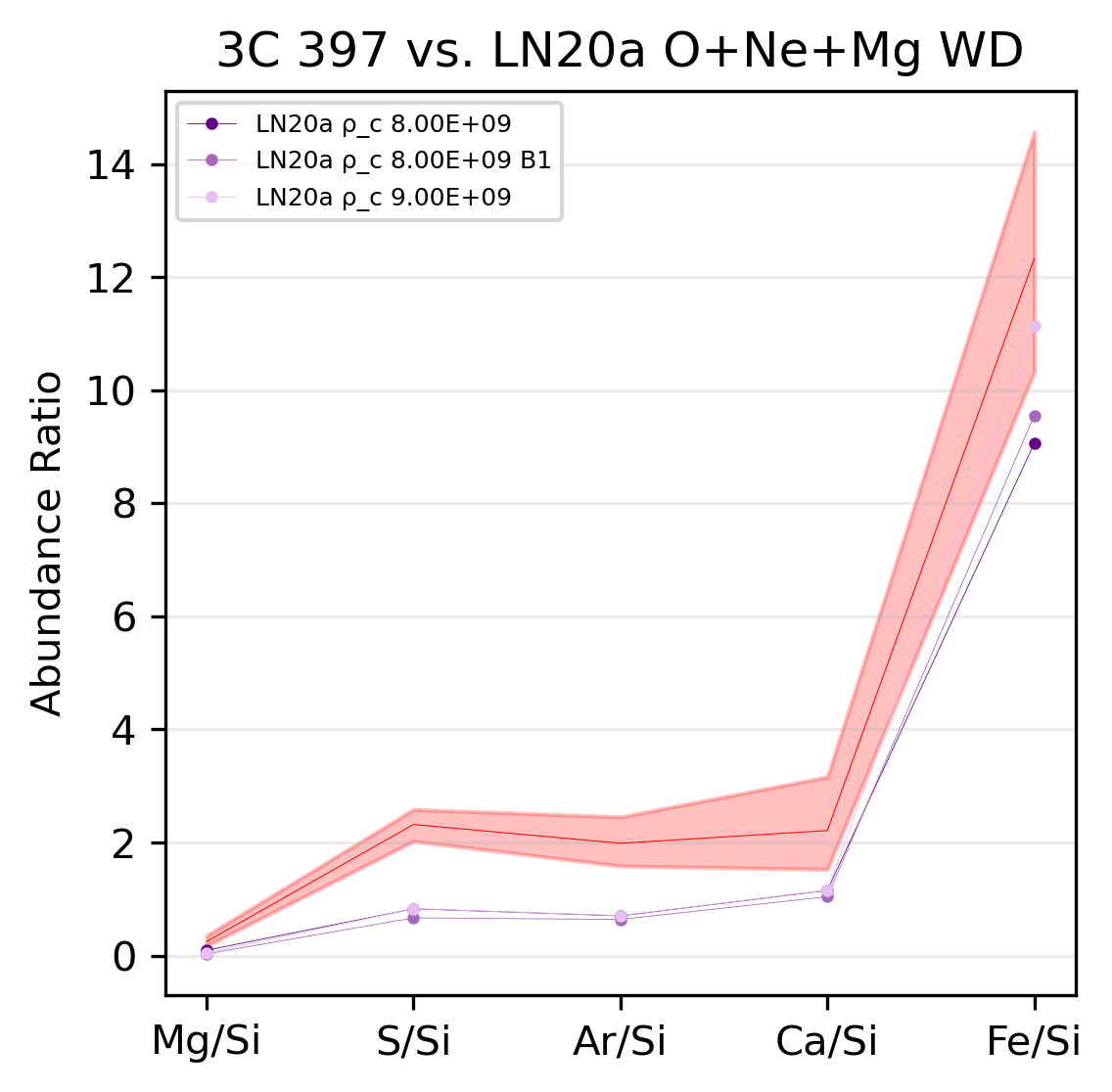}}
	\end{center}
    {Continued from above.}
\end{figure*}

\begin{figure*}\ContinuedFloat
	\begin{center}
		\subfloat{\includegraphics[angle=0,width=0.40\textwidth,scale=0.5]{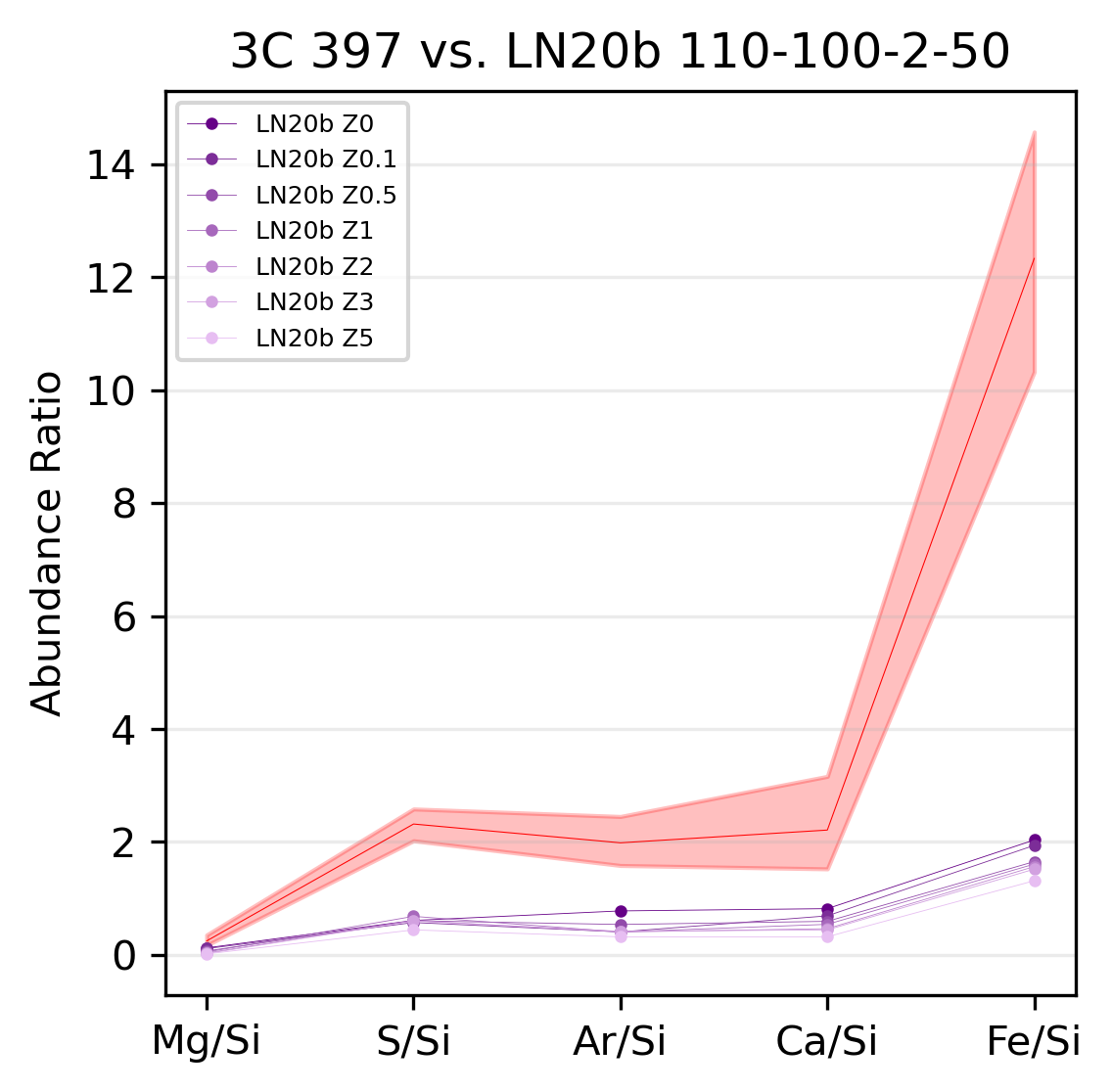}}
		\subfloat{\includegraphics[angle=0,width=0.40\textwidth,scale=0.5]{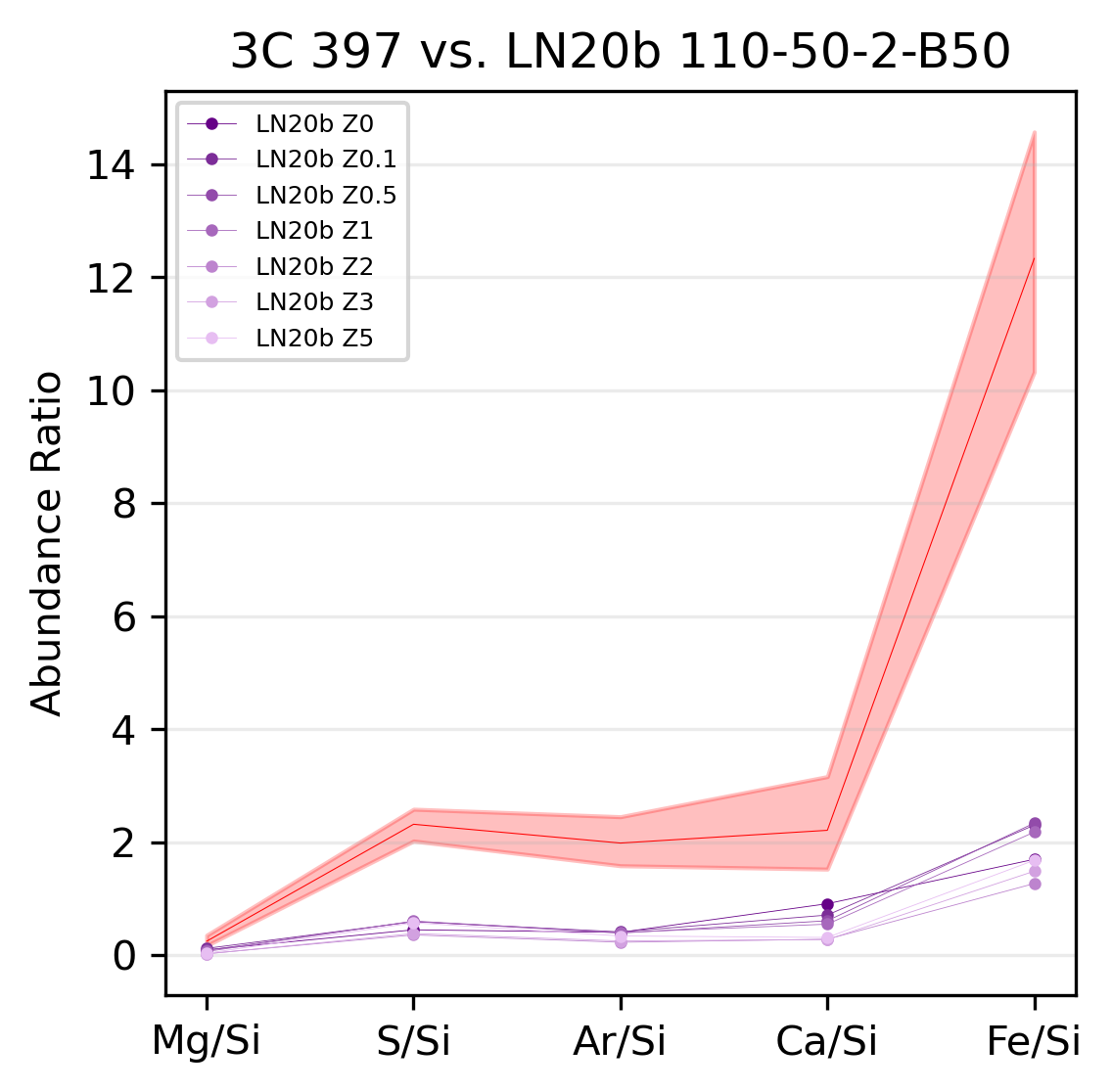}} \\
		\subfloat{\includegraphics[angle=0,width=0.40\textwidth]{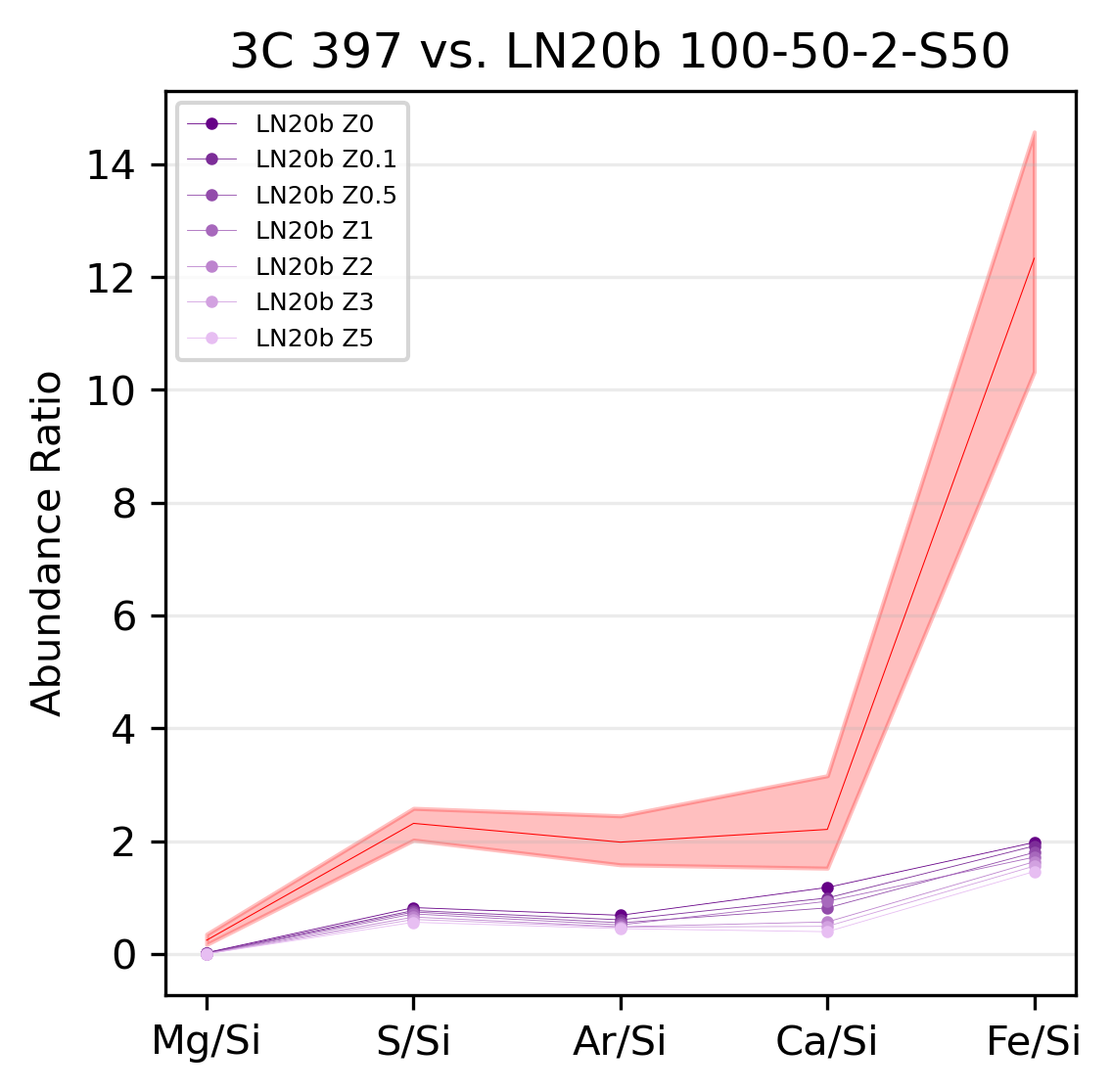}}
		\subfloat{\includegraphics[angle=0,width=0.40\textwidth]{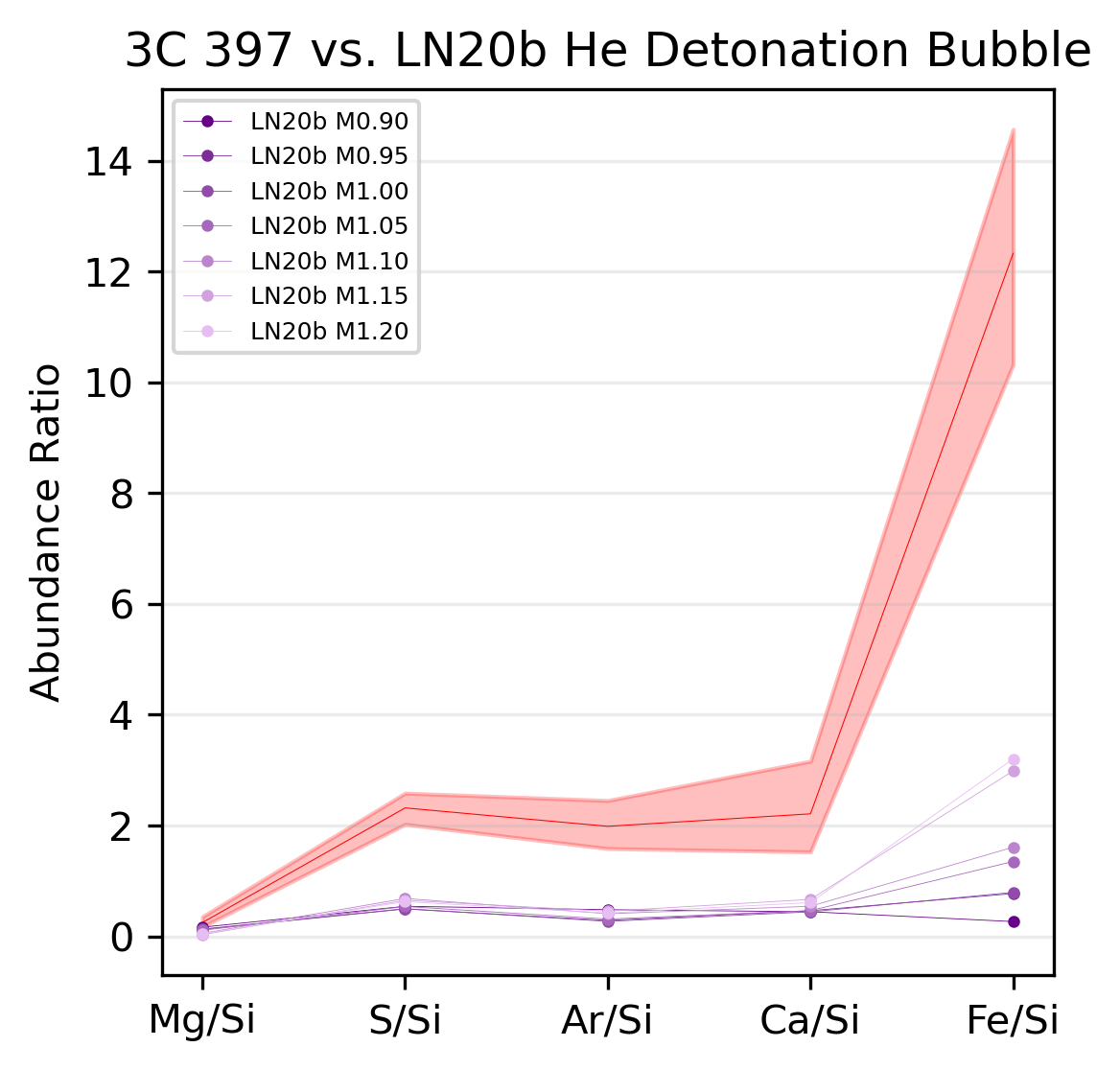}} \\
		\subfloat{\includegraphics[angle=0,width=0.40\textwidth,scale=0.5]{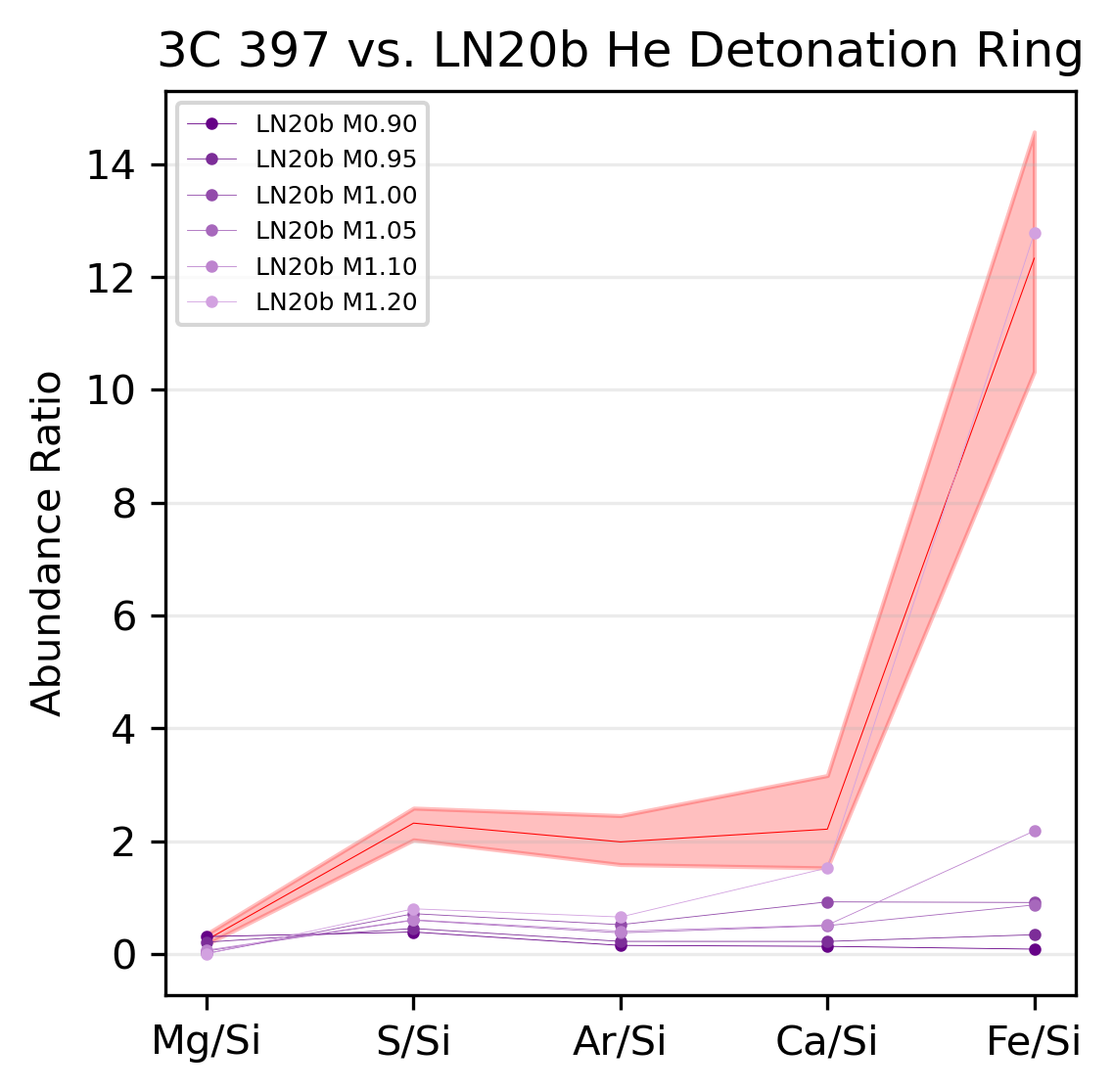}}
		\subfloat{\includegraphics[angle=0,width=0.40\textwidth]{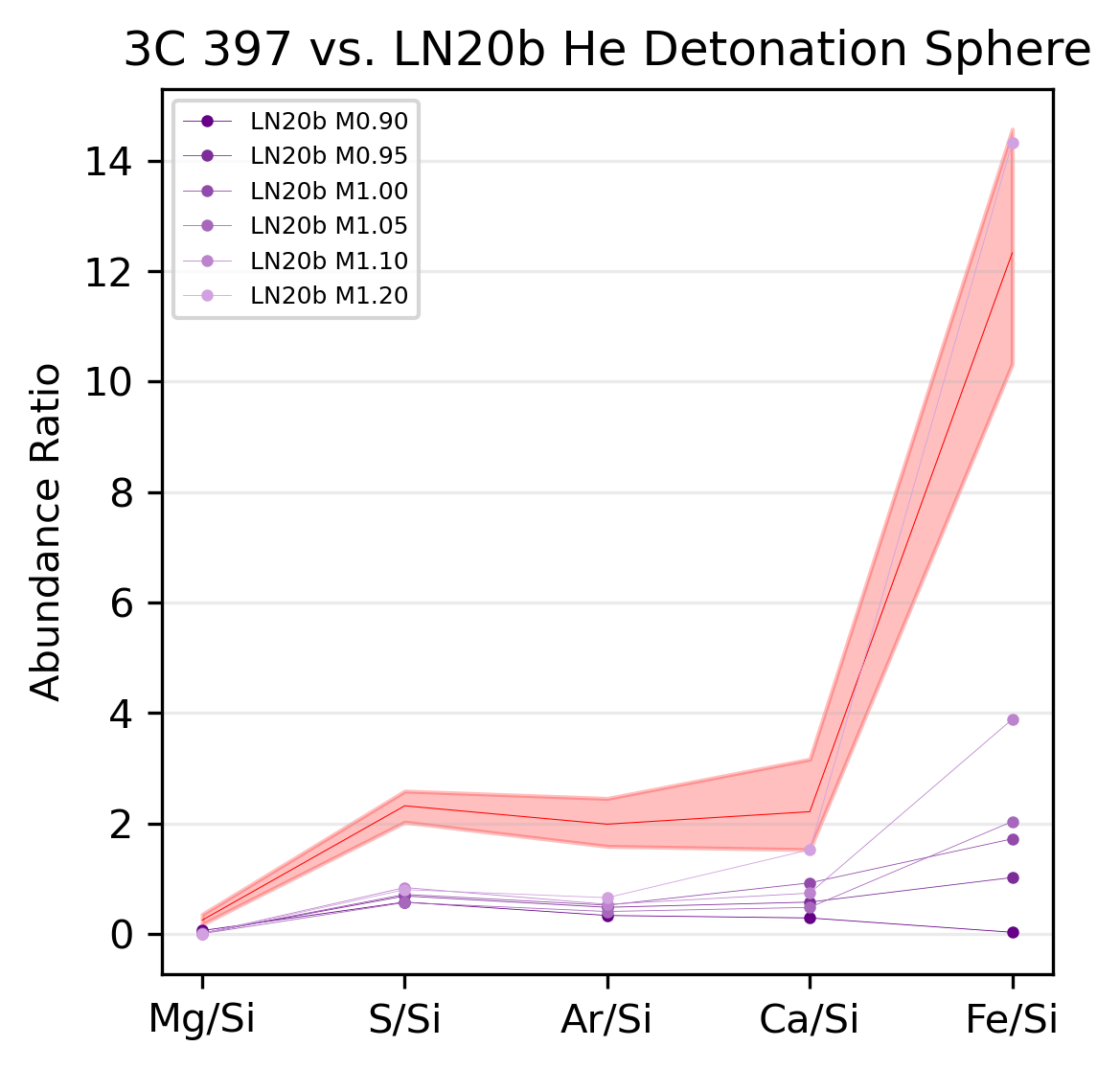}}
	\end{center}
    {Continued from above.}
\end{figure*}

\begin{figure*}
	\begin{center}
		\subfloat{\includegraphics[angle=0,width=0.40\textwidth,scale=0.5]{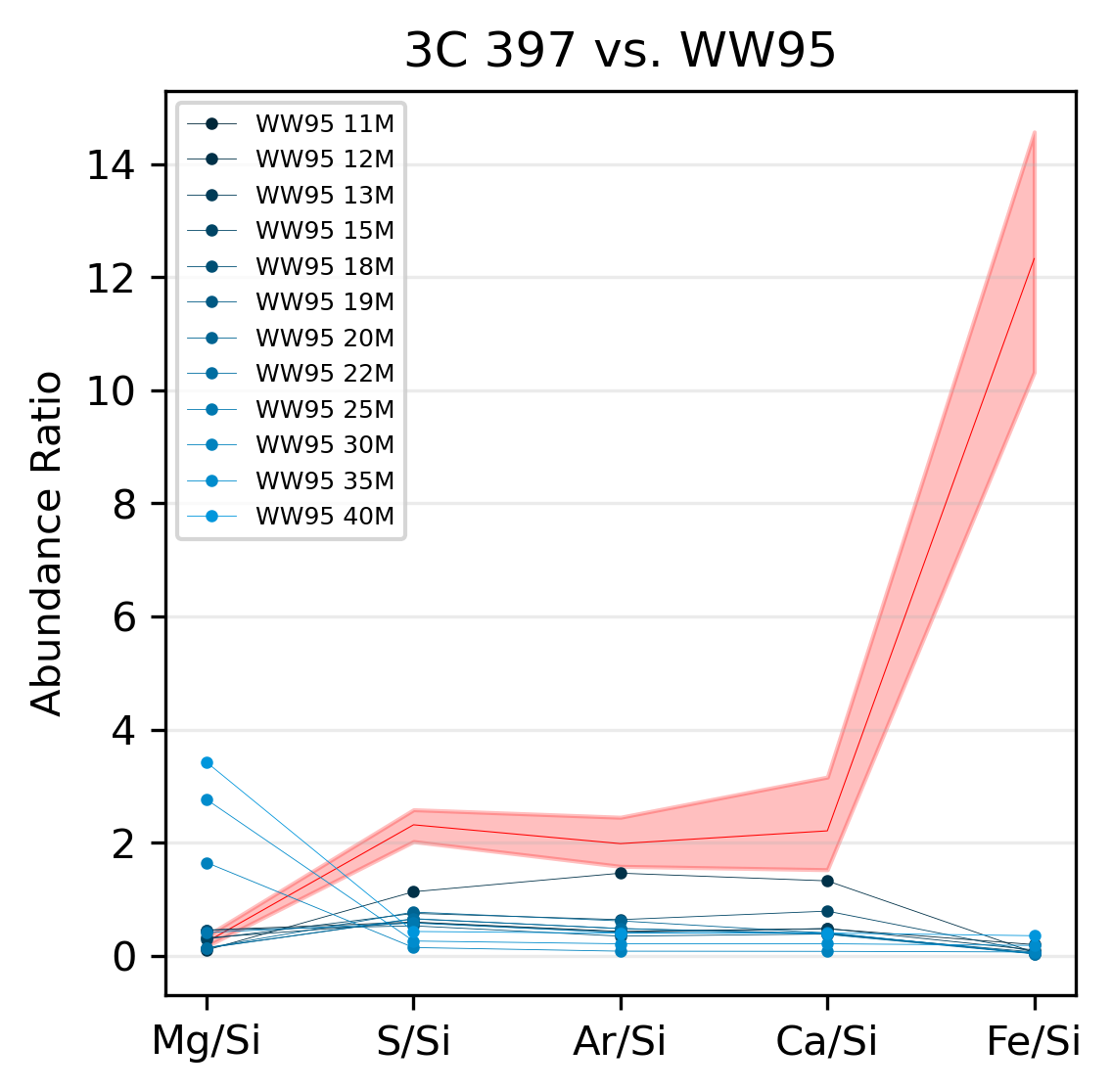}}
		\subfloat{\includegraphics[angle=0,width=0.40\textwidth,scale=0.5]{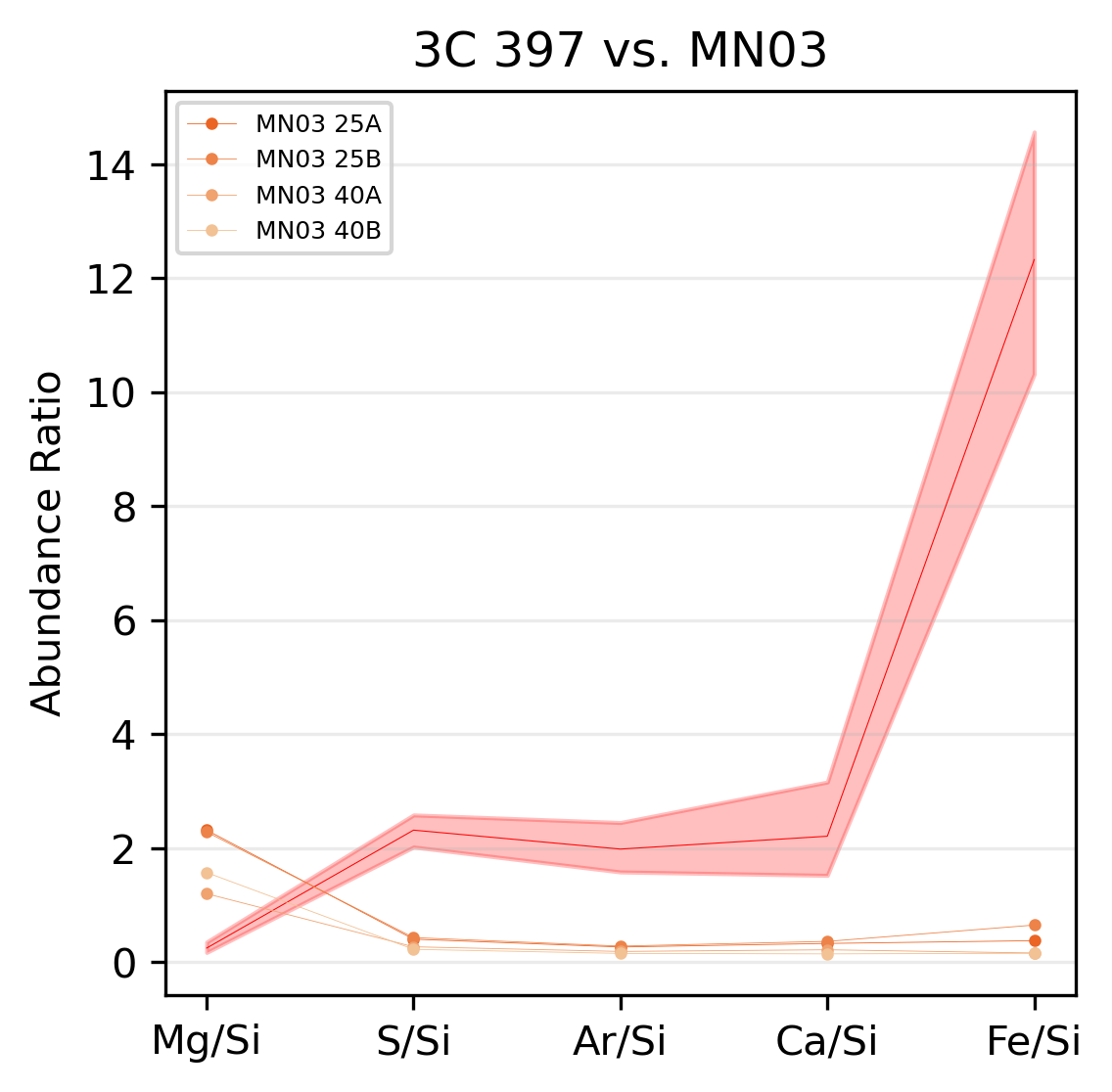}} \\
		\subfloat{\includegraphics[angle=0,width=0.40\textwidth]{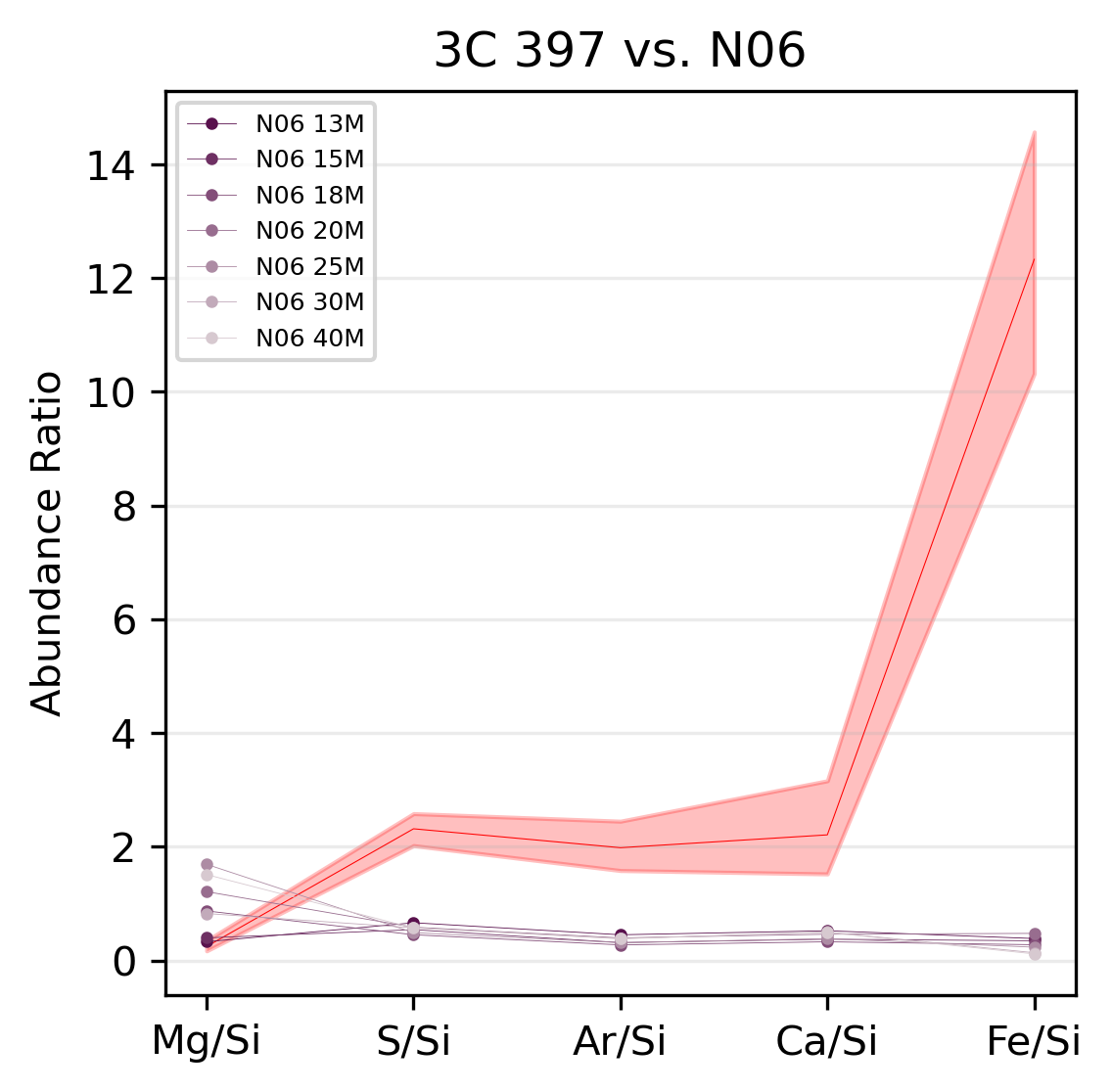}}
		\subfloat{\includegraphics[angle=0,width=0.40\textwidth]{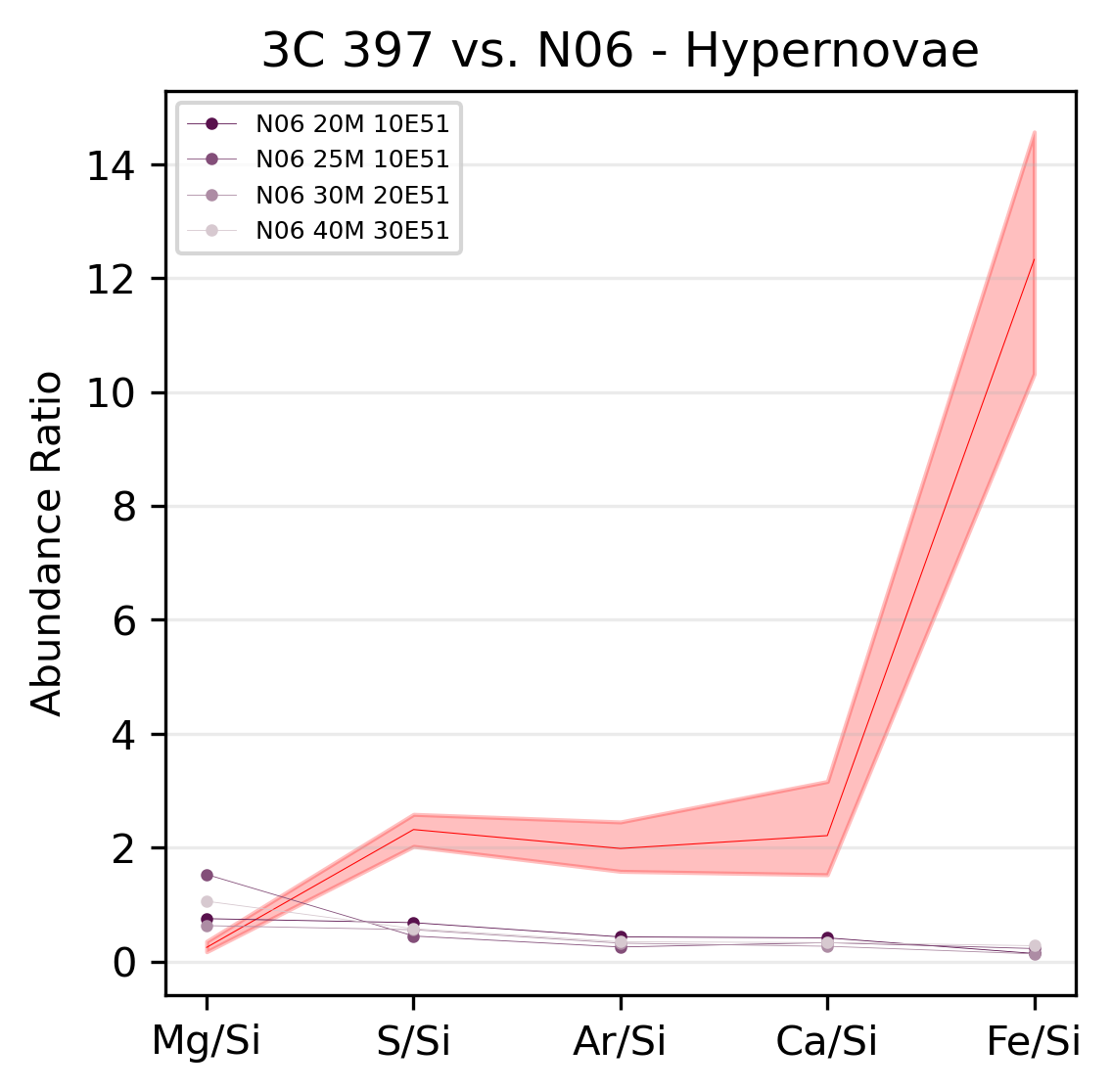}} \\
		\subfloat{\includegraphics[angle=0,width=0.40\textwidth,scale=0.5]{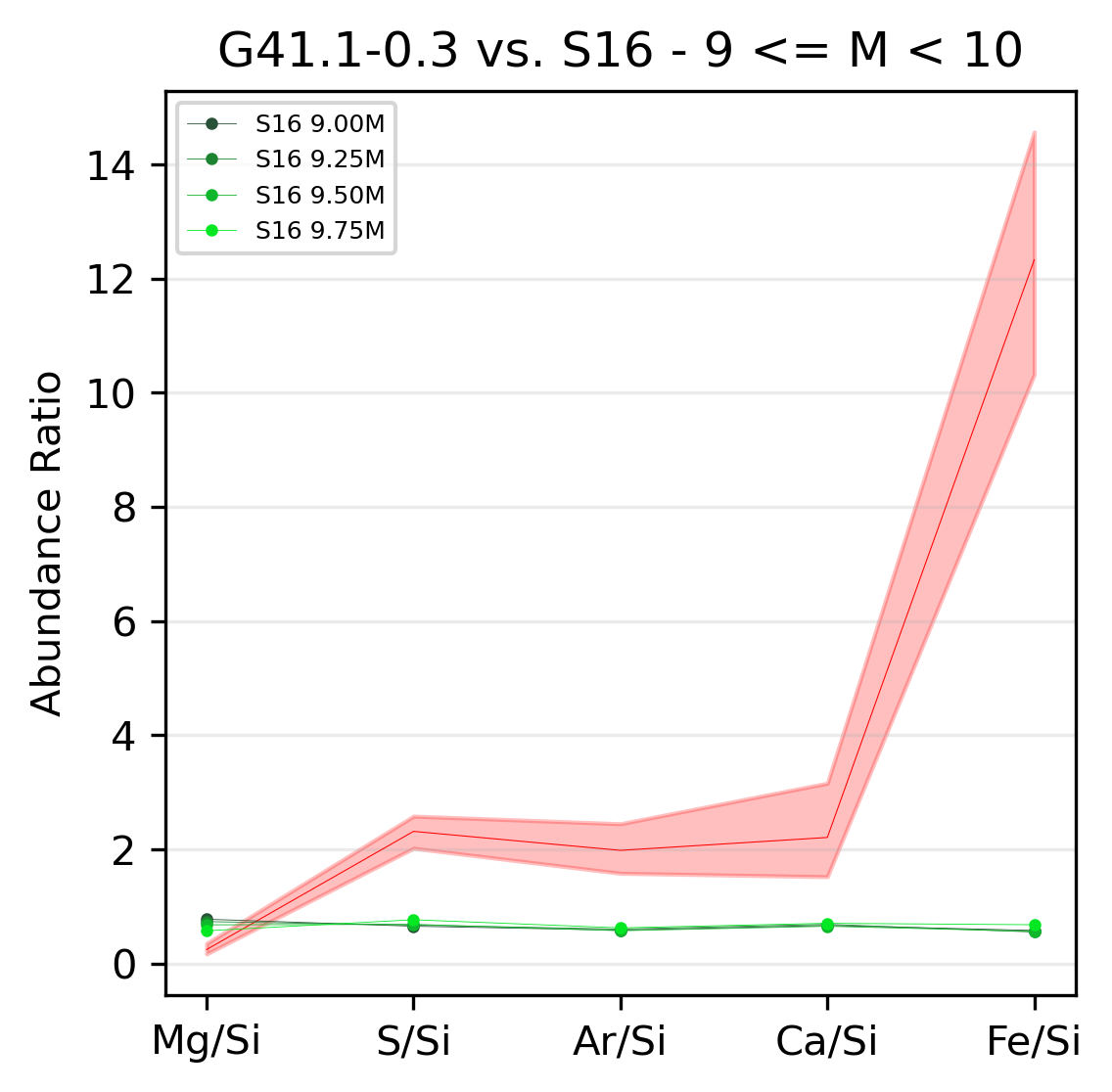}}
		\subfloat{\includegraphics[angle=0,width=0.40\textwidth]{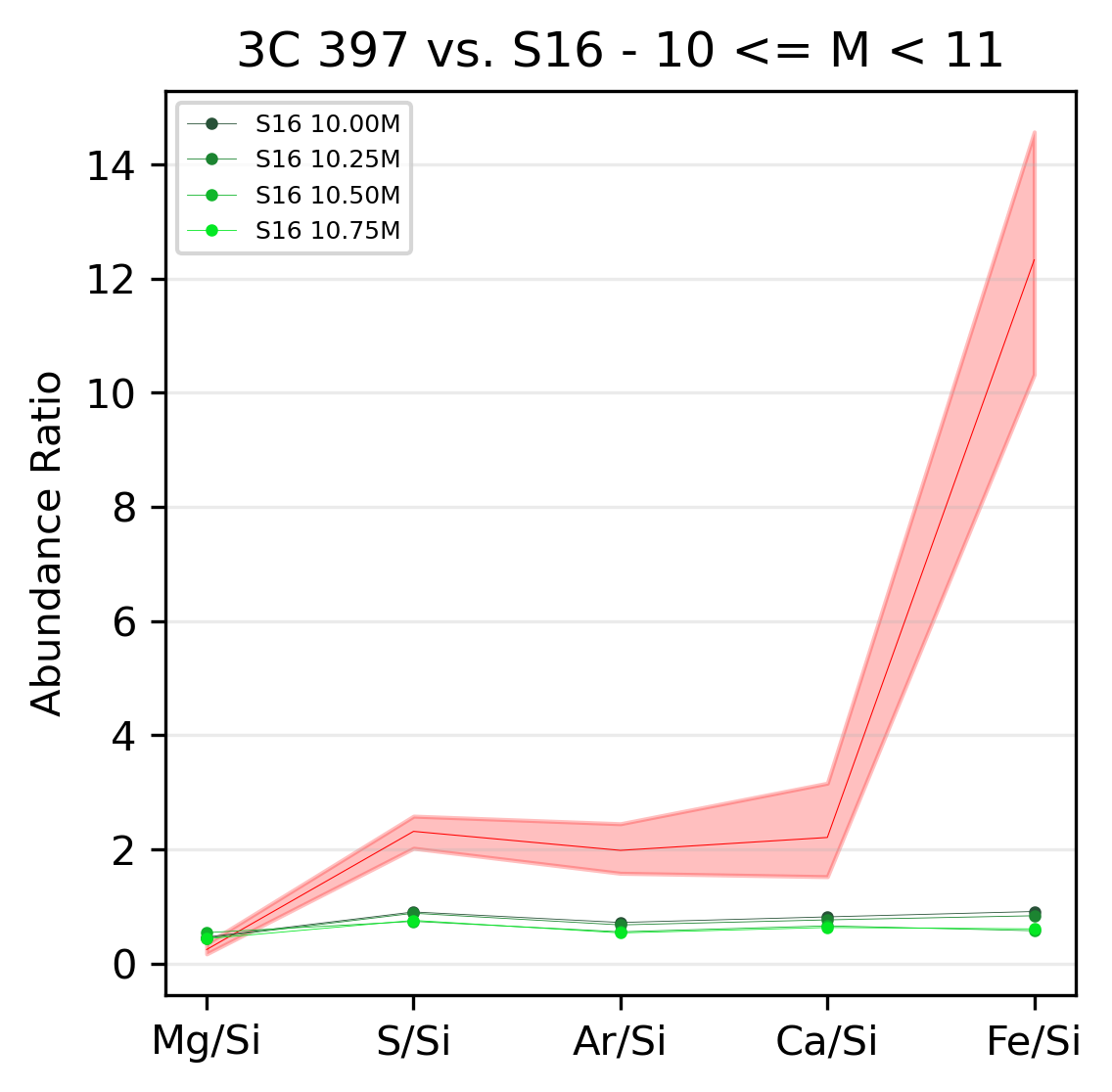}}
	\end{center}
    \caption{Nucleosynthesis comparisons between 3C 397 and the tested CC models.}
    \label{fig:3c397_cc}
\end{figure*}

\begin{figure*}\ContinuedFloat
	\begin{center}
		\subfloat{\includegraphics[angle=0,width=0.40\textwidth,scale=0.5]{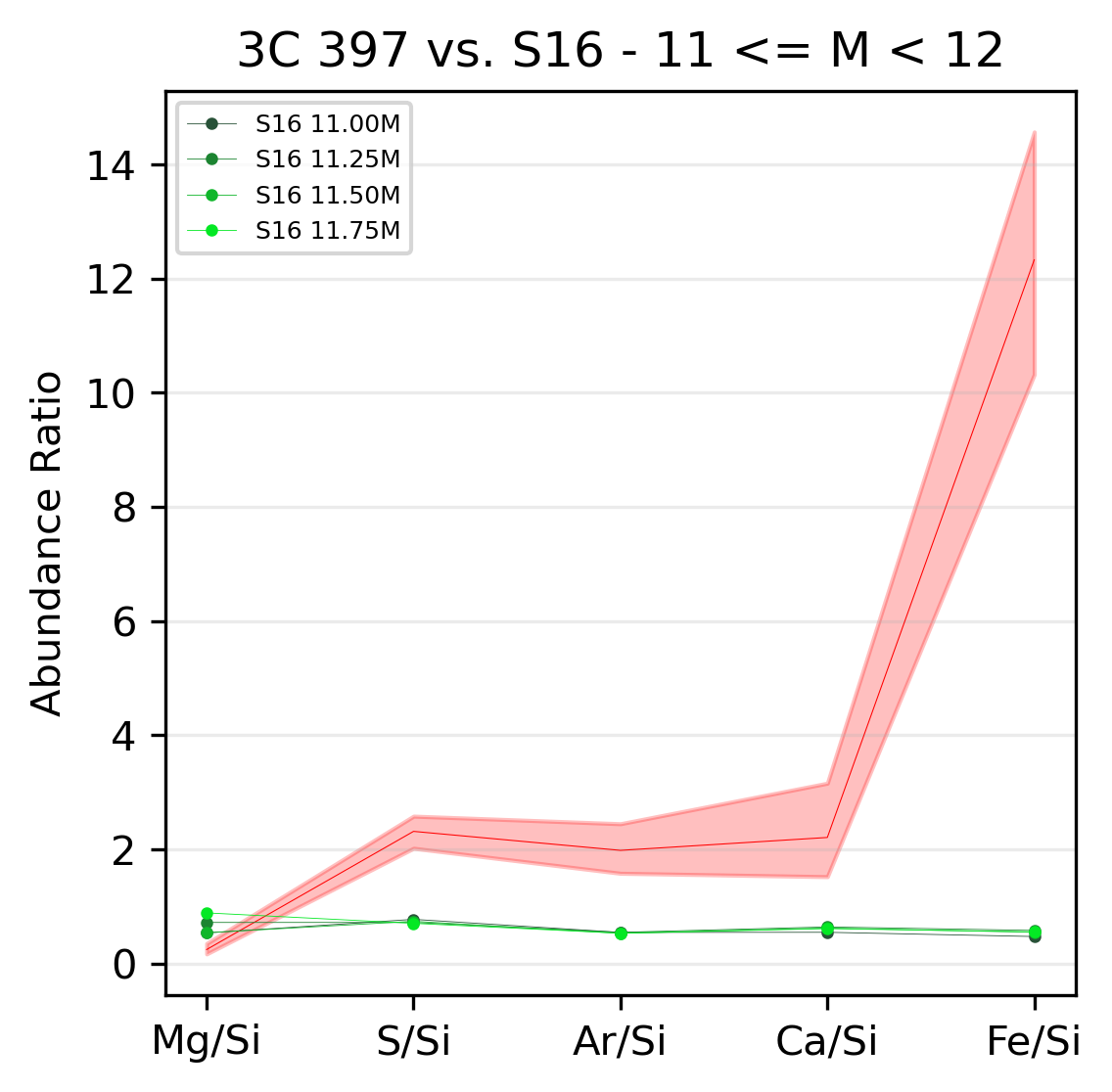}}
		\subfloat{\includegraphics[angle=0,width=0.40\textwidth,scale=0.5]{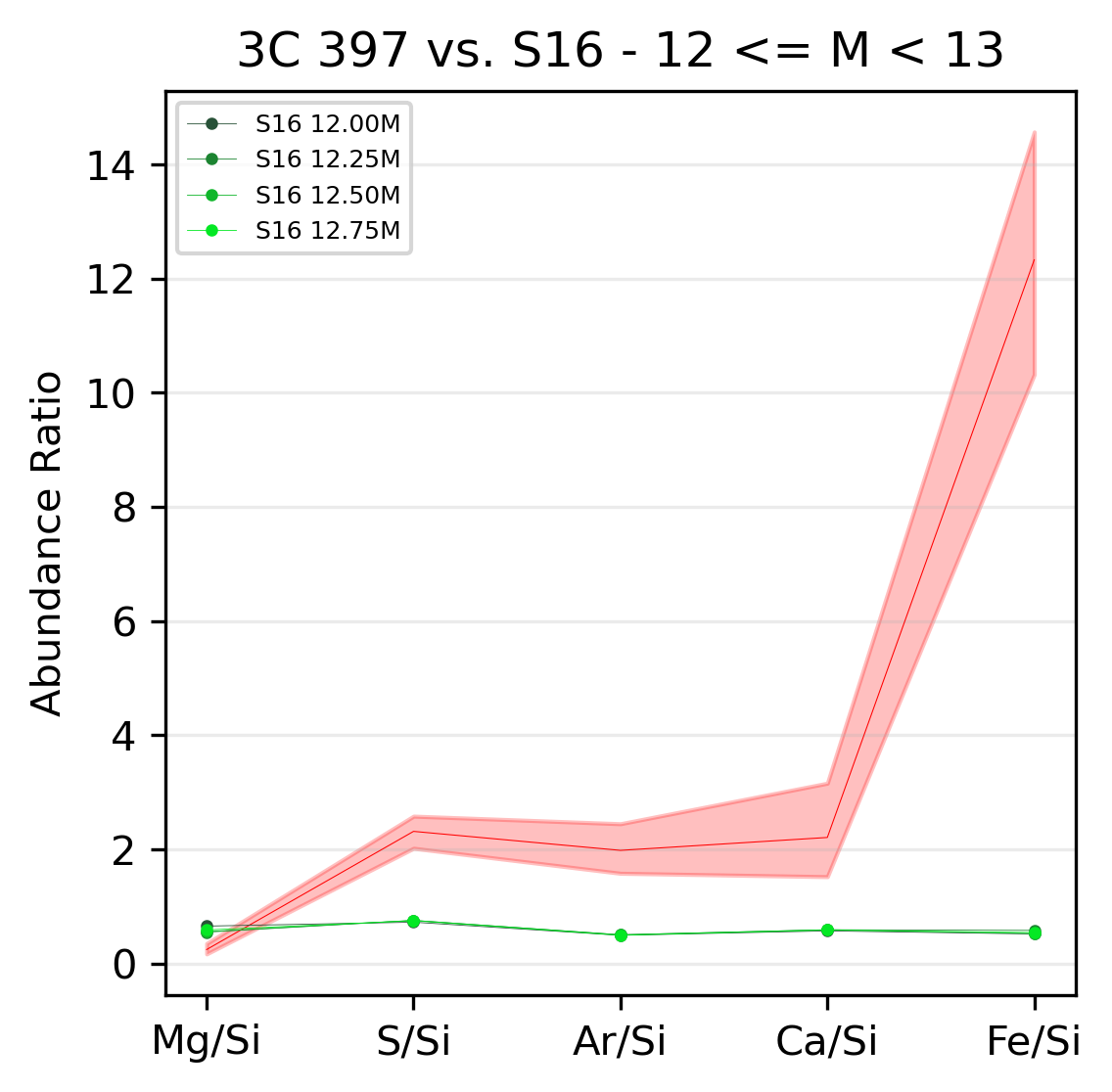}} \\
		\subfloat{\includegraphics[angle=0,width=0.40\textwidth]{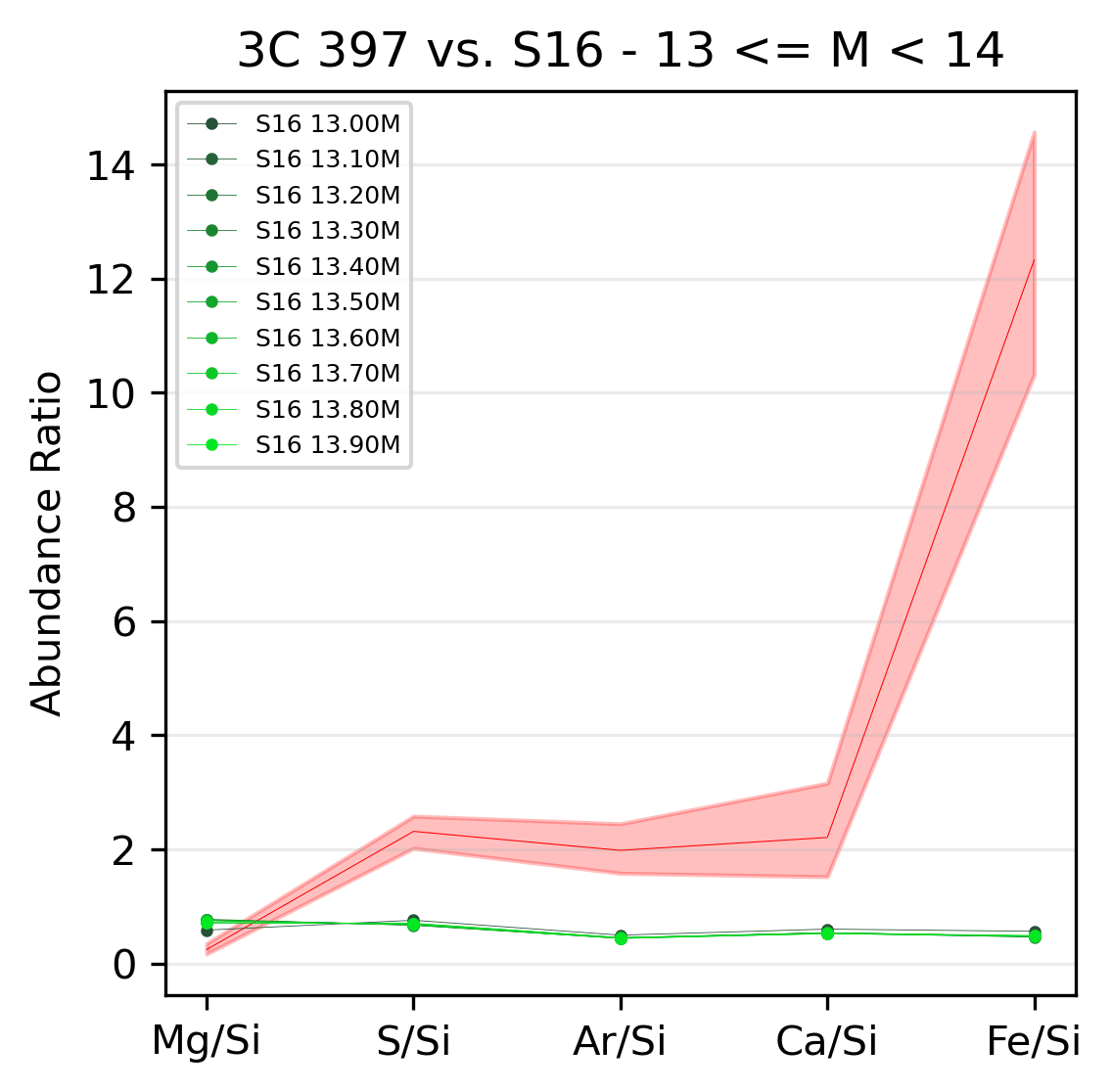}}
		\subfloat{\includegraphics[angle=0,width=0.40\textwidth]{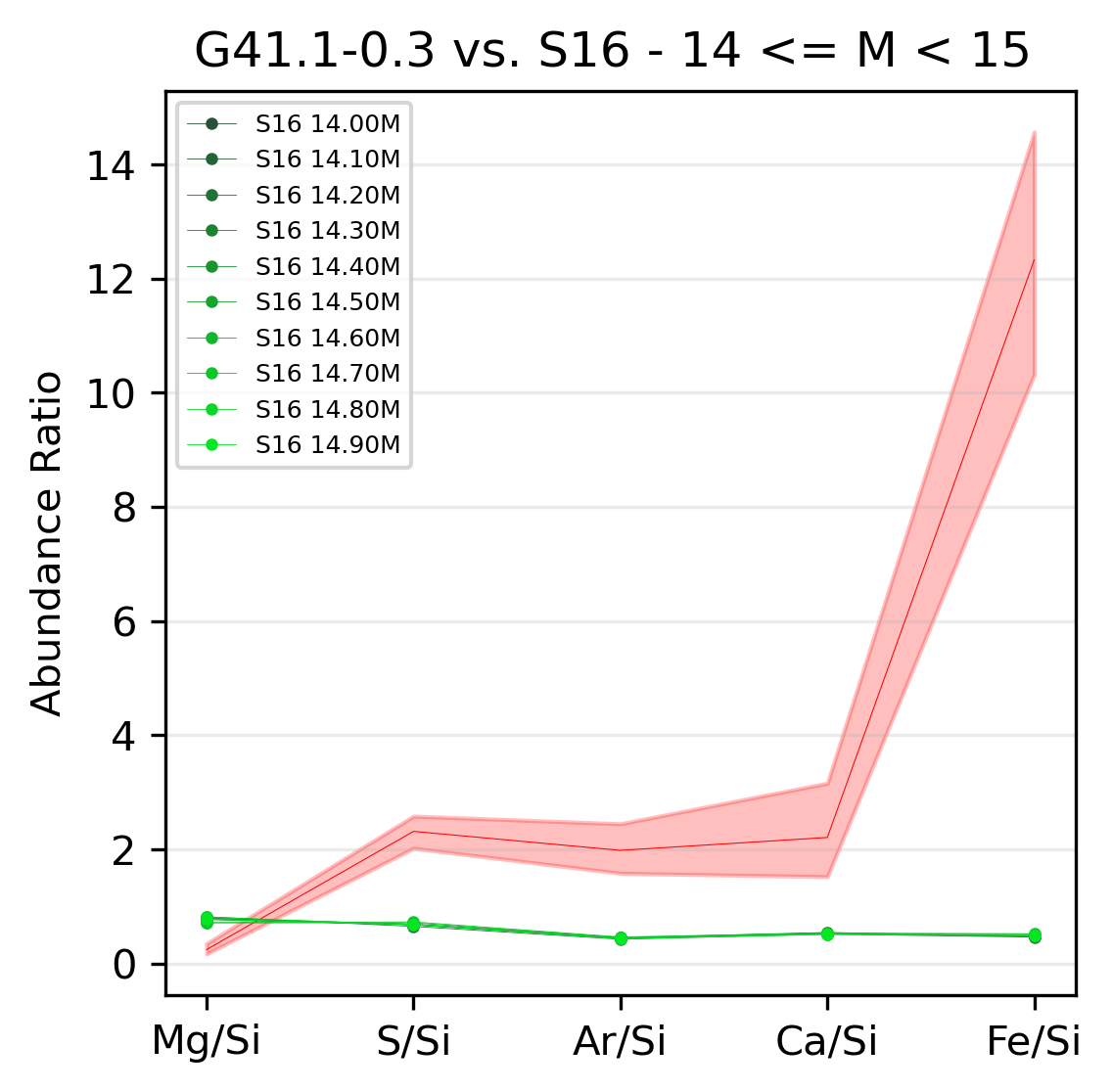}} \\
		\subfloat{\includegraphics[angle=0,width=0.40\textwidth,scale=0.5]{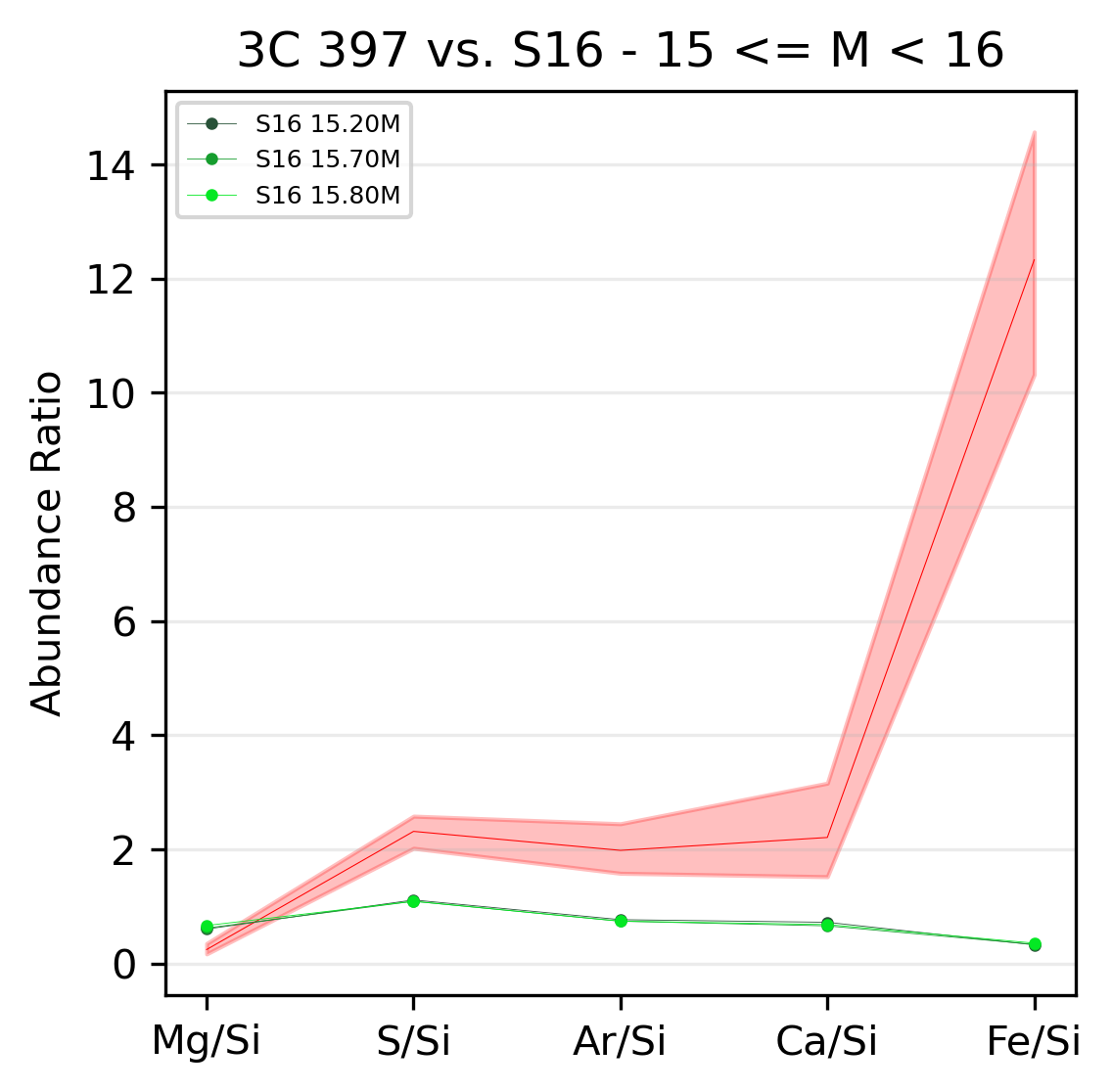}}
		\subfloat{\includegraphics[angle=0,width=0.40\textwidth]{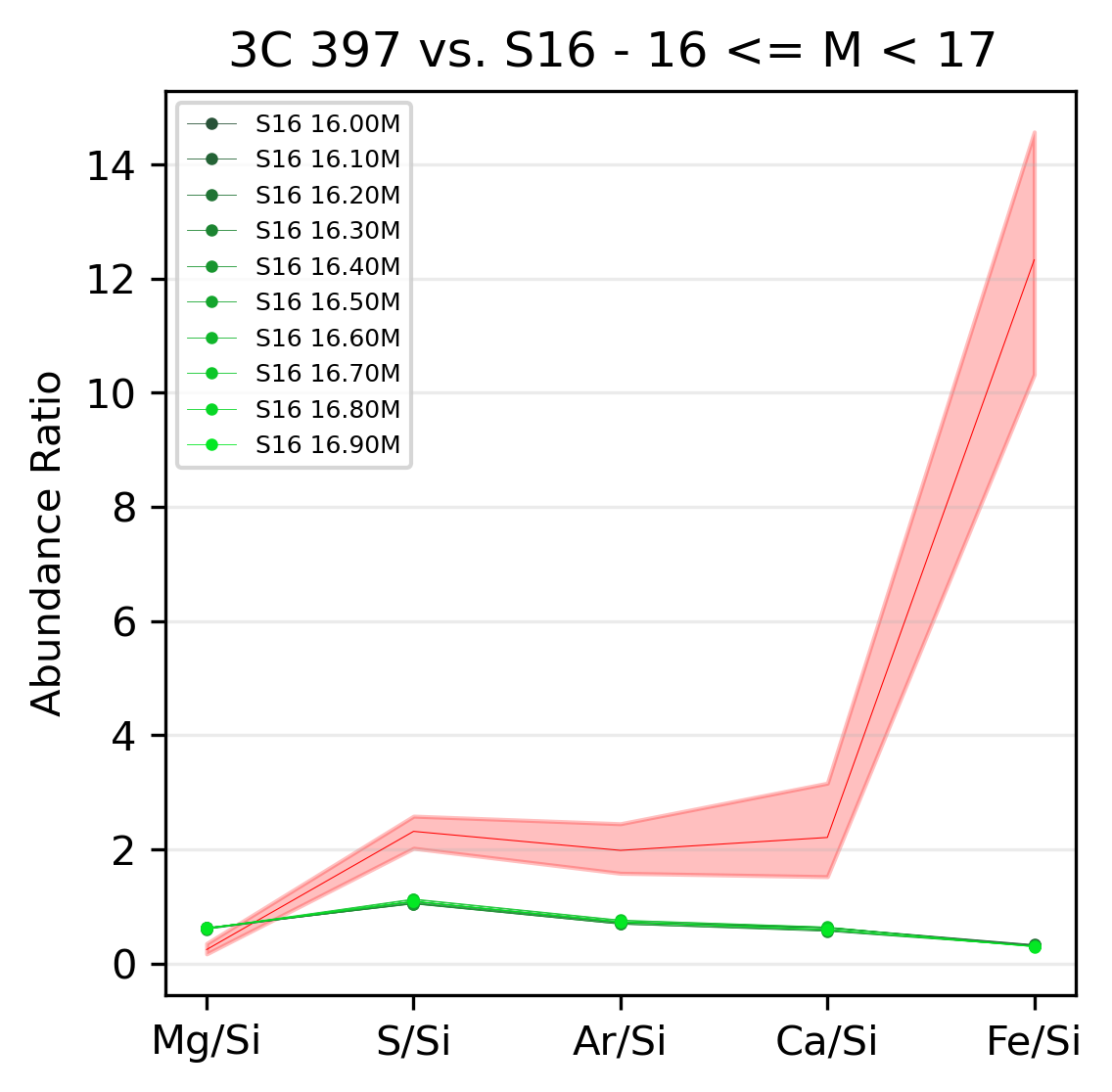}}
	\end{center}
    {Continued from above.}
\end{figure*}

\begin{figure*}\ContinuedFloat
	\begin{center}
		\subfloat{\includegraphics[angle=0,width=0.40\textwidth,scale=0.5]{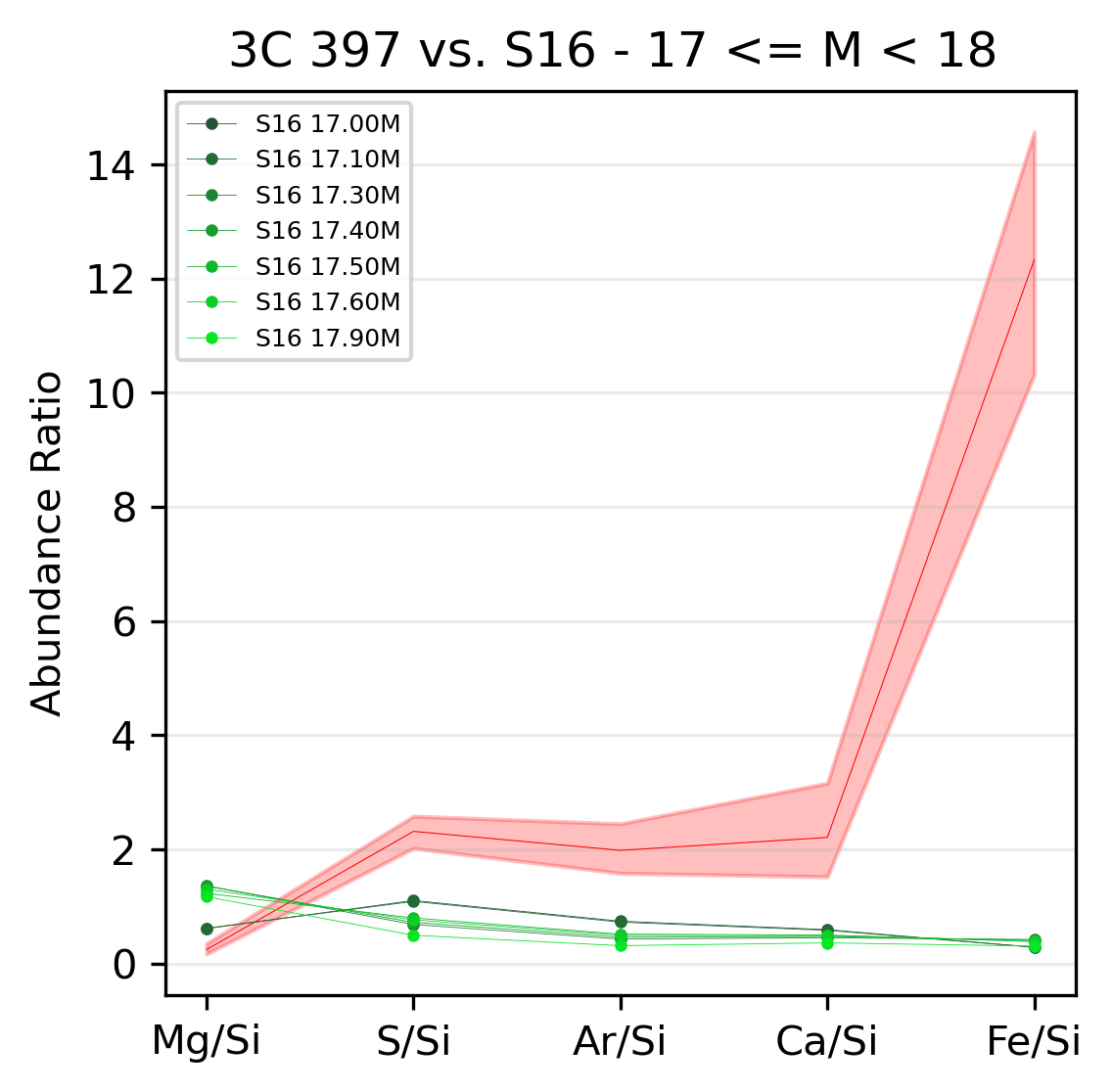}}
		\subfloat{\includegraphics[angle=0,width=0.40\textwidth,scale=0.5]{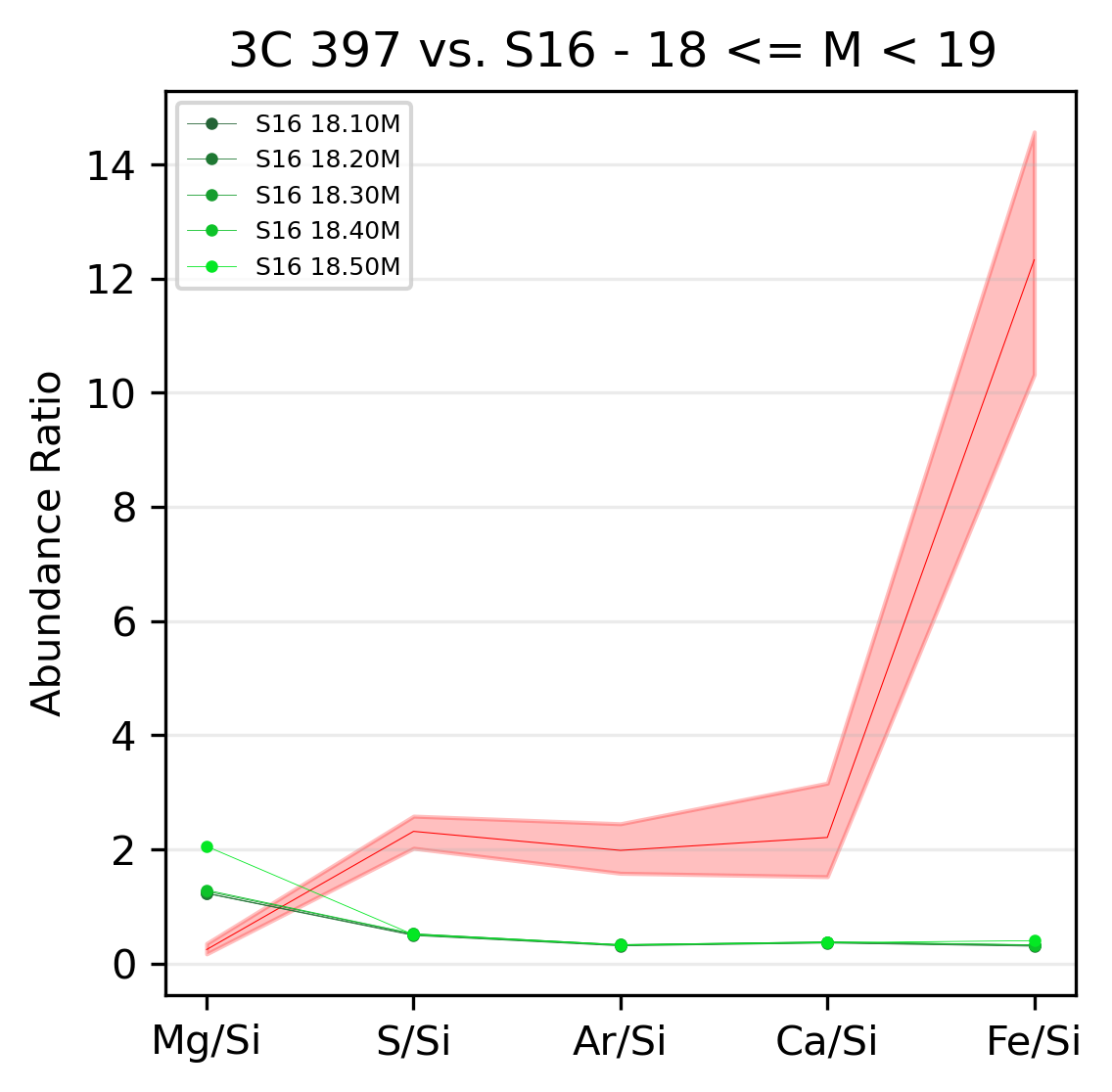}} \\
		\subfloat{\includegraphics[angle=0,width=0.40\textwidth]{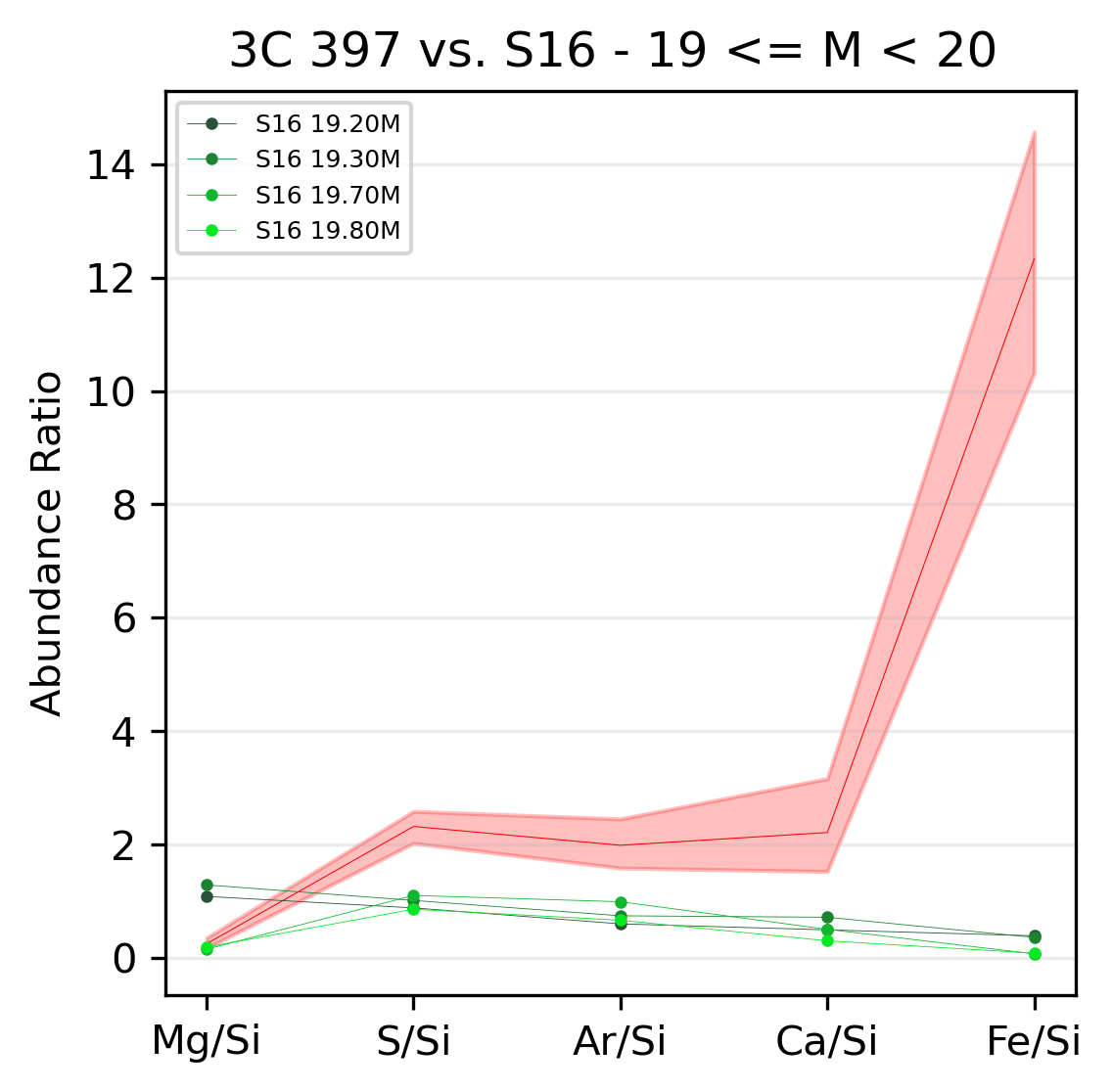}}
		\subfloat{\includegraphics[angle=0,width=0.40\textwidth]{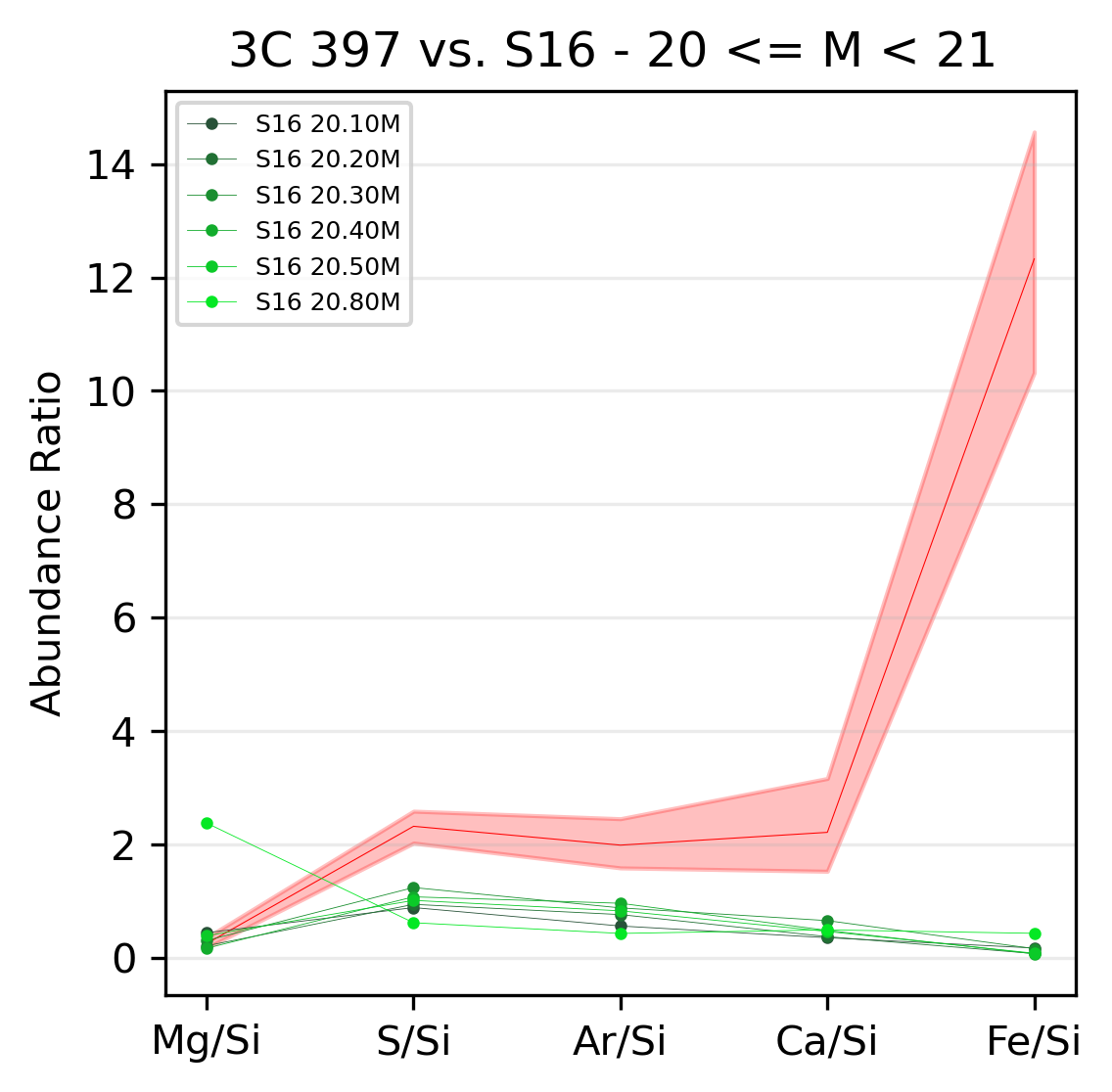}} \\
		\subfloat{\includegraphics[angle=0,width=0.40\textwidth,scale=0.5]{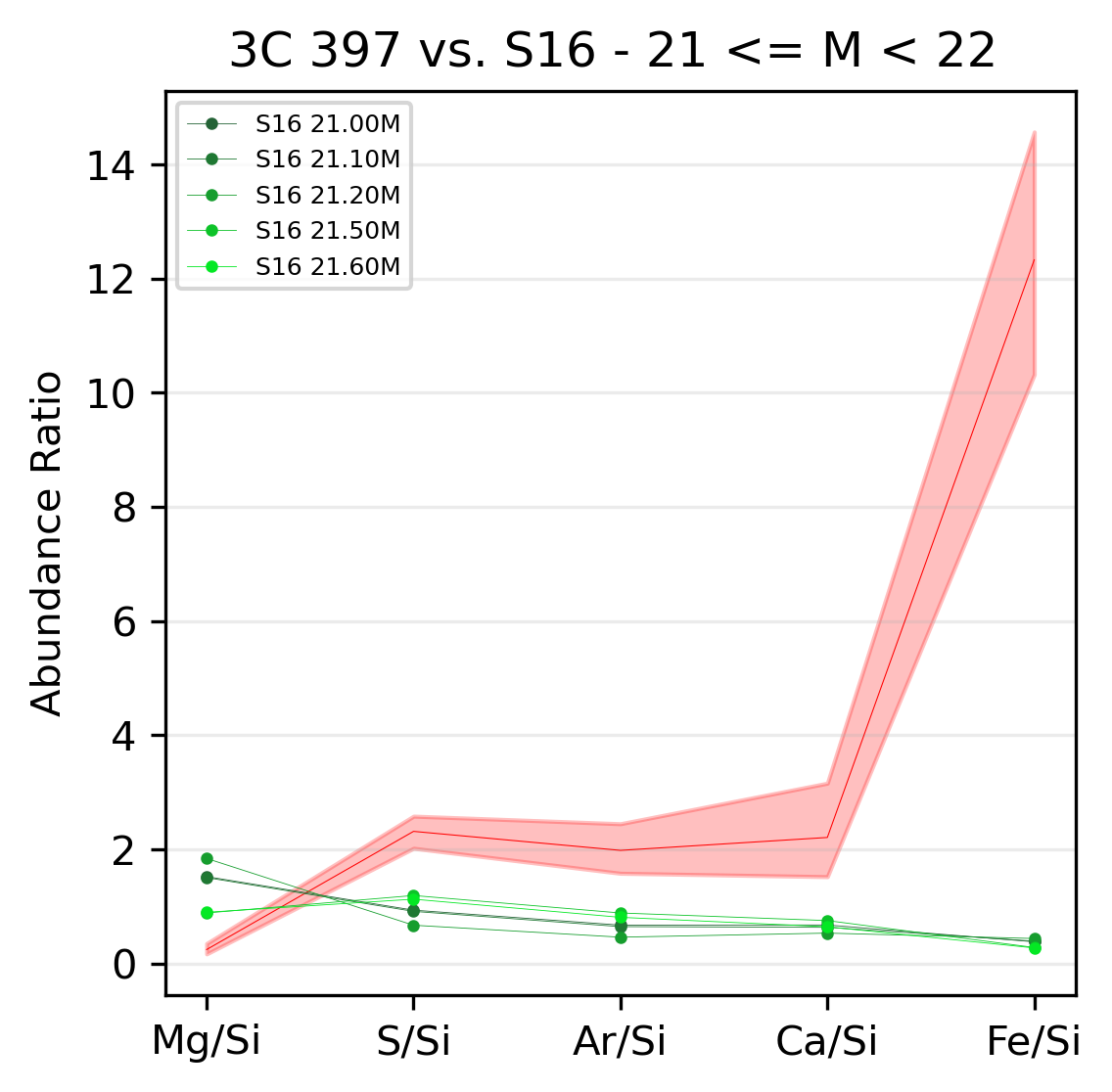}}
		\subfloat{\includegraphics[angle=0,width=0.40\textwidth]{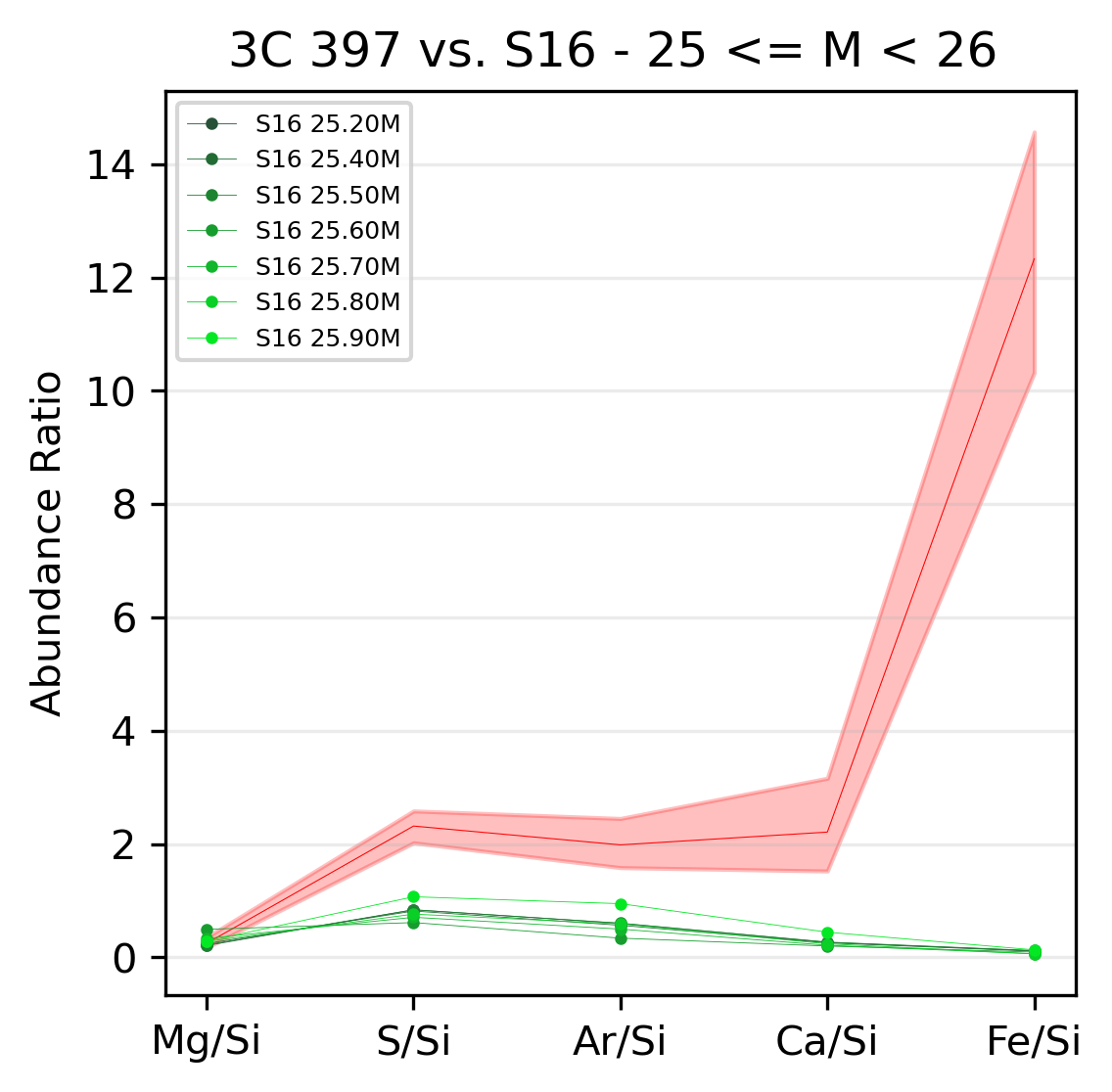}}
	\end{center}
    {Continued from above.}
\end{figure*}

\begin{figure*}\ContinuedFloat
	\begin{center}
		\subfloat{\includegraphics[angle=0,width=0.40\textwidth,scale=0.5]{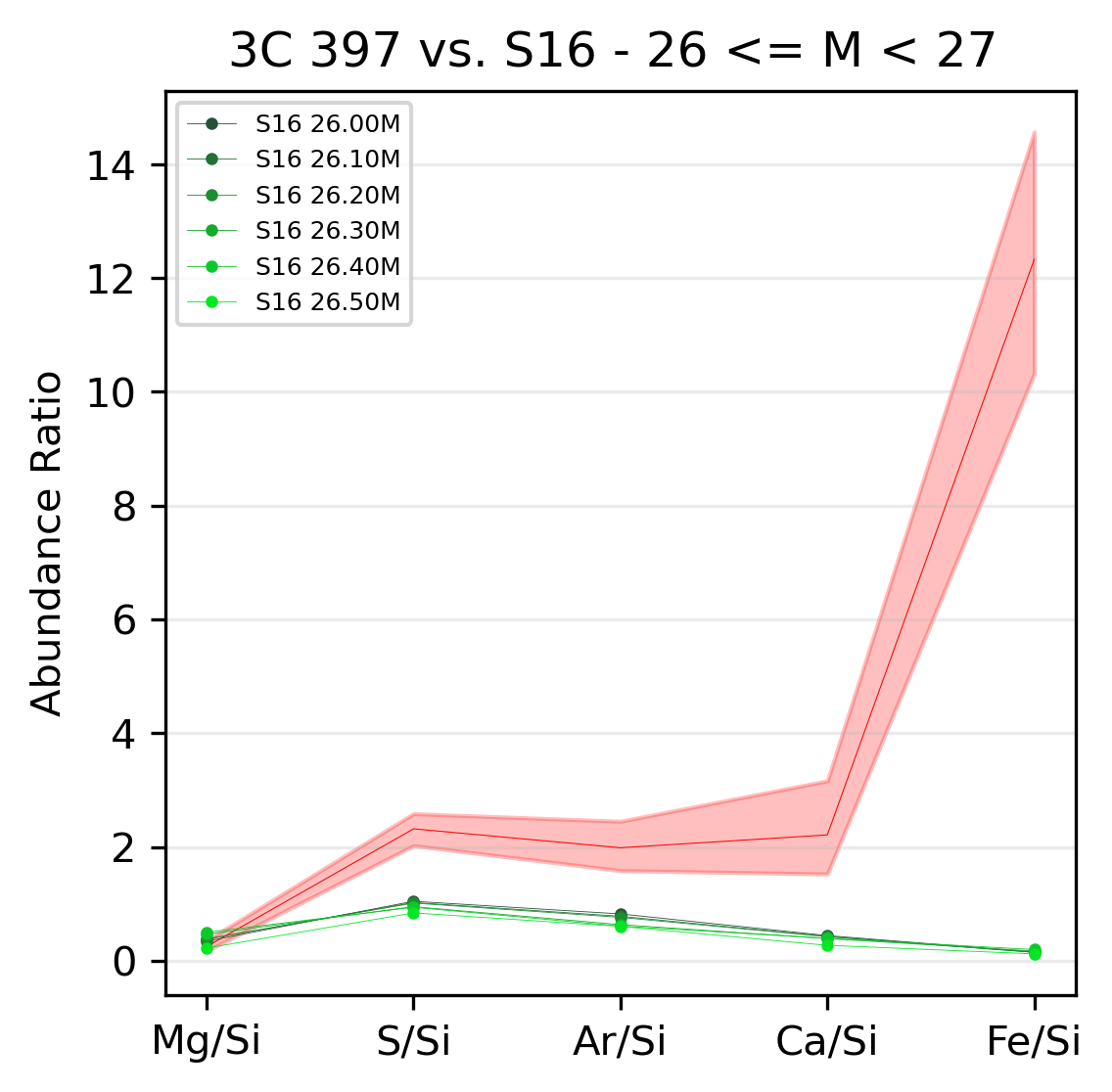}}
		\subfloat{\includegraphics[angle=0,width=0.40\textwidth,scale=0.5]{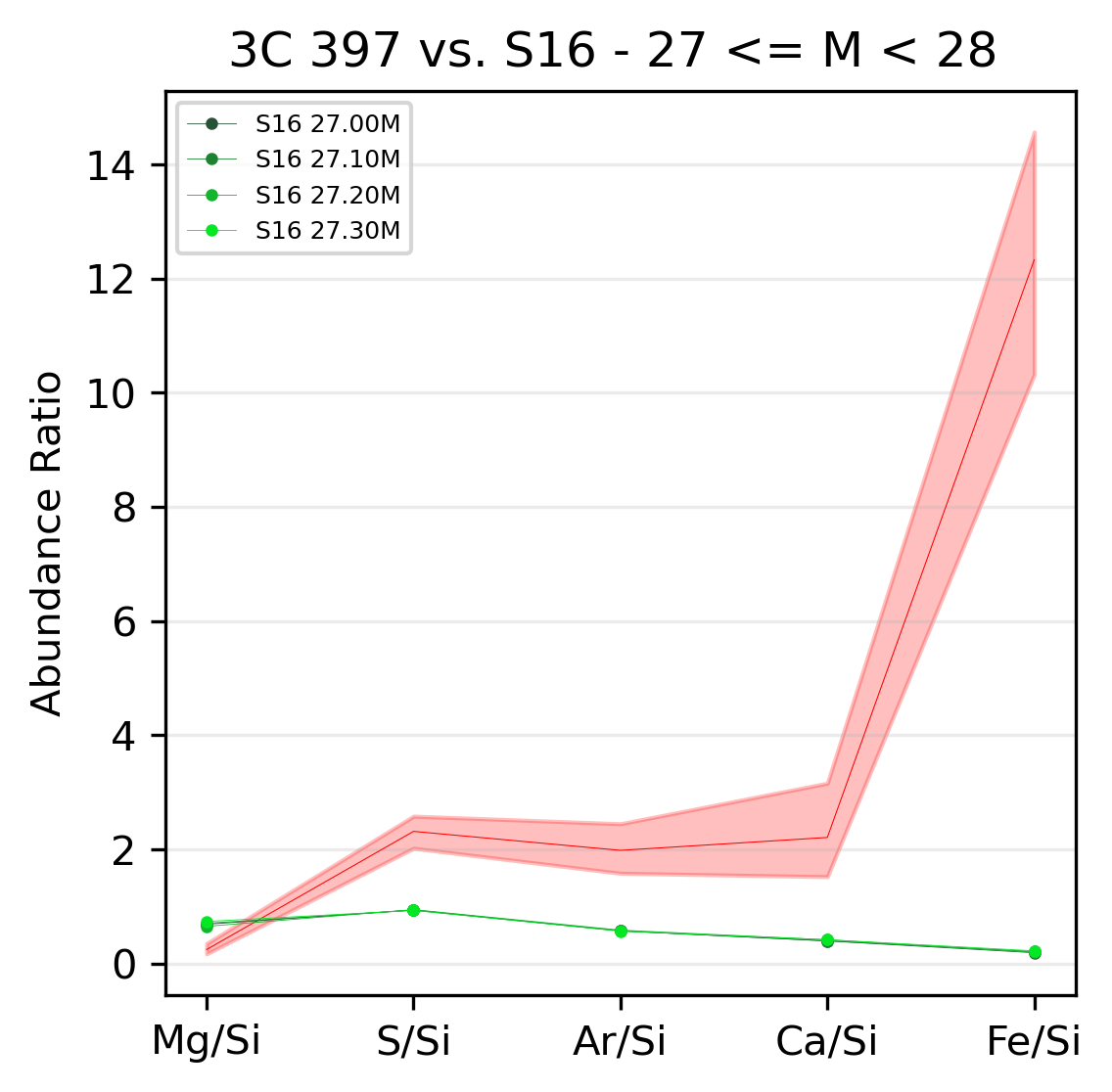}} \\
		\subfloat{\includegraphics[angle=0,width=0.40\textwidth]{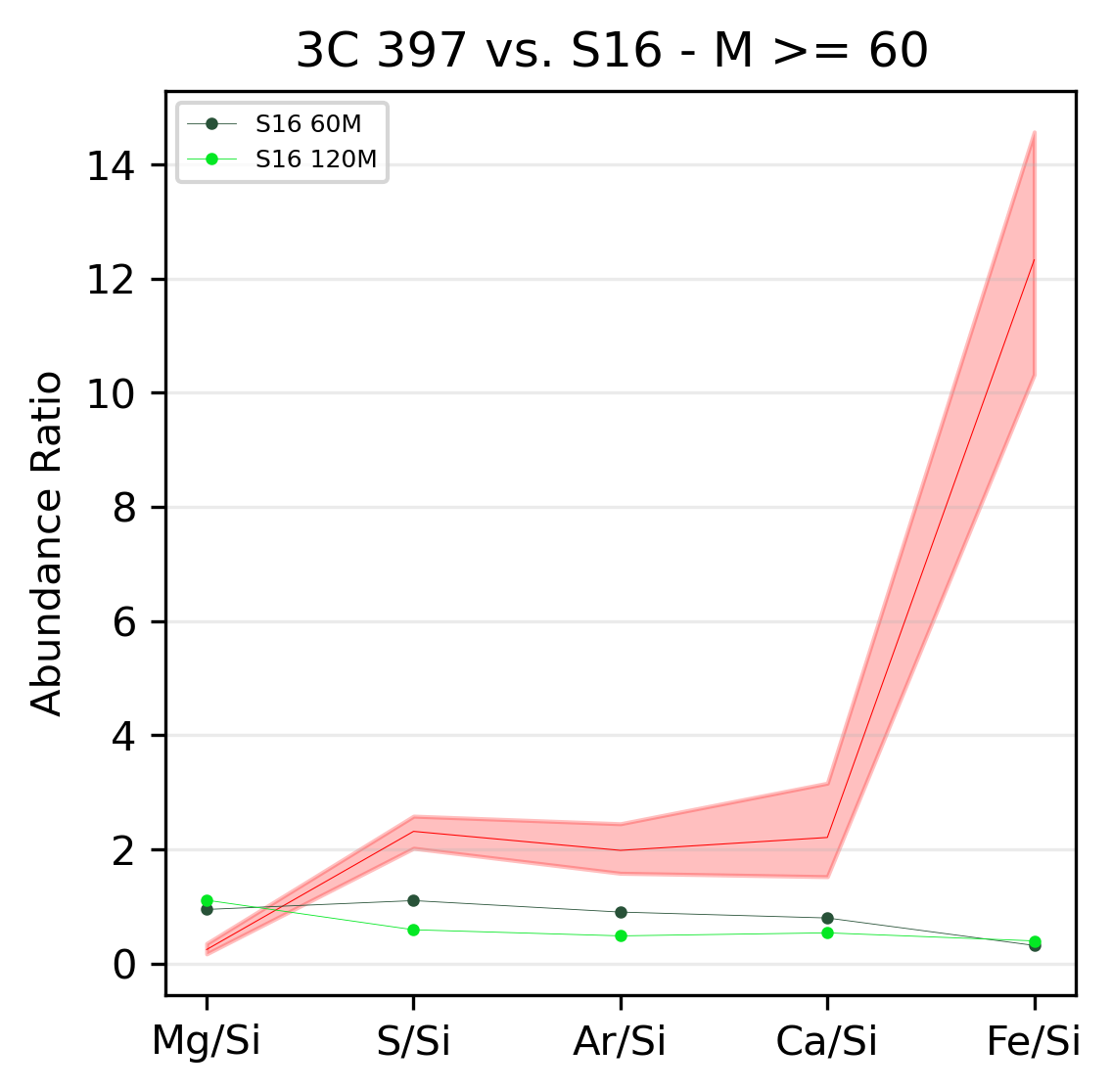}}
	\end{center}
    {Continued from above.}
\end{figure*}

\begin{figure*}
	\begin{center}
		\subfloat{\includegraphics[angle=0,width=0.40\textwidth,scale=0.5]{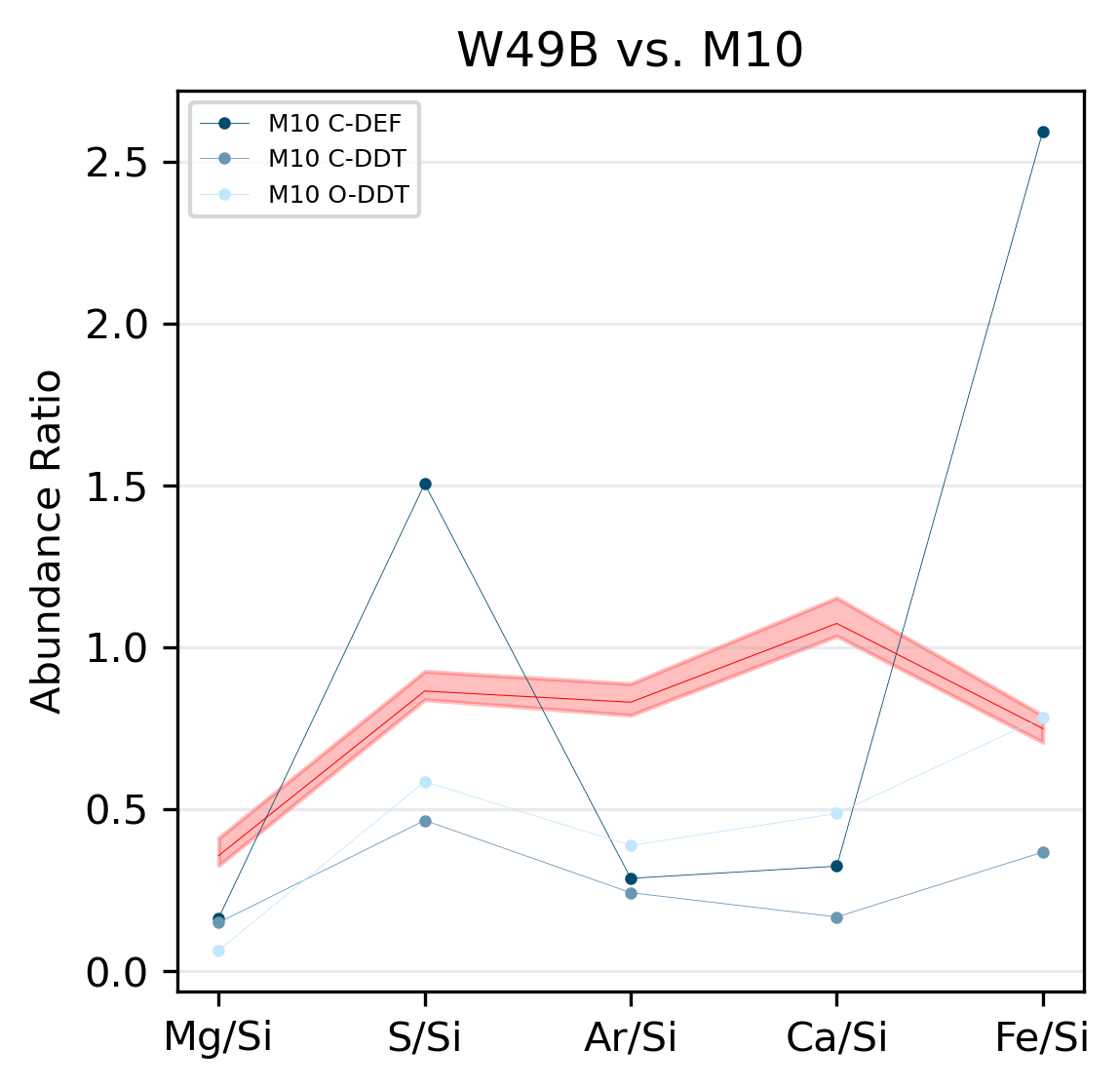}}
		\subfloat{\includegraphics[angle=0,width=0.40\textwidth,scale=0.5]{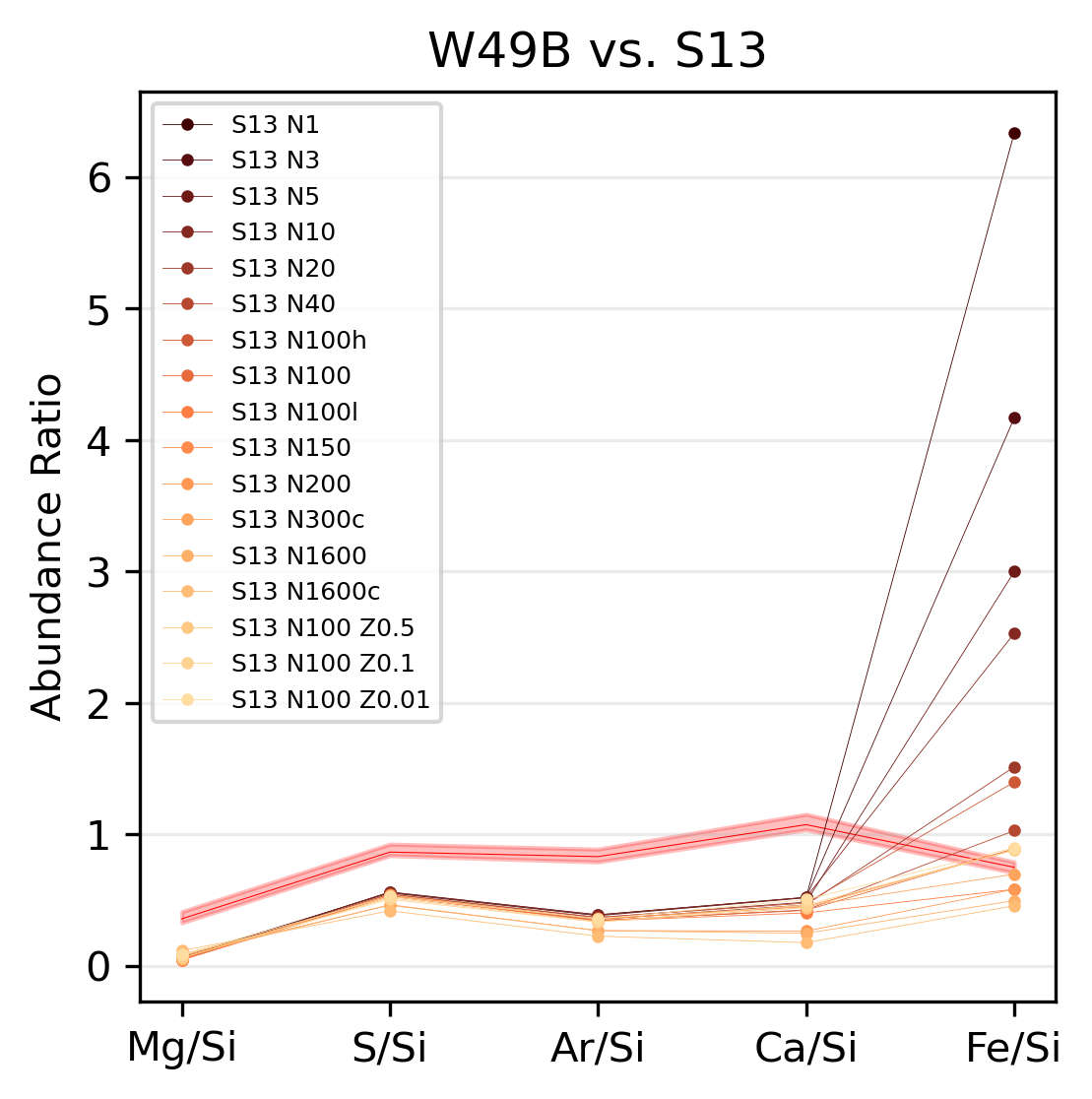}} \\
		\subfloat{\includegraphics[angle=0,width=0.40\textwidth]{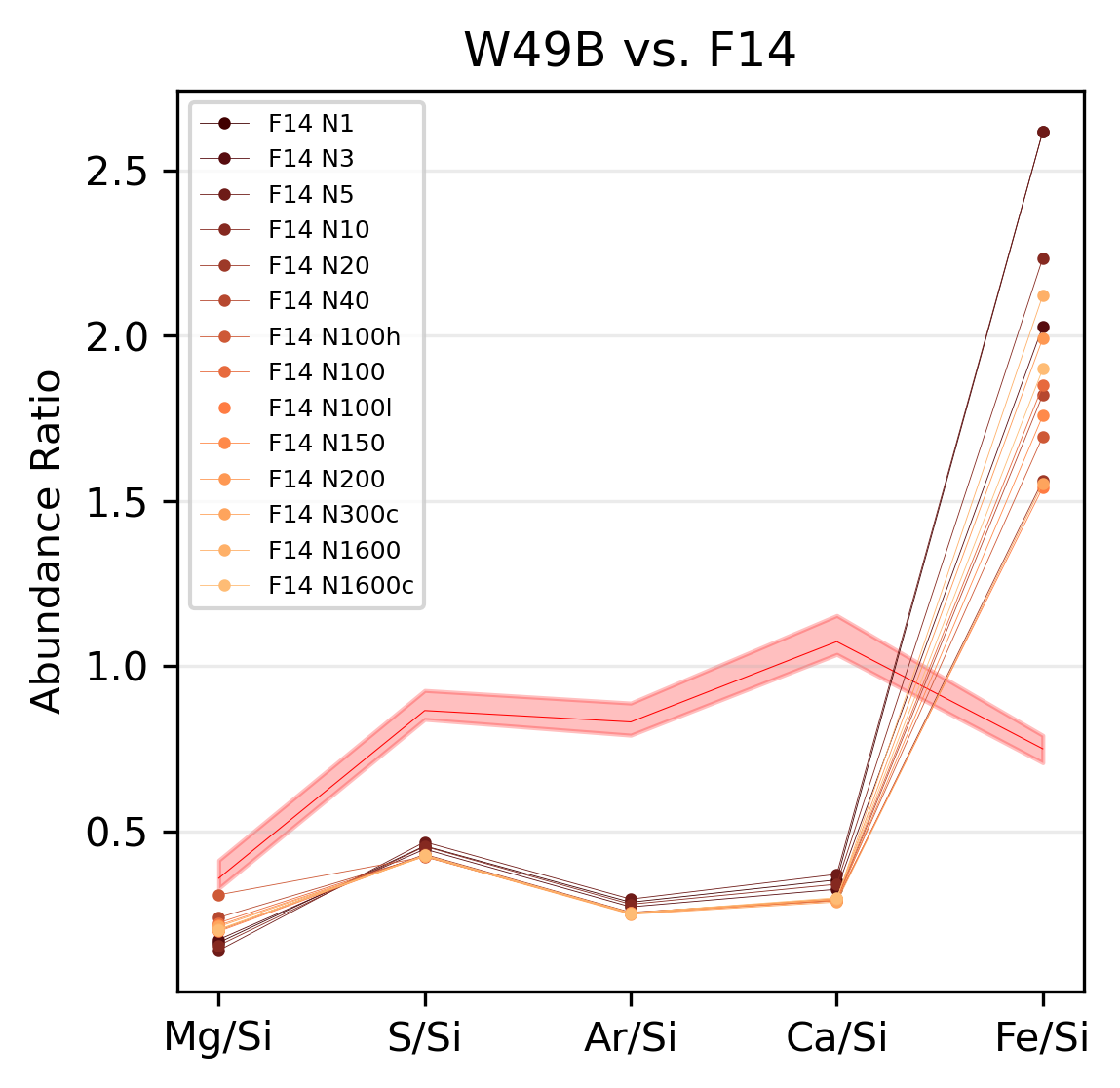}}
		\subfloat{\includegraphics[angle=0,width=0.40\textwidth]{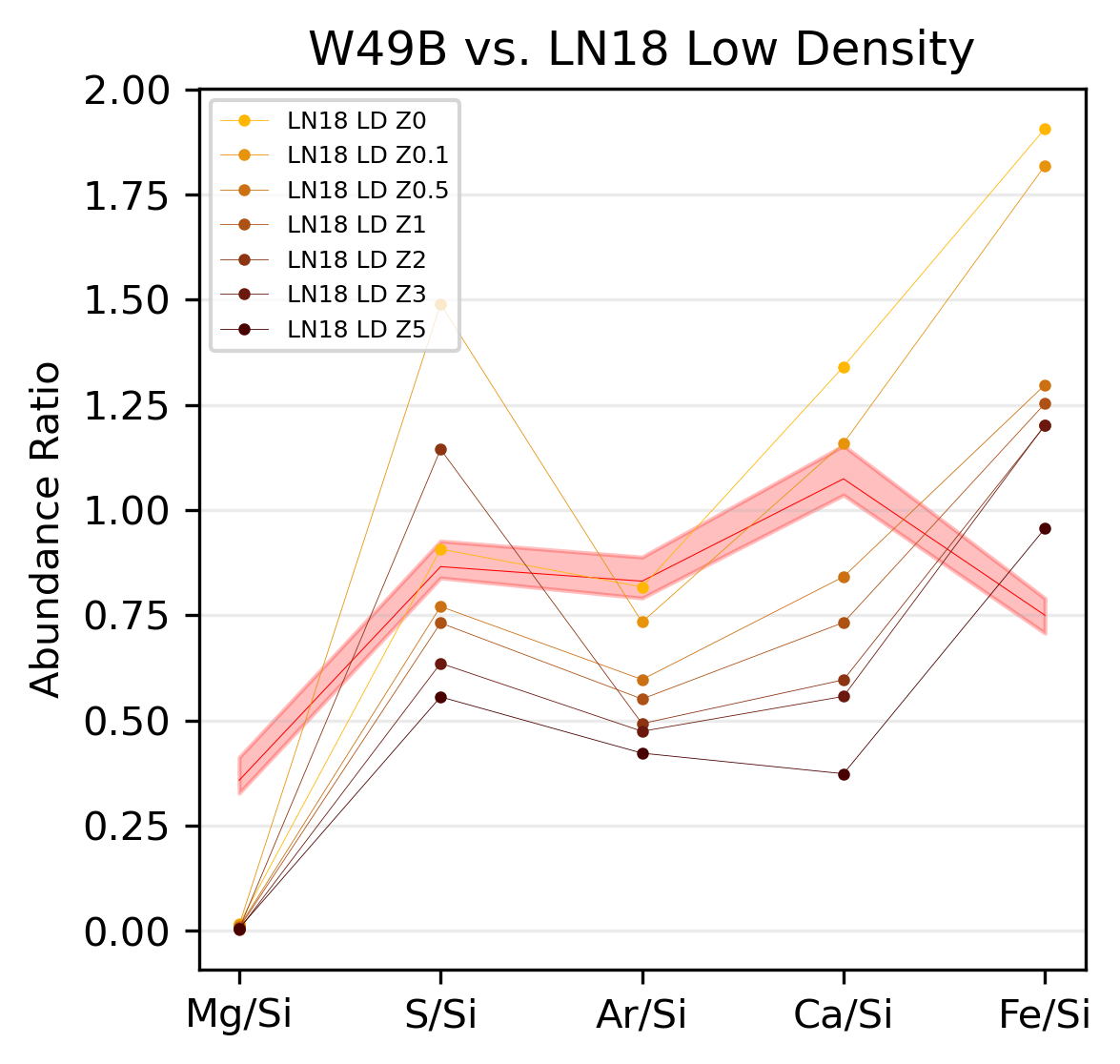}} \\
		\subfloat{\includegraphics[angle=0,width=0.40\textwidth,scale=0.5]{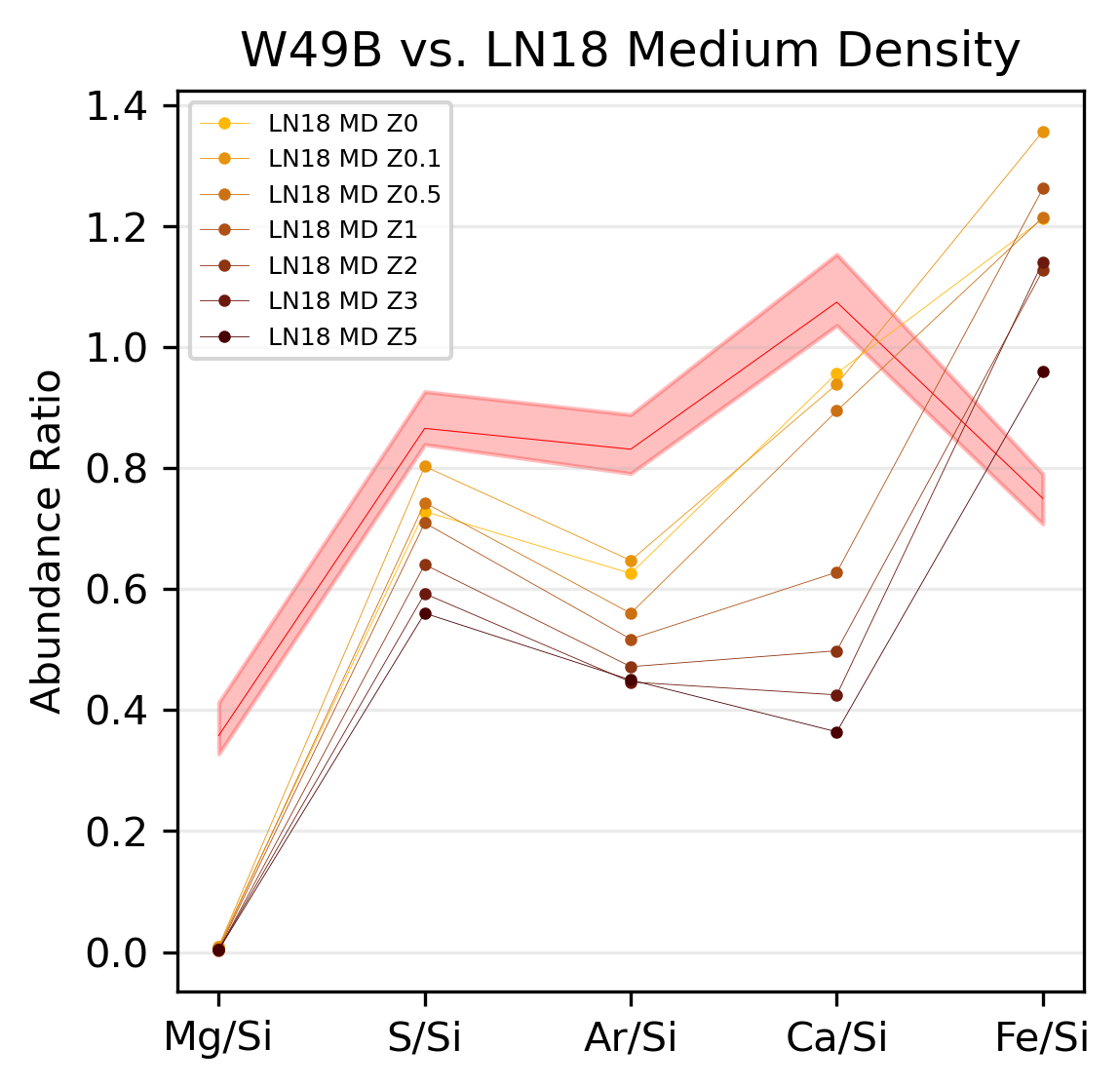}}
		\subfloat{\includegraphics[angle=0,width=0.40\textwidth]{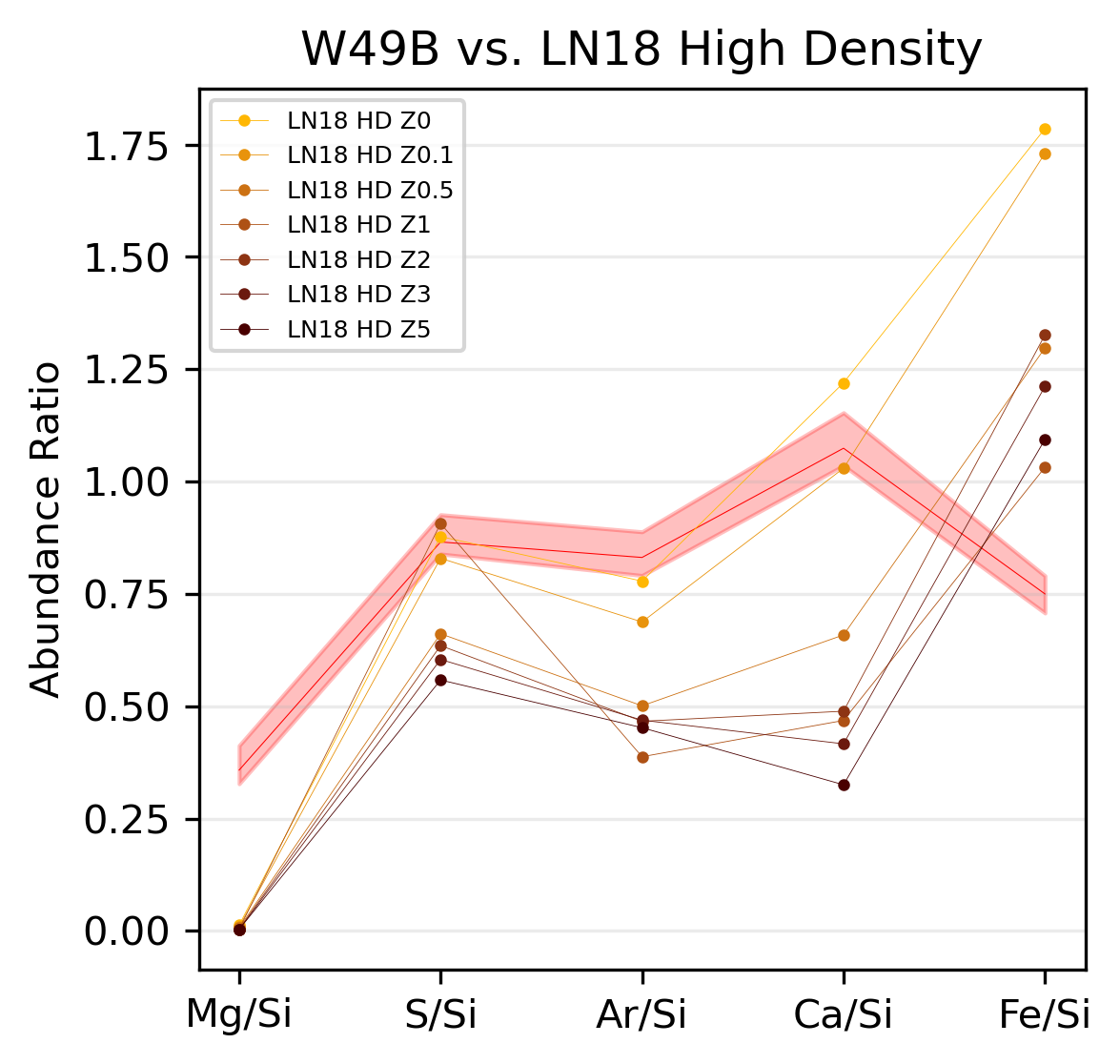}}
	\end{center}
    \caption{Nucleosynthesis comparisons between W49B and the tested Ia models.}
    \label{fig:w49b_ia}
\end{figure*}

\begin{figure*}\ContinuedFloat
	\begin{center}
		\subfloat{\includegraphics[angle=0,width=0.40\textwidth,scale=0.5]{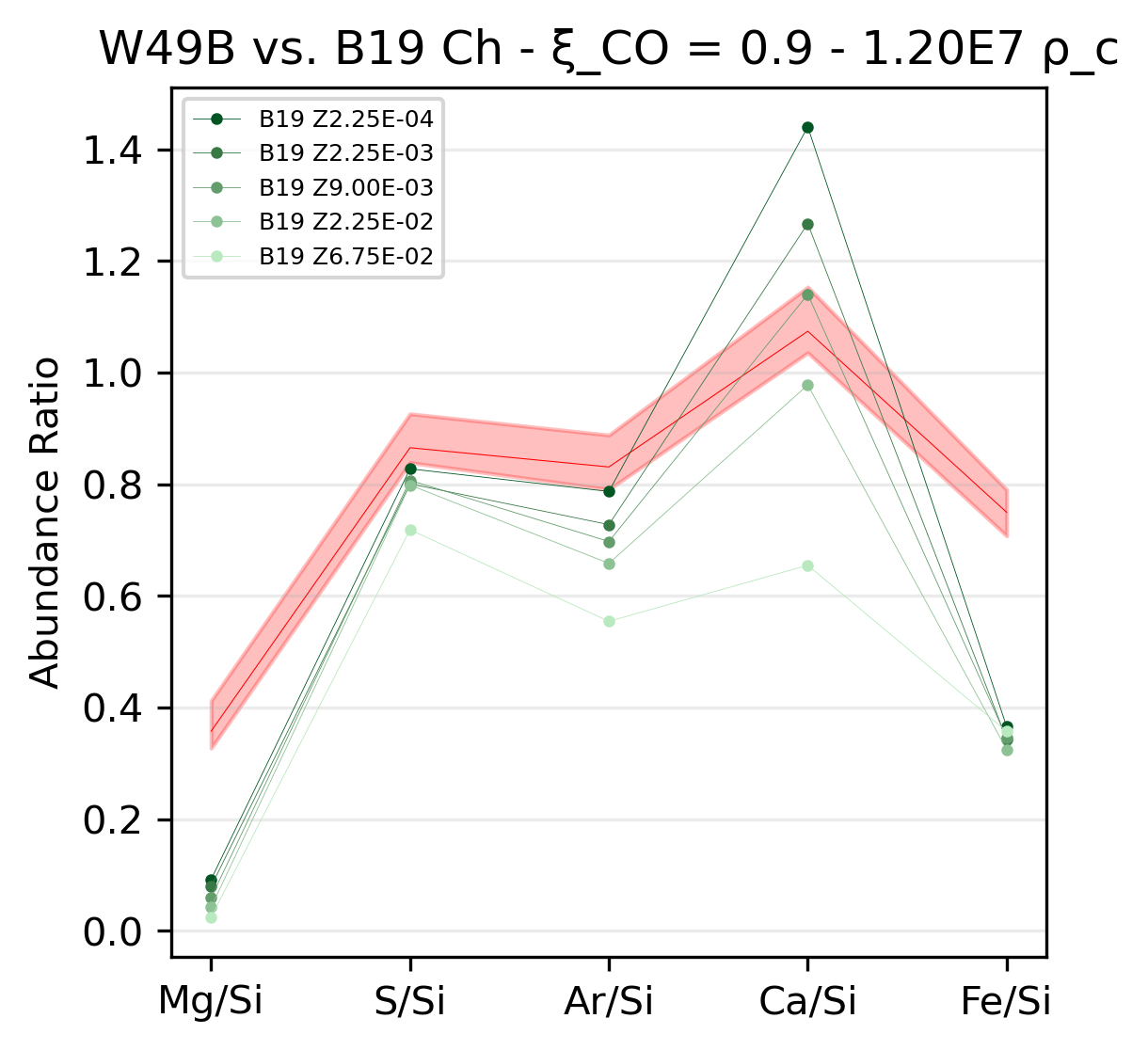}}
		\subfloat{\includegraphics[angle=0,width=0.40\textwidth,scale=0.5]{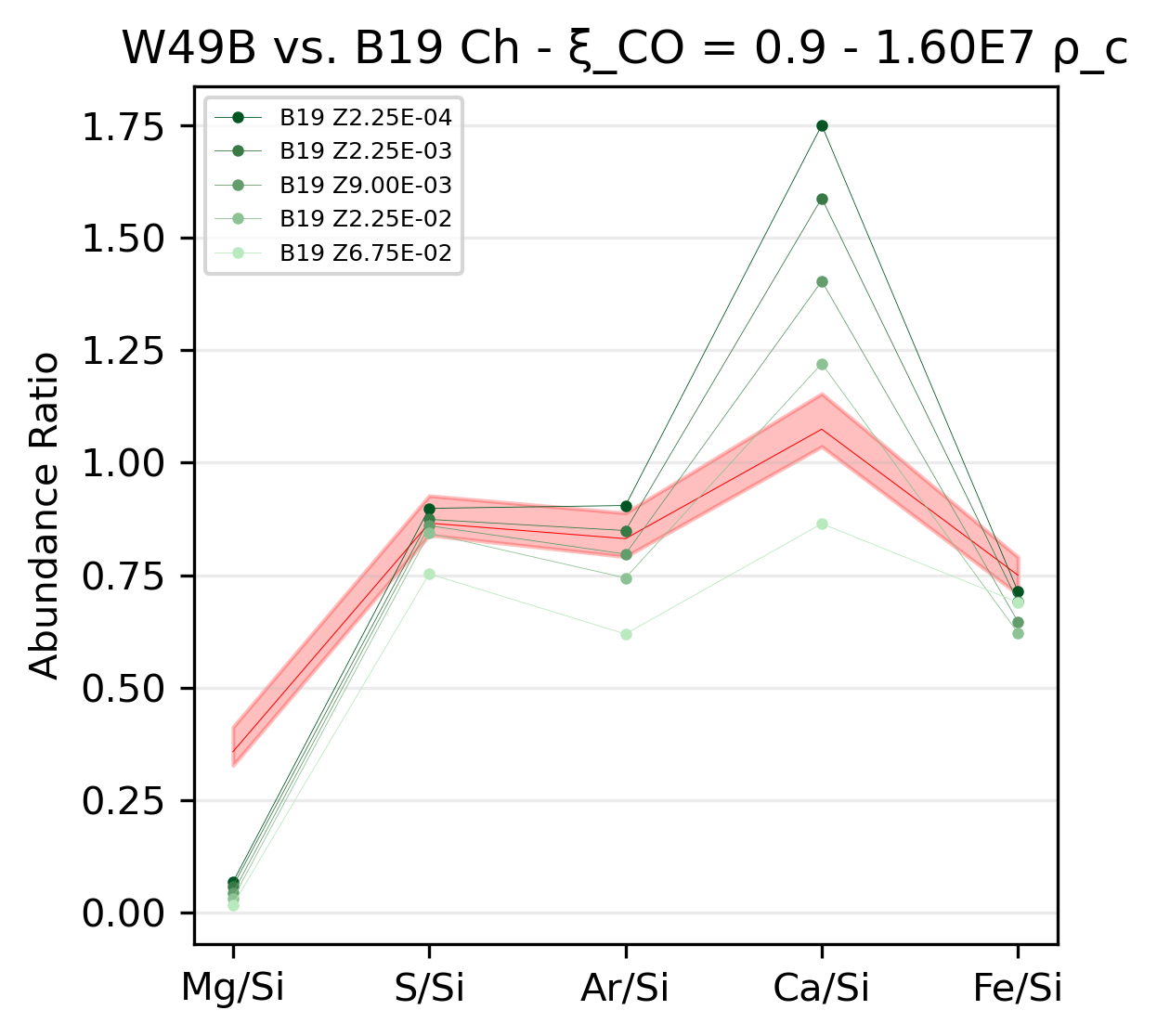}} \\
		\subfloat{\includegraphics[angle=0,width=0.40\textwidth]{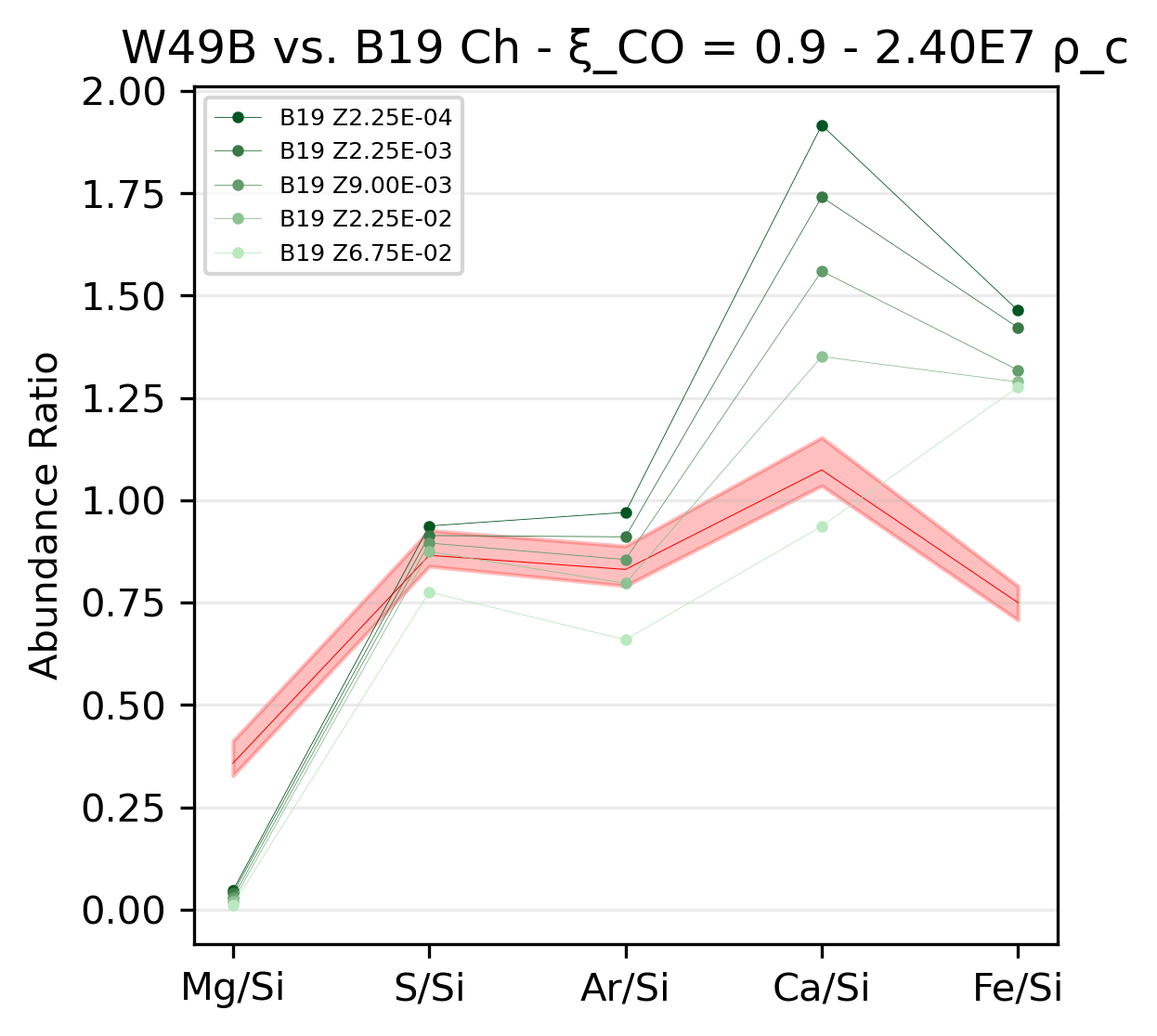}}
		\subfloat{\includegraphics[angle=0,width=0.40\textwidth]{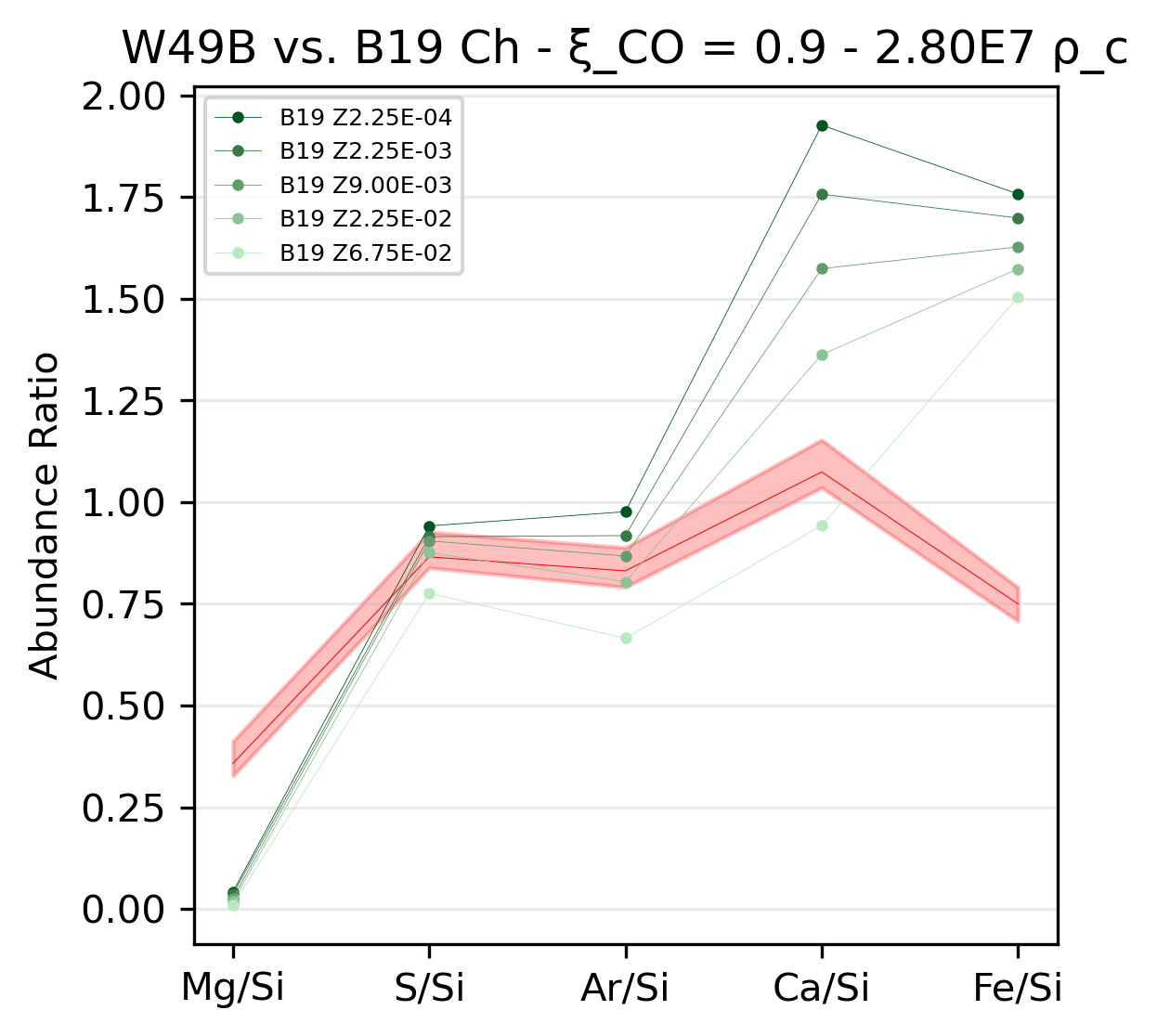}} \\
		\subfloat{\includegraphics[angle=0,width=0.40\textwidth,scale=0.5]{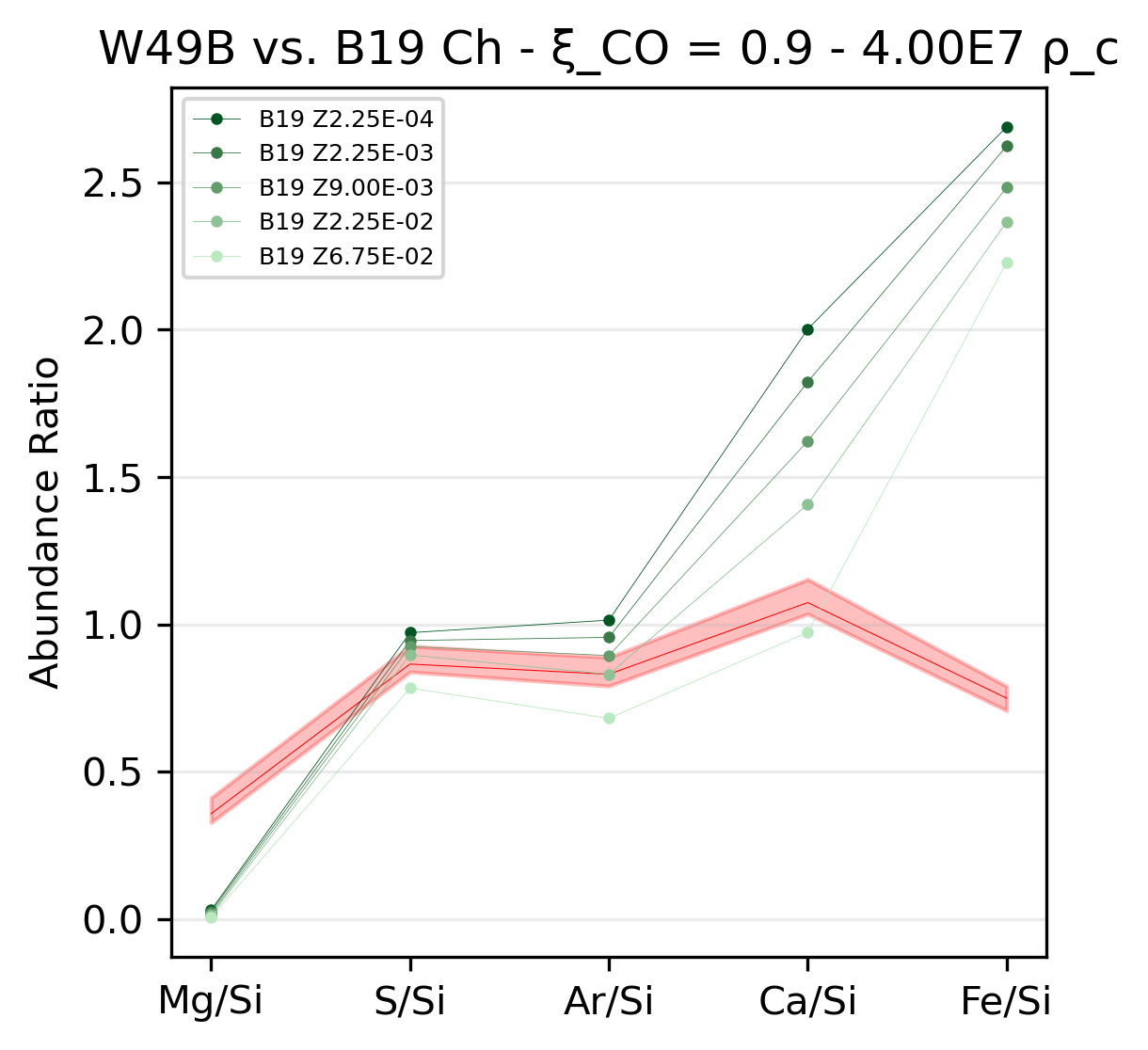}}
		\subfloat{\includegraphics[angle=0,width=0.40\textwidth]{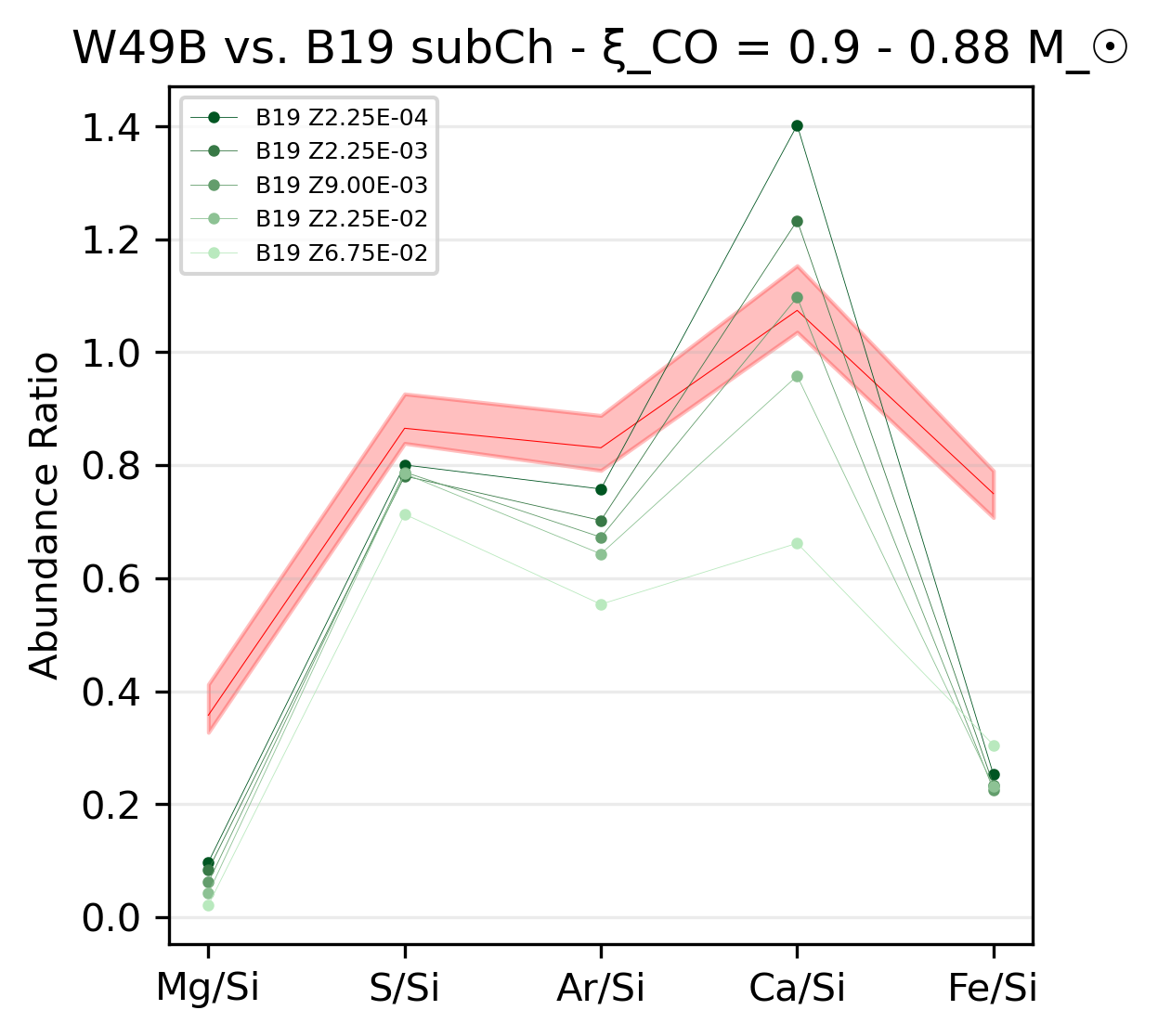}}
	\end{center}
    {Continued from above.}
\end{figure*}

\begin{figure*}\ContinuedFloat
	\begin{center}
		\subfloat{\includegraphics[angle=0,width=0.40\textwidth,scale=0.5]{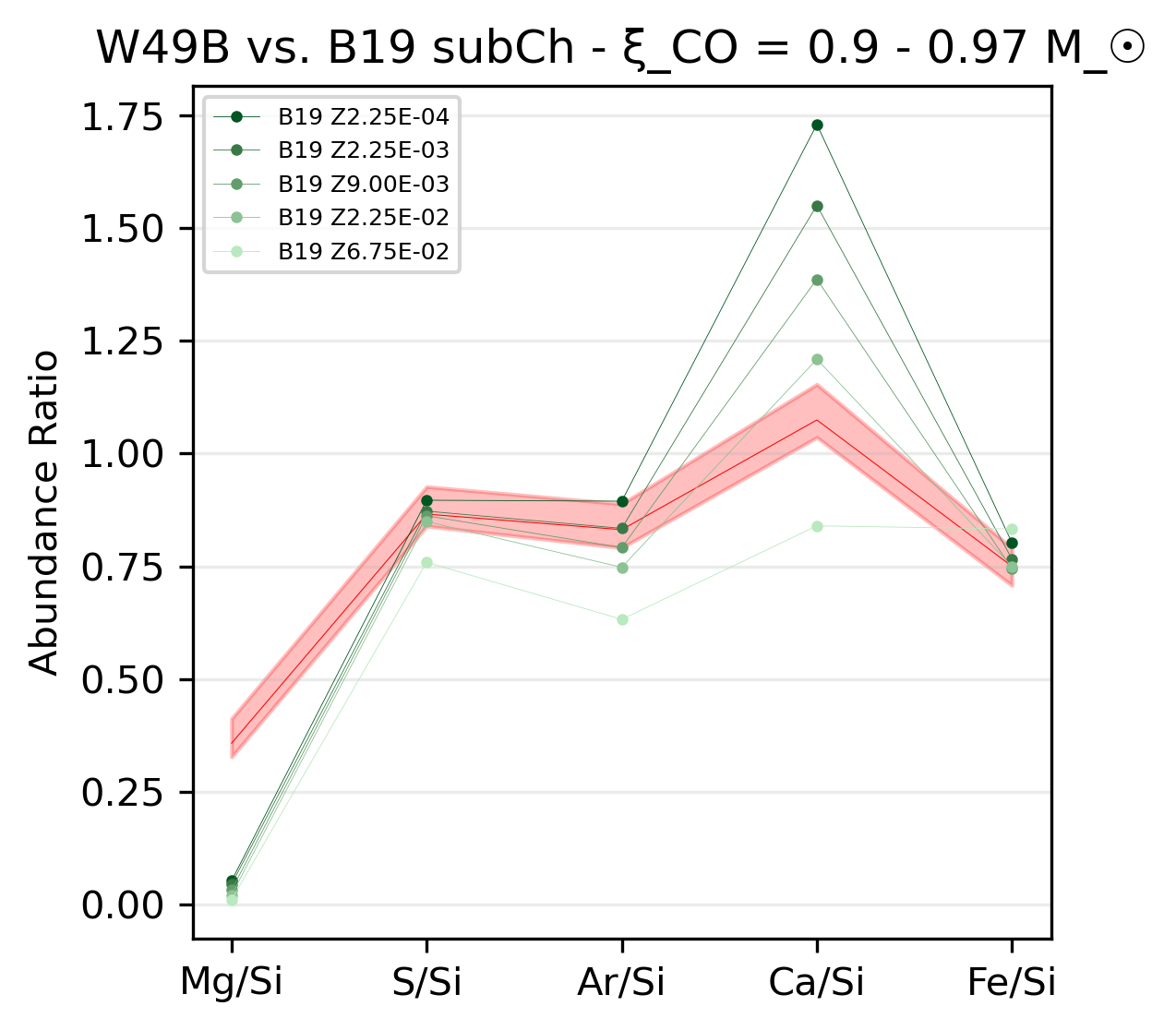}}
		\subfloat{\includegraphics[angle=0,width=0.40\textwidth,scale=0.5]{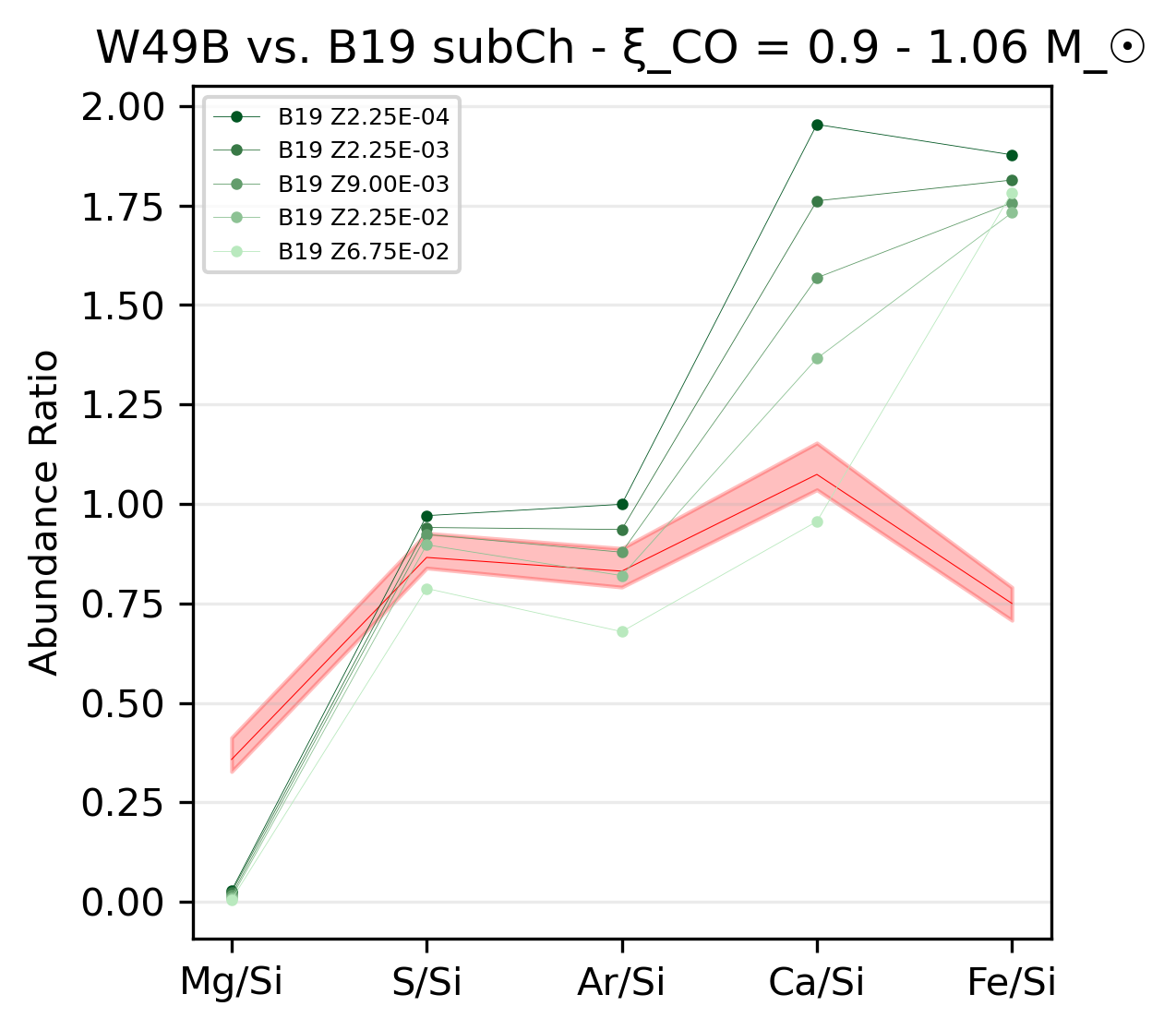}} \\
		\subfloat{\includegraphics[angle=0,width=0.40\textwidth]{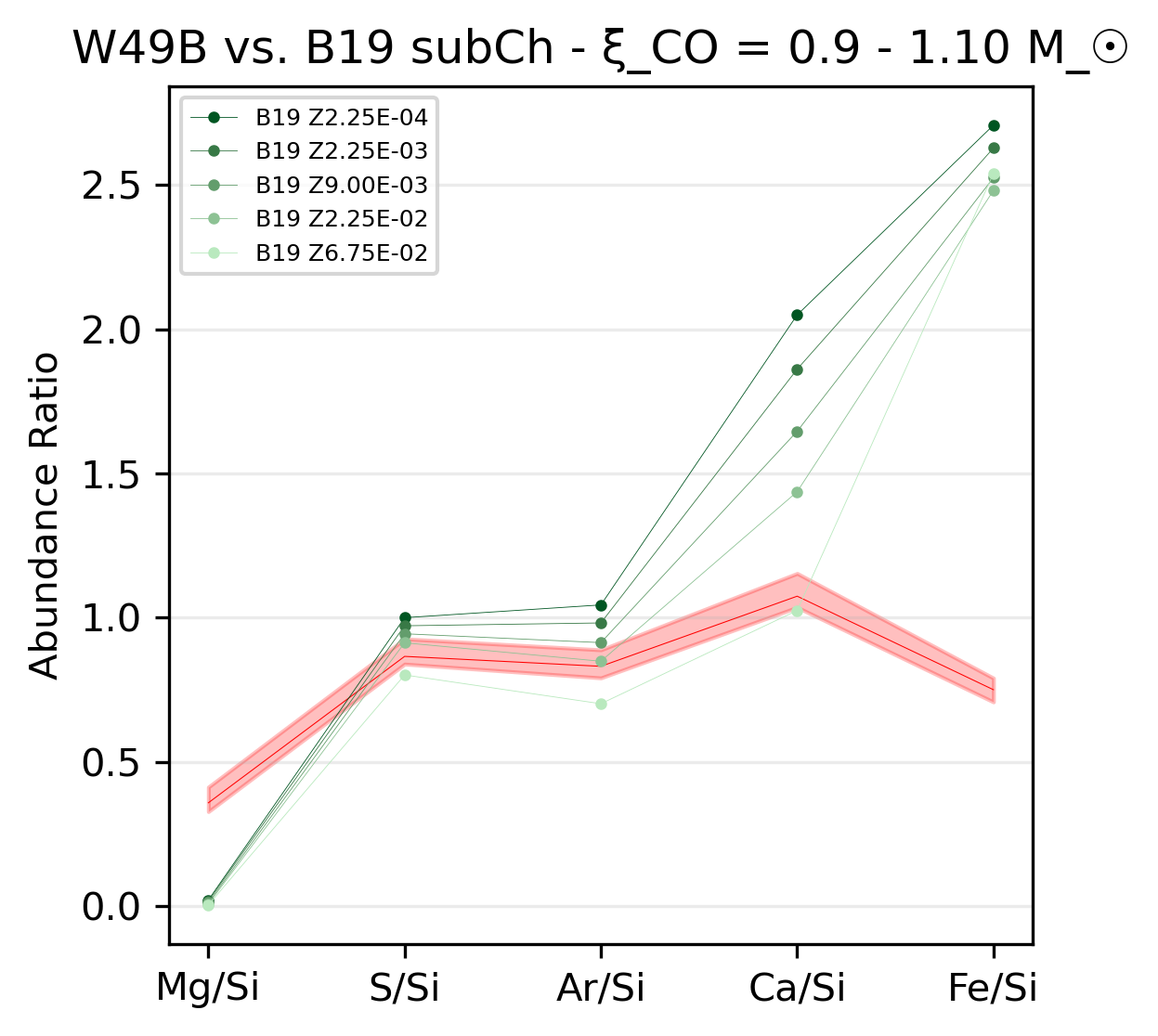}}
		\subfloat{\includegraphics[angle=0,width=0.40\textwidth]{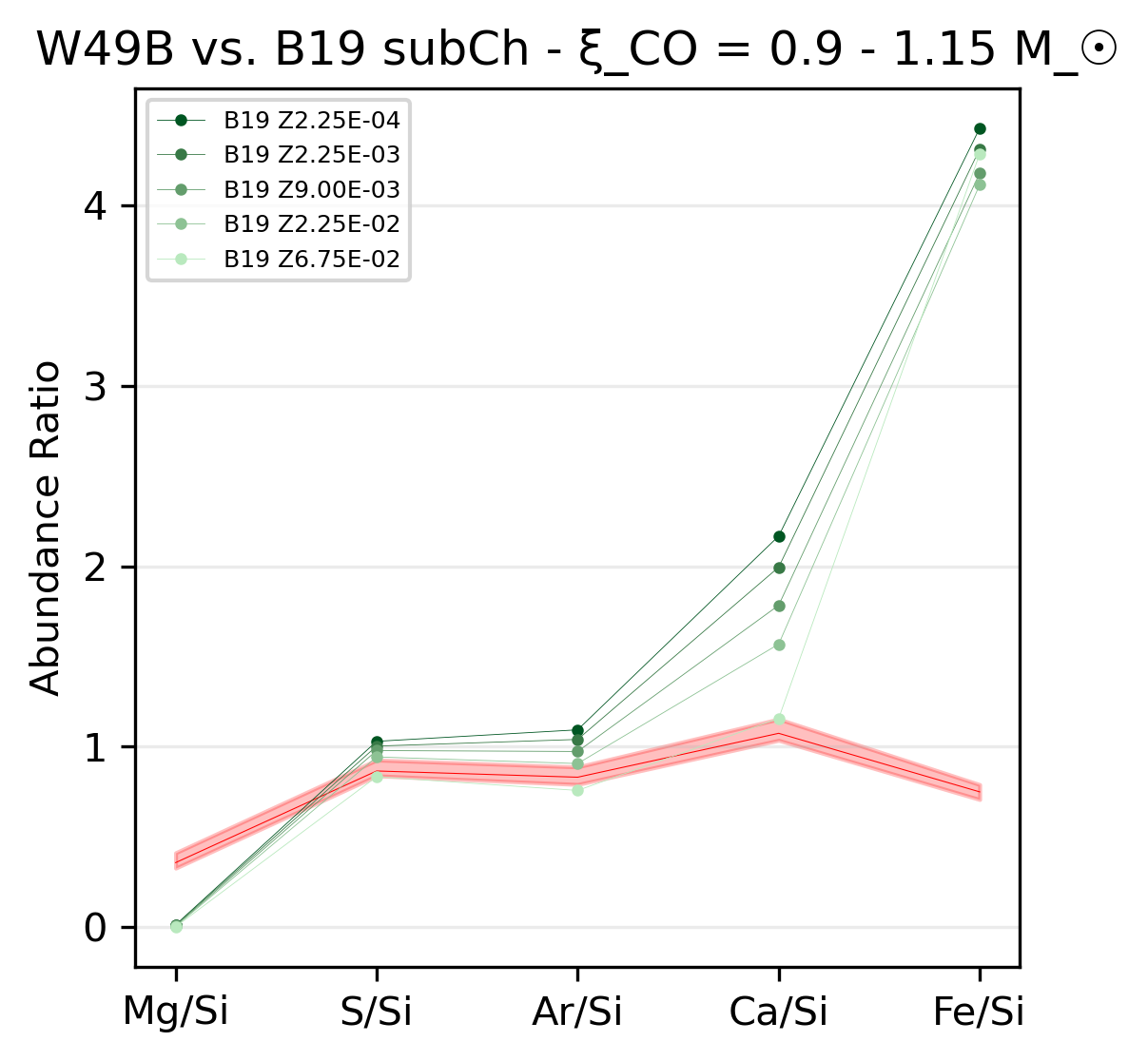}} \\
		\subfloat{\includegraphics[angle=0,width=0.40\textwidth,scale=0.5]{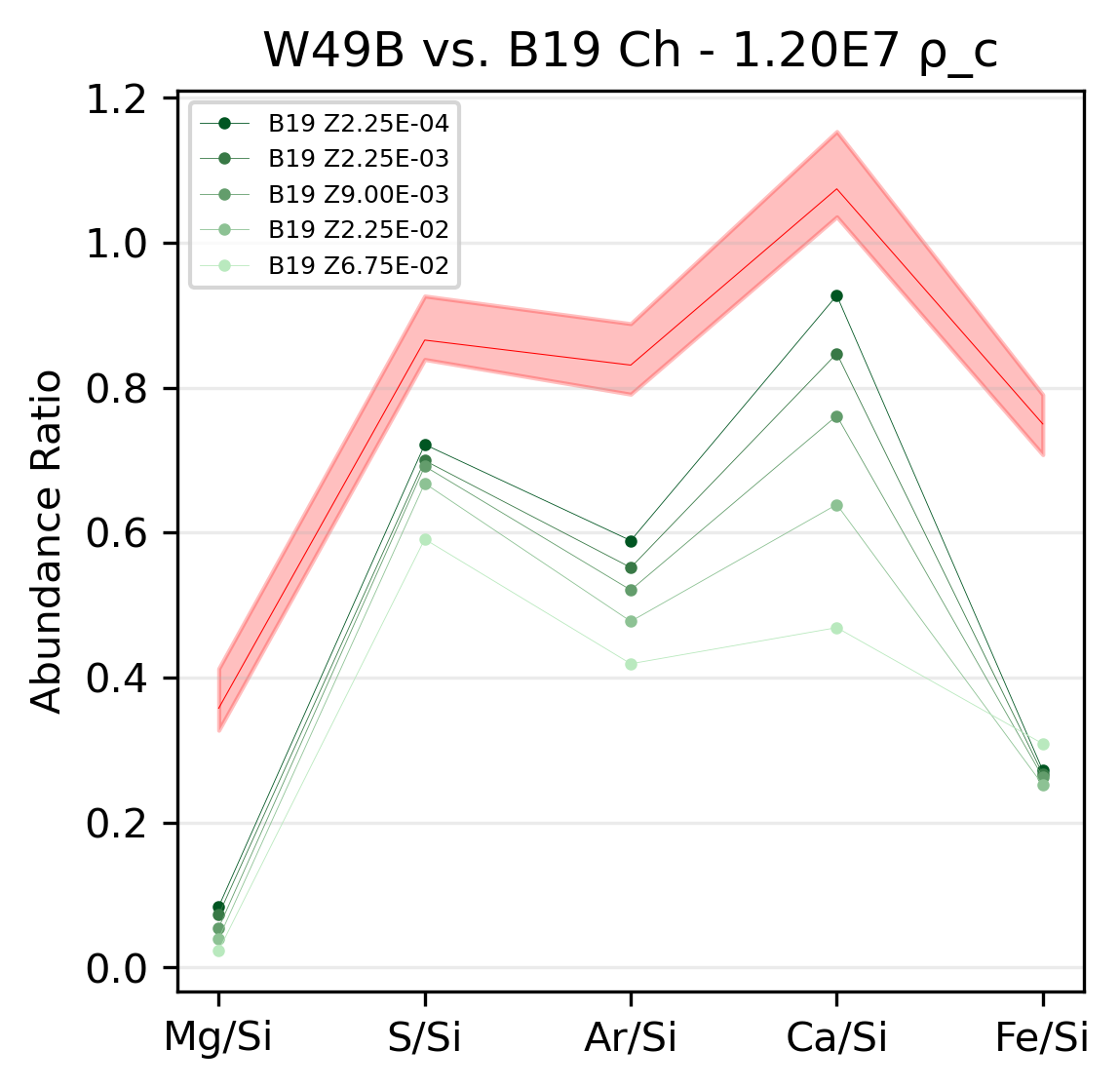}}
		\subfloat{\includegraphics[angle=0,width=0.40\textwidth]{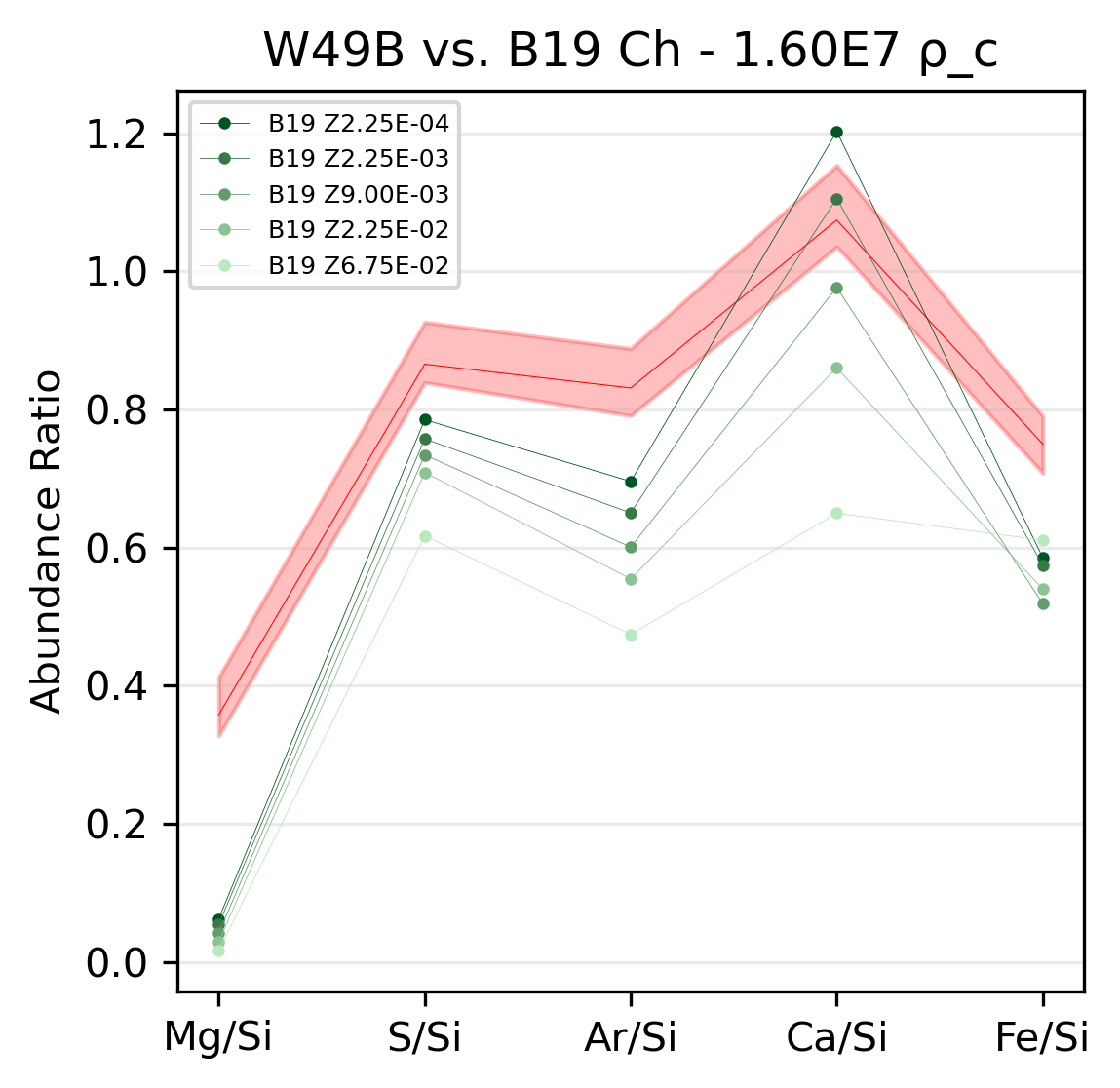}}
	\end{center}
    {Continued from above.}
\end{figure*}

\begin{figure*}\ContinuedFloat
	\begin{center}
		\subfloat{\includegraphics[angle=0,width=0.40\textwidth,scale=0.5]{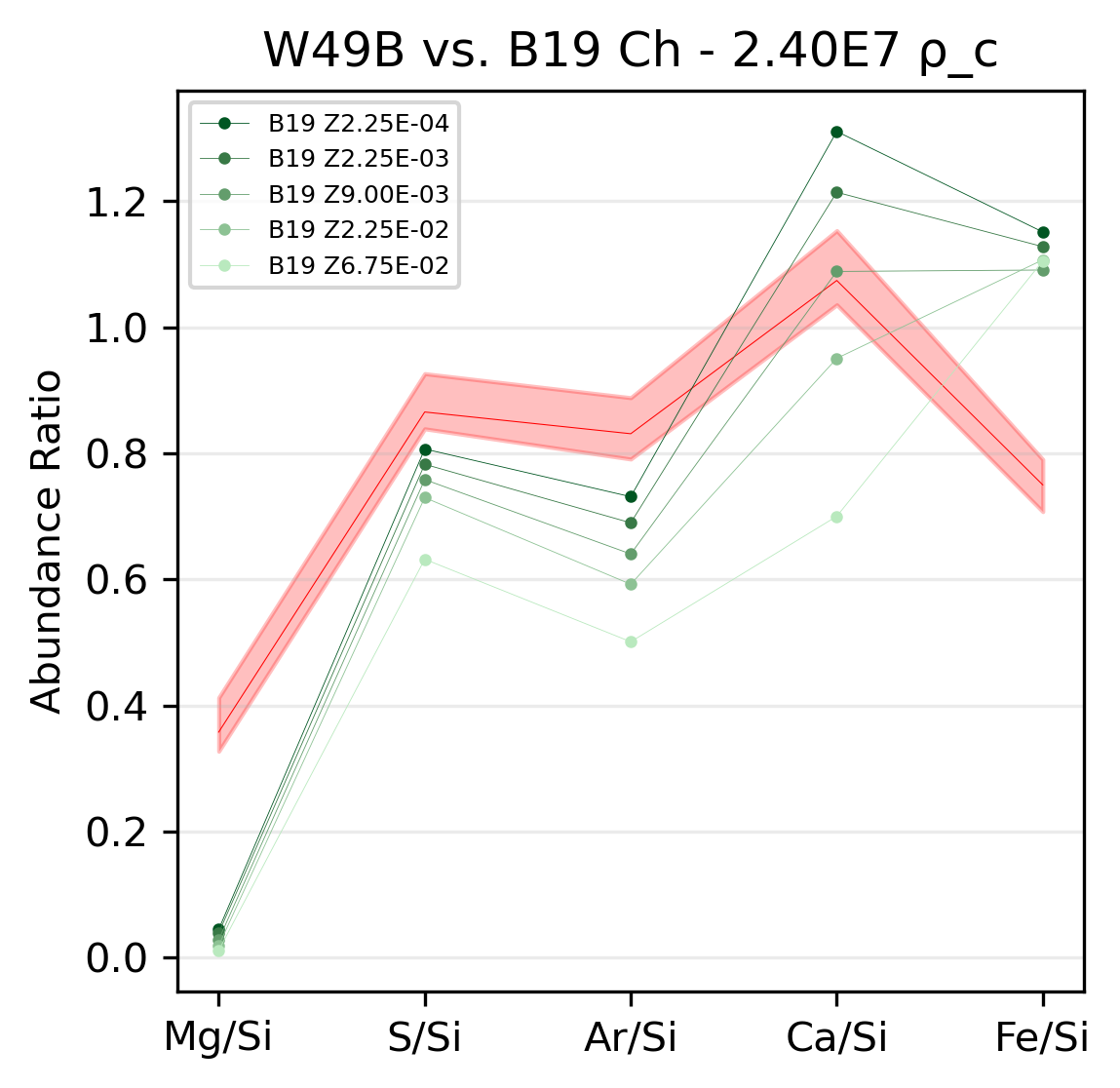}}
		\subfloat{\includegraphics[angle=0,width=0.40\textwidth,scale=0.5]{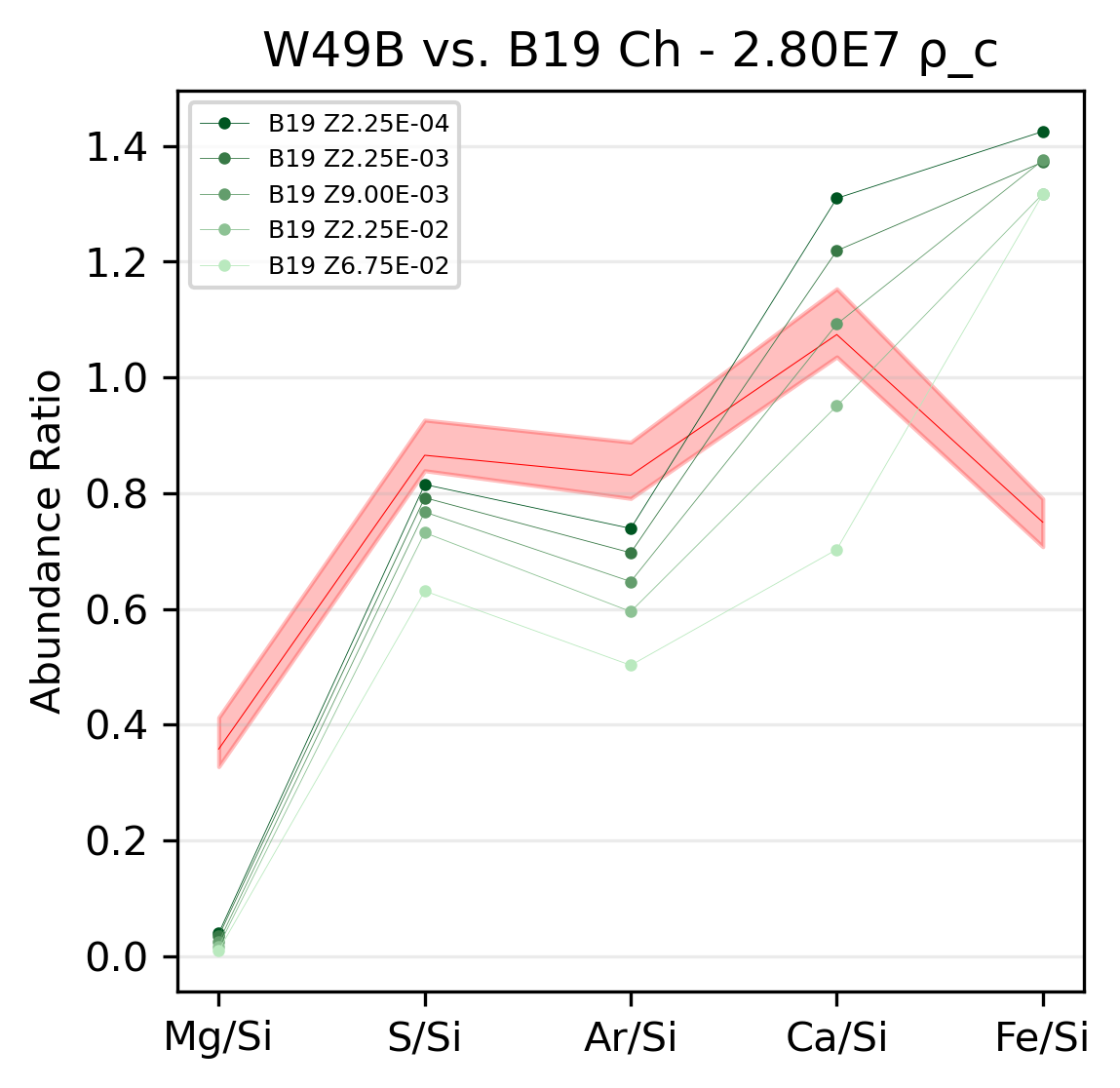}} \\
		\subfloat{\includegraphics[angle=0,width=0.40\textwidth]{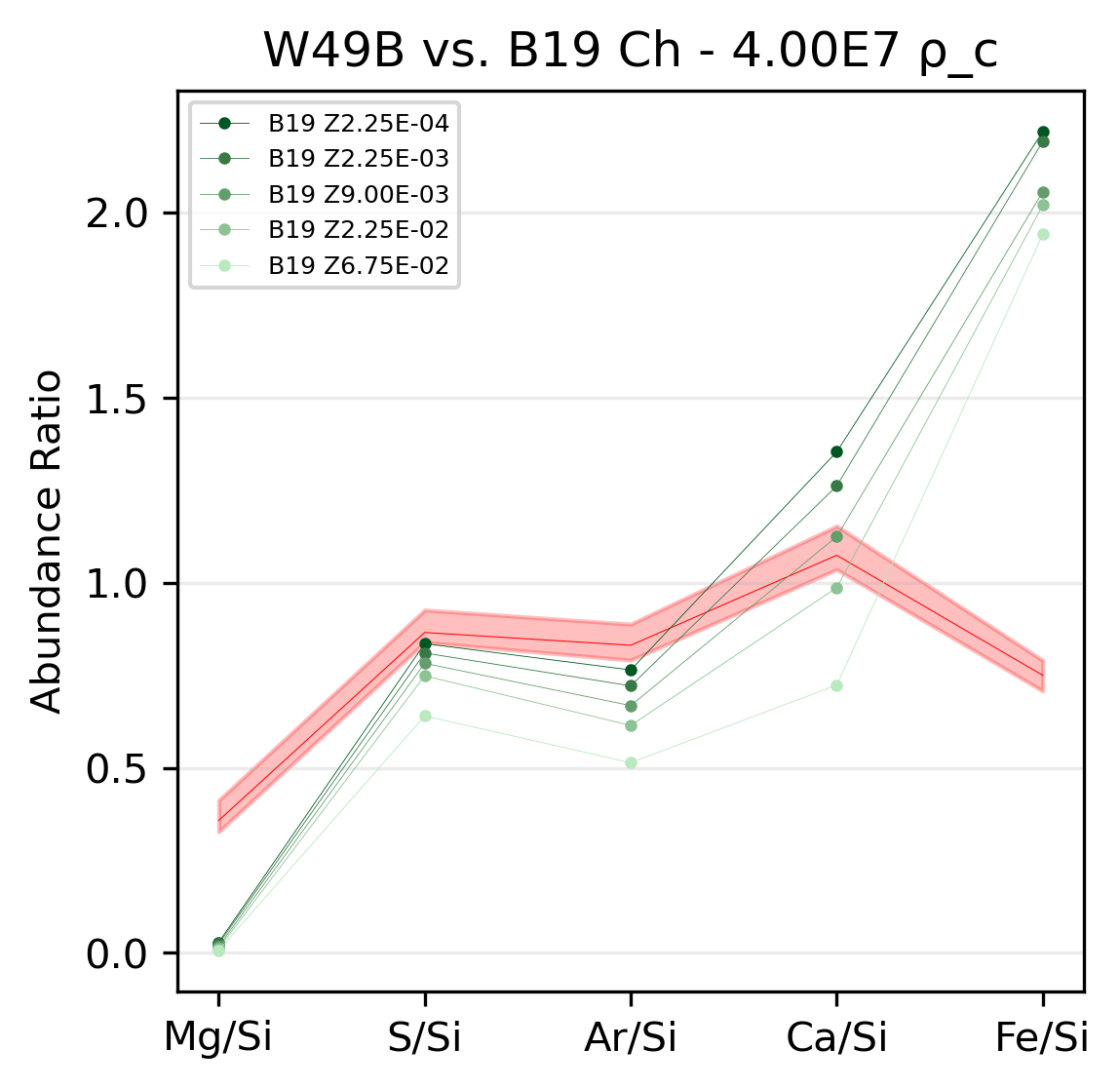}}
		\subfloat{\includegraphics[angle=0,width=0.40\textwidth]{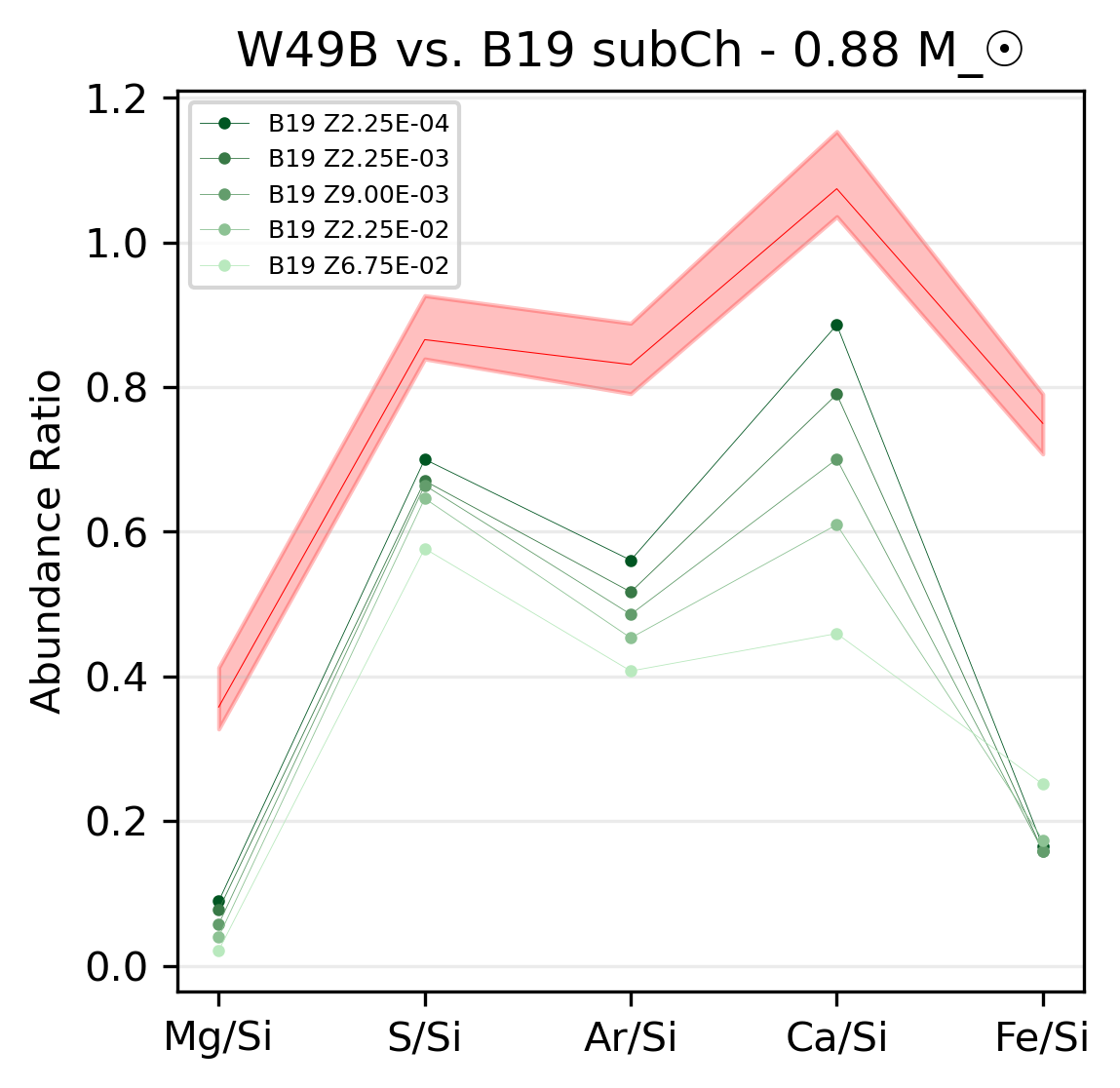}} \\
		\subfloat{\includegraphics[angle=0,width=0.40\textwidth,scale=0.5]{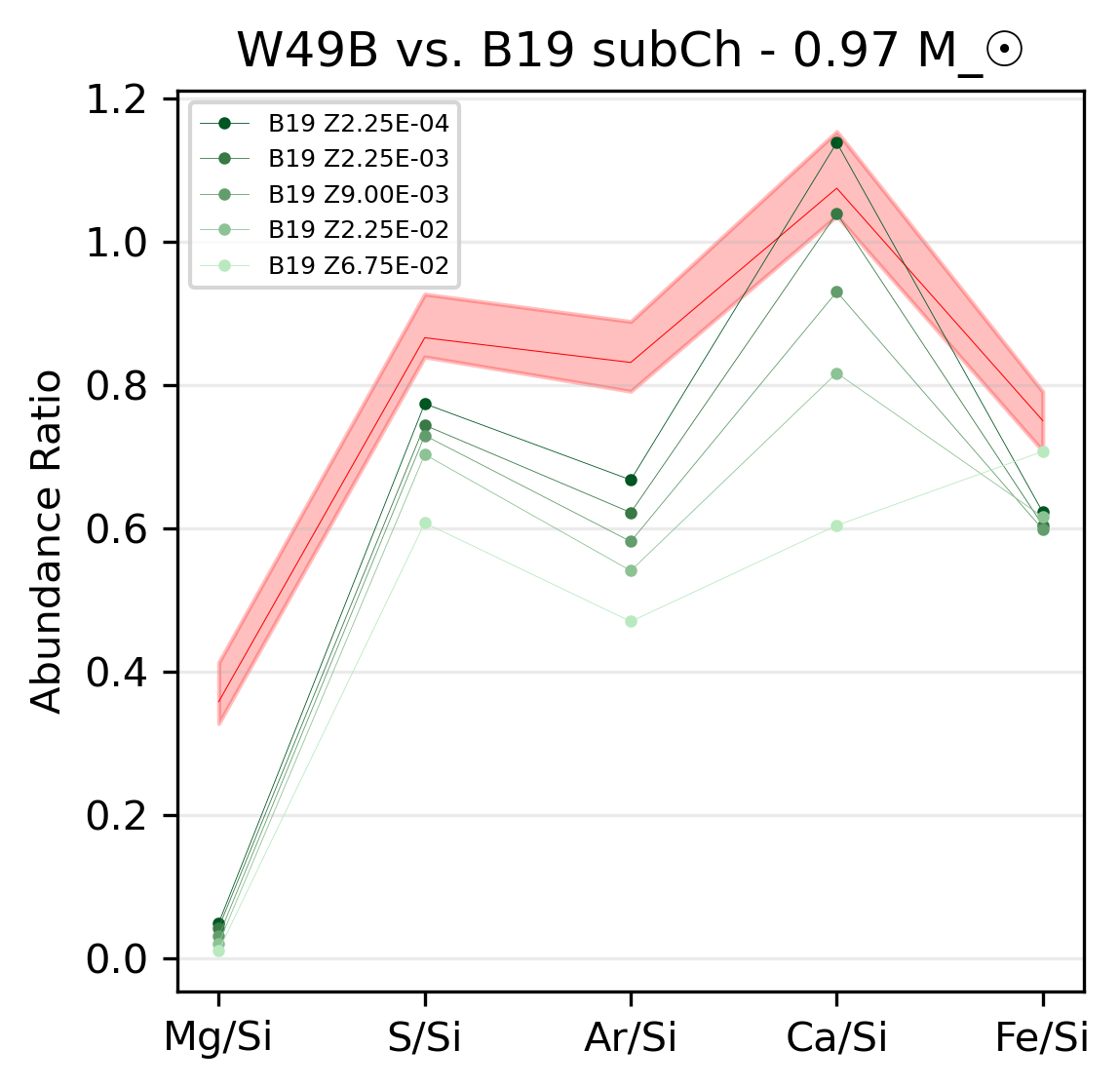}}
		\subfloat{\includegraphics[angle=0,width=0.40\textwidth]{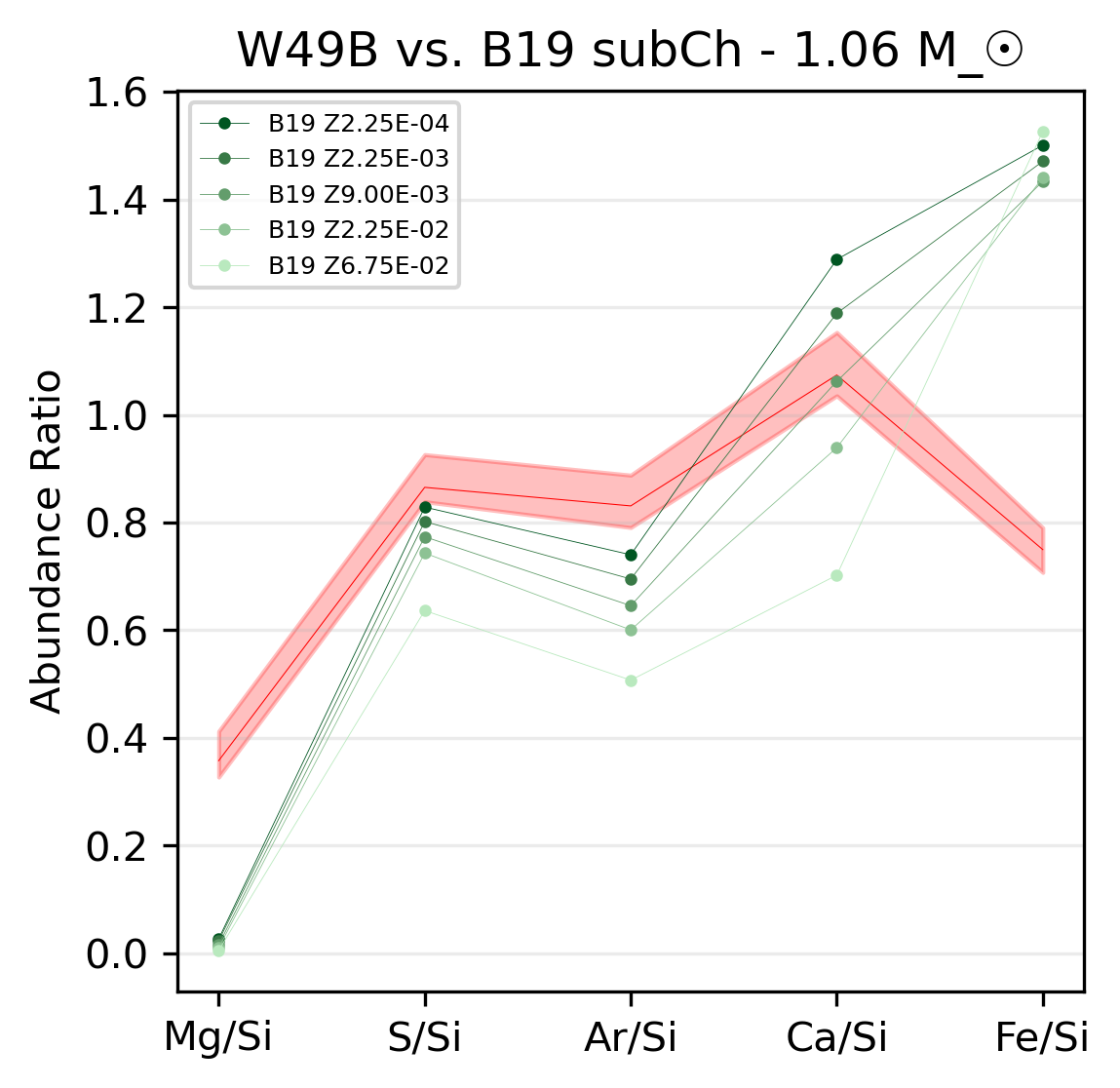}}
	\end{center}
    {Continued from above.}
\end{figure*}

\begin{figure*}\ContinuedFloat
	\begin{center}
		\subfloat{\includegraphics[angle=0,width=0.40\textwidth,scale=0.5]{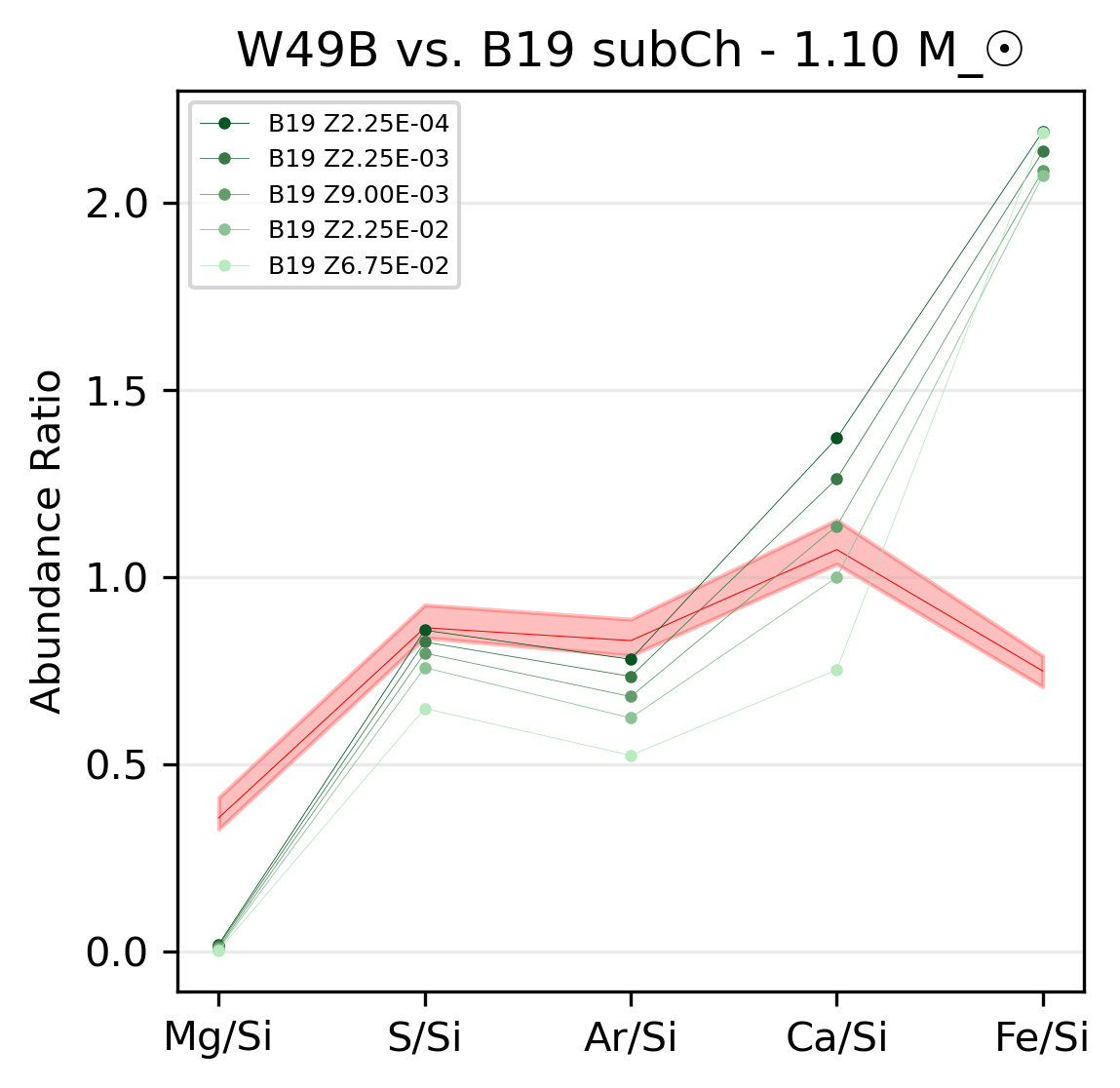}}
		\subfloat{\includegraphics[angle=0,width=0.40\textwidth,scale=0.5]{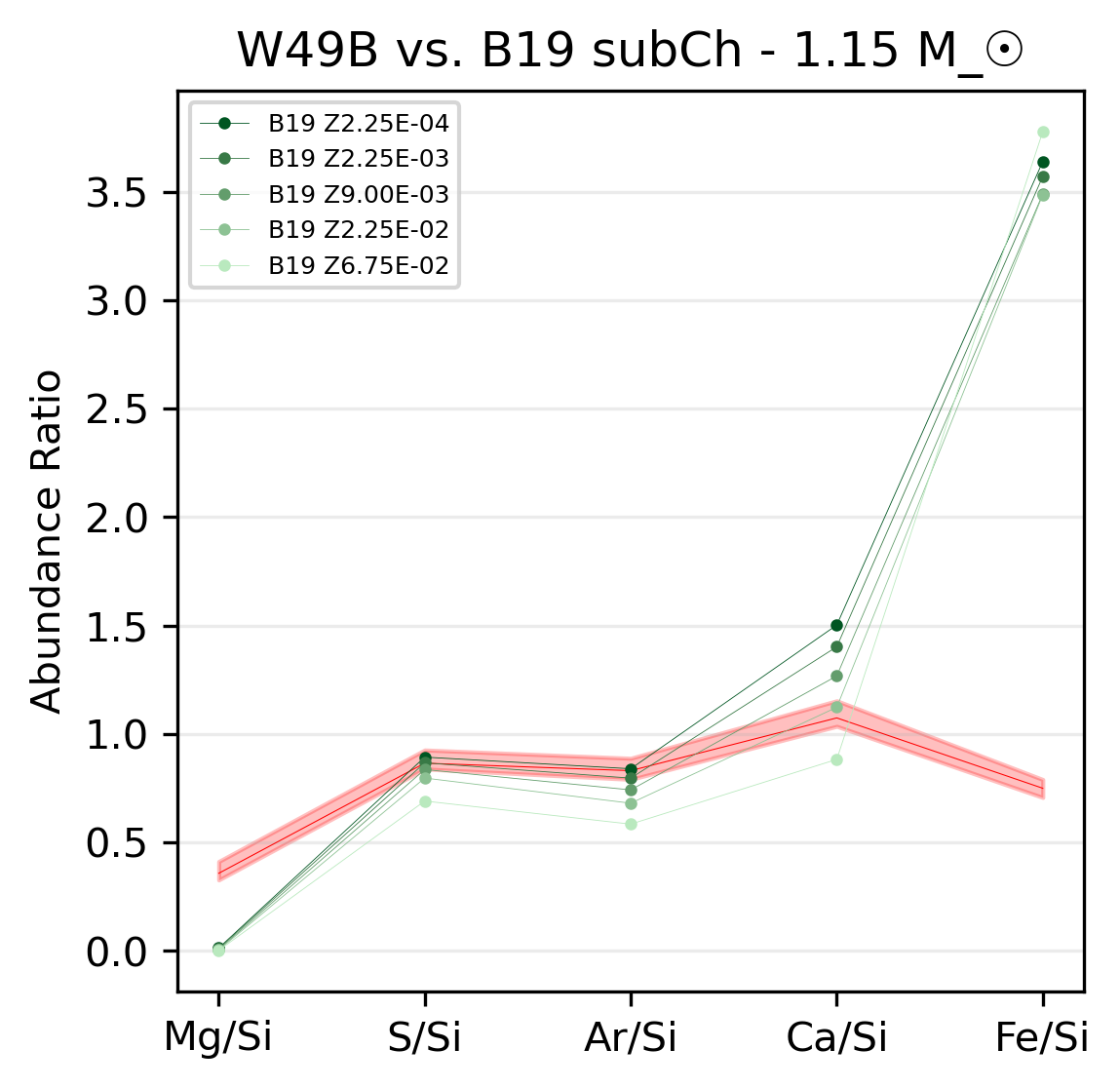}} \\
		\subfloat{\includegraphics[angle=0,width=0.40\textwidth]{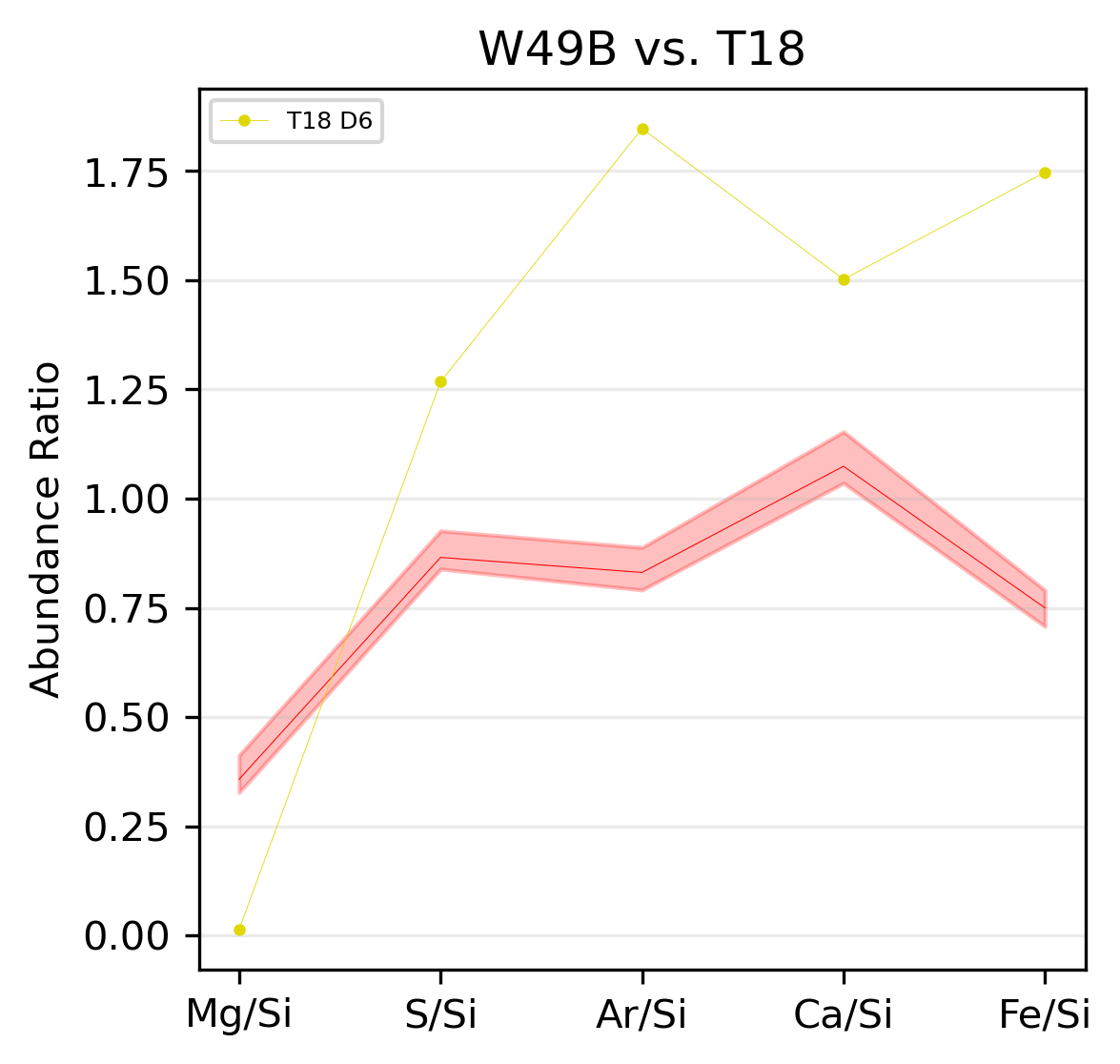}}
		\subfloat{\includegraphics[angle=0,width=0.40\textwidth]{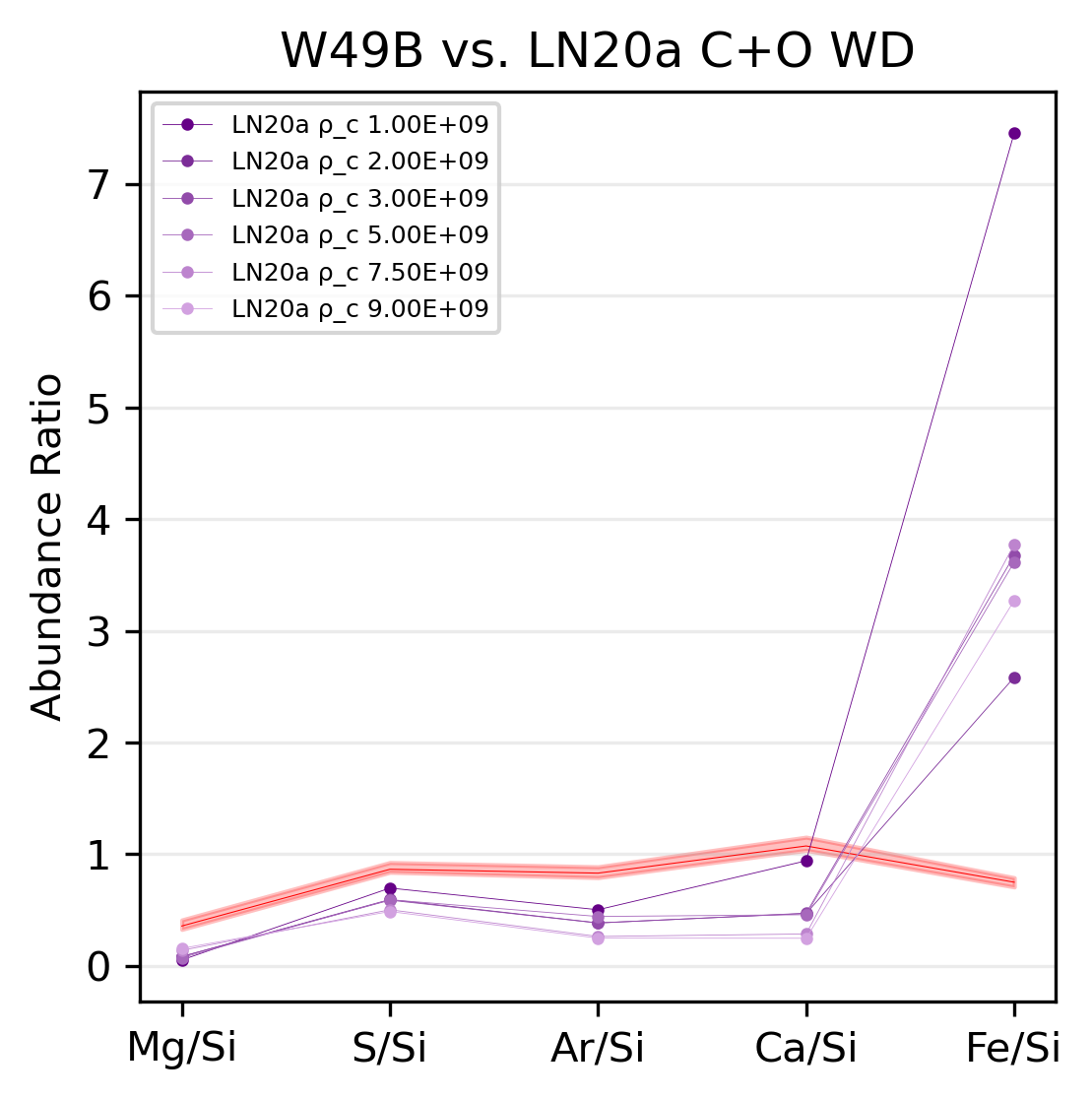}} \\
		\subfloat{\includegraphics[angle=0,width=0.40\textwidth,scale=0.5]{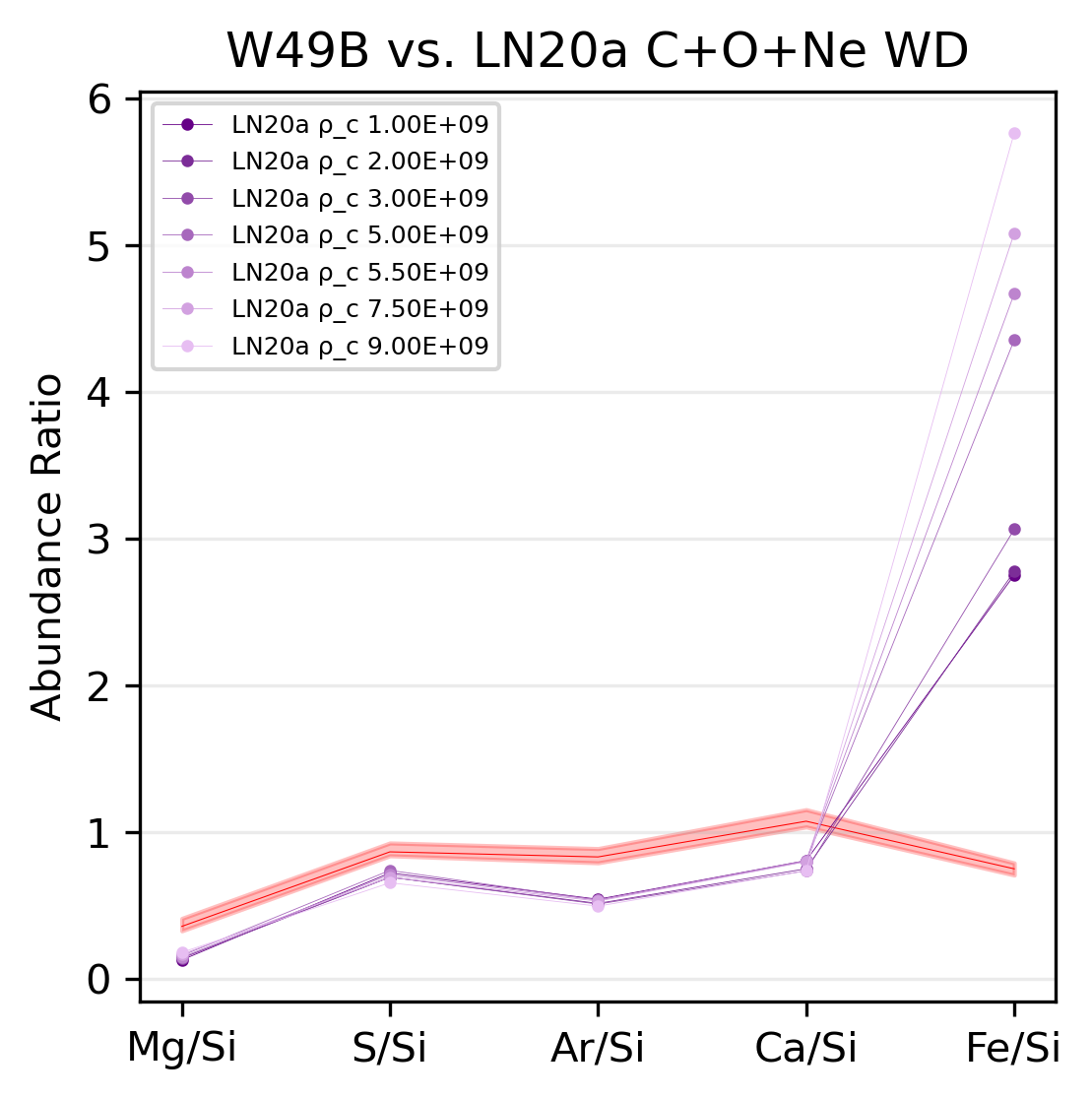}}
		\subfloat{\includegraphics[angle=0,width=0.40\textwidth]{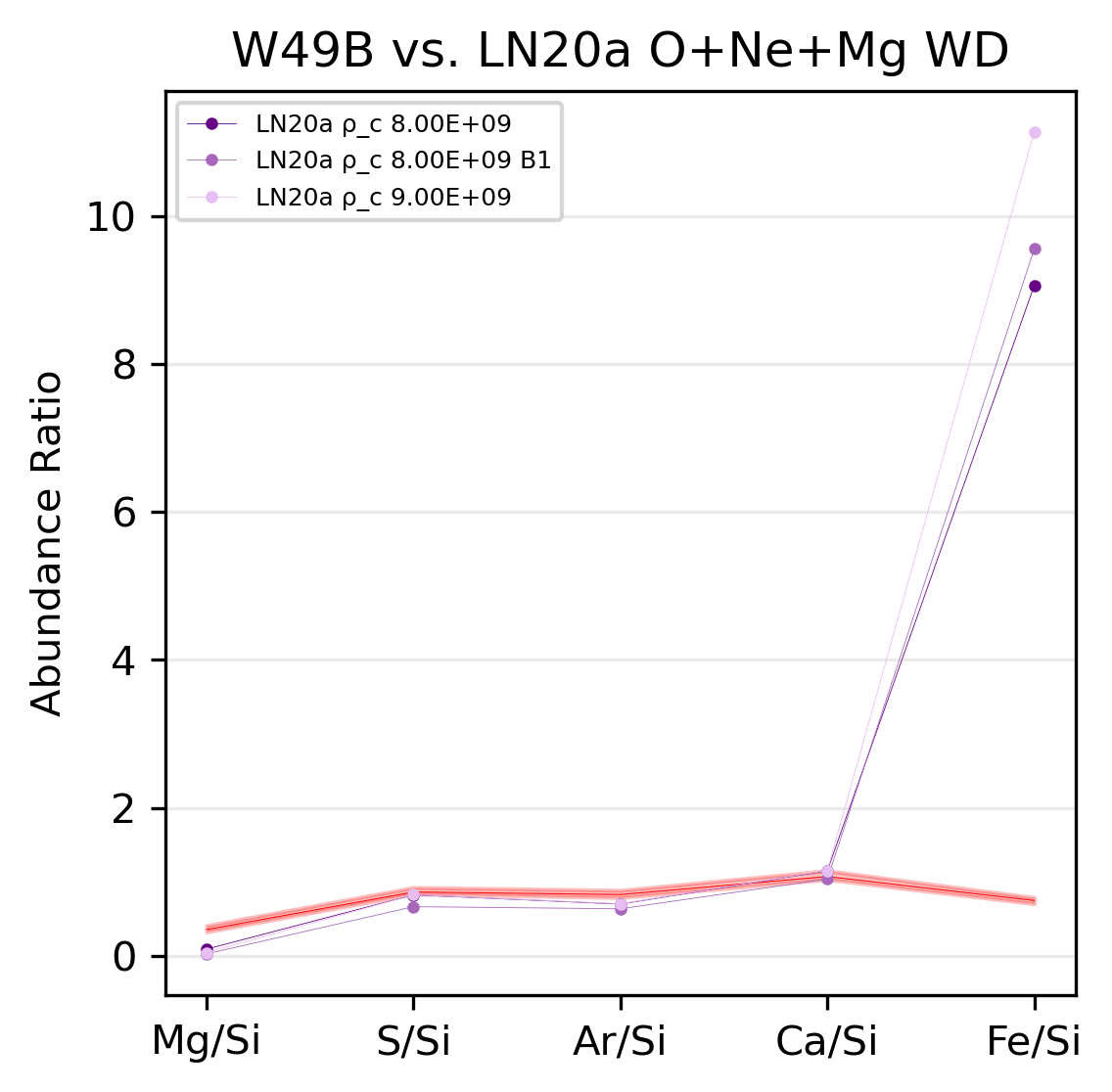}}
	\end{center}
    {Continued from above.}
\end{figure*}

\begin{figure*}\ContinuedFloat
	\begin{center}
		\subfloat{\includegraphics[angle=0,width=0.40\textwidth,scale=0.5]{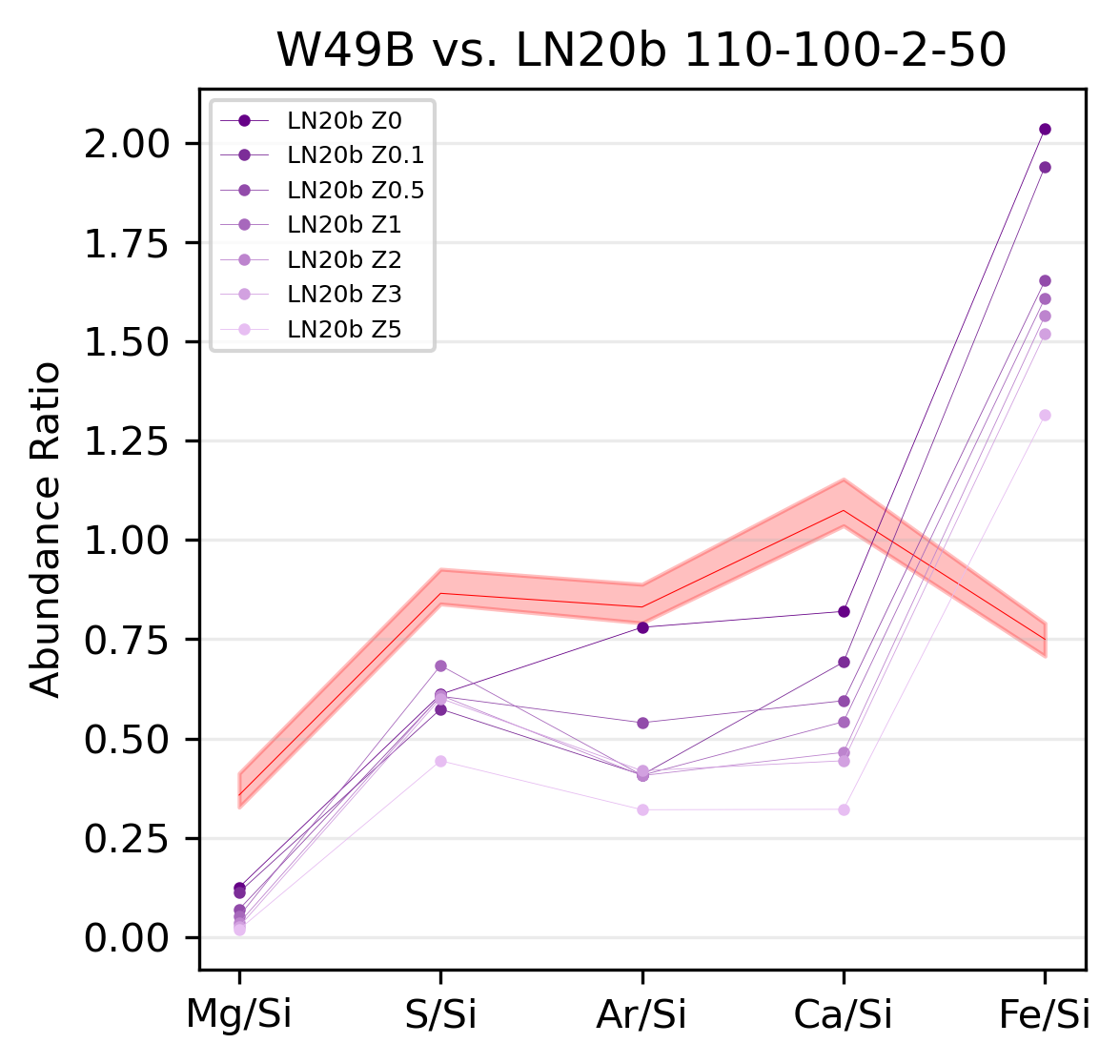}}
		\subfloat{\includegraphics[angle=0,width=0.40\textwidth,scale=0.5]{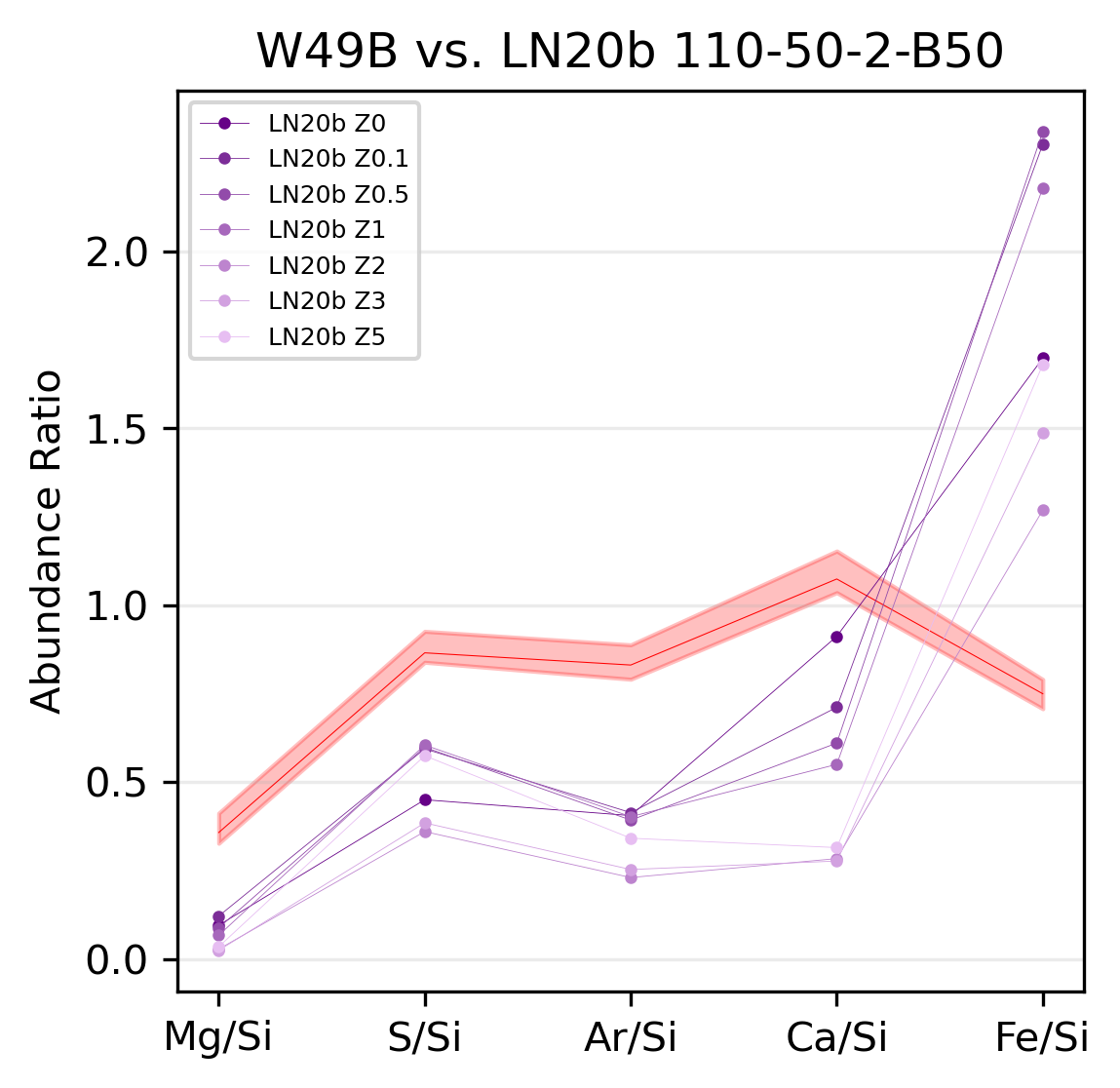}} \\
		\subfloat{\includegraphics[angle=0,width=0.40\textwidth]{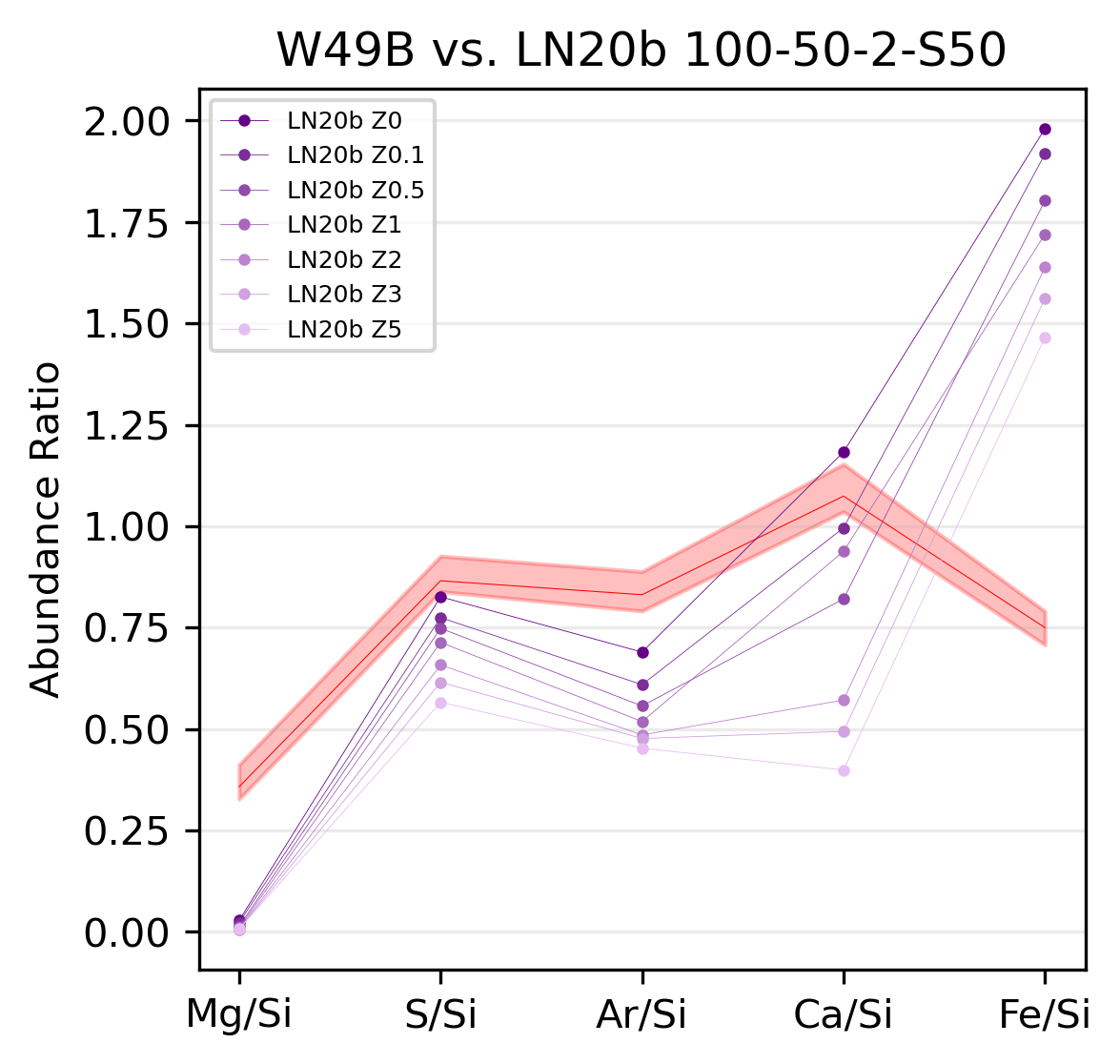}}
		\subfloat{\includegraphics[angle=0,width=0.40\textwidth]{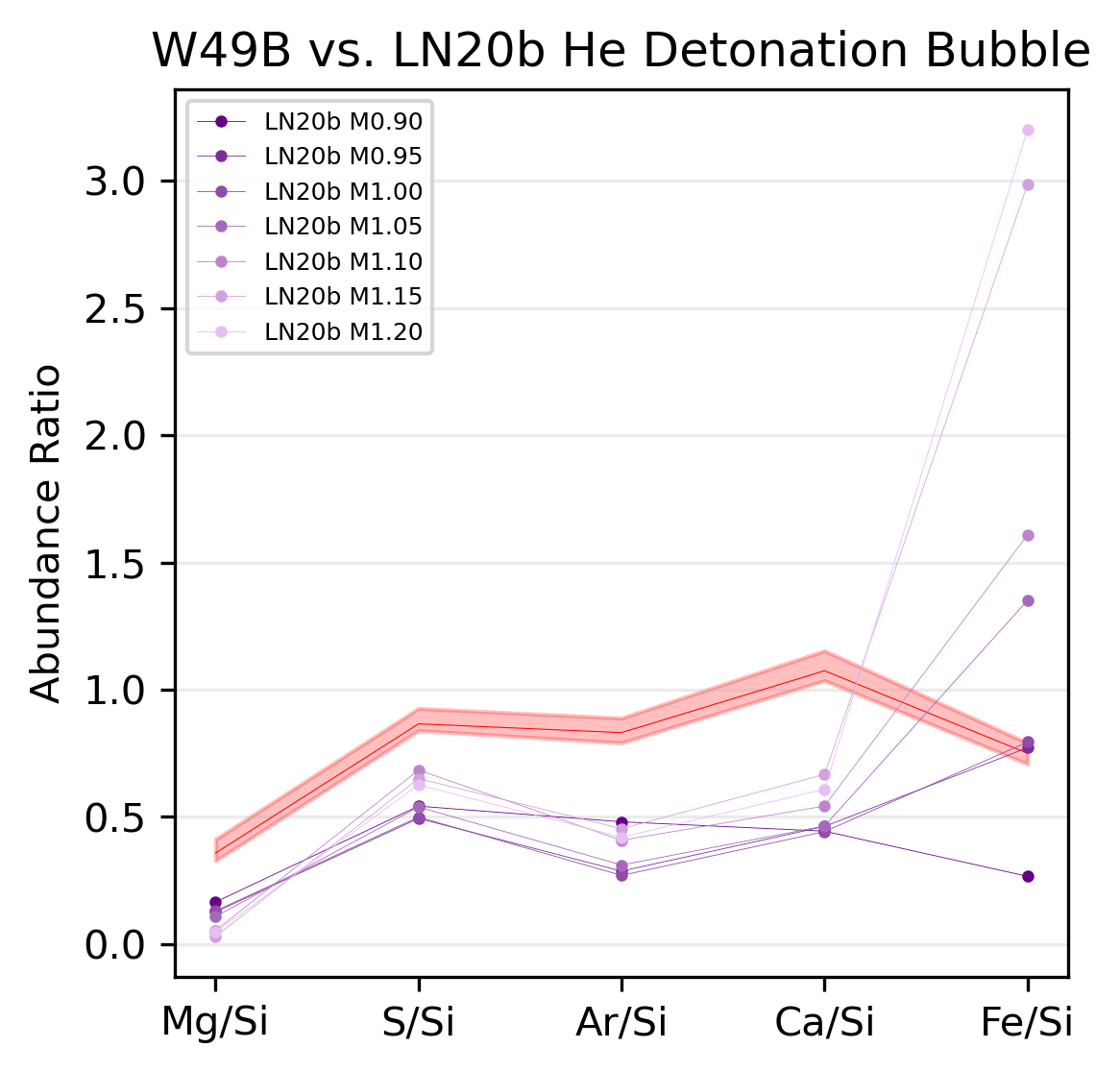}} \\
		\subfloat{\includegraphics[angle=0,width=0.40\textwidth,scale=0.5]{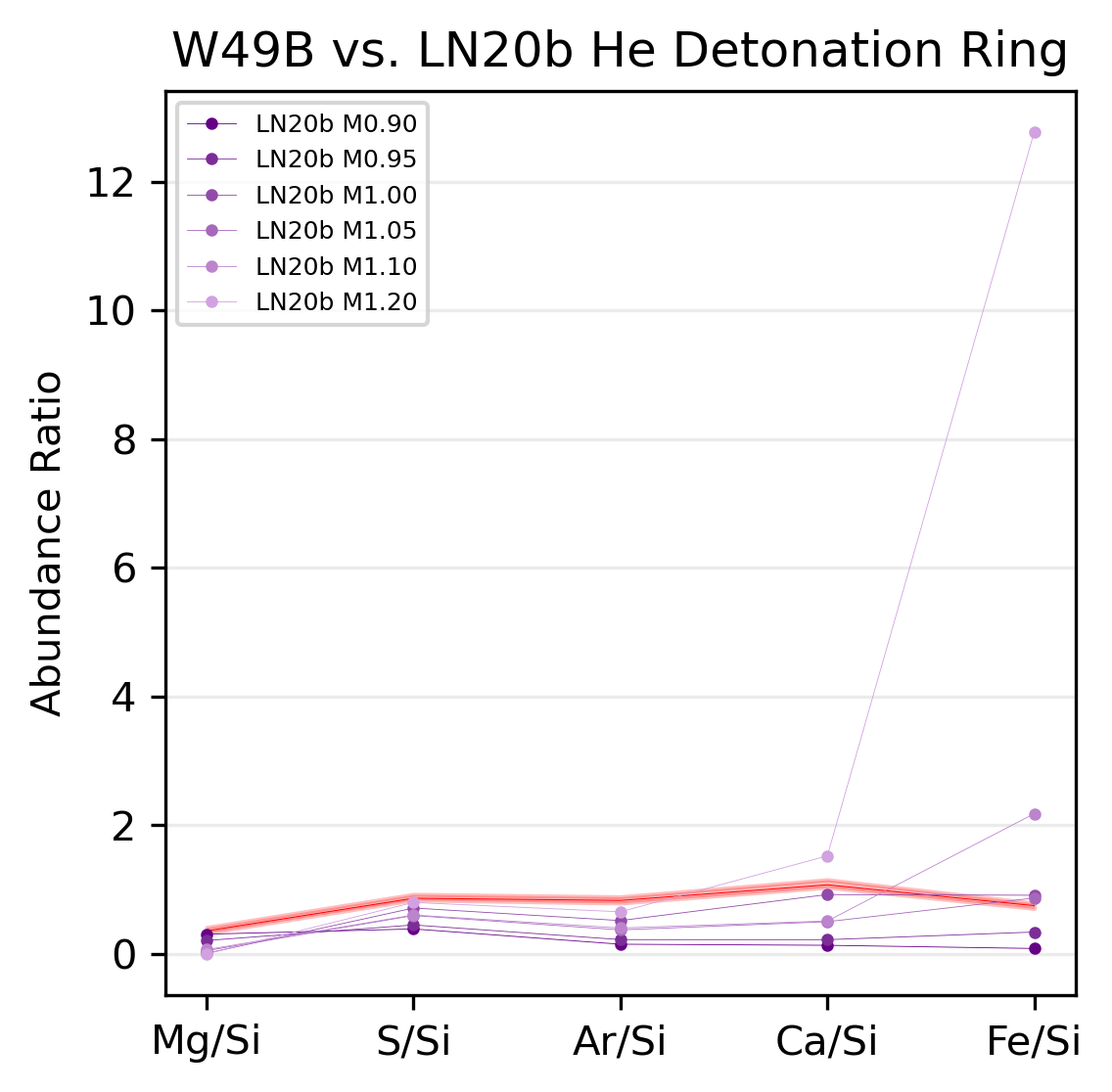}}
		\subfloat{\includegraphics[angle=0,width=0.40\textwidth]{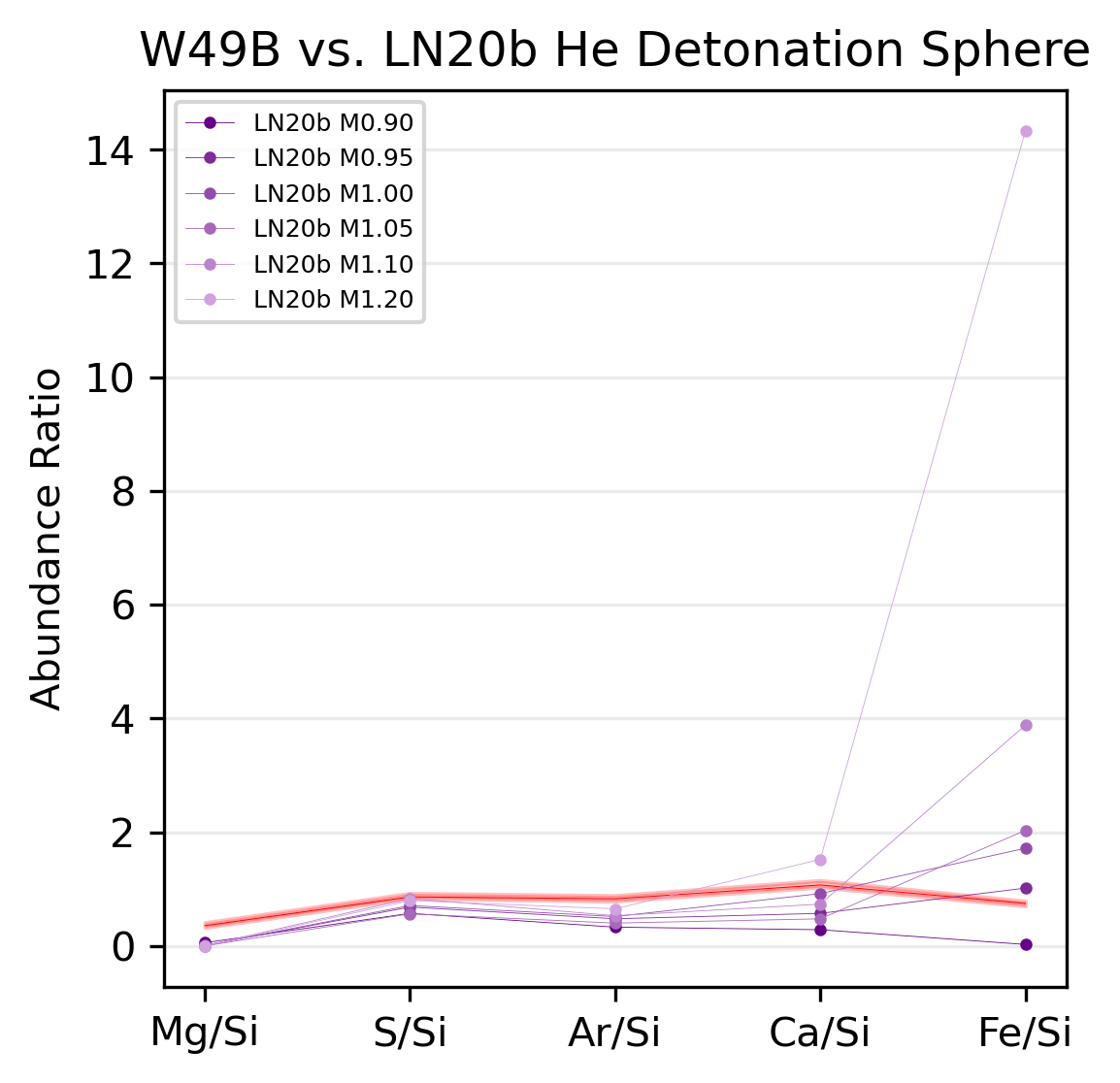}}
	\end{center}
    {Continued from above.}
\end{figure*}

\begin{figure*}
	\begin{center}
		\subfloat{\includegraphics[angle=0,width=0.40\textwidth,scale=0.5]{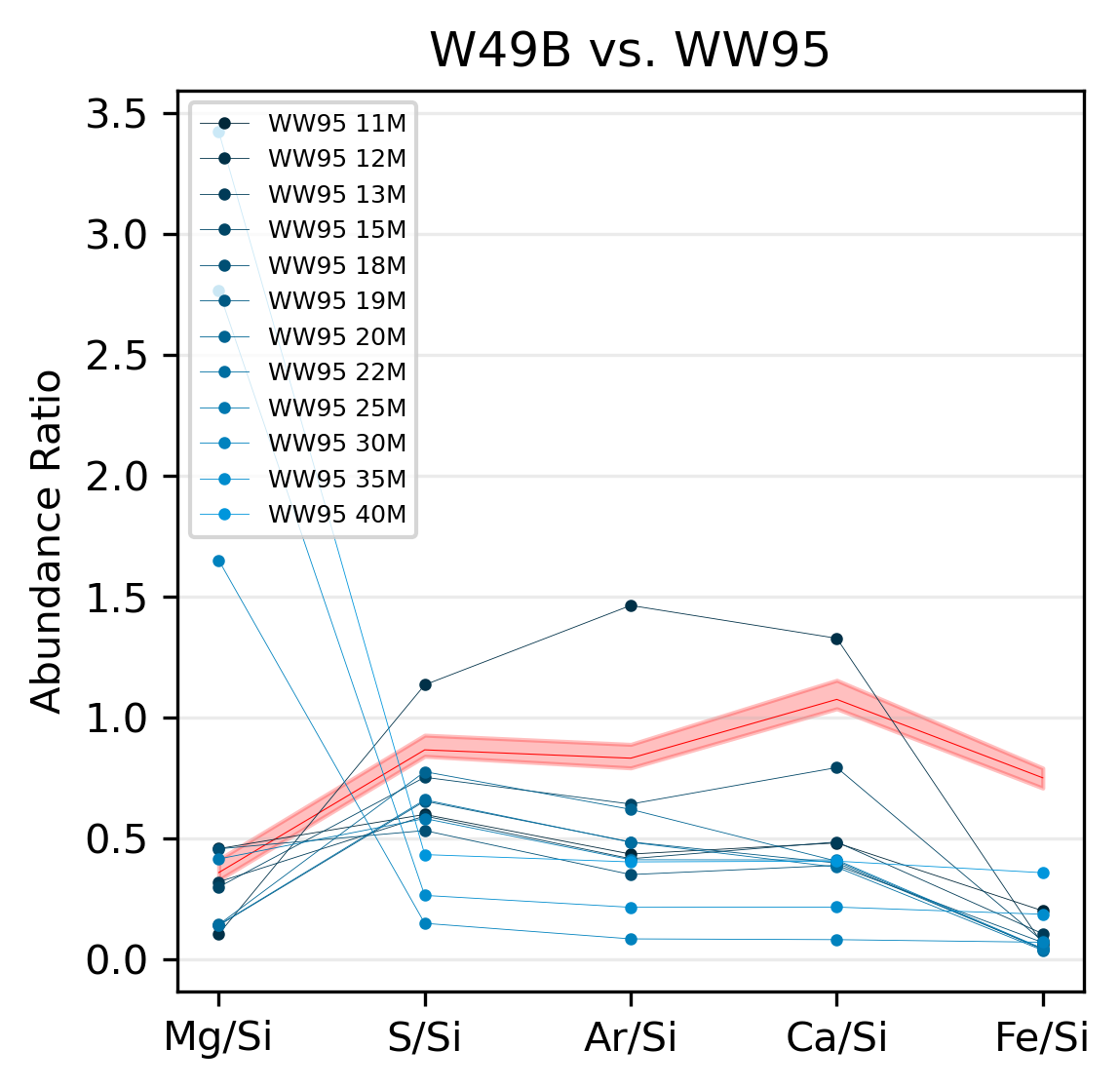}}
		\subfloat{\includegraphics[angle=0,width=0.40\textwidth,scale=0.5]{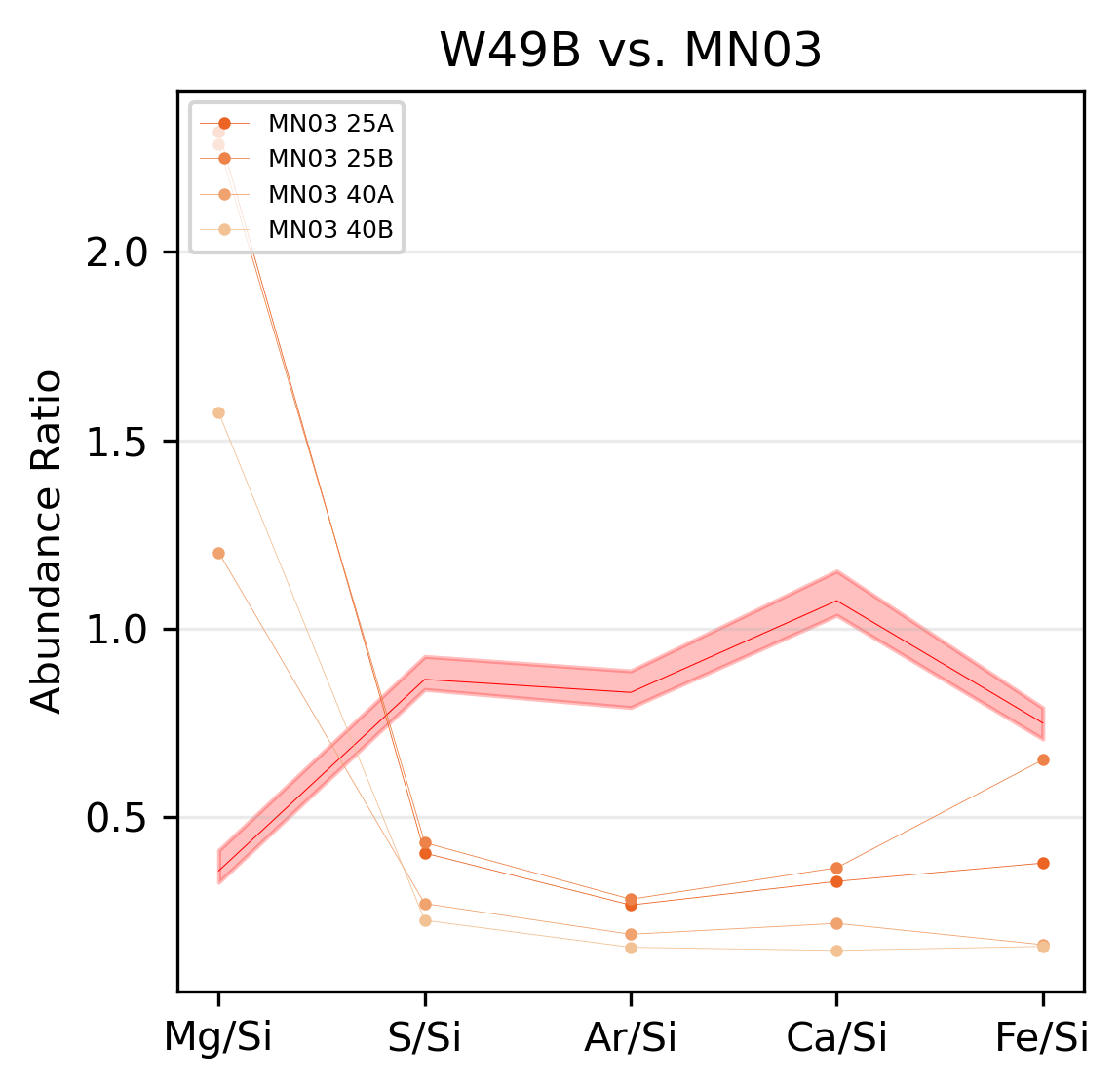}} \\
		\subfloat{\includegraphics[angle=0,width=0.40\textwidth]{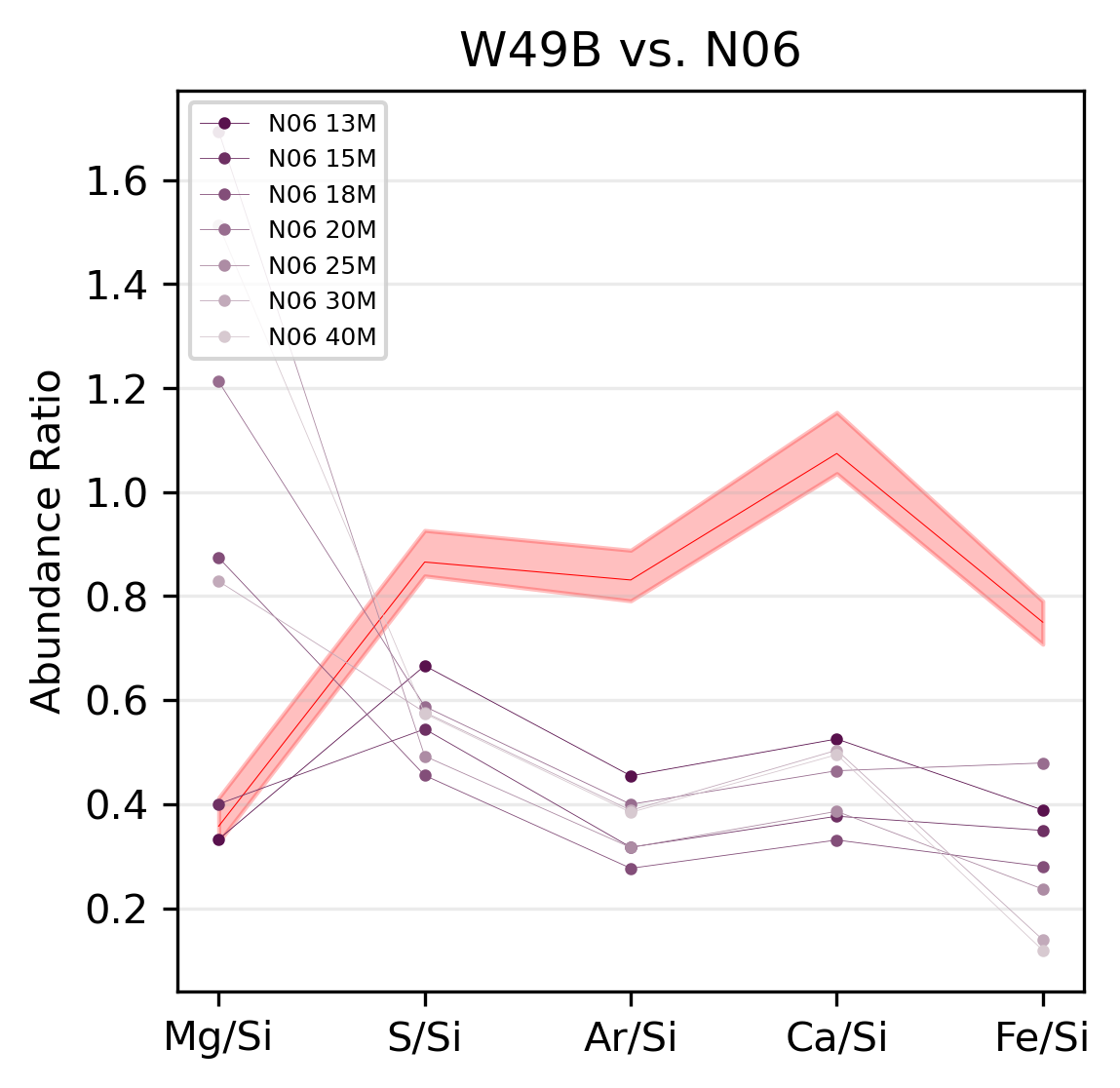}}
		\subfloat{\includegraphics[angle=0,width=0.40\textwidth]{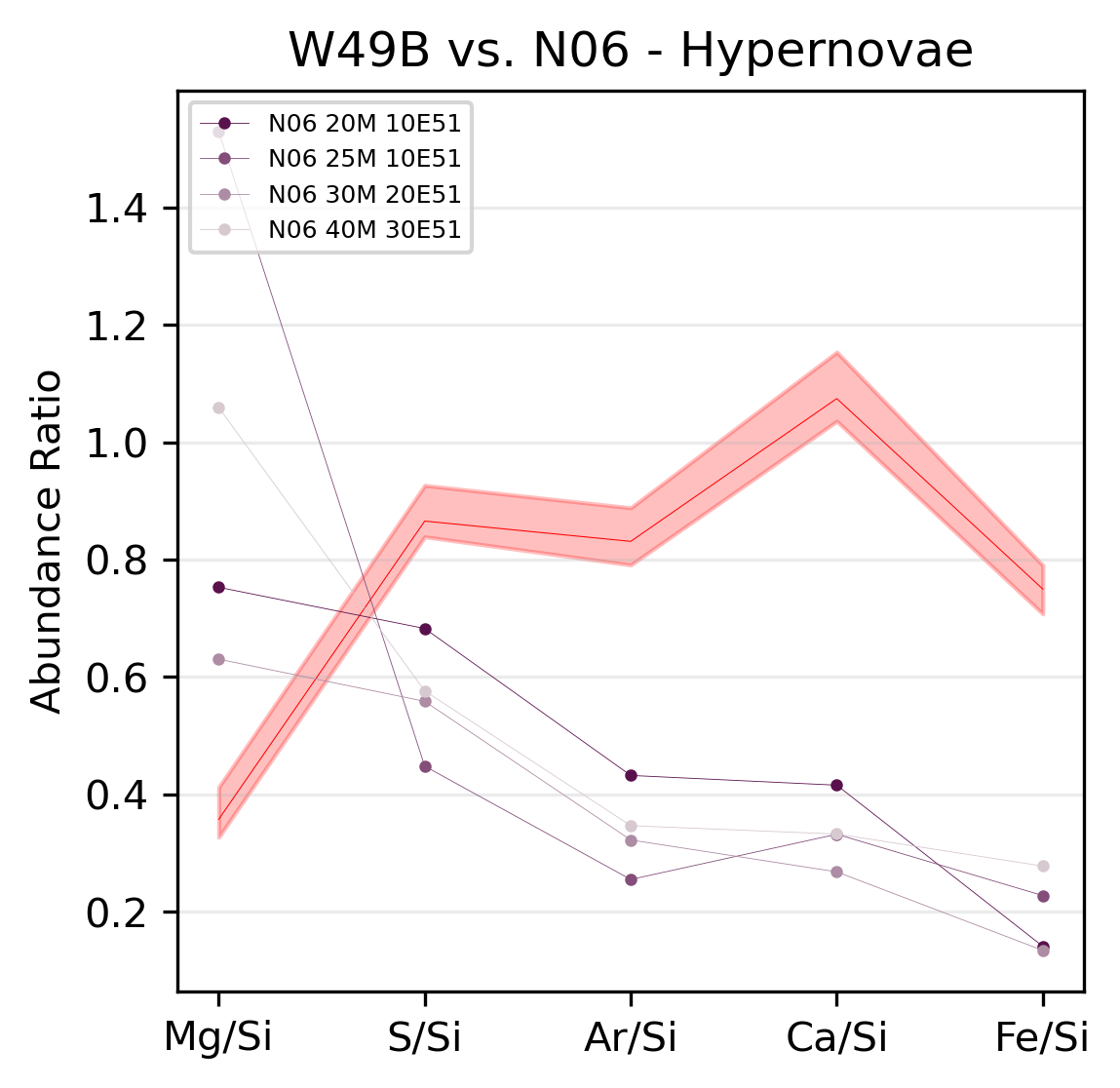}} \\
		\subfloat{\includegraphics[angle=0,width=0.40\textwidth,scale=0.5]{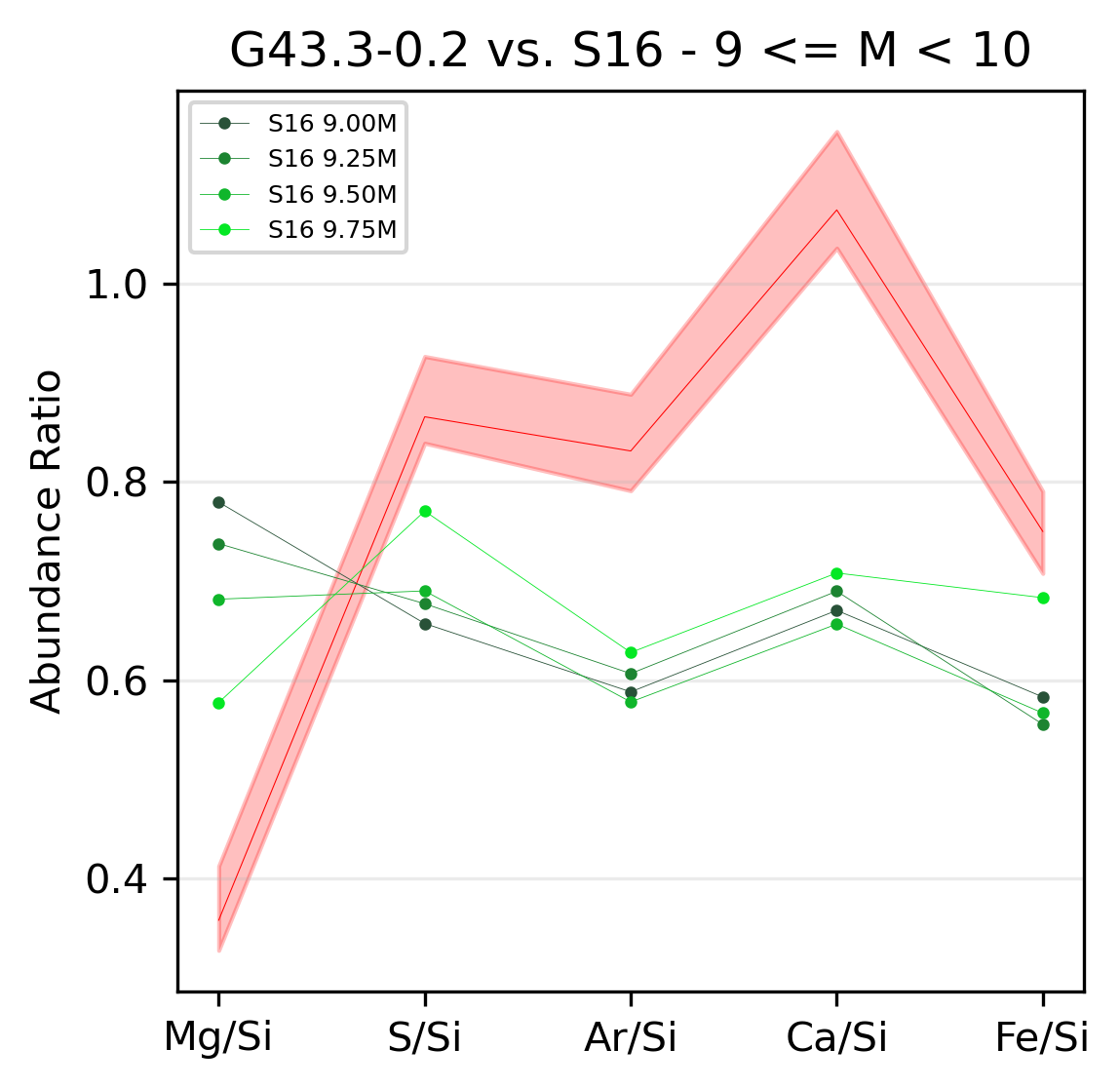}}
		\subfloat{\includegraphics[angle=0,width=0.40\textwidth]{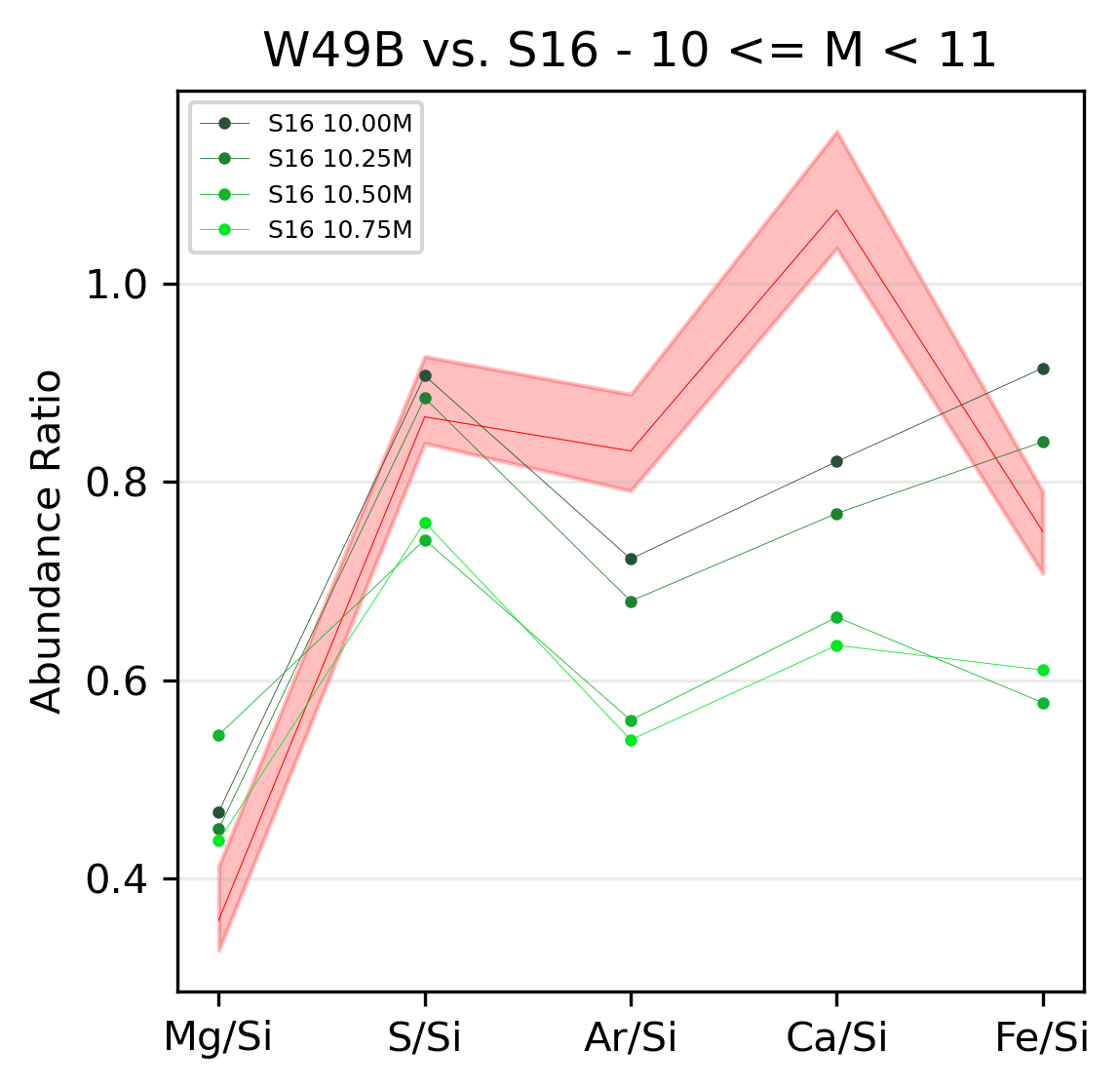}}
	\end{center}
    \caption{Nucleosynthesis comparisons between W49B and the tested CC models.}
    \label{fig:w49b_cc}
\end{figure*}

\begin{figure*}\ContinuedFloat
	\begin{center}
		\subfloat{\includegraphics[angle=0,width=0.40\textwidth,scale=0.5]{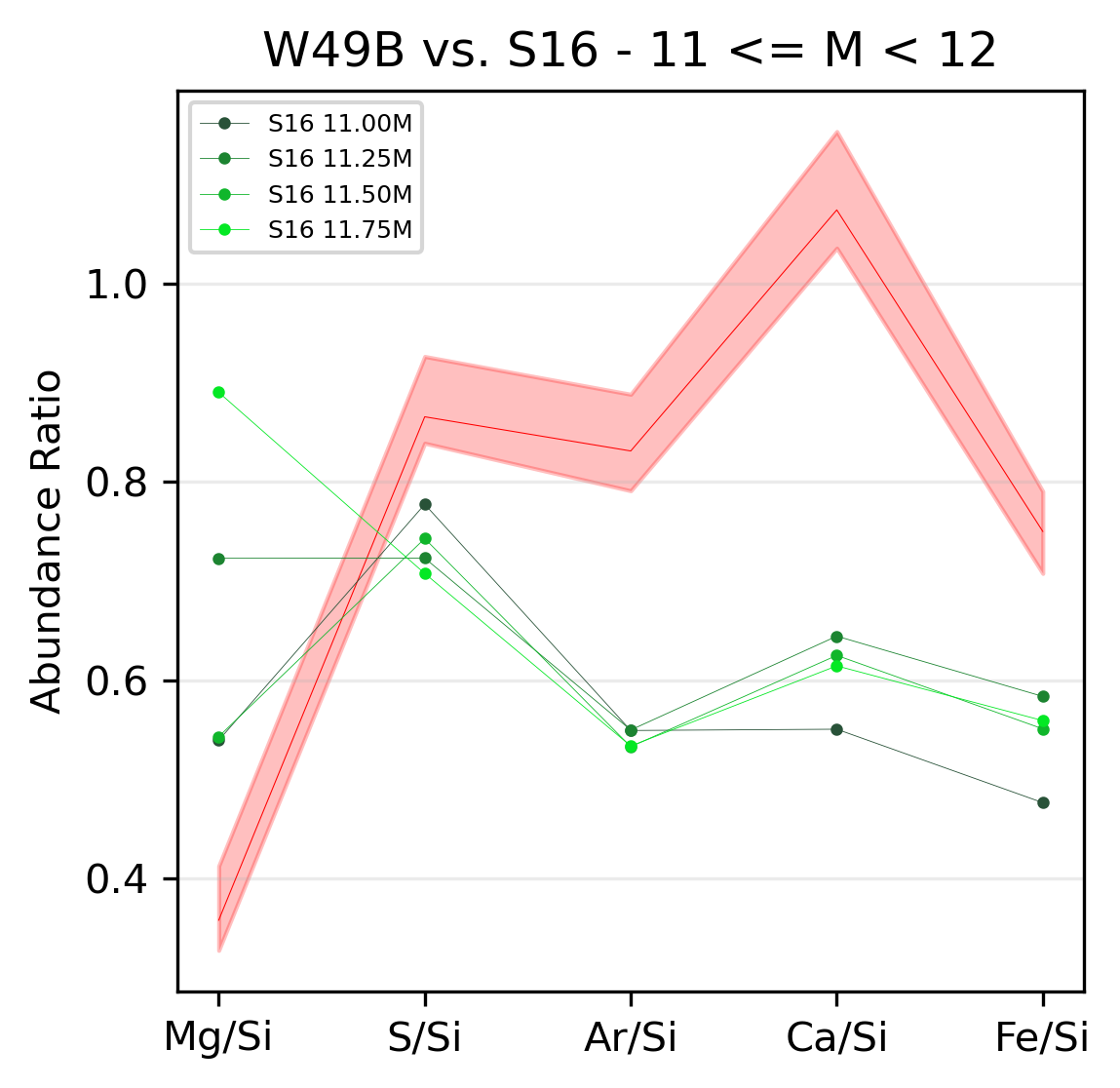}}
		\subfloat{\includegraphics[angle=0,width=0.40\textwidth,scale=0.5]{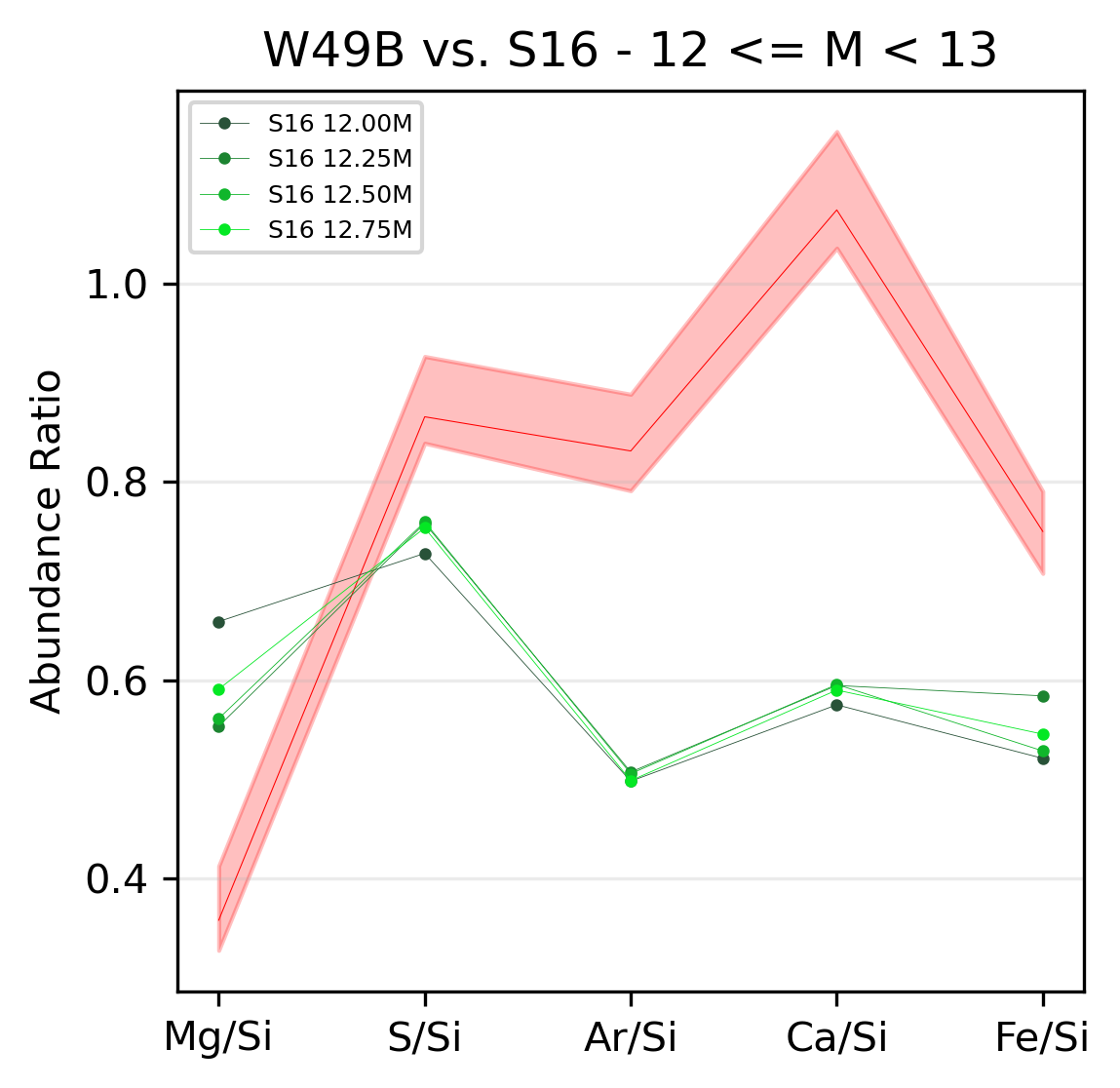}} \\
		\subfloat{\includegraphics[angle=0,width=0.40\textwidth]{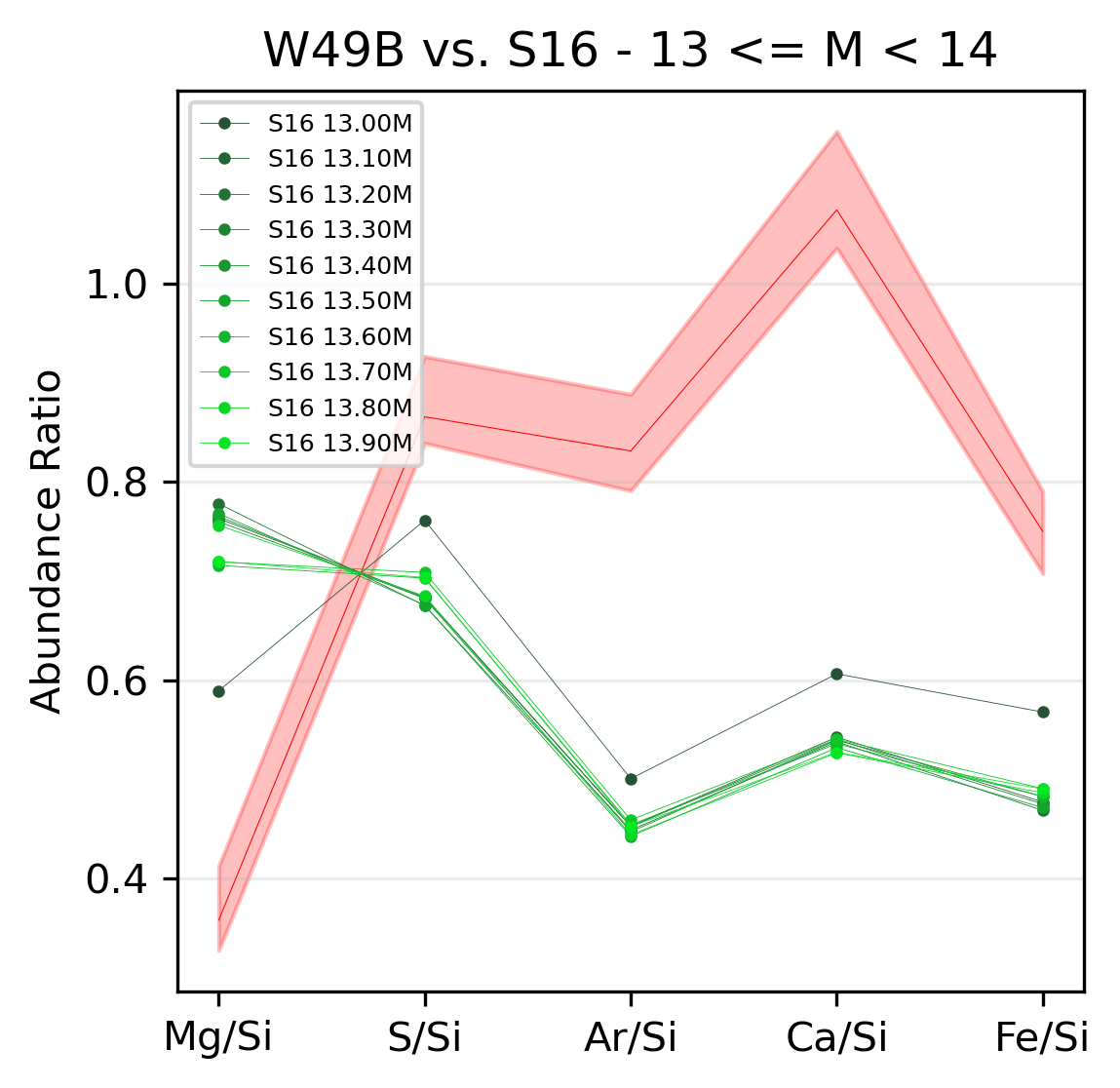}}
		\subfloat{\includegraphics[angle=0,width=0.40\textwidth]{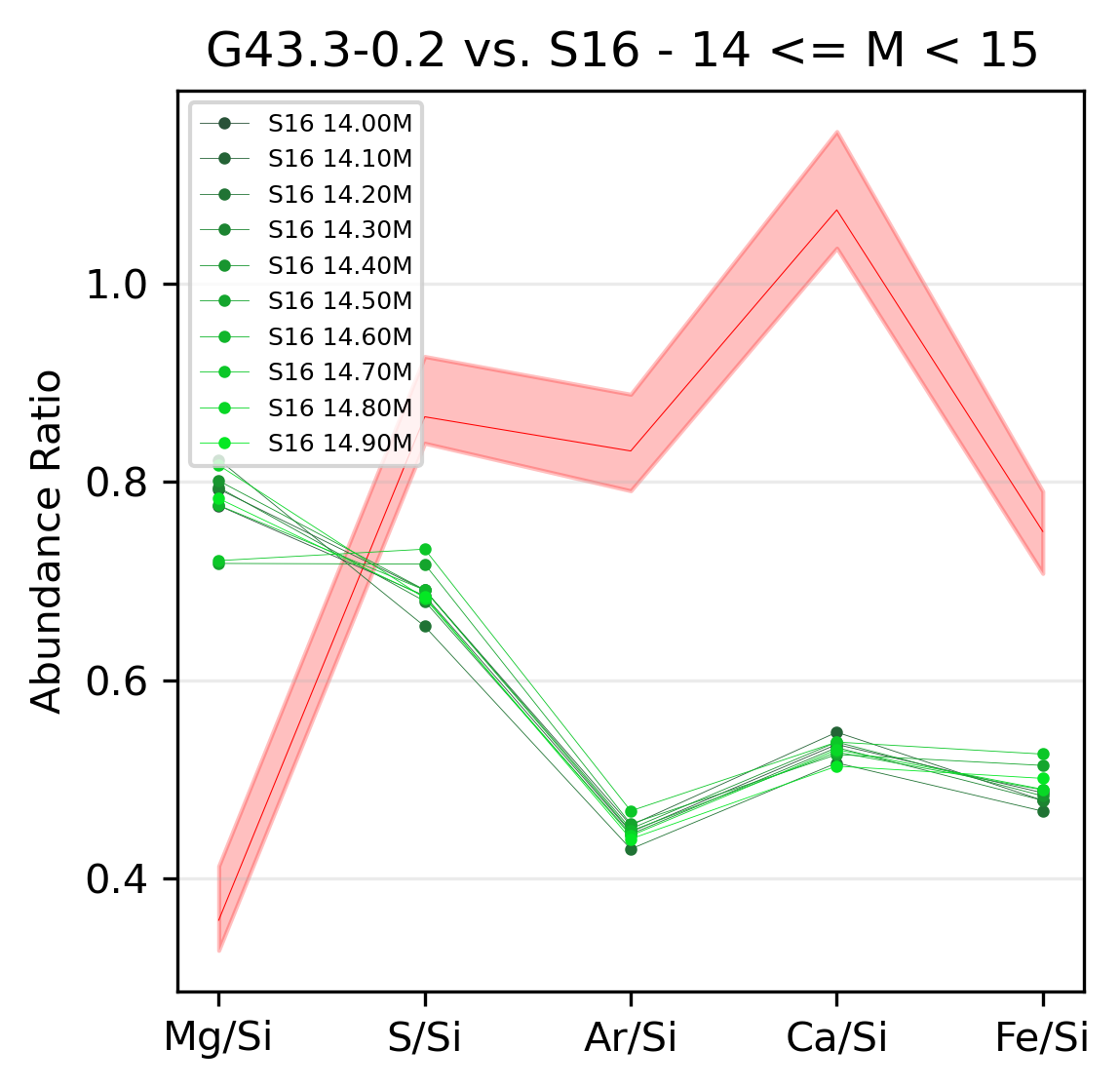}} \\
		\subfloat{\includegraphics[angle=0,width=0.40\textwidth,scale=0.5]{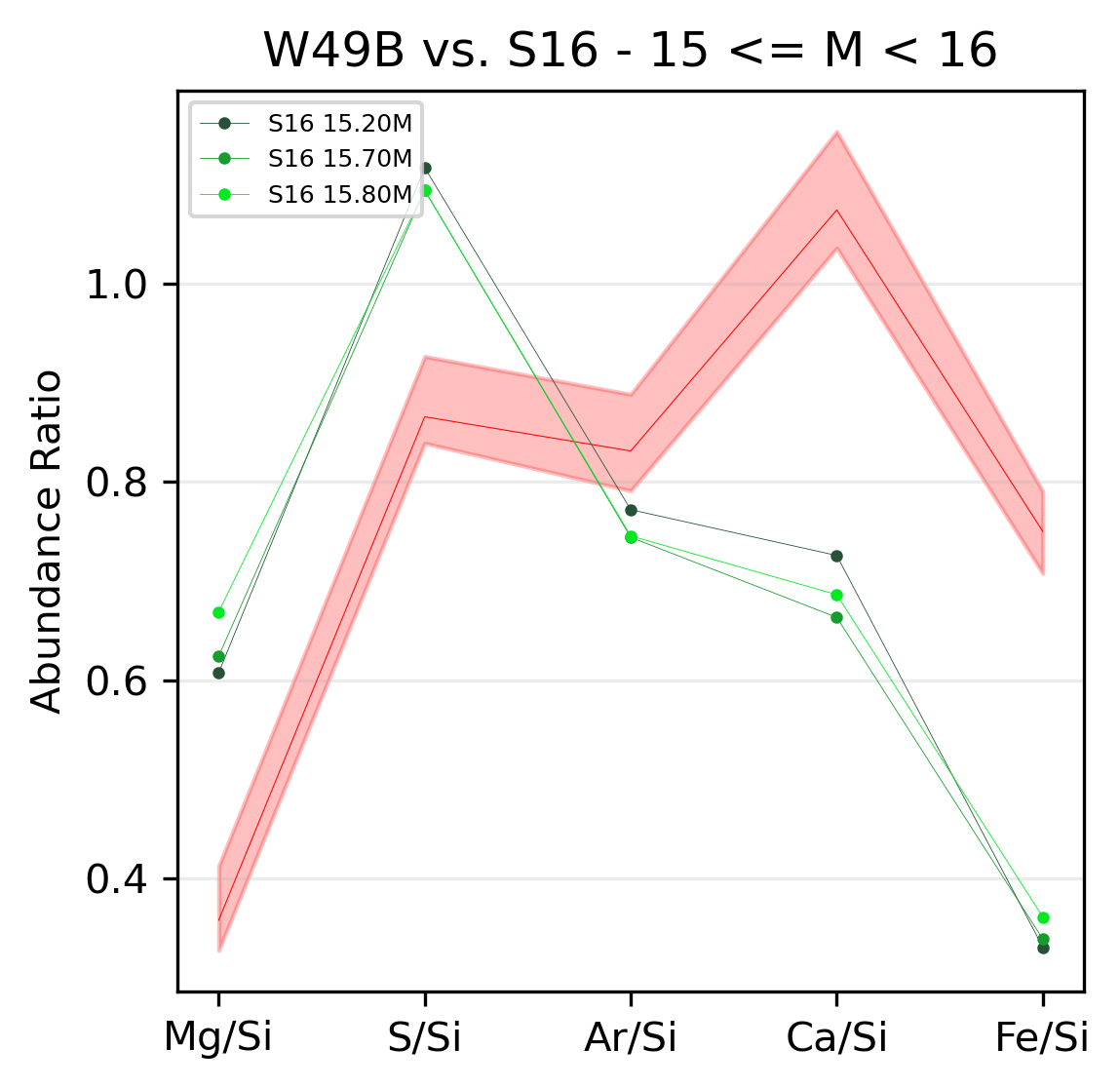}}
		\subfloat{\includegraphics[angle=0,width=0.40\textwidth]{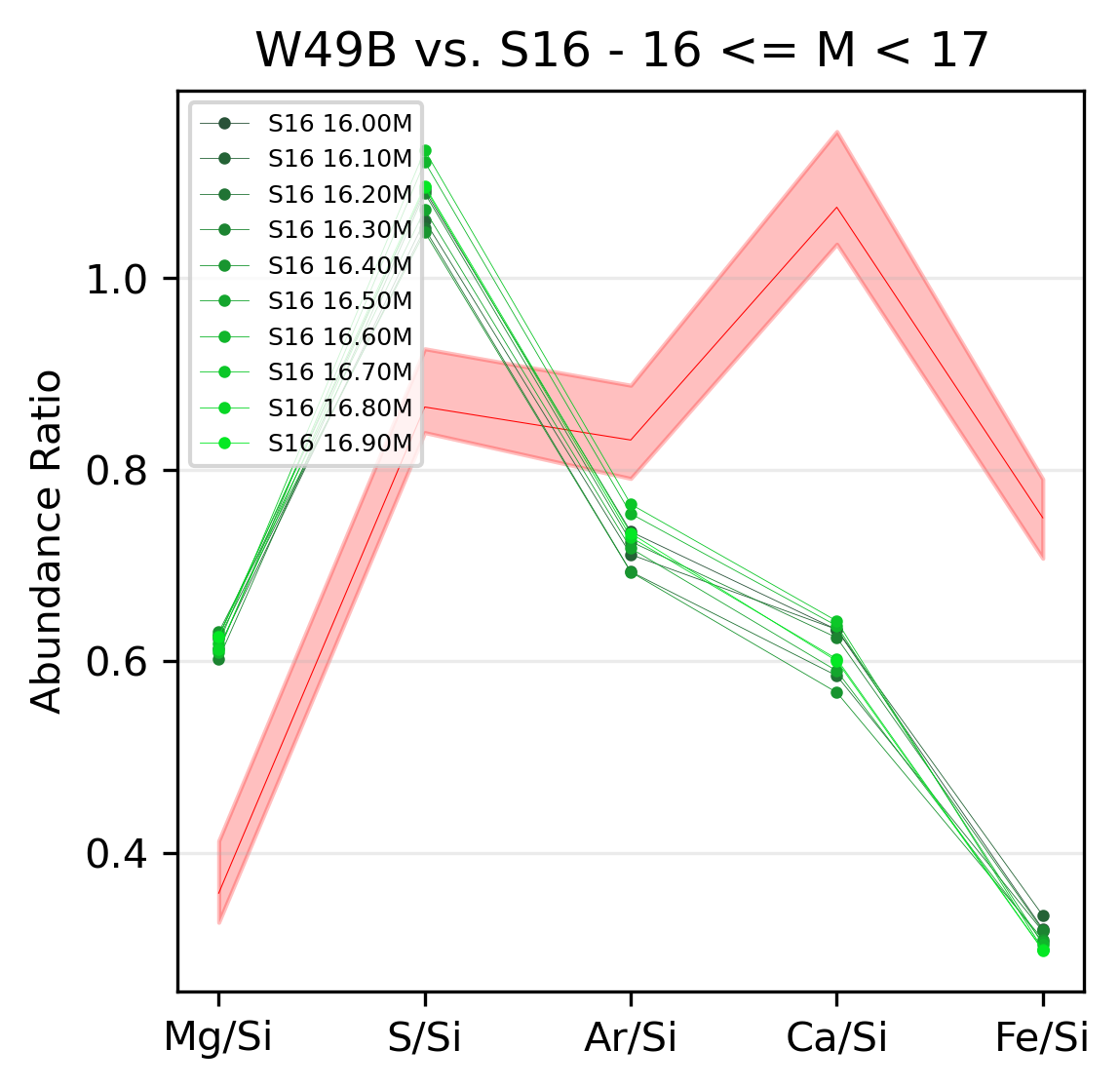}}
	\end{center}
    {Continued from above.}
\end{figure*}

\begin{figure*}\ContinuedFloat
	\begin{center}
		\subfloat{\includegraphics[angle=0,width=0.40\textwidth,scale=0.5]{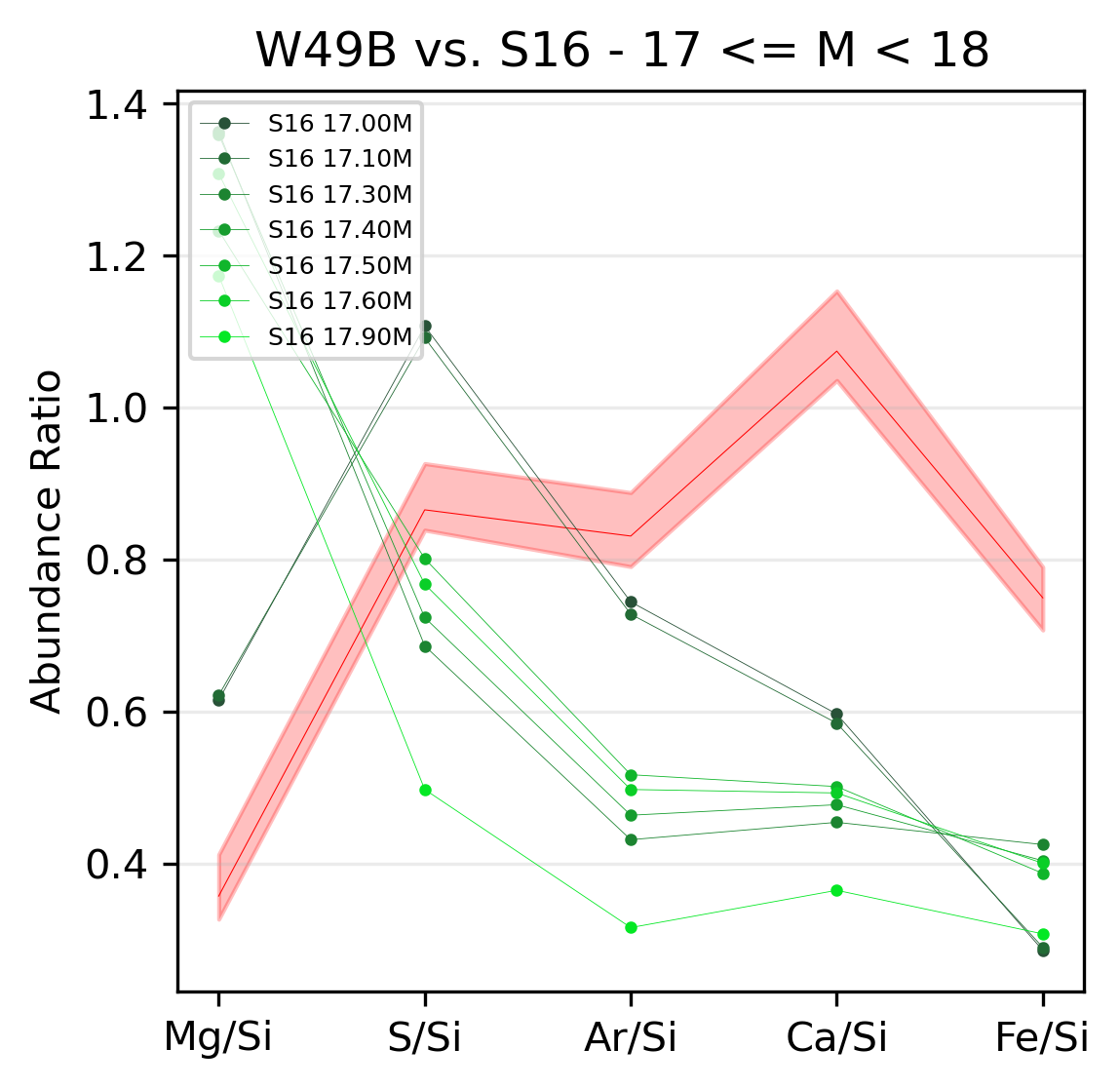}}
		\subfloat{\includegraphics[angle=0,width=0.40\textwidth,scale=0.5]{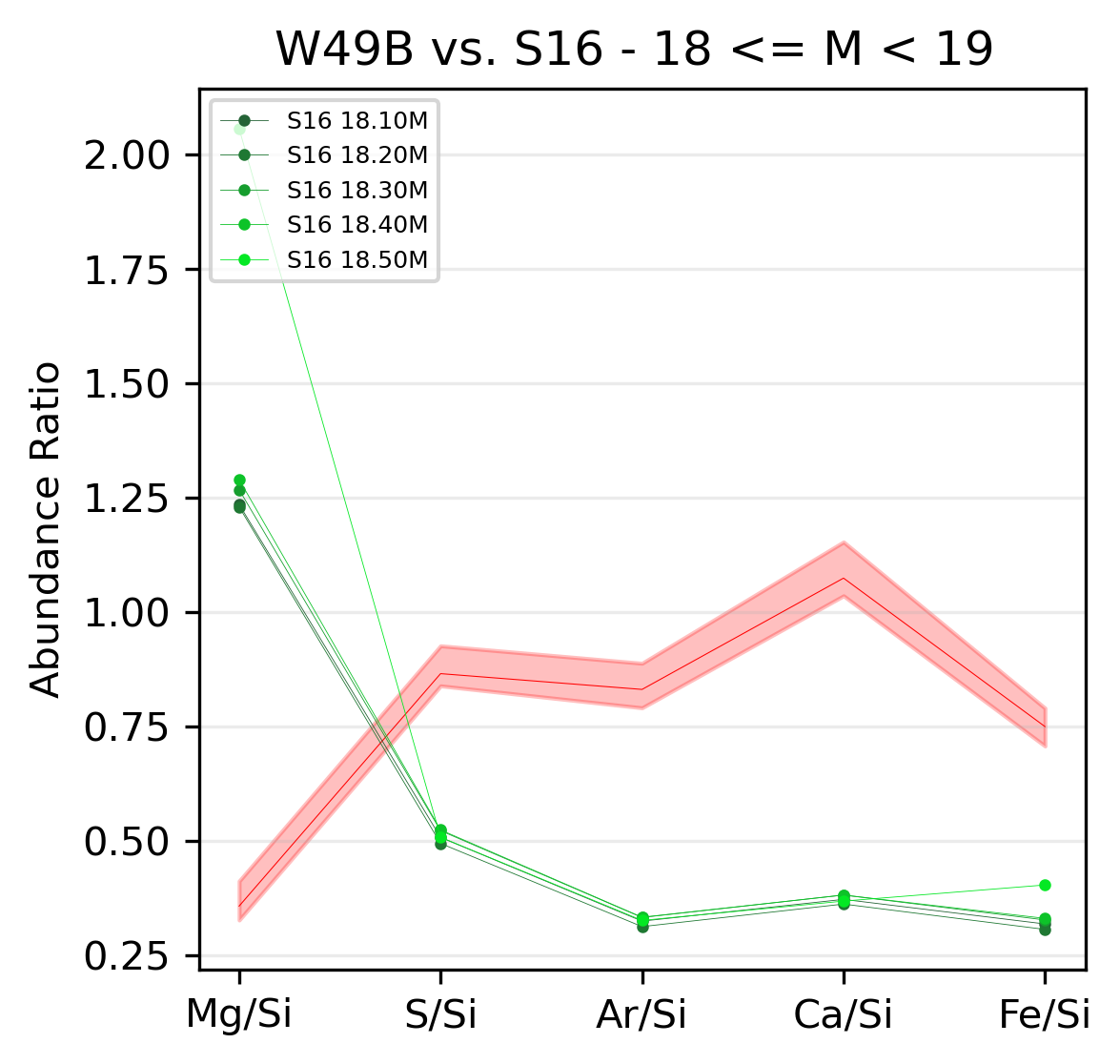}} \\
		\subfloat{\includegraphics[angle=0,width=0.40\textwidth]{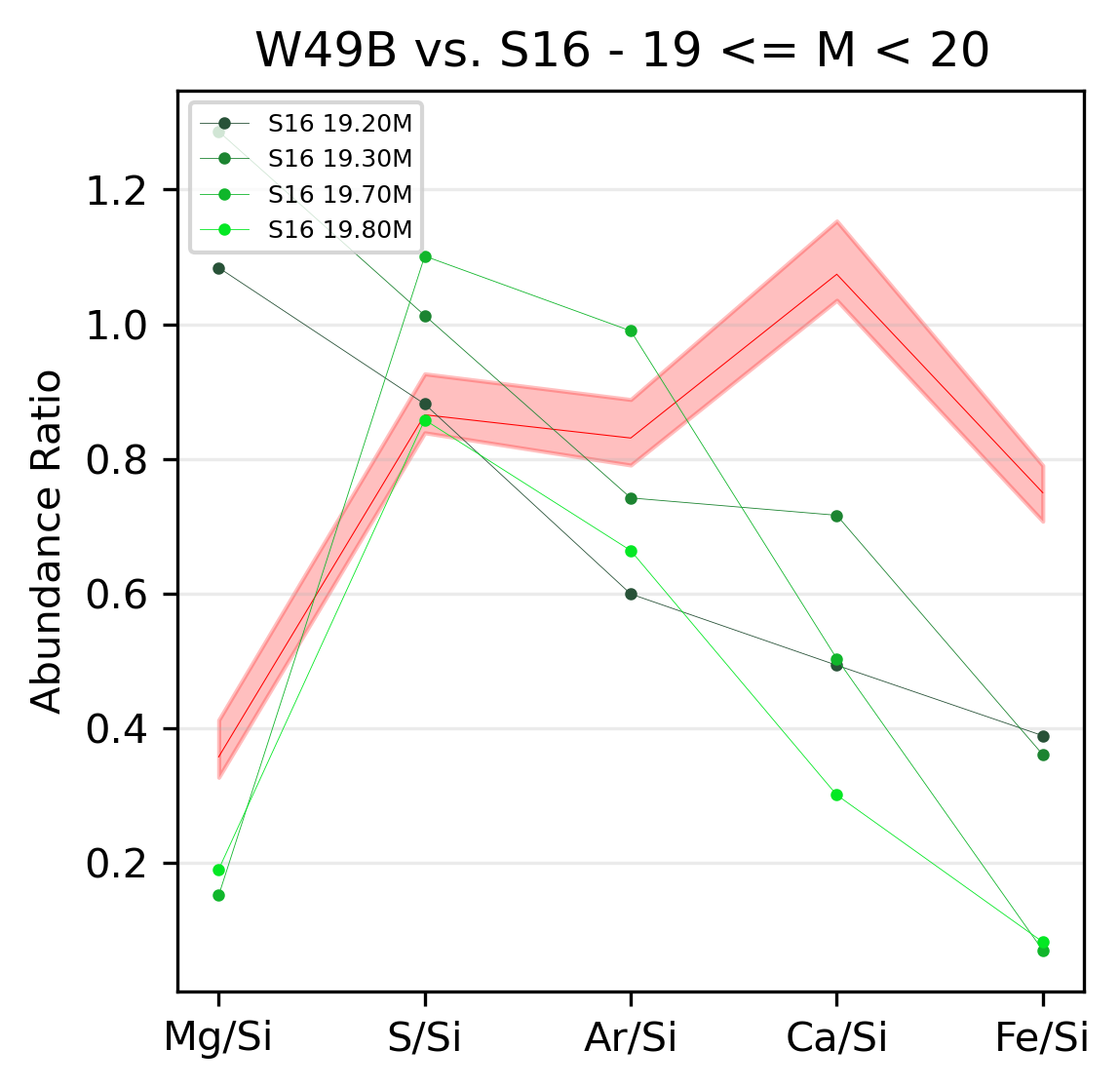}}
		\subfloat{\includegraphics[angle=0,width=0.40\textwidth]{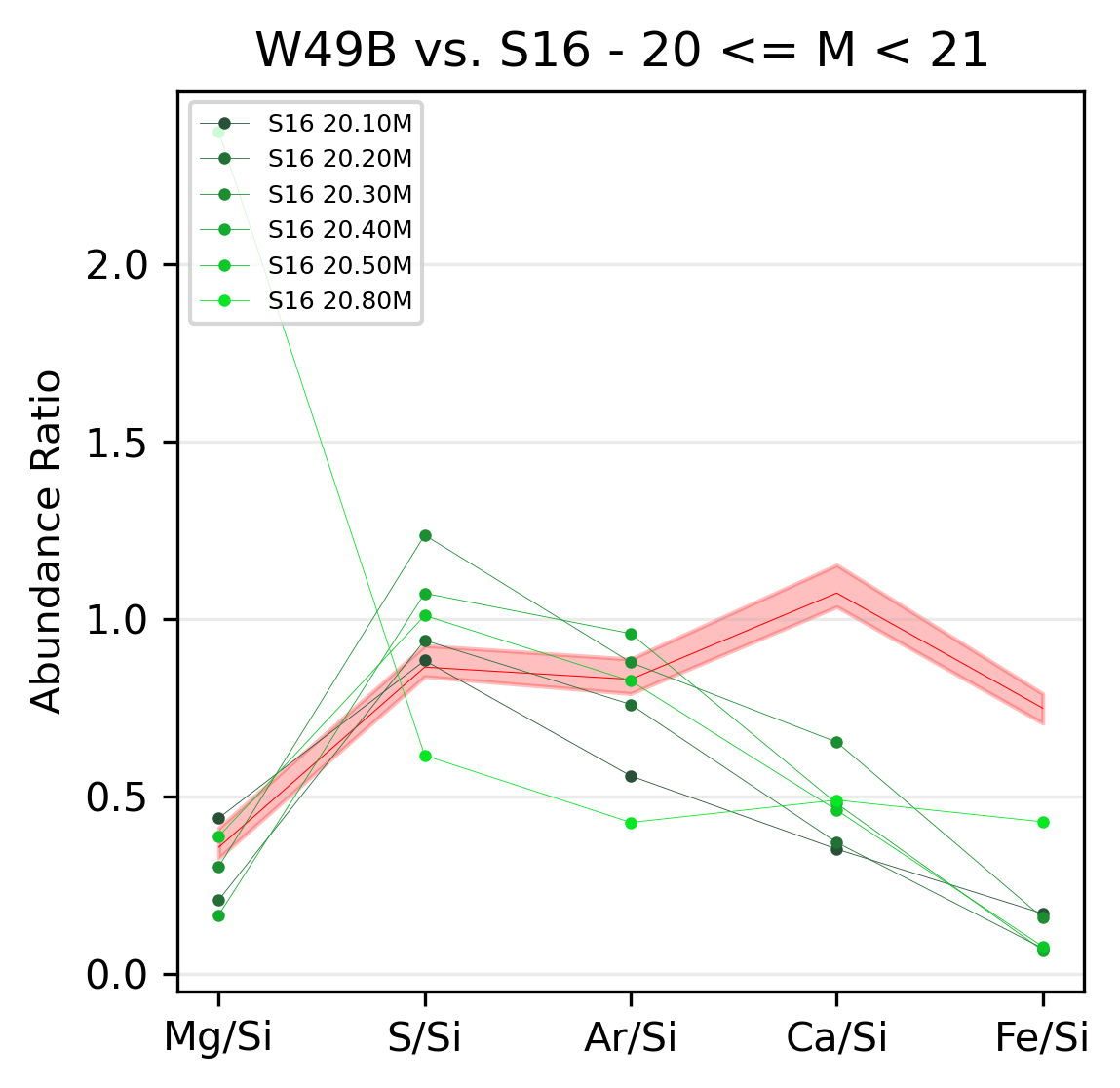}} \\
		\subfloat{\includegraphics[angle=0,width=0.40\textwidth,scale=0.5]{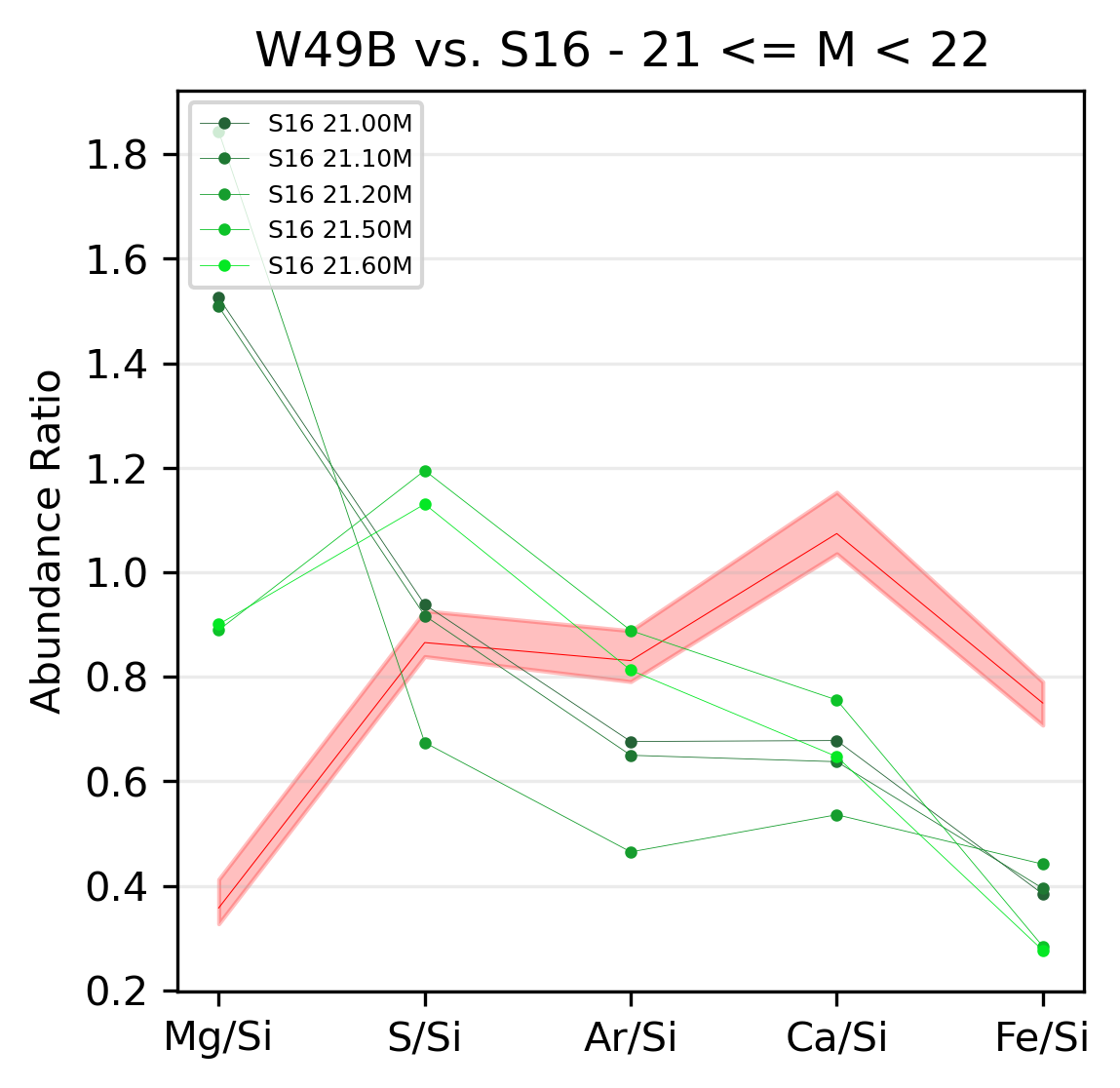}}
		\subfloat{\includegraphics[angle=0,width=0.40\textwidth]{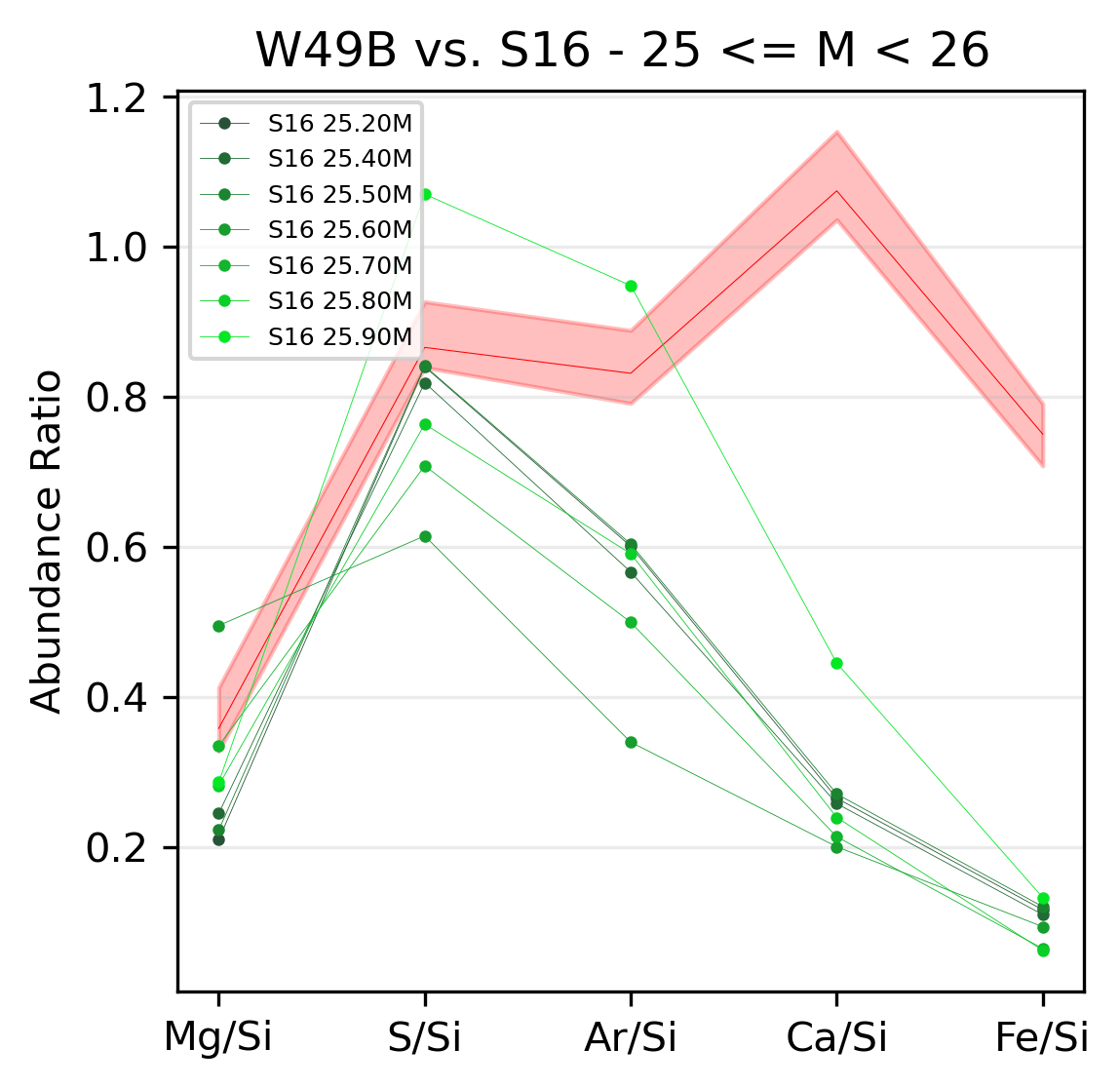}}
	\end{center}
    {Continued from above.}
\end{figure*}

\begin{figure*}\ContinuedFloat
	\begin{center}
		\subfloat{\includegraphics[angle=0,width=0.40\textwidth,scale=0.5]{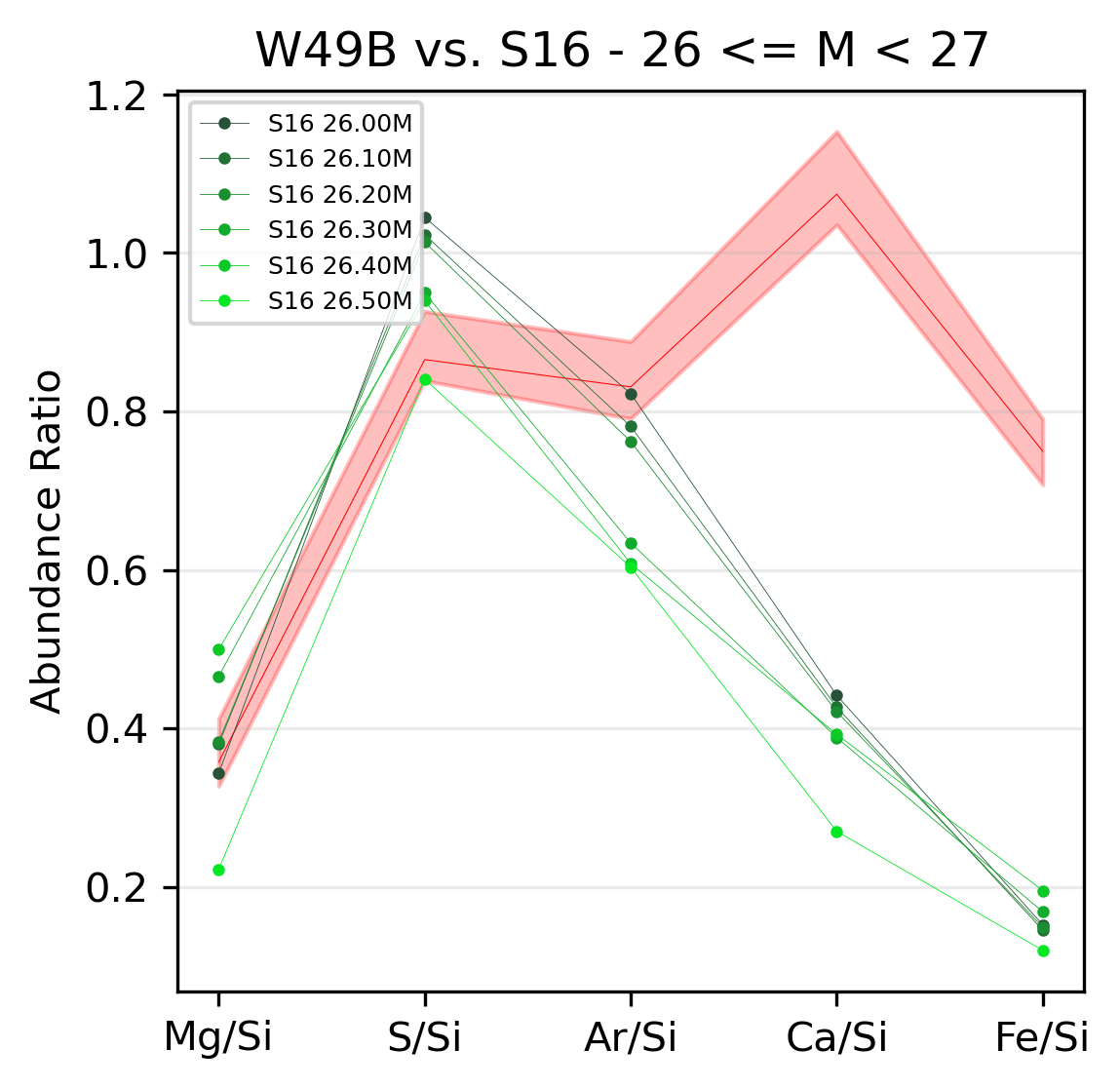}}
		\subfloat{\includegraphics[angle=0,width=0.40\textwidth,scale=0.5]{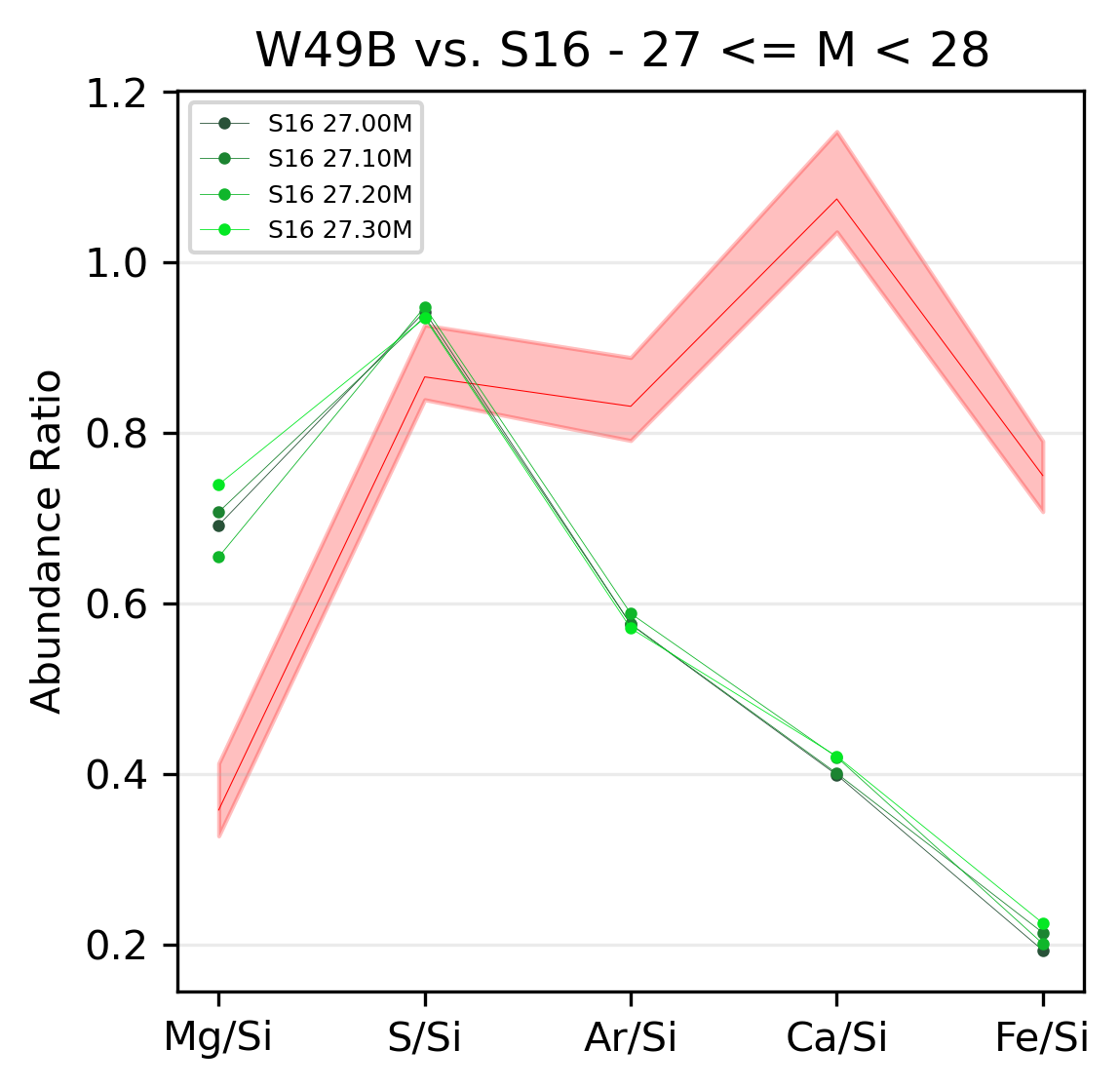}} \\
		\subfloat{\includegraphics[angle=0,width=0.40\textwidth]{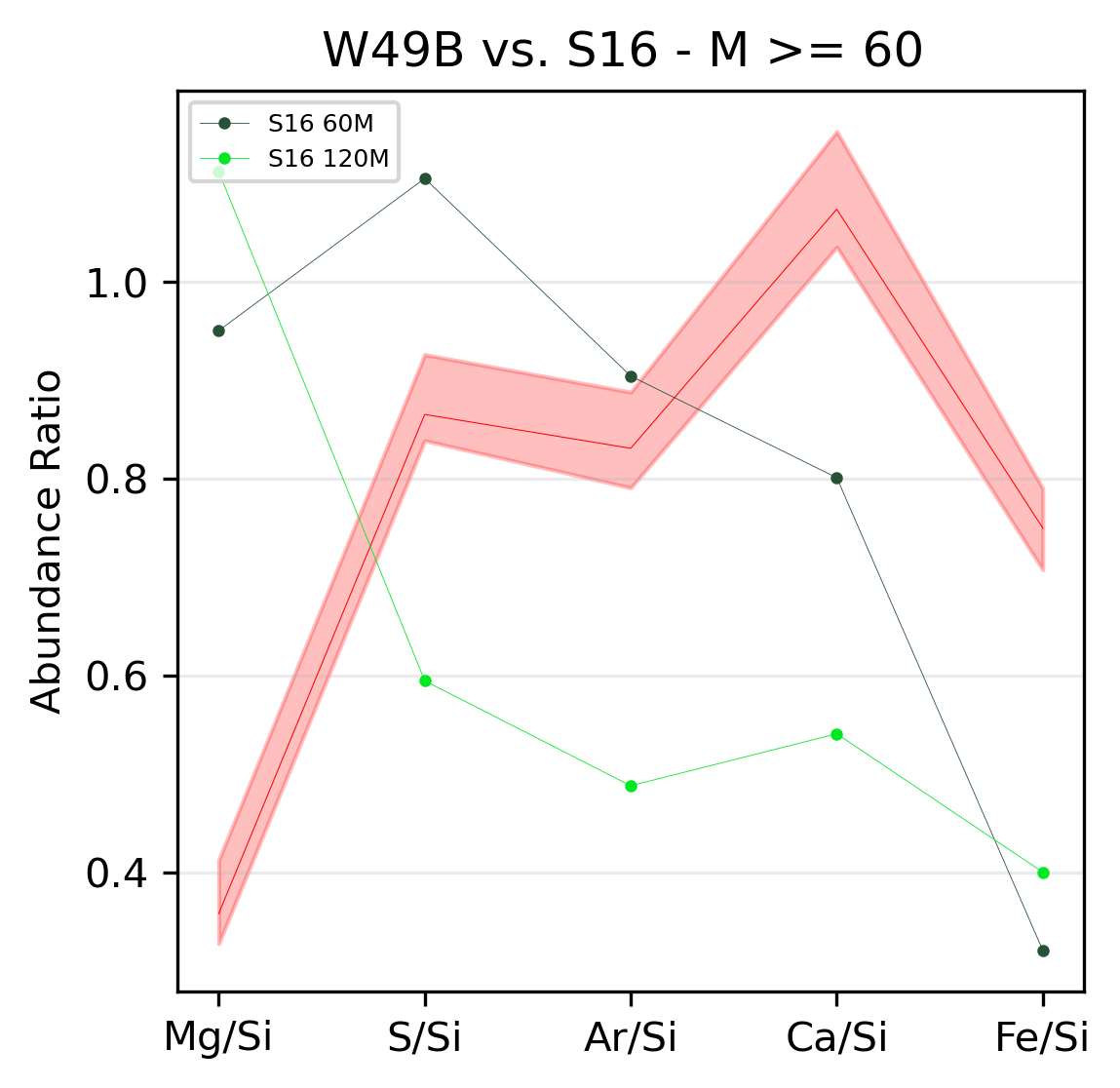}}
	\end{center}
    {Continued from above.}
\end{figure*}

\label{lastpage}
\end{document}